\RequirePackage{fix-cm}% fix the issues with T1-encoded CModern fonts
\documentclass[a4paper,11pt,twoside=off]{scrbook}%scrreprt%report
\usepackage{fixltx2e} %general LaTeX2e bugfixes: should be loaded before amsthm, subfig
\usepackage[ngerman,english]{babel}
\usepackage[utf8]{inputenc}%latin1
\usepackage{graphicx}
\usepackage{placeins}
\usepackage{bbm}
\usepackage{dsfont}
\usepackage{mathrsfs}
\usepackage{framed}
\usepackage{wrapfig}
\usepackage{amsmath}
\usepackage{amsfonts}
\usepackage{color}
\usepackage{array}
\usepackage{caption}
\usepackage{subcaption}
\usepackage{abbrevs}
\usepackage{amssymb}
\usepackage[numbers,sort&compress,merge]{natbib}
\usepackage{tikz}
\usepackage{hyperref}
\usepackage{dsfont}
\usepackage{bigints}
\usepackage[nottoc,notlot,notlof]{tocbibind}
\usepackage{mdwlist}
\usepackage{attrib}
\usepackage[T1]{fontenc}%T1 font encoding (for umlaute searching etc.)
\usepackage{booktabs}
\usepackage{multicol}
\usepackage{geometry}
\usepackage{layout}
\usepackage{appendix}
\usepackage[CaptionBefore]{fltpage}

\input{glyphtounicode}
\definecolor{darkblue}{rgb}{0.0,0.0,0.4}
\definecolor{darkgreen}{rgb}{0.0,0.4,0.0}
\definecolor{darkred}{rgb}{0.6,0.0,0.0}
\hypersetup{
    colorlinks=true,
        linkcolor=darkgreen, % fuer elektronische Version
        citecolor=darkred,
        urlcolor=darkblue,
    pdfpagemode=UseOutlines,% --> ob outlines oder thumbs links
    pdftitle={Functional renormalisation approach to driven dissipative dynamics},
    pdfauthor={Steven Mathey},
    pdfsubject={Statistical physics},
    pdfkeywords={turbulence,renormalisation,renormalization,super-fluid turbulence,statistical physics,scaling},
	pdfstartview 	={FitH}
}

\makeatletter
\DeclareRobustCommand*{\bfseries}{%
  \not@math@alphabet\bfseries\mathbf
  \fontseries\bfdefault\selectfont
  \boldmath
}
\makeatother

\renewcommand{\mathbf}[1]{{\boldsymbol{#1}}}
\renewcommand{\simeq}{\cong}
\newcommand{\op}{\omega,\mathbf{p}}
\newcommand{\hop}{\hat{\omega},\hat{\mathbf{p}}}
\newcommand{\mop}{-\omega,-\mathbf{p}}
\newcommand{\tx}{t,\mathbf{x}}
\newcommand{\Eq}[1]{Eq.~(\ref{#1})}
\newcommand{\Eqs}[1]{Eqs.~(\ref{#1})}
\newcommand{\eq}[1]{(\ref{#1})}
\newcommand{\eg}{e.g.~}
\newcommand{\ie}{i.e.~}

\newcommand{\Ie}{I.e.~}
\newcommand{\egc}{e.g.,~}
\newcommand{\iec}{i.e.,~}

\newcommand{\brv}{[\mathbf{v}]}
\newcommand{\nuk}[2]{\tilde{\nu}_{#1}^{#2}}
\newcommand{\nukk}{\nuk{k}{}}
\newcommand{\nukp}{\nuk{p}{}}
\newcommand{\nukps}{\nuk{p}{2}}
\newcommand{\nukq}{\nuk{q}{}}
\newcommand{\nukr}{\nuk{r}{}}
\newcommand{\T}{T(\hat{q}){}}
\newcommand{\twidlT}{\tilde{T}(\epsilon){}}
\newcommand{\hp}{\hat{p}}
\newcommand{\vhp}{\hat{\mathbf{p}}}
\newcommand{\cred}[1]{#1}
\newcommand{\pretitle}[1]{\section*{#1}}

\newabbrev\RG{Renormalisation Group (RG)}[RG]
\newabbrev\FRG{Functional Renormalisation Group (FRG)}[FRG]
\newabbrev\KPZ{Kardar--Parisi--Zhang (KPZ)}[KPZ]
\newabbrev\NS{Navier--Stokes (NS)}[NS]
\newabbrev\GPE{Gross--Pitaevskii equation (GPE)}[GPE]
\newabbrev\OnePI{1 Particle Irreducible (1PI)}[1PI]
\newabbrev\TwoPI{2 Particle Irreducible (2PI)}[2PI]
\newabbrev\IR{Infrared (IR)}[IR]
\newabbrev\UV{Ultraviolet (UV)}[UV]
\newabbrev\BMW{Blaizot--Mendez--Wschebor (BMW)}[BMW]
\newabbrev\SGPE{Driven-Dissipative Gross--Pitaevskii Equation (DDGPE)}[DDGPE]
\newabbrev\NTFPs{Non-Thermal Fixed Points (NTFP)\vphantom{\NTFP}}[NTFPs]
\newabbrev\NTFP{Non-Thermal Fixed Point (NTFP)\vphantom{\NTFPs}}[NTFP]
\newabbrev\EPCs{Exciton--Polariton Condensates (EPC)\vphantom{\EPC}}[EPCs]
\newabbrev\EPC{Exciton--Polariton Condensate (EPC)\vphantom{\EPCs}}[EPC]
\newabbrev\MSR{Martin--Siggia--Rose/Janssen--de Dominicis (MSR/JD)}[MSR/JD]

\makeatletter
\renewcommand\maybe@space@{%
  \maybe@ictrue % <= this is new
  \expandafter   \@tfor
    \expandafter \reserved@a
    \expandafter :%
    \expandafter =%
                 \nospacelist
                 \do \t@st@ic
  \ifmaybe@ic % <= this is new
    \space
  \fi
}
\makeatother

\begin{document}
%\hypersetup{pageanchor=false}
%\title{Universal scaling properties of randomly stirred fluids \\[1cm]
%Universal scaling properties of driven-dissipative dynamics \\[1cm]
%Functional renormalisation approach to driven dissipative dynamics \\[1cm]
%Scaling laws in stochastically driven fully developed turbulence and the Functional Renormalization Group \\[1cm]
%Burgers turbulence as a fixed point of the functional renormalisation group}
\frontmatter %use with scrbook
\newcommand{\HRule}{\rule{\linewidth}{0.7mm}}

\begin{titlepage}

\begin{center}

{\large

\bfseries

Dissertation\\[0.5em]

submitted to the\\[0.5em]

Combined Faculties of the Natural Sciences and Mathematics\\[0.5em]

of the Ruperto-Carola-University of Heidelberg. Germany\\[0.5em]

for the degree of\\[0.5em]

Doctor of Natural Sciences\\[0.5em]

\vfill

Put forward by\\[0.5em]

Steven, Mathey\\[0.5em]

born in: Geneva, Switzerland\\[0.5em]

Oral examination: 15th October 2014

}

\newpage

%\mbox{}\thispagestyle{empty}

%\newpage

\thispagestyle{empty}

\vspace*{0.4cm}

% Title

{

	\sffamily\Huge\bfseries

	\HRule \\ \mbox{} \\[0.6em]

	Functional renormalisation approach to driven dissipative dynamics \\[1.4em] 

	\mbox{} \\ \HRule

}

{	\large

  \bfseries

  \vfill

  \begin{tabular}{ll}

  Referees: & Prof.\,Dr.\,Thomas Gasenzer \\

            & Prof.\,Dr.\,J\"urgen Berges

  \end{tabular}

}
\end{center}

\newpage 

\selectlanguage{ngerman}
\pretitle{Zusammenfassung}
\ResetAbbrevs{All}

In der vorliegenden Arbeit werden die getrieben-dissipativen station\"aren skaleninvarianten Zust\"ande der Burgers- und Gross--Pitaevskii-Gleichungen (GPE) untersucht.\vphantom{\GPE}

Die Pfadintegral-Darstellung des station\"aren Zustands der stochastischen Burgers-Gleichung wird verwendet, um skaleninvariante L\"osungen des Systems an Fixpunkten der Renormierungsgruppe zu studieren.
Die funktionale Renormierungsgruppe wird genutzt, um die Physik in einer nicht-perturbativen N\"aherung zu beschreiben.
Eine Approximation, die Galileiinvarianz ber\"ucksichtigt und die die Frequenz- und Impulsabh\"angigkeit der zwei-Punkt Geschwindigkeits-Korrelationsfunktion beschreiben kann, wird konstruiert.
Ein System von Fixpunktgleichungen der Renormierungsgruppe f\"ur beliebige Frequenz und Impulsabh\"angigkeiten des inversen Propagators wird aufgestellt.
In allen betrachteten Raumdimensionen ergeben diese ein Kontinuum von Fixpunkten und einen isolierten Fixpunkt.
Diese Ergebnisse weisen eine sehr gute \"Ubereinstimmung mit den aus der Literatur bekannten Werten nur f\"ur $d=1$ auf.
In der Literatur werden jedoch fast ausschliesslich wirbelfreie L\"osungen behandelt, w\"ahrend die in der vorliegenden Arbeit verwendete N\"aherung exklusiv f\"ur L\"osungen mit Vortizit\"at anwendbar ist. Dadurch ist diese \"ahnlicher zur Navier--Stokes-Turbulenz.

Station\"are Nichtgleichgewichtszust\"ande ultrakalter Bose-Gase gekoppelt an externe Energie- und Teilchenreservoirs, wie zum Beispiel Exziton-Polariton Kondensate, stehen mit der stochastischen \KPZ Gleichung durch die Dichte- und Phasenzerlegung der gemittelten komplexen Wellenfunktion in Beziehung.
Diese Ergebnisse legen nahe, dass die skaleninvarianten L\"osungen, die in diesem Kontext festgestellt werden k\"onnen, auch g\"ultig sind f\"ur quasi-station\"are Zust\"ande des konservativen Systems fern des Gleichgewichts (nicht-thermische Fixpunkte), welche mit Hilfe der GPE beschrieben werden k\"onnen.
F\"ur das KPZ-Modell bekannte Ergebnisse werden auf ultrakalte Bose-Gase angewandt.
Auf diese Weise l\"a\ss t sich eine neue Skalenrelation herleiten, welche dazu verwendet werden kann, Kolmogorovs Skalenexponent $-5/3$ der turbulenten Energieverteilung in einer inkompressiblen Fl\"ussigkeit zu bestimmen. Dar\"uberhinaus erh\"alt man eine anomale Korrektur zum inkompressiblen wie auch zum kompressiblen Anteil des Energiespektrums eines verd\"unnten Bose-Gases.

\selectlanguage{english}

\newpage

\pretitle{Abstract}
\ResetAbbrevs{All}

In this thesis we investigate driven-dissipative stationary scaling states of Burgers' and Gross--Pitaevskii equations (GPE)\vphantom{\GPE}.

The path integral representation of the steady state of the stochastic Burgers equation is used in order to investigate the scaling solutions of the system at renormalisation group fixed points.
We employ the functional renormalisation group in order to access the non-perturbative regime.
We devise an approximation that respects Galilei invariance and is designed to resolve the frequency and momentum dependence of the two-point velocity correlation function.
We establish a set of renormalisation group fixed point equations for effective inverse propagators with an arbitrary frequency and momentum dependence.
In all spatial dimensions they yield a continuum of fixed points as well as an isolated one.
These results are fully compatible with the existing literature for $d=1$ only.
For $d\neq1$ however results of the literature focus almost exclusively on irrotational solutions while the solutions that our approximation can capture contain necessarily vorticity and are closer to Navier-Stokes turbulence.

Non-equilibrium steady states of ultra-cold Bose gases coupled to external reservoirs of energy and particles such as exciton--polariton condensates are related to the stochastic \KPZ equation by the density and phase decomposition of the average complex wave function.
We postulate that the scaling that we obtain in this context applies as well to far-from-equilibrium quasi-stationary steady states (non-thermal fixed points) of the corresponding closed system described by the \GPE.
We translate results found in the \KPZ literature to their corresponding dual in the ultra-cold Bose gas set-up.
We find that this provides a new scaling relation which can be used to analytically identify the classical Kolmogorov $-5/3$ exponent and its anomalous correction.
Moreover we estimate the anomalous correction to the scaling exponent of the compressible part of the kinetic energy spectrum of the Bose gas which is confirmed by numerical simulations of the \GPE.
\newpage

\begin{flushleft}

\begin{minipage}{.5\textwidth}

  \normalsize

  Steven Mathey\\[0.4em]\small

  Institut f\"ur Theoretische Physik \\ 

  Philosophenweg 16 \\

  D-69120 Heidelberg\\

  Deutschland					

\end{minipage}

\end{flushleft}

\vspace{1em}

\begin{flushleft}

\begin{minipage}{\textwidth}

  \normalsize

  \textsl{Primary advisor}\vspace*{-0.7em}\\ \rule{\linewidth}{0.2mm}

\end{minipage}

\end{flushleft}

\begin{flushleft}

\begin{minipage}{.45\textwidth}

  \normalsize

  Prof.\,Dr.\,Thomas Gasenzer \\[0.4em] \small

  Institut f\"ur Theoretische Physik \\ 

  Philosophenweg 16 \\

  D-69120 Heidelberg\\

  Deutschland					

\end{minipage}

\end{flushleft}

\vspace{1em}

\begin{flushleft}

\begin{minipage}{\textwidth}

  \normalsize

  \textsl{Secondary advisor}\vspace*{-0.7em}\\ \rule{\linewidth}{0.2mm}

\end{minipage}

\end{flushleft}

\begin{flushleft}

\begin{minipage}{.45\textwidth}

  \normalsize

  Prof.\,Dr.\,Jan M. Pawlowski \\[0.4em] \small

  Institut f\"ur Theoretische Physik \\ 

  Philosophenweg 16 \\

  D-69120 Heidelberg\\

  Deutschland					

\end{minipage}

\end{flushleft}

\end{titlepage}

\pretitle{Publication}

This thesis contains discussions and results from the following paper which is currently under review at Physical Review A.
\begin{itemize*}
 \item S. Mathey, T. Gasenzer, J. M. Pawlowski, Anomalous Scaling at Non-thermal Fixed Points of Burgers' and Gross--Pitaevskii Turbulence, \href{http://arxiv.org/abs/1405.7652}{arXiv:1405.7652 [cond-mat.quant-gas]}
\end{itemize*}
I conducted this work myself under the supervision of T.~Gasenzer and J.~M.~Pawlowski.
%The results about Non-thermal fixed points of ultra-cold Bose gases where obtained in collaboration with T. Gasenzer.
~\\[0.5cm]

\pretitle{Declaration by author}

This thesis is composed of my original work, and contains no material previously published or written by another person except where due reference has been made in the text. I have clearly stated the contribution by other authors to jointly-authored works that I have included in my thesis. The content of my thesis is the result of work I have carried out since the commencement of my graduate studies at the Heidelberg Graduate School of Fundamental Physics, Institut f\"ur Theoretische Physik, Universit\"at Heidelberg and does not include material that has been submitted by myself to qualify for the award of any other degree or diploma in any university or other tertiary institution.\\[0.5cm]

\newpage

\pretitle{Acknowledgments}
%\medskip\bigskip

My first and biggest thanks go to my advisors Thomas Gasenzer and Jan M. Pawlowski.
They placed me at the fascinating interface of turbulence and renormalisation.
They assembled enough funds to send me to conferences all over Europe.
They offered challenging guidance that made me improve by myself and straightforward advice when it was needed.
They questioned my work when I was not able to recognize that it was questionable.
%every one of my statements even when I didn't realise that they where doubtful.
They were patient and helped me improve.
They made me feel welcome in their respective research groups.

\smallskip

I would like to acknowledge the Heidelberg Graduate School for Fundamental Physics for holding together an excellent graduate program and for financing my month in Les Houches, the University of Heidelberg for providing all the necessary infrastructure and the Landesgraduiertenf\"orderungsgesetz for the financial support.

\smallskip

I would like to thank as well Sebastian Bock, L\'eonie Canet, Isara Chantesana, Sebastian Erne, Thomas Gasenzer, Markus Karl, Alexander Liluashvili, David Mesterh\'azy, Mario Mitter, Boris Nowak, Jan M. Pawlowski, Nikolai Philipp, Andreas Samberg, Martin Trappe, Gilles Tarjus and Nicolas Wschebor for stimulating scientific discussions.

\smallskip

Thank you to Sebastian Bock, Isara Chantesana, Nicolai Christiansen, Martin G\"arttner, Sebastian Heupts, Markus Karl, Kevin Falls, Boris Nowak and Andreas Samberg, for enduring and even trying to answer my random questions, for saving me when my computer was not being cooperative and/or for the warm welcome in the institute.

\smallskip

Thanks to my special proofreading team: Sebastian Bock, Isara Chantesana, Markus Karl, Gabriela Loza, Maureen Mathey, David Mesterh\'azy, Mario Mitter and Andreas Samberg.
Be it by pointing out typos or telling me to redo whole sections, they made me improve this work more than I could ever have by myself.
Of course, any remaining errors or mistakes are solely due to my shortcoming.
I thank as well Boris Nowak and Jan Schole who made their numerical data easily available to me.

\smallskip

I thank my parents who supported me during my studies and encouraged me along a road that I could choose for myself.

\smallskip

I thank my Ph.D. examination committee for taking an interest in my work and the time to learn about it. Thomas Gasenzer and J\"urgen Berges who will read this thesis and Markus Oberthaler and Karlheinz Meier who will take part in the oral examinations.
%(I hope that they like it.)

\smallskip

I thank the staff of the Institute for theoretical physics. They are always helpful and in a good mood. They are as well very patient with my poor German skills.

\smallskip

I thank as well the very wise users of the \LaTeX{} \href{http://tex.stackexchange.com/}{Stack Exchange}. They saved my life countless times.

\smallskip

My final thanks go to Gabriela Loza for the continued encouragement, support, patience and advice during the years that lead the conclusion of this work.
Thank you for filling my life with more than equations.

\clearpage
\thispagestyle{plain}
\par\vspace*{.35\textheight}{\centering \huge Para mi Amorcita\par}
%\chapter*{Acknowledgements}
%A lot of people helped me.

%\hypersetup{pageanchor=true}

\mainmatter %use with scrbook
\tableofcontents

\chapter{Introduction}
\ResetAbbrevs{All}

\begin{quote}
"A process cannot be understood by stopping it. Understanding must move with the flow of the process, must join it and flow with it."

\attrib{The First Law of Mentat, \cite{herbert1965dune}}
\end{quote}~\newline

%\cite{Altman2013a}
%cited somewhere else
%\cite{Berges:2000ew} Non-perturbative renormalization flow in quantum field theory and statistical physics
%\cite{Berges2012a} Introduction to the nonequilibrium functional renormalization group
%\cite{Mesterhazy2013a} Dynamic universality class of Model C from the functional renormalization group
%\cite{Adams:1995cv} Solving nonperturbative flow equations
%\cite{Berges:2004yj} Introduction to nonequilibrium quantum field theory
%possible liste
%http://inspirehep.net/record/1084868 Nonlinear amplification of instabilities with longitudinal expansion
%http://inspirehep.net/record/945590 Out of equilibrium dynamics of coherent non-abelian gauge fields
%http://inspirehep.net/record/882964 Strong versus weak wave-turbulence in relativistic field theory
%http://inspirehep.net/record/818348 fermion production qft
%http://inspirehep.net/record/725085 ultracold
%http://inspirehep.net/record/1264281 non abelian plasma
%http://inspirehep.net/record/1184493
%http://inspirehep.net/record/683003 thermalisation
%http://inspirehep.net/record/757639

\noindent Physical systems may look very different when observed at different scales \cite{powersof10,*powersof10huang}. In most cases it appears that processes which occur on very large scales are decoupled from those happening on small ones and can be described almost independently from each other.
Even if it is technically true that a butterfly flapping its wings can affect distant weather patterns, this is never included in tropical cyclone forecast models.
There are however systems realised in nature where correlations can propagate across a large range of scales. This typically leads to scale invariance of observables.
It is well known that such phenomena occur for specific values of the thermodynamic parameters of equilibrium systems. These are critical points of the phase diagram where fluctuations occur on all spatial scales and the correlation length is infinite \cite{Huang1987a,Goldenfeld1992a}. In thermal equilibrium critical states are not the norm though. \cred{Indeed}, the parameters of the system have to be precisely tuned in order to observe criticality.

The situation is however different outside of thermal equilibrium. There, many systems spontaneously evolve to a critical steady state (see \eg \cite{Kardar1986a,Bak1987a,Bak1990a,frette1996avalanche}). One of the most famous examples of such a far-from-equilibrium critical state is turbulence \cite{reynolds1883experimental,*reynolds1894dynamical,Kolmogorov1941a,*Kolmogorov1941b,*Kolmogorov1941c,Frisch2004a,monin2007statisticalI,*monin2007statisticalII,Nazarenko2011a,zakharov2012kolmogorov}.
\cred{Indeed}, it seems to appear almost spontaneously in many fundamentally different systems ranging from classical hydrodynamics to high-energy heavy-ion collisions all the way to ultra-cold Bose gas dynamics.
It can either be sustained by an appropriate driving mechanism or be a transient before the thermalisation of the system.
One of the hallmarks of turbulence is that, when it is realised, conserved charges undergo cascades.
The charge is transported either from large to small scales or in the other direction in a way that is local in Fourier space.
A stationary, unidirectional and local transport of charge is established.
For example in the case of a direct cascade of energy dissipation happens at small spatial scales while energy is injected at the large ones.
Then a steady state is established with energy flowing from large to small scales while keeping its spectral distribution constant.
Because they involve many different scales such processes depend on the dynamics of many interacting degrees of freedom and are still poorly understood and under heavy investigation \cite{anderson1972more}.

Considerable progress in our understanding of turbulence in quantum field theories has been made recently because of the use \cite{Berges:2008wm,Berges:2008sr,Scheppach:2009wu} of non-perturbative methods such as $1/N$ re-summations of \TwoPI effective field equations 
\cite{Berges:2001fi,Aarts:2001yn,Aarts:2002dj,Berges:2002cz,Gasenzer:2005ze,Temme2006,Berges:2007ym,Gasenzer2009a,Kronenwett:2010ic} and semi-classical methods \cite{Nazarenko2001a,Micha:2004bv,Vinen2006a,Berges:2007ym,bradley2008a,Berges:2008wm,Tsubota2008a,Weiler2008a,Blakie2008a,Tsubota2010a,Nowak:2010tm,Berges:2010ez,Gasenzer:2011by,Nowak:2011sk,Schmidt:2012kw,Bradley2012a,Schole:2012kt,Nowak:2012gd,davis2013a,reeves2013a,Karl:2013mn,Nowak:2013juc,Karl:2013kua,Barenghi2014a} as well as recent experimental realisations of quantum turbulence \cite{Henn2009a,neely2010a,seman2011a,Neely2012a}.
Applications of \FRG methods to dynamical evolution, non-thermal fixed points, and turbulence have received increasing attention recently \cite{Gasenzer:2008zz,Berges:2008sr,Canet2010a,Gasenzer:2010rq,Canet2011a,Kloss2012a,Sieberer2013a,Sieberer2013b,Kloss2013a}.
In particular the dynamics of ultra-cold Bose gases \cite{Nazarenko2001a,Vinen2006a,bradley2008a,Tsubota2008a,Weiler2008a,Tsubota2010a,Nowak:2010tm,Nowak:2011sk,Schmidt:2012kw,Bradley2012a,Schole:2012kt,Nowak:2012gd,reeves2013a,Karl:2013mn,Nowak:2013juc,Karl:2013kua,Barenghi2014a}
 and gauge systems such as relativistic heavy-ion collisions \cite{Fukushima:2011nq,Berges:2012us,Berges:2012ks,Berges:2012ev,Schlichting:2012es,Kurkela:2012hp,Berges:2013eia,Blaizot:2013lga,Fukushima:2013dma,Gasenzer:2013era,Berges:2013lsa} have received a great deal of attention.
 
\begin{figure}[t]
\center
\includegraphics[width=\textwidth]{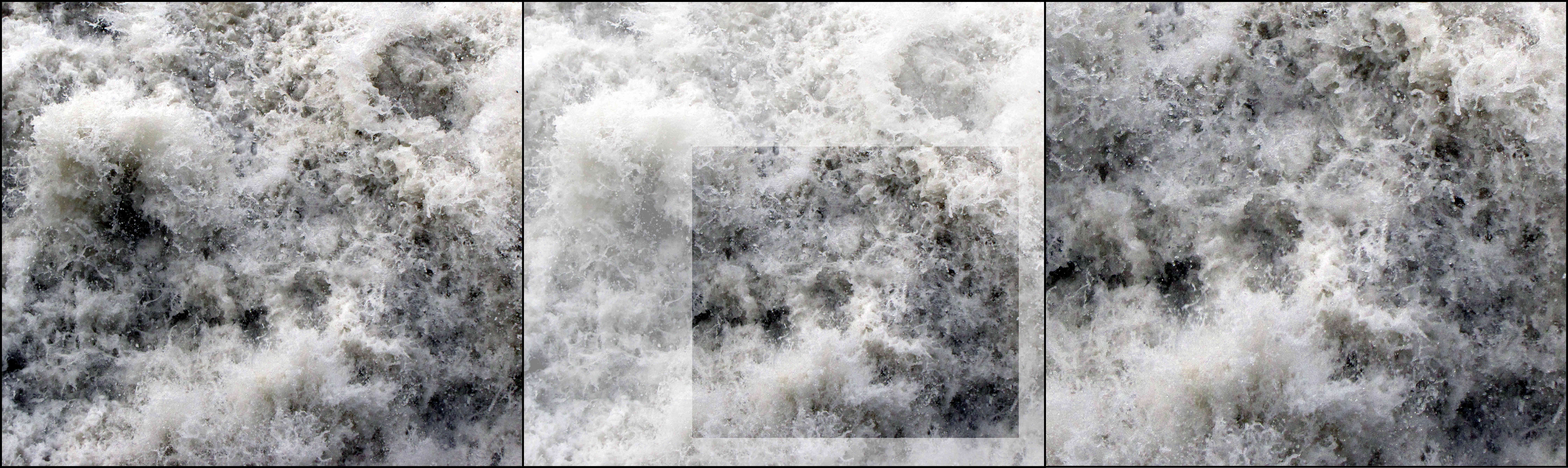}
\caption{Self-similar turbulence on the Neckar (Heidelberg)}
\label{fig:photo}
\end{figure}

In this work we investigate stationary classical and quantum turbulence. We focus on driven-dissipative dynamics which provide a natural mechanism to establish out-of-equilibrium steady states.
It is well known from experimental
\cite{chen1980vertical,castaing1990velocity,bramwell1998universality,hof2003scaling,xu2007,ictr2008,avila2011onset,eckhardt2011critical,Ouellette2011,%} and \cite{
Maurer1998,Walmsley2007a,Walmsley2008a},
numerical \cite{ictr2008,Takashi20009a,Schumacher2014a,%} and \cite{
Araki2002a,Kobayashi2005a,Numasato2010a} as well as analytical
\cite{Kolmogorov1941a,*Kolmogorov1941b,*Kolmogorov1941c,Forster1977a,Collina1997a,Adzhemyan1999a,Berera2001a,Frisch2004a,Barbi2010a,Chevillard2012,Mejiamonasterio2012a,ruelle2012a,ruelle2014,%} and \cite{
Volovik2004a,Kozik2009a} studies that both classical and quantum turbulence exhibits scale invariant observables. Moreover in the case of classical hydrodynamic turbulence this is as well evident form everyday life (see Figure (\ref{fig:photo})).
Based on this observation many authors have proposed analogies between classical turbulence and critical systems (see \eg \cite{nelkin1974turbulence,Eyink1994a,bramwell1998universality,eckhardt2011critical}). This points towards the \RG \cite{kadanoff1966a,wilson1971renormalizationI,*wilson1971renormalizationII,wilson1972critical,wilson1975a} and the idea that turbulence is realised as one of its fixed points.

Although it was introduced as a method to handle the divergences appearing in quantum field theory (see \cite{schweber1994qed,huang2013a} and references therein), the \RG has proven to be a tremendously effective tool in statistical physics.
It relates different effective descriptions of a system under increasing levels of coarse-graining of the degrees of freedom.
It has enabled us to understand the reason for the universality of critical physics. Why critical systems only come in a small number of different kinds even though the underlying microscopic physics can be infinitely varied. Moreover it enables the calculation of the critical properties of the system such as scaling exponents and dimensionless ratios and explains the infinite correlation length at a critical point as being the only non-zero correlation length that one can find at a fixed point of the \RG transformation (see \eg \cite{Goldenfeld1992a,Bagnuls:2000ae,Delamotte:2007pf} and references therein for overviews).

The \RG has also been applied very successfully outside of thermal equilibrium. Examples of such applications can be found in \eg \cite{Hohenberg1977a,Forster1977a,Medina1989a,Mathey2010a,mitra2012a} and in \cite{Forster1977a,Yakhot1986a,Yakhot1986b,Berera2001a} in the case of stationary hydrodynamic turbulence. See also \cite{Adzhemyan1999a} for an overview and references to the older literature and \cite{Gasenzer:2008zz,Berges:2008sr,Canet2010a,Gasenzer:2010rq,Canet2011a,Kloss2012a,Sieberer2013a,Sieberer2013b,Kloss2013a} for applications of the \FRG.
In the case of hydrodynamic turbulence its applications are, however, limited because the system is fundamentally non-perturbative.
\cred{Indeed}, at an \RG fixed point one finds perfect scale invariance. The cascade dynamics are realised from infinitely large to infinitely small scales and all the degrees of freedom take part in them. Then the corresponding Reynolds number (or coupling constant) which determines the extent of the scaling range must be infinite.

In this work we turn to the \FRG which is a non-perturbative version of the \RG. It expresses the \RG directly in terms the flow of a functional (typically a coarse-grained effective action or a generating functional) and contains by construction the full interacting theory. Note that the (perturbative) \RG is defined in terms of the flow of a few parameters of the system which makes it only effective close to fixed points.
The \FRG was first introduced directly as a flow equation for the effective Lagrangian of the theory in \cite{polchinski1984}.
Many \FRG flow equations have been written since then (see \eg \cite{Pawlowski:2005xe,Rosten:2010vm} for overviews).
The flow equation of Wetterich \cite{Wetterich:1991be,Wetterich:1992yh} which acts on the \IR coarse-grained \OnePI effective action has proven to be particularly useful. It provides an intuitive framework to compute \RG transformations because it relates directly physical observables and bare quantities by a continuous tuning of the cut-off scale.
See Refs.~\cite{Berges:2000ew,Polonyi:2001se,Rosten:2010vm} for general and \cite{Delamotte:2003dw,Litim:2006ag,Gies:2006wv,Pawlowski:2005xe,Igarashi:2009tj,Scherer:2010sv,Metzner:2011cw,canet2011b,Braun:2011pp,Berges2012a,Boettcher:2012cm} for more specific reviews.
Wetterich's \FRG has proven to be a powerful and versatile tool in many areas of physics.
It has recently gained a lot of interest in a wide range of non-equilibrium phenomena \cite{Gasenzer:2008zz,Gasenzer:2010rq}, including reaction-diffusion processes \cite{Canet:2003yu}, quantum decoherence \cite{Latorre:2004pk,Kehrein2004a}, open quantum systems \cite{Mitra2006a,Zanella:2006am}, critical dynamics \cite{Canet:2006xu,Berges:2008sr,Mesterhazy2013a}, transport in quantum systems \cite{Gezzi2007a,Jakobs2007a,Korb2007a}, cosmology \cite{Matarrese:2007wc}, strongly correlated transport in solids and Kondo physics \cite{Karrasch2008a,Schoeller2009a,Jakobs2009b}, disordered systems \cite{Balog2013a} and driven-dissipative dynamics \cite{Sieberer2013a,Sieberer2013b}.

The \FRG has been applied to classical steady state \NS turbulence in \cite{Collina1997a,Barbi2010a,Mejiamonasterio2012a}. The results of the classical perturbative approaches \cite{Forster1977a,Yakhot1986a,Yakhot1986b,Adzhemyan1999a} were recovered but the truly non-perturbative turbulent fixed point is still out of reach. However the work of the authors of \cite{Canet2010a,Canet2011a,Kloss2012a,Kloss2013a} who apply the \FRG to the similar although simpler problem of the stochastic \KPZ equation has been very successful and gives hope that the \FRG can be of use in the context of hydrodynamic turbulence.

\section{Summary}

In this work we study stationary driven-dissipative dynamics. As an intermediate step in between \KPZ and \NS dynamics we study the stochastic Burgers equation \cite{Burgers1939a} in arbitrary number of spatial dimension. Burgers' equation is formally similar to \NS apart from the fact that it does not contain a pressure term.
This simplifies the problem greatly as compared to the usual incompressible fluid dynamics because in the latter case the pressure is used to enforce the incompressibility condition. It is a slave to the dynamics of the velocity field and is determined by an inhomogeneous Laplace equation.
This introduces a non-local component to the equations of motion of the velocity field when this equation is inverted to express pressure in terms of the velocity field \cite{guyon2001a}.

Burgers' equation has received a lot of attention over the years because, for a given set of initial conditions, is can be solved implicitly in the limit of vanishing viscosity with the method of characteristics (see \cite{Frisch2000a,Bec2007a} and references therein).
Despite this apparent simplicity Burgers' equation still contains interesting and complex features.
Its solutions may acquire discontinuities \ie shocks, after a finite time evolution and these in turn produce a rich scaling behaviour when stochastic components are included in the dynamics. Moreover for spatial dimension $d>1$ the literature almost exclusively concentrates on solutions where the velocity field is the gradient of a potential $\mathbf{v} = \boldsymbol{\nabla}\theta$. To the best of our knowledge only \cite{Neate2011a,Choquard2013} does not make this simplification. This case is equivalent to \KPZ equation and can even be linearised by expressing the potential as\footnote{$\nu$ is the kinematic viscosity of the fluid. The equation of motion of $Z$ is linear.} $\theta = -2\nu\log(Z)$.

\paragraph{Section \ref{sec:the_functional_renormalisation_group} - The Functional Renormalisation Group}\hspace{0pt}\\

The second part of this work is about the \RG and its application to critical phenomena. In particular we focus on the \FRG. We emphasise the properties of \RG fixed point and explain how to compute them.

We start with a short qualitative introduction to the \RG and \FRG. Precise definitions are given in Section (\ref{sec:defgamma}). We turn to \FRG fixed point equations in Section (\ref{sec:FRG_fixed_point_equations}). We first report on a calculation that was made as an introductory exercise to the \FRG, Section (\ref{sec:Local_potential_approximation}). We reproduce results that can be found in Ref.~\cite{Bervillier:2007rc} and explain in detail how they are obtained. We consider the critical $\lambda^4$ scalar field theory \cite{weinberg1995quantum,kleinert2001critical,maggiore2005modern} and apply the local potential approximation \cite{Golner1986a,Wetterich:1991be,Morris:1994ie} to find the non-Gaussian \RG fixed point of the theory and compute its scaling exponents for $d=3$. Section (\ref{sec:Local_potential_approximation}) serves as well as an introduction to the \RG method since it is relatively simple and contains the most important ideas.

Section (\ref{sec:momentum_dependent_propagator}) represents \textbf{original research} work. We establish a set of \FRG fixed point equations that can be used when the space and time dependence of the two-point correlation function is an essential property of the system. \cred{Indeed}, the advective non-linearity of fluid mechanics contains a spatial derivative which produces a rich momentum dependence of correlation functions.
For this reason the local potential approximation does not provide good results in the context of hydrodynamic turbulence.
This section generalises the method first used in \cite{Pawlowski:2003XX} in the context of thermal equilibrium Yang-Mills theory \cite{weinberg1995quantum,maggiore2005modern} to an out-of-equilibrium set-up.

\paragraph{Section \ref{sec:burgers_turbulence} - Burgers Turbulence}\hspace{0pt}\\

The third part of this work is devoted to the stationary state of the stochastic Burgers equation.
The homogeneous Burgers equation is supplemented with a Gaussian random forcing which exhibits a stationary two-point correlation function.
Then even though individual solutions of the forced hydrodynamic equation have a strong time dependence the problem becomes stationary when the average over the different realisations of the forcing is taken.

We start with a brief introduction of the physics that we wish to describe and set up its mathematical formulation. In Section (\ref{sec:scaling_and_correlation_functions}) we review the results of the literature with a focus on the scaling properties of different correlation functions. Next we establish a path integral formulation of the steady state generating functional in Section (\ref{sec:Burgers_functionals}). In particular we compute its action (see \Eqs{eq:Burgersaction} or \eq{eq:Burgersaction_fourier}).

Section (\ref{sec:frg_calculation}) is the heart of the present work and contains exclusively \textbf{original research}. We apply the \FRG and the fixed point equations established in Section (\ref{sec:momentum_dependent_propagator}) to compute the two-point correlation function of the stochastic Burgers equation. We establish an approximation scheme in Section (\ref{sec:approximation_scheme}) and write \RG flow equations for its parameters in Section (\ref{sec:flow_equations}). Next we adapt the fixed point equations of Section (\ref{sec:momentum_dependent_propagator}) to our truncation in Section (\ref{sec:fixed_point_equations}). The different quantities that we have introduced in Sections (\ref{sec:approximation_scheme}) and (\ref{sec:fixed_point_equations}) are related to physically observable quantities in Section (\ref{sec:computing_observables}). Finally the \RG fixed point equations are analysed in detail in Section (\ref{sec:asymptotic_properties_of_the_flow_integrals}) and fixed points as well as their 
properties are discussed.

In Section (\ref{sec:scaling_limit_p_ll_k}) we solve the \RG flow equations in the asymptotic regime of vanishing re-scaled momentum. We find in all dimensions a continuum of fixed points supplemented with an additional isolated fixed point (see Figure (\ref{fig:alphas})). Along the continua of fixed points the value of the scaling exponents are extracted (see \Eq{eq:eta12alpha1}) from the asymptotic form of the \RG fixed point equations. In Section (\ref{sec:scaling_limit_p_gg_k}) we study the opposite asymptotic regime of infinite re-scaled momentum. We identify a range of values of the scaling exponents for which the fixed point theories are \UV convergent (see Figure (\ref{fig:eta_range})).
Finally in Section (\ref{sec:cascades}) we give a physical interpretation for the boundaries of this range in terms of the locality properties of the applied forcing and the transport of kinetic energy. In particular we find that the fixed point theories become \UV divergent when a cascade of energy to the \UV (direct cascade) sets in.

For $d=1$ our results are in good agreement with known results. Based on the values of the scaling exponents the fixed point that appear in the perturbative calculation of \cite{Medina1989a} are contained within our continuum of fixed points. Moreover, in accord with \cite{Kloss2013a} we find that the continuum of fixed points extends further into the region where the perturbative calculation breaks down.
For $d\neq1$ the perturbative calculation can only access the Gaussian fixed points which are trivially contained in our calculation since we write fixed point equations without analysing their stability. We try however to compare our results to those of \cite{Kloss2013a} which is an \FRG calculation of the scaling properties of the stochastic \KPZ equation.
Based on the comparison of the obtained scaling exponents the authors of \cite{Kloss2013a} find more fixed points than we do. We attribute this discrepancy to the fact that for $d\neq1$ Burgers' and \KPZ equations are only equivalent when the velocity field of Burgers' equation is the gradient of a potential and conclude that the fixed points that we find describe physics that is closer that of \NS equation which contains vorticity.
The isolated fixed point that we find in all dimensions appears nowhere else in the literature. Assuming that it is not an artefact of our approximation scheme we postulate that it is relevant in the case of forcing mechanisms that are strongly non local.

\paragraph{Section \ref{sec:ultracold_bose_gases} - Ultra-cold Bose Gases}\hspace{0pt}\\

In the fourth part of this work we change gears and consider the dynamics of dilute Bose gases.
We start by summarising relevant results concerning out-of-equilibrium steady states in closed systems. In particular in the regime where occupation numbers are large and the \GPE and the truncated Wigner approximation apply.
We introduce the concept of \NTFPs and their scaling properties in Section (\ref{sec:super_fluid_turbulence_and_non_thermal_fixed_points}). In Section (\ref{sec:driven_dissipative_gross_pitaevskii_equation}) we introduce the \SGPE as a model for a dilute gas of Bosons in contact with external reservoirs of particles and energy. The inclusion of driving and dissipation into the closed system enables its mapping to the stochastic Burgers equation which was introduced in the Part \ref{sec:burgers_turbulence}. 

We now go on to the \textbf{original research}.% These results where obtained in collaboration with Thomas Gasenzer.
In Section (\ref{sec:resultsSWT}) we assume that both approaches, the closed \GPE and the open \SGPE, describe the same out-of-equilibrium steady state and employ the mapping in between the \SGPE and Burgers' equation to apply the results of Section (\ref{sec:frg_calculation}) and extract non-trivial scaling relations in the context of out-of-equilibrium Bose gases at a \NTFP. This analysis produces two important results.

In the context of classical hydrodynamics a scaling relation in between the two exponents of the stochastic Burgers equation \eq{eq:chiofeta1andorz} emerges because of the Galilei invariance of the hydrodynamic theory.
The first important result, which is detailed in Section (\ref{sec:gallilee_invariance_and_kolmogorov_scaling}), comes from the translation of this scaling relation to a dual relation in between the exponents of the Bose gas (see \Eq{eq:kappa_and_z}).
In conjuncture with the scaling relations of \cite{Scheppach:2009wu} this reduces the number of unknown exponents characterising the \NTFP of the ultra-cold Bose gas to one.
In particular this makes it possible to identify the scaling exponent of the kinetic energy spectrum of the energy cascade as containing a canonical Kolmogorov scaling exponent of $-5/3$ with an unknown anomalous correction.

Section (\ref{sec:resultsAcoustic}) contains the second important result. It comes from translating the values of the scaling exponents of Burgers' equation computed in the \KPZ literature for $d=1$, $2$ and $3$ to the ultra-cold Bose gas set-up.
We find an interesting difference between the dynamics of the phase of the Bose gas wave function and the traditional \KPZ equation. Indeed, even though they are formally identical, the latter can not be applied to the former because it describes an unbounded field which can not represent a phase angle.
We compute anomalous corrections to the scaling of the compressible kinetic energy spectrum of the Bose gas. See \Eqs{eq:anomalousLit} where these exponents are shown and Figure (\ref{fig:anomalous_simulations}) where they are compared to the results of a numerical simulation.

\paragraph{Appendices \ref{sec:Local_potential_approximation_fixed_point_coefficients} to \ref{sec:list_of_abbreviations}}\hspace{0pt}\\

Most of the technical details are given in the appendices.
Appendix \ref{sec:Local_potential_approximation_fixed_point_coefficients} contains the derivation of a recursion relation \eq{eq:local_pot_recursion} which is necessary to compute the fixed point properties of the $\lambda^4$ scalar field theory in Section (\ref{sec:local_pot_fixed_point}).
Appendices \ref{sec:explicit_expressions_for_the_flow_integrals} to \ref{sec:equations_for_ai} provide many details on the fixed point analysis of Burgers' equation. The derivation of the \RG flow equations is discussed in Appendix \ref{sec:explicit_expressions_for_the_flow_integrals}. Their re-scaled form as well as the computation of their asymptotic behaviour is given in detail in Appendix \ref{sec:rescaled_flow_integrals}. Appendix \ref{sec:equations_for_ai} contains the derivation of sets of equations which are discussed in detail in Section (\ref{sec:scaling_limit_p_ll_k}) and used to constrain the scaling exponents of the fixed points.
In Appendix \ref{sec:kinetic_energy_spectrum_decomposition} we give details on the decomposition of the kinetic energy spectrum of the ultra-cold Bose gas in order to give a precise definition to the compressible kinetic energy spectrum which is discussed in Section (\ref{sec:resultsAcoustic}).
Appendix \ref{sec:notations_and_conventions} contains a list notation and conventions that are use through out this work and may be a little fuzzy in the main text.
Finally Appendix \ref{sec:list_of_abbreviations} contains a reminder of the different acronyms and short-hand notations that we use throughout this work.
%\layout
\chapter{The Functional Renormalisation Group}
\label{sec:the_functional_renormalisation_group}
\ResetAbbrevs{All}

In this section we briefly review the \RG and its application to critical phenomena \cite{kadanoff1966a,wilson1971renormalizationI,*wilson1971renormalizationII,wilson1972critical,wilson1975a}. In particular we focus on the \FRG \cite{Wetterich:1991be,Wetterich:1992yh}. We emphasise the properties of \RG fixed points and explain how to compute them.

We start with a short qualitative introduction to the \RG and \FRG. Precise definitions are kept for Section (\ref{sec:defgamma}). We turn to \FRG fixed point equations in Section (\ref{sec:FRG_fixed_point_equations}). We start by reporting on a calculation that was made as an introductory exercise to the \FRG, Section (\ref{sec:Local_potential_approximation}). We reproduce results that can be found in Ref.~\cite{Bervillier:2007rc} and explain in detail how the they are obtained. Section (\ref{sec:Local_potential_approximation}) serves as well as an introduction to the \RG method since it is relatively simple and contains the most important ideas. Finally we establish in Section (\ref{sec:momentum_dependent_propagator}), a set of \FRG fixed point equations that can be used when the space and time dependence of the propagator is an essential property of the fixed point and can not be neglected as it is with the local potential approximation. This generalises the method first used in \cite{Pawlowski:2003XX} in the case of thermal equilibrium Yang-Mills theory \cite{weinberg1995quantum,maggiore2005modern} to an out-of-equilibrium set-up.

In most of this work we consider the theory of stochastic hydrodynamics. The dynamical variables are described by a real field of $d$ components $\mathbf{v}(t,\mathbf{x})$ and depending on $d+1$ space-time variables $(\tx)$. $d$ is the dimension of space. We consider a particular unit system where time has the dimension of space squared\footnote{This implies that the kinematic viscosity is dimensionless. See Section (\ref{sec:burgers_turbulence}).}. Then the canonical dimension of the velocity field $\mathbf{v}(t,\mathbf{x})$, is one over space.

\RG transformations provide a link between effective descriptions of our system at different levels of coarse-graining. We consider here field theories that are characterised by a generating functional that can be expressed as a path integral
\begin{align}
Z[\mathbf{J}] & = \int \mathcal{D}\left[\mathbf{v}\right] \, \text{e}^{-S[\mathbf{v}]+\int \, \mathbf{J} \cdot \mathbf{v}},
\label{eq:generating_functional}
\end{align}
with a given action functional $S[\mathbf{v}]$. We use the notation $\mathcal{D}\left[\mathbf{v}\right] = \Pi_{\op} \text{d}^dv(\op)$. As usual correlation functions of the velocity field are obtained by taking derivatives of $Z[\mathbf{J}]$ and setting $\mathbf{J}=0$,
\begin{align}
& \langle {v}_{i_1}(t_1,\mathbf{x}_1) \cdot \,..\,\cdot  {v}_{i_n}(t_n,\mathbf{x}_n) \rangle = \frac{\delta^{n} Z[\mathbf{J}=0]}{\delta J_{i_1}(t_1,\mathbf{x}_1)..\delta J_{i_{n}}(t_{n},\mathbf{x}_{n})}.
\label{eq:computing_correlation_functions}
\end{align}
Note that more precise definitions are given in Appendix \ref{sec:notations_and_conventions}.
We introduce a momentum cut-off scale $k$ and write
\begin{align}
Z[\mathbf{J}] & =  \int \underset{\omega,p<k}{\Pi} \text{d}^d{v}(\omega,\mathbf{p}) \, \text{e}^{-S_k[\mathbf{v}]+\int \, \mathbf{J} \cdot \mathbf{v}}.
\label{eq:def_gamma_k}
\end{align}
The functional integration over momentum scales larger than the cut-off $k$ is absorbed into the exponential of the Wilson effective action, $S_k[\mathbf{v}]$ \cite{wilson1971renormalizationI,*wilson1971renormalizationII}.
The cut-off dependence of $S_k[\mathbf{v}]$ is defined by \Eq{eq:def_gamma_k} in such a way that the integration over momentum scales larger than $k$ does not have to be performed any more. By definition the \RG transformation does not affect the observable physics. In order for $Z[\mathbf{J}]$ to be independent of the cut-off scale we need to change the action with $k$ in just the right way.
$S_k[\mathbf{v}]$ can be interpreted physically as an effective action for the theory coarse-grained at the cut-off scale. For momentum scales smaller than $k$ correlation functions are computed as in \Eq{eq:computing_correlation_functions} except that the momenta that enter the calculation are restricted to be smaller than $k$ and $S_k[\mathbf{v}]$ is used instead of $S[\mathbf{v}]$. On the other hand, for momentum scales larger than $k$ no functional integration is necessary any more. Correlation functions are directly given by the derivatives of $S_k[\mathbf{v}]$.
%We use a sharp cut-off in \Eq{eq:def_gamma_k}.

Note that we do not cut off the frequency dependence. Spatial fluctuations of the velocity field are truncated if they are smaller than $1/k$ while temporal fluctuations over time are left completely free. Since the coarse-graining is done under the functional integration the physical quantities do not depend on the way in which we implement the cut-off. This does however make a difference when approximations are made.
One great advantage of the \RG technique is that fluctuations are included gradually in such a way that we do not encounter divergences at as we go from one level of coarse-graining to another. It was found that only cutting off the spatial fluctuations does not spoil this property (see \eg \cite{Medina1989a,Janssen1999a,frey1999scaling,Frey1994a,Collina1997a,Canet2010a,Barbi2010a,Canet2011a,Mejiamonasterio2012a,Kloss2012a,Sieberer2013a,Sieberer2013b,Kloss2013a}). We will see in Section (\ref{sec:frg_calculation}) that in the case of stochastic hydrodynamics described by Burgers' equation this is not necessary as well.

Before we discuss how $S_k[\mathbf{v}]$ is computed in practice, let us introduce the re-scaled dimensionless variables
\begin{align}
& \hat{\mathbf{x}} = \mathbf{x}\, k, && \vhp = \mathbf{p} \,k^{-1}, \nonumber \\
& \hat{t} =  t \, k^2 Z_1, && \hat{\omega} = \omega \, /(k^{2} Z_1) , \nonumber \\
& \hat{\mathbf{v}} =  \mathbf{v} \, 1/(k\sqrt{Z_2}), && \hat{\mathbf{J}} =  \mathbf{J} \, k^{-d-1} \sqrt{Z_2}.
%& \hat{t} =  t \, k^z && \hat{\omega} = \omega \, k^{-z}, \nonumber \\
%& \hat{\mathbf{v}} =  \mathbf{v} \, k^{1-z} && \hat{\mathbf{J}} =  \mathbf{J} \, k^{-d-1}.
\label{eq:rescaled_variables_intro}
\end{align}
The $k$-dependent rescaling factors $Z_i$ will be introduced shortly.
With these our generating functional takes the form\footnote{We use the short-hand notations $\int_{t,\mathbf{x}}=\int\mathrm{d}t\,\mathrm{d}^{d}x$ and $\int_{\omega,\mathbf{p}}=(2\pi )^{-d-1}\int\mathrm{d}\omega\,\mathrm{d}^{d}p$.}
\begin{align}
Z[\mathbf{J}] & = \int \underset{\hat{\omega},\hp<1}{\Pi} \text{d}^d\hat{{v}}(\hat{\omega},\vhp) \, \text{e}^{-\hat{S}_k[\hat{\mathbf{v}}] +\int_{\hat{t},\hat{\mathbf{x}}} \, \hat{\mathbf{J}}(\hat{t},\hat{\mathbf{x}}) \cdot \hat{\mathbf{v}}(\hat{t},\hat{\mathbf{x}})}.
\label{eq:def_gamma_k_rescaled}
\end{align}
With the re-scaled variables the dependence on $k$ disappears from the integral measure and is entirely confined to $\hat{S}_k[\hat{\mathbf{v}}]$. Then there is no reason to single out $S[\mathbf{v}]$ as compared to all the other effective theories given by $\hat{S}_k[\hat{\mathbf{v}}]$ for different values of $k$.
$\hat{S}_k[\hat{\mathbf{v}}]$ can be used instead of $S[\mathbf{v}]$ as an effective action and leads to the same generating functional $Z[\mathbf{J}]$.
%This enables the computation of correlation functions through the computation of the change of the parameters of $\hat{S}_k[\hat{\mathbf{v}}]$ instead of the computation of the whole path integral. \cred{Indeed}, any value of $k$ can be used instead of the original cut-off scale (used to define the original generating functional, \Eq{eq:generating_functional}) so that many different (and possibly leading to simpler a computation of $Z[\mathbf{J}]$) effective actions become available. The flow of $\hat{S}_k[\hat{\mathbf{v}}]$ is then required to relate the parameters of the computed observable to the parameters of the bare action $S[\mathbf{v}]$.

Note that there is a certain freedom in the re-scaling of $S_k[\mathbf{v}]$. \cred{Indeed}, since we use a momentum cut-off space must be re-scaled with the cut-off in order for $\hp$ to be always smaller than one, but the time and the field can be re-scaled with arbitrary cut-off dependent factors.
This is natural and usually implicit in the case of the original action $S\brv $.
Whatever the context $S\brv$ always contains some operators with no associated coupling. These have been set to one.
Compare \eg the terms containing time derivatives in \Eqs{eq:Burgersaction} and \eq{eq:action_gpe} with the terms containing a Laplace operator.
These couplings are set to one because if they were not we could simply go to variables where they are. In other words theories defined by $S\brv $ and $\tilde{S}\brv = S[\lambda \mathbf{v}]$ describe the same physics. If we want to have equivalent theories to be all represented by a single $\hat{S}_k[\hat{\mathbf{v}}]$ along the \RG flow we need to impose \RG conditions. Typically we set to one the first two terms of the Taylor expansion of the momentum dependent inverse propagator of $S_k\brv$, \ie $S_k^{(2)}(0,\mathbf{0}) = 1$ and $\partial^2_pS_k^{(
2)}(0,\mathbf{0}) = 1$. Such conditions make it possible to define $Z_1$ and $Z_2$ and re-scale the field and the time variables unambiguously. Once this is done the \RG flow is a property of the re-scaled quantities $S_k[\mathbf{\hat{v}}]$ only. The re-scaling factors $Z_i$ are "spectators" and can be expressed in terms of $S_k[\mathbf{\hat{v}}]$. Their logarithmic cut-off derivative $k\partial_k \log(Z_i) = \eta_i(k)$ are the anomalous dimensions of the field and time. They account for the anomalous scaling of correlation functions at fixed points.

In the theory of renormalisation, fixed points of the \RG flow play a special role. \cred{Indeed}, if $\hat{S}_k[\hat{\mathbf{v}}]$ reaches a fixed point it looses its dependence on $k$. Then a change of the cut-off scale only affects the dimension-full quantities of $S_k\brv$ by re-scaling them. A typical correlation function $O(\tx;\mathbf{g})$ will depend on the space-time variables as well as on the different parameters (or couplings) of the system $\mathbf{g}$. By construction when it is expressed in terms of re-scaled variables the re-scaled expression cancels out the $k$-dependence with cut-off dependent couplings,
\begin{align}
 O(\tx;\mathbf{g}) = k^{d_O} (Z_2)^{n/2} \, \hat{O}(\hop;\mathbf{\hat{g}}(k)).
\end{align}
$k^{d_O}$ carries the dimension of $O(\tx,\mathbf{g})$ and $n$ is the number of fields that it contains. At an \RG fixed point the parameters are numbers $\mathbf{\hat{g}}(k) = \mathbf{\hat{g}}^*$ and the re-scaling factors are simple scaling functions of the cut-off scale\footnote{This is because the re-scaled equations for $Z_i$ only depend on $\mathbf{\hat{g}}(k)$, $k\partial_k Z_i/Z_i = \eta_i(\mathbf{\hat{g}}(k))$.} $Z_i = Z_{i0}\,k^{\eta_i}$. Then we find that the correlation functions satisfy
\begin{align}
 O(\tx;\mathbf{g}) = (Z_{20})^{n/2} \, k^{d_O+n\eta_2/2} \, \hat{O}(t k^{2+\eta_1},\mathbf{x} k;\mathbf{\hat{g}}^*), && \text{for all }k>0.
\end{align}
The freedom that we have to choose the cut-off scale is expressed as a scale invariance of correlation functions. One can see this as a theory which describes a system that is invariant under simultaneous coarse-graining and re-scaling of the microscopic degrees of freedom. One averages over the finest details and tries to cancel this by scaling them down to a size where they would not be observable any way. If one can do this perfectly, then the system contains details on all scales which are all identical to each other.

The computation of the \RG flow of $S_k[\mathbf{v}]$ gives access to all the information contained in the generating functional. \cred{Indeed}, it is apparent from \Eq{eq:def_gamma_k} that in the limit $k\to \infty$ the flowing effective action reduces to the original action $S_{\infty}[\mathbf{v}] = S[\mathbf{v}]$. On the other hand, in the limit $k\to 0$ all the fluctuations have been included.
This feature is particularly explicit when the \RG flow is expressed in terms of the flowing effective action $\Gamma_k\brv$ which can be computed fro $S_k\brv$ by a Legendre transform of the short scale variable features of $S_k\brv$. See Section (\ref{sec:flowing_effective_action}) for details.
$\Gamma_k\brv$ represents physical observables on scales smaller than the inverse cut-off $1/k$ and acts as an effective action on larger scales.
As $k$ is lowered it changes continuously from being equal to the original action for $k\to\infty$ to the $\OnePI$ effective action once the cut-off is sent to zero.
It can be defined through the Legendre transform of the coarse-grained Schwinger functional $W_k[\mathbf{J}] = \log(Z_k[\mathbf{J}])$,
\begin{align}
\Gamma_k[\mathbf{v}] = \sup_{\mathbf{J}}\left[-\log\left(Z_k[\mathbf{J}]\right) + \int \mathbf{J} \cdot \mathbf{v}\right] - \Delta S_k\brv.
\label{eq:k_to_zero}
\end{align}
$Z_k[\mathbf{J}]$ is computed in the same way as $Z[\mathbf{J}]$ in \Eq{eq:generating_functional} with the difference that the action is supplemented with a cut-off term,
\begin{align}
S\brv \to S\brv + \Delta S_k\brv && \Delta S_k\brv = \frac{1}{2}\int_{\omega,\mathbf{p}} \mathbf{v}(\omega,\mathbf{p}) \cdot \mathbf{v}(-\omega,-\mathbf{p}) \, R_k(p).
\end{align}
$R_k(p)$ is a positive function that is very large for $p\ll k$ and very small otherwise.
See section \ref{sec:1pi_effective_action} or \cite{Zinnjustin2002a} for more details.

In practice, the exact computation of an \RG transformation is almost always impossible.
An appropriate approximation scheme must be devised.
Typical \RG transformations project $S_k[\mathbf{v}]$ on a finite set of flowing parameters that characterise the theory at different scales. In its \FRG formulation this projection is done rather late in the calculation. \cred{Indeed}, an exact differential equation \cite{Wetterich:1992yh} can be written for $\Gamma_k[\mathbf{v}]$,
\begin{align}
k \partial_k \Gamma_k[\mathbf{v}] 
= \frac{1}{2} \text{Tr}\left[ \left(\Gamma_k^{(2)}[\mathbf{v}]+R_k\right)^{-1} k\partial_k R_k \right].
\label{eq:wetterich}
\end{align}
\Eq{eq:wetterich} is then used to project the \RG flow on an appropriate set of parameters and can be used to include as much non-perturbative effects as possible. $R_k(p)$ is a positive function that is zero in the limit $k/p \to 0$ and diverges when $k/p\to \infty$. It acts as a momentum dependent mass term that cuts off field fluctuations with momentum smaller than $k$. See Section (\ref{sec:def_flow_equation}) for precise definitions.

The \RG approach to scaling and critical phenomena was first introduced in \cite{kadanoff1966a,wilson1971renormalizationI,*wilson1971renormalizationII,wilson1972critical,wilson1975a}. See \eg \cite{Goldenfeld1992a,Bagnuls:2000ae,Delamotte:2007pf} for introductory texts. The \FRG provides a non-perturbative framework to implement the coarse-graining inherent to the \RG. As opposed to the usual field theoretical perturbative \RG it takes into account irrelevant operators and provides physical information far away from fixed points.
The flow equation for the effective action \Eq{eq:wetterich} was introduced in \cite{Wetterich:1991be,Wetterich:1992yh} and provides an intuitive framework to compute \RG transformations since the derivatives of $\Gamma_{k\to 0}[\mathbf{v}]$ are directly related to physical observables while they can be interpreted as effective couplings at a finite cut-off value.
Moreover the flow of $\Gamma_k\brv$ is reversible. Even though the small scale fluctuations are coarse-grained infinitely high order correlation functions are taken into account such that no memory is lost as $k$ is decreased.

\section{Effective action}
\label{sec:defgamma}

In this section we give some details on the definition, interpretation and use of $\Gamma_{k}[\mathbf{v}]$ and its $k\to 0$ limit.
We relate $W_k[\mathbf{J}]$ and $\Gamma_k\brv$ to $S_k\brv$ through different Legendre transformations.
This clarifies the use of $W_k[\mathbf{J}]$ and $\Gamma_k\brv$ as coarse-grained quantities.
As in the previous section we use a sharp cut-off because it makes the coarse-graining procedure more intuitive. We will however arrive at expressions that do not depend on the particular choice of the cut-off and are valid for smooth cut-off's as well.
See \eg \cite{aoki2000introduction,Rosten:2010vm} and references therein for overviews.

We start by giving a precise meaning to \Eq{eq:def_gamma_k}. Our starting point is the generating functional of velocity correlation functions defined in \Eq{eq:generating_functional}. In order to implement the coarse-graining of small scale fluctuations, we introduce the cut-off scale $k$ and break the functional integration in two parts. The small scale (as compared to $1/k$) velocity fluctuations are integrated first while the large scale features of $\mathbf{v}(\tx)$ are kept as parameters. Then the integration over the remaining field variables is performed,
\begin{align}
Z[\mathbf{J}] & = \int \underset{\omega,p<k}{\Pi} \, \text{d}^d{v}(\omega,\mathbf{p}) \, \underset{\omega,p>k}{\Pi} \, \text{d}^d\mathbf{v}(\omega,\mathbf{p}) \, \text{e}^{-S[\mathbf{v}]+\int \, \mathbf{J} \cdot \mathbf{v}} \nonumber \\
& \equiv \int \underset{\omega,p<k}{\Pi} \, \text{d}^d{v}(\omega,\mathbf{p}) \, \text{e}^{-S_k[\mathbf{v}_<,\mathbf{J}_>]+\int \, \mathbf{J}_{<} \cdot \mathbf{v}_{<}}.
\label{eq:Sk}
\end{align}
We have used the notation $\mathbf{X}_{>,<}$ defined in Fourier space,
\begin{align}
\mathbf{X}_{>}(\omega,\mathbf{p}) = \theta(p-k) \, \mathbf{X}(\omega,\mathbf{p}), &&
 \mathbf{X}_{<}(\omega,\mathbf{p}) = \theta(k-p) \, \mathbf{X}(\omega,\mathbf{p}).
\end{align}
It denotes the Fourier truncated field. $\mathbf{X}_{<}$ is a coarse-grained version of $\mathbf{X}$. It is identical to $\mathbf{X}$ on large scales but is smooth on scales smaller than $1/k$. $\mathbf{X}_{>}$ is the opposite. It contains all the fine details of $\mathbf{X}$ but none of its overall features.
$S_k[\mathbf{v}_<,\mathbf{J}_>]$ depends the smooth (over scales smaller than $k$) features of $\mathbf{v}$ and the sharp details of $\mathbf{J}$. It is an effective action where all the fluctuations over scales smaller than $1/k$ are already included,
\begin{align}
\text{e}^{-S_k[\mathbf{v}_<,\mathbf{J}_>]} & = \int \underset{\omega,p>k}{\Pi} \, \text{d}^d{v}(\omega,\mathbf{p}) \, \text{e}^{-S[\mathbf{v}]+\int \, \mathbf{J}_> \cdot \mathbf{v}_>}.
\label{eq:Sk2}
\end{align}
Since $\mathbf{v}$ and $\mathbf{J}$ are conjugate variables, we can take the Legendre transformation of $S_k[\mathbf{v}_<,\mathbf{J}_>]$ with respect to one of its field variables and recover a functional of the full, not truncated, other.
We show in the following two Sections that the Legendre transform of $S_k[\mathbf{v}_<,\mathbf{J}_>]$ with respect to $\mathbf{J}_>$ is the flowing effective action which interpolates from the original action $S\brv$ for $k\to\infty$ and the \OnePI effective action for $k\to 0$ and that its Legendre transform with respect to $\mathbf{v}_<$ is closely related to the flowing Schwinger functional.

\subsection{Flowing effective action}
\label{sec:flowing_effective_action}

We start by changing variables from $\mathbf{J}_>$ to $\mathbf{v}_>$. We take the Legendre transform of $S_k[\mathbf{v}_<,\mathbf{J}_>]$ with respect to $\mathbf{J}_>$ and define the flowing effective action $\Gamma_k\brv$,
\begin{align}
&\Gamma_k[\mathbf{v}] = \sup_{\mathbf{J}_>}\left[S_k[\mathbf{v}_<,\mathbf{J}_>] + \int \, \mathbf{\mathbf{J}_>} \cdot \mathbf{v}_>\right], 
\label{eq:def_gamma_legendre}
\end{align}
When this is inserted in \Eq{eq:Sk} we get,
\begin{align}
Z[\mathbf{J}] & =  \int \underset{\omega,p<k}{\Pi} \, \text{d}^d{v}(\omega,\mathbf{p}) \, \text{e}^{ -\Gamma_k[\mathbf{v}] + \int \, \mathbf{J} \cdot \mathbf{\mathbf{v}}}.
\label{eq:gammak}
\end{align}
Note that the functional integration is only performed on $\mathbf{v}_<$ because of the cut-off. On the other hand $\Gamma_k\brv$ depends on the full $\mathbf{v}$ field.
On the right hand side of \Eq{eq:gammak} $\Gamma_k\brv$ is actually an implicit function of $\mathbf{J}_>$ through the relation ${\delta \Gamma_k[\mathbf{v}]}/{\delta \mathbf{v}_{>}} = \mathbf{J}_>$.
\Eq{eq:gammak} can be reformulated in such a way that the integration measure is not restricted,
\begin{align}
Z[\mathbf{J}] & = \int  \mathcal{D}\brv \, \exp\left[ -\Gamma_k[\mathbf{v}] + \int \, \mathbf{J} \cdot \mathbf{\mathbf{v}} - \frac{1}{2} \int_{\op} \left|J(\op)-\frac{\delta \Gamma_k[\mathbf{v}]}{\delta \mathbf{v}(\op)} \right|^2 \, \tilde{R}_k(p)\right].
\label{eq:gammak_2}
\end{align}
We have introduced $\tilde{R}_k(p)$ which is zero for $p<k$ and very large otherwise.
It makes it possible to enforces the cut-off without restricting the integration measure through the identity
\begin{align}
 \delta\left[ J_>-\frac{\delta \Gamma_k[\mathbf{v}]}{\delta \mathbf{v}_>}\right] = \lim_{\sigma \to 0} \frac{1}{N(\sigma)} \exp\left[-\frac{1}{2 \sigma^2} \int_{\op} \left| J(\op)-\frac{\delta \Gamma_k[\mathbf{v}]}{\delta \mathbf{v}(\op)}\right|^2 \theta(p-k)\right].
 \label{eq:delta_funct}
\end{align}
and the identification $\tilde{R}_k(p) = \lim_{\sigma \to 0} \theta(p-k)/\sigma^2$. $N(\sigma)$ is a normalisation factor which can be reabsorbed into the path integral measure. Note that the integrand inside the exponential of \Eq{eq:delta_funct} actually only contains the Fourier truncated quantities $J_>-{\delta \Gamma_k[\mathbf{v}]}/{\delta \mathbf{v}_>}$ since it is multiplied by $\theta(p-k)$.

One can see from \Eq{eq:gammak_2} what happens with $\Gamma_k\brv$ in both limits $k\to 0$ and $k\to \infty$. If the cut-off $k$ is sent to infinity, we have $\tilde{R}_{k\to\infty}(p)=0$. Then all the coarse-graining is gone and we recover \Eq{eq:generating_functional}. We have $\Gamma_{k\to \infty}[\mathbf{v}] = S[\mathbf{v}]$. On the other hand, if $k \to 0$ we have $\tilde{R}_{k\to 0}(p)\to \infty$ for all $p$ and the delta distribution of \Eq{eq:delta_funct} acts on the full field $\mathbf{v}(\tx)$ instead of its short scale features only. Then there are no fluctuations of $\mathbf{v}(\tx)$ left in the path integral and we have instead ${\delta \Gamma_{k\to 0}[\mathbf{v}]}/{\delta \mathbf{v}} = \mathbf{J}$ for all momenta. Only the average velocity field remains. $\Gamma_{k\to 0}\brv = \Gamma\brv$ is the \OnePI effective action.

\Eq{eq:gammak_2} shows as well that all the fluctuations of $\mathbf{v}_<$ are present in the path integral while $\mathbf{v}_>$ is fixed to its average value $\mathbf{v}_> = \delta S_k[\mathbf{v}_<,\mathbf{J}_>]/\delta \mathbf{J}_>$.
$\Gamma_k[\mathbf{v}]$ generates physical correlation functions on scales smaller than $1/k$. On large scales however, it acts as an effective action and can be used to replace $S[\mathbf{v}]$ in \Eq{eq:generating_functional} if the fluctuations with momentum larger than $k$ are cut off.

\subsection{Flowing Schwinger functional}

We now go back to \Eq{eq:Sk2}. This time we change variables from $\mathbf{v}_<$ to $\mathbf{J}_<$,
\begin{align}
\tilde{W}_k[\mathbf{J}] = \sup_{\mathbf{v}_<}\left[-S_k[\mathbf{v}_<,\mathbf{J}_>] + \int \, \mathbf{\mathbf{J}_<} \cdot \mathbf{v}_<\right].
\end{align}
When this is inserted in \Eq{eq:Sk2} we get
\begin{align}
\text{e}^{\tilde{W}_k[\mathbf{J}]} =  \int \underset{\omega,p>k}{\Pi} \, \text{d}^d{v}(\omega,\mathbf{p}) \, \text{e}^{-S\brv+\int \, \mathbf{J} \cdot \mathbf{v}}.
\label{eq:wk}
\end{align}
This time it is $\mathbf{v}_<$ in the integrand that implicitly depends on $\mathbf{J}_<$ through $\delta \tilde{W}_k[\mathbf{J}]/\delta \mathbf{J}_{<} = \mathbf{v}_<$. Note that the Legendre transform is not defined with the same sign as the transform with respect to $\mathbf{J}_>$. We make this choice of the definition of $\tilde{W}_k[\mathbf{J}]$ so that it is the Legendre of $\Gamma_k[\mathbf{v}]$. As before we can shift the cut-off from the integration measure to the integrand
\begin{align}
\text{e}^{\tilde{W}_k[\mathbf{J}]} =  \int \mathcal{D}\brv \, \exp\left[-S\brv+\int \, \mathbf{J} \cdot \mathbf{v}-\frac{1}{2}\int_{\op} \left|\mathbf{v}(\op)-\frac{\delta \tilde{W}_k}{\delta \mathbf{J}(\op)}\right|^2 \, R_k(p)\right].
\label{eq:wk2}
\end{align}
$R_k(p)$ is zero for $p>k$ and infinite otherwise. It is an \IR cut-off. Note the difference with the \UV cut-off, $\tilde{R}_k(p)$. 

The flowing Schwinger functional is defined by subtracting the term
\begin{align}
\Delta S_k[\mathbf{v}] & = \frac{1}{2}\int_{\omega,\mathbf{p}} \mathbf{v}(\omega,\mathbf{p}) \cdot \mathbf{v}(-\omega,-\mathbf{p}) \, R_k(p),
\end{align}
to the action $S\brv$ in \Eq{eq:Sk} and taking the logarithm of the obtained coarse-grained generating functional,
\begin{align}
\text{e}^{ W_k[\mathbf{J}]} = \int D\brv \, \text{e}^{-S[\mathbf{v}]- \Delta S_k[\mathbf{v}] +\int \, \mathbf{J} \cdot \mathbf{v}}.
\label{eq:wk3_preliminary}
\end{align}
We see that $\tilde{W}_k[\mathbf{J}]$, the functional that we obtain through the Legendre transform of $S_k[\mathbf{v}_<,\mathbf{J}_>]$, is slightly different from ${W}_k[\mathbf{J}]$. It is however clear that $\tilde{W}_k[\mathbf{J}]$ is the Legendre transform of $\Gamma_k\brv$. Then it is no surprise that $\tilde{W}_k[\mathbf{J}]\neq {W}_k[\mathbf{J}]$. Indeed, in order to recover the Schwinger functional from the flowing effective action an additional factor of $\Delta S_k[\mathbf{v}]$ must be added to $\Gamma_k\brv$ before its Legendre transformation. See \Eqs{eq:ek4} and \eq{eq:gammak_3}. One can write $\tilde{W}_k[\mathbf{J}]$ as a double Legendre transform of ${W}_k[\mathbf{J}]$ with a factor of $\Delta S_k\brv$ inserted in between the two transformations and recover \Eq{eq:wk2} from \Eq{eq:wk3_preliminary},
\begin{align}
&\tilde{W}_k[\mathbf{J}] = \sup_{\mathbf{v}} \left[ -\Gamma_k\brv + \int \mathbf{v} \cdot \mathbf{J} \right] \nonumber \\
&\Gamma_k\brv = \sup_{\mathbf{J}}\left[ -W_k[\mathbf{J}] + \int \mathbf{v} \cdot \mathbf{J} \right] - \Delta S_k\brv.
\end{align}

$\Gamma_k\brv$ and $\tilde{W}_k[\mathbf{J}]$ were both defined with a sharp cut-off. Note that the cut-off procedure has been entirely shifted into the definitions of $\tilde{R}_k(p)$ and ${R}_k(p)$ in \Eqs{eq:gammak_2} and \eq{eq:wk2}. We are actually free to choose smooth functions instead of $\tilde{R}_k(p)$ and ${R}_k(p)$. It is only important that they are positive, that $\tilde{R}_k(p<k)$ and ${R}_k(p>k)$ be very small and $\tilde{R}_k(p>k)$ and ${R}_k(p<k)$ be very large.

The sharp cut-off makes the coarse-graining procedure more intuitive and helps to understand what is going on.
Since it was introduced as an intermediate step in the computation of $Z[\mathbf{J}]$, it will in principle not affect the outcome of the computation of physical quantities.
However it does not interact well approximation schemes because it is highly non-analytic.
When an approximation is made in the computation of $Z[\mathbf{J}]$ the error depends on the cut-off that we choose and may be large with a sharp cut-off.
In practice it is better to use smooth cut-off functions instead of $\tilde{R}_k(p)$ and ${R}_k(p)$.
In the following we will simply keep $R_k(p)$ and $\tilde{R}_k(p)$ as free parameters until it becomes necessary to choose a specific form for one of them in Sections (\ref{eq:renormalisation_group_flow_equation}) and (\ref{sec:frg_calculation}).

\subsection{Flow equation}
\label{sec:def_flow_equation}

We have used \Eqs{eq:Sk2} and \eq{eq:def_gamma_legendre} to define $\Gamma_k\brv$. A somewhat simpler, although equivalent, definition is written in terms of the flowing Schwinger functional $W_k[\mathbf{J}]$.
We introduce a positive cut-off function $R_k(p)$ that satisfies with $R_k(p \gg k) = 0$ and $R_k(p \ll k) = \infty$ and define $W_k[\mathbf{J}]$ as
\begin{align}
\text{e}^{ W_k[\mathbf{J}]} = \int D\brv \, \text{e}^{-S[\mathbf{v}]- \Delta S_k[\mathbf{v}] +\int \, \mathbf{J} \cdot \mathbf{v}},&&
\Delta S_k[\mathbf{v}] = \frac{1}{2}\int_{\omega,\mathbf{p}} \mathbf{v}(\omega,\mathbf{p}) \cdot \mathbf{v}(-\omega,-\mathbf{p}) \, R_k(p).
\label{eq:wk3}
\end{align}
As we discussed in the previous section, $\Gamma_k[\mathbf{v}]$ is not exactly the Legendre transform of $W_k[\mathbf{J}]$. Instead we have
\begin{align}
& \tilde{\Gamma}_k[\mathbf{v}] =\sup_{\mathbf{J}}\left[- W_k[\mathbf{J}] + \int \mathbf{J}\cdot \mathbf{v}\right],
\label{eq:ek4}
\end{align}
and the flowing effective action is
\begin{align}
\Gamma_k[\mathbf{v}] = \tilde{\Gamma}_k[\mathbf{v}] - \Delta S_k[\mathbf{v}].
\label{eq:gammak_3}
\end{align}
On can check from this definition of $\Gamma_k\brv$ that $\Gamma_{k\to\infty}\brv = S\brv$ and $\Gamma_{k\to0}\brv = \Gamma\brv$. Moreover we have seen in Section (\ref{sec:flowing_effective_action}) that $\Gamma_k\brv$ is a coarse-grained effective action.
This definition can be used to write a differential equation for the flowing effective action. \cred{Indeed}, $\Gamma_k\brv$ changes continuously with $k$. The cut-off derivative of $W_k[\mathbf{J}]$ can be computed directly form \Eq{eq:wk3}. Then this can be inserted into \Eq{eq:gammak_3} in order to write,
\begin{align}
k \partial_k \Gamma_k[\mathbf{v}] 
= \frac{1}{2} \text{Tr}\left[ \left(\Gamma_k^{(2)}[\mathbf{v}]+R_k\right)^{-1}k\partial_k R_k \right].
\label{eq:wetterich2}
\end{align}
See \eg \cite{Wetterich:1992yh,Berges:2000ew,Polonyi:2001se,Rosten:2010vm} for a detailed derivations and overviews.
$\Gamma_k^{(2)}[\mathbf{v}]+R_k$ is the inverse of the propagator computed from $W_k[\mathbf{J}]$. It satisfies
\begin{align}
 W_k^{(2)}[\mathbf{J}] \left(\Gamma_k^{(2)}[\mathbf{v}]+R_k\right) = \mathds{1},
\label{eq:w_times_G}
\end{align}
with
\begin{align}
& W_{k,ij}^{(2)}[\mathbf{J}](\tx;t',\mathbf{x}') = \frac{\delta^2 W_k[\mathbf{J}]}{\delta J_i(\tx) \delta J_j(t',\mathbf{x}')}, && \Gamma_{k,ij}^{(2)}\brv (\tx;t',\mathbf{x}') = \frac{\delta^2 \Gamma_k\brv}{\delta v_i(\tx) \delta v_j(t',\mathbf{x}')},
\label{eq:proper_definitions1}
\end{align}
and
\begin{align}
& R_{k,ij}^{(2)}(\tx;t',\mathbf{x}') = \delta_{ij} \int_{\op} \text{e}^{i\left[\omega (t-t') - \mathbf{p}\cdot(\mathbf{x}-\mathbf{x})\right]} \, R_k(p),\nonumber \\
& \mathds{1}_{ij}(\tx;t',\mathbf{x}') = \delta_{ij} \delta(t-t') \delta(\mathbf{x}-\mathbf{x}').
\label{eq:proper_definitions2}
\end{align}
The product and the trace of two operators are defined straightforwardly as
\begin{align}
 & {AB}(\tx;t',\mathbf{x}') = \int_{t'',\mathbf{x}''} {A}(\tx;t'',\mathbf{x}'') {B}(t'',\mathbf{x}'';t',\mathbf{x}'),
&& \text{Tr}\left({A}\right) =  \int_{\tx} A(\tx;\tx).
\label{eq:proper_definitions_2}
\end{align}
\Eq{eq:wetterich2} is a functional differential equation for $\Gamma_k\brv$. It contains the full dependence of $\Gamma_k\brv$ on $\mathbf{v}(\tx)$ and can be used to extract equations for all of the derivatives of the flowing effective action. Note that it couples $\Gamma_k\brv$ to its second field derivative. The $n$-th derivative of \Eq{eq:wetterich2} will contains derivatives of $\Gamma_k\brv$ up to order $n+2$. We have an infinite hierarchy of equations for all the $\Gamma_k^{(n)}$.

In practice \Eq{eq:wetterich2} can only be approximately solved. Typically one writes an ansatz for $\Gamma_k\brv$ in terms of unknown parameters which can be extracted by taking appropriate derivatives of the ansatz (see \eg \Eqs{eq:effective_pot_ansatz} and \eq{eq:trunc}). Then one can obtain differential equations for the parameters by taking the same derivative on both sides of \Eq{eq:wetterich2}. See \Eqs{eq:flow_local_pot_not_rescaled} and \eq{eq:flow1} (with \Eqs{eq:explicit_flow_1} and \eq{eq:explicit_flow_2}). The different parameters of the ansatz are couples to each other through the coupling of the different derivatives of $\Gamma_k\brv$ and obey complex non-linear differential equations because of the structure of \Eq{eq:wetterich2}. In most cases these equations need to be treated numerically.

\subsection{One particle irreducible effective action}
\label{sec:1pi_effective_action}

We can see from \Eq{eq:wk3} that $\Gamma_{k\to 0}[\mathbf{v}] \equiv \Gamma[\mathbf{v}]$ is the Legendre transform of the Schwinger functional, $W[\mathbf{J}]$ which generates all the connected correlation functions of the velocity field. This information is equivalently contained in $\Gamma[\mathbf{v}]$. Here, we briefly show how correlation functions are computed from $\Gamma[\mathbf{v}]$. See \eg \cite{Zinnjustin2002a} for a detailed exposition.

Once one has obtained a good estimation for $\Gamma\brv$ the first step is to determine the average field, $\langle \mathbf{v}(t,\mathbf{x})\rangle$ by solving the equation, $\delta \Gamma[\mathbf{v}=\langle{\mathbf{v}}\rangle] /\delta \mathbf{v}=0$. This is equivalent to $\langle{\mathbf{v}}\rangle= \delta W[\mathbf{J}=0]/\delta \mathbf{J}$. In the case of stochastic hydrodynamics, we get $\langle{\mathbf{v}}\rangle=0$ because of rotational symmetry.

The two-point correlation function is the inverse of the second derivative of $\Gamma[\mathbf{v}]$ evaluated at $\langle{\mathbf{v}}\rangle$,
\begin{align}
\langle v_i(t,\mathbf{x}) v_j(t',\mathbf{x}') \rangle_{\text{c}} = \left( \frac{\delta^2 \Gamma[\mathbf{v}=\langle \mathbf{v} \rangle]}{\delta v_i(t,\mathbf{x}) \delta v_j(t',\mathbf{x}') } \right)^{-1}.
\label{eq:inverse_gamma}
\end{align}
Note that this is a equivalent to \Eq{eq:w_times_G} in the limit $k\to0$. We include a $c$ index to differentiate the disconnected correlation functions, generated by $Z[\mathbf{J}]$, from the connected ones which are generated by $W[\mathbf{J}]$.

Finally, higher order correlation functions can be computed from higher order derivatives of $\Gamma[\mathbf{v}]$. \cred{Indeed}, we can express $n$-point correlation functions using $n-2$ functional derivatives of the two-point function with respect to $\mathbf{J}$,
\begin{align}
& \langle \mathbf{v}_{i_1}(t_1,\mathbf{x}_1) \, \mathbf{v}_{i_2}(t_2,\mathbf{x}_2) \dots \mathbf{v}_{i_n}(t_n,\mathbf{x}_n) \rangle_{\text{c}} \nonumber \\
& \qquad = \frac{\delta^{n-2}}{\delta J_{i_1}(t_1,\mathbf{x}_1)..\delta J_{i_{n-2}}(t_{n-2},\mathbf{x}_{n-2})} \frac{\delta^2W[\mathbf{J}=0]}{\delta J_{i_{n-1}}(t_{n-1},\mathbf{x}_{n-1})\delta J_{i_{n}}(t_{n},\mathbf{x}_{n})} \nonumber \\
& \qquad = \frac{\delta^{n-2}}{\delta J_{i_1}(t_1,\mathbf{x}_1)..\delta J_{i_{n-2}}(t_{n-2},\mathbf{x}_{n-2})} \left( \frac{\delta^2 \Gamma[\mathbf{v}=\langle \mathbf{v} \rangle]}{\delta v_{i_{n-1}}(t_{n-1},\mathbf{x}_{n-1}) \delta v_{i_n}(t_n,\mathbf{x}_n) } \right)^{-1}.
\label{eq:high_corr_fct}
\end{align}
This can be expressed in terms of the derivatives of $\Gamma\brv$ by using the chain rule for derivatives,
\begin{align}
 \frac{\delta A^{-1}(\tx;t',\mathbf{x}')}{\delta J(t'',\mathbf{x}'')} = - \left(A^{-1} \frac{\delta A}{\delta J(t'',\mathbf{x}'')} A^{-1}\right)(\tx;t',\mathbf{x}').
\end{align}
Note that the last line of \Eq{eq:high_corr_fct} contains an additional internal derivative. \cred{Indeed}, $\Gamma\brv$ depends on $\mathbf{J}(\tx)$ only implicitly through $\mathbf{v}(\tx)$ and the relation ${\delta {\Gamma}[\mathbf{v}]}/{\delta \mathbf{v}} = \mathbf{J}$. This produces an additional propagator multiplying the whole equation.
In particular we have
\begin{align}
& \langle \mathbf{v}_{i_1}(t_1,\mathbf{x}_1) \, \mathbf{v}_{i_2}(t_2,\mathbf{x}_2) \mathbf{v}_{i_3}(t_3,\mathbf{x}_3) \rangle_{\text{c}} \nonumber \\
& \quad = - \int_{\tau_1,\mathbf{z}_1;\tau_2,\mathbf{z}_2;\tau_3,\mathbf{z}_3} \frac{\delta^3 \Gamma[\mathbf{v}=\langle \mathbf{v} \rangle]}{\delta v_{j_1}(\tau_1,\mathbf{z}_1) \delta v_{j_2}(\tau_2,\mathbf{z}_2)\delta v_{j_3}(\tau_3,\mathbf{z}_3)} \left( \frac{\delta^2 \Gamma[\mathbf{v}=\langle \mathbf{v} \rangle]}{\delta v_{i_{1}}(t_{1},\mathbf{x}_{1}) \delta v_{j_1}(\tau_1,\mathbf{z}_1) } \right)^{-1} \nonumber \\
& \quad \times  \left( \frac{\delta^2 \Gamma[\mathbf{v}=\langle \mathbf{v} \rangle]}{\delta v_{i_{2}}(t_{2},\mathbf{x}_{2}) \delta v_{j_2}(\tau_2,\mathbf{z}_2) } \right)^{-1}
 \left( \frac{\delta^2 \Gamma[\mathbf{v}=\langle \mathbf{v} \rangle]}{\delta v_{i_{3}}(t_{3},\mathbf{x}_{3}) \delta v_{j_3}(\tau_3,\mathbf{z}_3) } \right)^{-1}.
 \label{eq:three_point_function}
\end{align}

One can apply the same procedure to $\Gamma_{k}[\mathbf{v}]$ and extract scale dependent correlation functions. These already represent physics on spatial scales larger than the inverse cut-off. We append a $k$ index to such correlation functions,
\begin{align}
\langle v_i(t,\mathbf{x}) v_j(t',\mathbf{x}') \rangle_{k,\text{c}} = \left( \frac{\delta^2 \Gamma_k[\mathbf{v}=\langle \mathbf{v} \rangle_k]}{\delta v_i(t,\mathbf{x}) \delta v_j(t',\mathbf{x}') } \right)^{-1}.
\end{align}
Note that because of the additional cut-off term these do not correspond to correlation functions computed from $W_k[\mathbf{J}]$ (see \Eq{eq:gammak_3}).

\section{Functional renormalisation group fixed point equations}
\label{sec:FRG_fixed_point_equations}

The greatest strength of the \FRG is that its starting point \Eq{eq:wetterich2}, is exact and can be used as a starting point to make non-perturbative approximations. In practice the flowing effective action must be truncated since \Eq{eq:wetterich2} generates an infinite hierarchy of differential equations for all of the derivatives of $\Gamma_k[\mathbf{v}]$. Typically one uses physical intuition and symmetry constraints to write an ansatz for the flowing effective action, $\Gamma_k[\mathbf{v}]$. Such an ansatz should contain unknown parameters which depend on the cut-off scale, $k$. Then the exact functional equation \Eq{eq:wetterich2} is projected on the subset of effective actions that the ansatz is capable of generating. This procedure leads to a set of differential equations for the parameters of the ansatz which can be handled with standard techniques \cite{numerical_recipes,hairer2008solving,*hairer1996solving,iserles2009first}.

It is clear that the quality of the outcome of any \FRG calculation will strongly depend on the quality of the chosen ansatz. $\Gamma_k[\mathbf{v}]$ should not explicitly break the symmetries of the problem at hand. If this is the case the symmetry breaking terms will lead to either trivial or wrong flow equations since they do not appear in the exact solution. However within the symmetry constraints $\Gamma_k[\mathbf{v}]$ should be as general as possible. Then the ansatz is usually simplified based on the physical process one is modelling. The truncated flowing effective action should be completely general (at least) in the physical sector which one hopes to resolve.

In choosing an appropriate ansatz, there are two different paths one can take. If one is mainly interesting in a precise evaluation of the correlation functions at equal positions and times,
\begin{align}
 \lambda_n = \langle v_{i_1}(\tx) v_{i_2}(\tx) \dots v_{i_n}(\tx) \rangle,
\end{align}
the momentum dependence of the parameters of $\Gamma_k[\mathbf{v}]$ is not so important. One can make a derivative expansion. \Ie the derivatives of the truncated effective action are assumed to be polynomials in the momentum and only the smallest powers of $p$ are kept. Within this approximation it is possible to model correlation functions with very large number of fields by including very large powers of $\mathbf{v}(\tx)$ into the ansatz. This will be discussed in Section \ref{sec:Local_potential_approximation}.

On the other hand, if one is more interested in the dependence of the correlation functions on space and time,
\begin{align}
F(t-t',\mathbf{x}-\mathbf{x}') = \langle v_i(\tx) v_j(t',\mathbf{x}') \rangle,
\end{align}
such an approximation will not produce good results. In this case it is better to use an ansatz with as much momentum dependence as possible.
It then becomes necessary to consider only a small numbers of powers of $\mathbf{v}(\tx)$ and the approximation becomes bad when high order correlation functions are computed. This is discussed in Section \ref{sec:momentum_dependent_propagator}.

\RG fixed point can of course be studied within both approaches. We show how this can be done the two following sections.
The \BMW approximation scheme \cite{Blaizot:2005xy,Blaizot:2005wd,Blaizot:2006vr,benitez2012} tries to bridge these two approaches by truncating the momentum dependence of correlation functions of order larger than a given order $s$ while considering the full momentum dependence of lower order order correlation functions. The \RG flow equations are evaluated at a an arbitrary constant field $\mathbf{v}(\tx) = \mathbf{w}$. They are closed by equating the field derivatives of order $s+1$ and $s+2$ of the effective action to the derivative with respect to the constant field of the derivative of order $s$,
\begin{align}
& \Gamma_k^{(s+1)}[\mathbf{w}](\omega_1,\mathbf{p}_1;..;\omega_s,\mathbf{p}_s) \to \frac{\partial}{\partial w_{i_{s+1}}}\Gamma_k^{(s)}[\mathbf{w}](\omega_1,\mathbf{p}_1;..;\omega_{s-1},\mathbf{p}_{s-1}), \nonumber \\
& \Gamma_k^{(s+2)}[\mathbf{w}](\omega_1,\mathbf{p}_1;..;\omega_{s+1},\mathbf{p}_{s+1}) \to \frac{\partial^2}{\partial w_{i_{s+1}}\partial w_{i_{s+2}}} \Gamma_k^{(s)}[\mathbf{w}](\omega_1,\mathbf{p}_1;..;\omega_{s-1},\mathbf{p}_{s-1}).
\end{align}
Such an identification is exact when the additional momentum variables on the left-hand side are set to zero. Functions of at most $s$ different momentum variables are taken into account.
One extracts \RG flow equations for correlation functions of order smaller than $s$ which are fully momentum dependent while the higher order derivatives of $\Gamma_k\brv$ are taken into account by the fact that the lower order derivatives still depend on a constant field.
Such a procedure has been applied successfully for O(N) models at thermal equilibrium and $s=2$. See \cite{Blaizot:2011ra,benitez2012} for detailed discussions. To the best of our knowledge it has not been generalised to a non-equilibrium set-up yet.

\subsection{Local potential approximation}
\label{sec:Local_potential_approximation}

The local potential approximation \cite{Golner1986a,Wetterich:1991be,Morris:1994ie} is the outcome of lowest order truncation of the derivative expansion of the effective action with the additional approximation of setting the anomalous dimension of the field to zero. In situations where it is acceptable to neglect the full momentum dependence of the field derivatives of the effective action, these can be Taylor expanded in their momentum variables around $p = 0$. The flowing effective action can be expanded in powers of the velocity field,
\begin{align}
 \Gamma_k[\mathbf{v}] = \sum_{n = 1}^\infty \int_{\omega_1,\mathbf{p}_1;..;\omega_n,\mathbf{p}_n} \frac{v_{i_1}(\omega_1,\mathbf{p}_1) .. v_{i_n}(\omega_n,\mathbf{p}_n)}{n!} \, \Gamma_{k,i_1i_2..i_n}^{(n)}(\omega_1,\mathbf{p}_1;..;\omega_n,\mathbf{p}_n).
\end{align}
$\Gamma_{k,i_1i_2..i_i}^{(n)}(\omega_1,\mathbf{p}_1;..;\omega_n,\mathbf{p}_n)$ are the field derivatives of $\Gamma_k[\mathbf{v}]$ evaluated at zero field and multiplied by $(2\pi)^{n(d+1)}$. They contain, if computed exactly, all the information contained in $\Gamma_k[\mathbf{v}]$. In a system that is invariant under translations we have
\begin{align}
& \Gamma_{k,i_1i_2..i_n}^{(n)}(\omega_1,\mathbf{p}_1;..;\omega_n,\mathbf{p}_n) = (2\pi)^{d+1} \, \delta\left(\mathbf{p}_1+ .. + \mathbf{p}_n\right) \delta\left(\omega_1+ .. + \omega_n\right) \nonumber\\
& \qquad \times \Gamma_{k,i_1i_2..i_n}^{(n)}(\omega_1,\mathbf{p}_1;..;\omega_{n-1},\mathbf{p}_{n-1}).
\end{align}
The derivative expansion consists in approximating $\Gamma_{k,i_1i_2..i_n}^{(n)}(\omega_1,\mathbf{p}_1;..;\omega_{n-1},\mathbf{p}_{n-1})$ by their truncated Taylor expansion in frequency and momentum around $(\op)=0$.

In this section we show how to compute fixed point properties to lowest order in the derivative expansion. This work was performed as an introduction to the \FRG. The results of \cite{Bervillier:2007rc} were reproduced as an exercise. Extensive discussions on the derivative expansion as well as very precise calculations can be found in \cite{Golner1986a,Wetterich:1991be,Morris:1994ie,Bagnuls:2000ae,Berges:2000ew,Litim:2001dt}.
% See arXiv:1110.2665 [cond-mat.stat-mech], refs 4-6 and 18 for applications of the derivative expansion.

Here we focus on the equilibrium critical properties of a scalar field in $d$ dimensions $\phi(\mathbf{x})$. We leave aside for a while our space and time dependent velocity field because, as will be discussed in Section \ref{sec:approximation_scheme}, the derivative expansion is not the right way to go in the case of a hydrodynamic theory. We consider the following action,
\begin{align}
 S[\phi] = \frac{1}{2}\int_{\mathbf{p}} \left(p^2+m^2\right) \phi(\mathbf{p}) \phi(-\mathbf{p}) + \lambda \int_{\mathbf{x}} \phi(\mathbf{x})^4.
 \label{eq:phi4_action}
\end{align}
It corresponds to the $\phi^4$ model which is a well known toy model in quantum field theory \cite{
weinberg1995quantum,kleinert2001critical,maggiore2005modern}. It can be used to describe the thermodynamics of the bosonic scalar field $\phi(\mathbf{x})$ and has many physical realisations ranging from the Higgs mechanism to water. It contains a phase transition from an ordered $\langle \phi \rangle = 0$ to a disordered $\langle \phi \rangle \neq 0$ state which is in the same universality class as the Ising model \cite{Ising1925}. See for example \cite{kleinert2001critical,kadanoff2000statics,maggiore2005modern} and references therein. Because of the symmetry of $S[\phi]$ under spatial rotations, the lowest order in the derivative expansion is the second one. The corresponding ansatz for the flowing effective action is
\begin{align}
 \Gamma_k[\phi] = \int_{\mathbf{x}} \frac{1}{2} Z_k(\phi(\mathbf{x})) \, \boldsymbol{\nabla}\phi(\mathbf{x}) \cdot \boldsymbol{\nabla}\phi(\mathbf{x}) + U_k(\phi(\mathbf{x})).
 \label{eq:effective_pot_ansatz}
\end{align}
$Z_k(\phi)$ and $U_k(\phi)$ are two unknown functions of the field and the cut-off scale which can be determined by inserting the ansatz into \Eq{eq:wetterich2}. Before we continue we will make the further approximation that $Z_k(\phi) = 1$. This is justified because the derivative expansion works best in situations where the anomalous dimension of the scalar field, $\eta$ is small. We can neglect the changes of $Z_k(\phi)$ with the cut-off scale since they are of higher order in $\eta$ as compared to the changes of $U_k(\phi)$ \cite{Litim:2001dt}.

\subsubsection{Renormalisation group flow equation}
\label{eq:renormalisation_group_flow_equation}

We are now ready to write an \RG flow equation for $U_k(\phi)$. This is done by evaluating the flow equation \Eq{eq:wetterich2} at a constant field, $\phi(\mathbf{x}) = \phi$. \cred{Indeed}, the second field derivative of the ansatz, \Eq{eq:effective_pot_ansatz} reads
\begin{align}
\frac{\delta^2 \Gamma_k}{\delta \phi(\mathbf{p}) \delta \phi(\mathbf{q})}[\phi] = \frac{1}{(2\pi)^d} p^2 \delta(\mathbf{p}+\mathbf{q})+\frac{1}{(2\pi)^{2d}}\int_{\mathbf{x}} \frac{\partial^2 U_k}{\partial^2\phi}(\phi(\mathbf{x})) \, \text{e}^{-i\mathbf{x}\cdot\left(\mathbf{p}+\mathbf{q}\right)}.
\end{align}
Then choosing a field that does not depend on the spatial variable and adding the cut-off term provides,
\begin{align}
\Gamma_k^{(2)}[\phi](\mathbf{p},\mathbf{q}) + R_k(p) \delta(\mathbf{p}+\mathbf{q}) = \left[ p^2 + \frac{\partial^2 U_k}{\partial^2\phi}(\phi)+R_k(p)\right] \delta(\mathbf{p}+\mathbf{q}).
\label{eq:regulated_prop}
\end{align}
We use the Litim cut-off \cite{Litim:2000ci,Litim:2002cf},
\begin{align}
 R_k(p) = \left(k^2-p^2\right) \theta\left(k^2-p^2\right),
 \label{eq:optimal_cut_off}
\end{align}
which minimises the length of the paths that the couplings follow, and therefore the error they accumulate because of the truncation, as the cut-off scale is sent to zero \cite{Pawlowski:2005xe}. $\theta(x)$ is the usual step function. It is one if its argument is positive and zero otherwise. Within the local potential approximation this cut-off function also has the remarkable property of removing all the momentum dependence from \Eq{eq:regulated_prop} when $p\leq k$.

Finally the flow equation reads
\begin{align}
\int_{\mathbf{x}} k\partial_k U_k(\phi) = \delta(\mathbf{0}) \int_{\mathbf{p}} k^2 \theta(k^2-p^2) \frac{(2\pi)^d}{k^2+\partial_\phi^2U_k(\phi)}.
\end{align}
Three remarks are in order here. First, the $k^2 \theta(k^2-p^2)$ term in the numerator on the right hand side is a result of taking the cut-off derivative of $\left(k^2 - p^2\right)$ in $R_k(p)$. There is in principle an additional term coming from taking the derivative of the theta function in \Eq{eq:optimal_cut_off}. This term however vanishes since it is the product of a delta function and its argument. Secondly, the trace on the right-hand side of \Eq{eq:wetterich2} contains a momentum integration which is regulated by the $\theta(k^2-p^2)$ term. It leads to a simple volume factor,
\begin{align}
\int_{\mathbf{p}} \theta(k^2-p^2) \equiv \Omega_d \, \frac{k^{d}}{d} = \frac{k^d \pi^{d/2}}{(2\pi)^{d}\Gamma(d/2+1)},
\end{align}
since by virtue of the Litim cut-off its integrand does not depend on the loop momentum. Finally, there are two infinities that cancel each other. On the left hand side, evaluating $\Gamma_k[\phi]$ at a constant field produces an integrand which is independent of $\mathbf{x}$ in the last term of \Eq{eq:effective_pot_ansatz}. Then we are left with $U_k(\phi)$ multiplied by an integral over all space. This 
volume factor is cancelled by 
the $\delta(\mathbf{0})$ on the right hand side. The latter term arises because of invariance of our system under spatial translations. \cred{Indeed}, when such a symmetry holds, any function of multiple momentum variables is proportional to $\delta(\mathbf{p}_1+..+\mathbf{p}_n)$. In our case the trace on the right-hand side of \Eq{eq:wetterich2} does not have any momentum argument and leads to $\delta(\mathbf{0}) = (2\pi)^{-d} \int_{\mathbf{p}}$. Taking these two remarks into account we write
\begin{align}
k\partial_k U_k(\phi) = \frac{k^{d}\Omega_d/d}{k^2+\partial_\phi^2U_k(\phi)}.
\label{eq:flow_local_pot_not_rescaled}
\end{align}
We emphasize here that, local potential approximation has enabled us to reduce the functional differential equation \Eq{eq:wetterich2} to a partial differential equation for an arbitrary potential $U_k(\phi)$.

In order to look for \RG fixed points, the next step is to go to re-scaled variables,
\begin{align}
\hat{\mathbf{x}} = \mathbf{x} k, && \hat{\phi}(\hat{\mathbf{x}}) = k^{(2-d)/2} \phi(\hat{\mathbf{x}}/k), && u_k(\rho) = k^{-d} U_k(k^{(d-2)/2}\sqrt{2\rho}).
\label{eq:rescaled_variables}
\end{align}
We have introduced the variable $\rho = \hat{\phi}^2/2 = k^{2-d} \phi^2/2$ which will simplify the following calculations since our system is symmetric under the transformation $\phi(\mathbf{x}) \rightarrow -\phi(\mathbf{x})$, see \Eq{eq:phi4_action}. With $u_k'(\rho) = \text{d}u_k(\rho)/\text{d}\rho$, the \RG flow equation for $u_k(\rho)$ finally reads
\begin{align}
k \partial_k u_k(\rho) = -d u_k(\rho) - (2-d) u_k'(\rho) \rho + \frac{\Omega_d/d}{2u_k''(\rho)\rho+ u_k'(\rho)+1}.
\label{eq:flow_local_pot}
\end{align}

\subsubsection{Renormalisation group fixed point}
\label{sec:local_pot_fixed_point}

Now that we have written down the \RG flow equation for the re-scaled potential $u_k(\rho)$ we can look for fixed points. At an \RG fixed point the effective potential must satisfy $k \partial_k u(\rho) = 0$, \ie
\begin{align}
d \, u(\rho) = (d-2) u'(\rho) \rho + \frac{\Omega_d/d}{2u''(\rho)\rho+ u'(\rho)+1}.
\label{eq:effective_pot_fp}
\end{align}
In order to solve \Eq{eq:effective_pot_fp} we expand $u(\rho)$ in a power series around its minimum,
\begin{align}
& u(\rho) = \lambda_0 + \sum_{n = 2} \frac{\lambda_n}{n!} (\rho-\rho_0)^n,
&& \left. \frac{\text{d}u}{\text{d}\rho} \right|_{\rho = \rho_0} = 0.
\label{eq:def_potential}
\end{align}
We have introduced the set of couplings $\lambda_n$, the mean field expectation value $\rho_0$, as well as the minimum of the potential $\lambda_0$. All of these couplings can be computed from \Eq{eq:effective_pot_fp} by taking the appropriate number of derivatives with respect to $\rho$ and evaluating the equation at $\rho = \rho_0$. First let us note that $\lambda_0$ is completely determined by all the other couplings but does not enter their determination. When \Eq{eq:effective_pot_fp} is evaluated at $\rho = \rho_0$ one gets
\begin{align}
\lambda_0 = \frac{\Omega_d}{d^2} \frac{1}{2\lambda_2 \rho_0 +1}.
\end{align}
On the other hand, the right hand side of \Eq{eq:effective_pot_fp} does not depend on $\lambda_0$ since it contains only derivatives of $u(\rho)$. Then $\lambda_0$ drops out when \Eq{eq:effective_pot_fp} is differentiated.

Equations for $\lambda_n$ and $\rho_0$ are written recursively. We start by taking the first and second derivative of \Eq{eq:effective_pot_fp} and evaluating them at $\rho = \rho_0$,
\begin{align}
& 0 = (d-2) \lambda_2 \rho_0 - \frac{\Omega_d}{d} \frac{2\lambda_3 \rho_0 + 3 \lambda_2}{\left(2 \lambda_2 \rho_0 +1 \right)^2}, \label{eq_fixed_point_couplings_3} \\[0.2cm]
& d \lambda_2 =  (d-2) \left( \lambda_3 \rho_0 + 2 \lambda_2 \right) + \frac{\Omega_d}{d} \left[ 2 \frac{\left(2\lambda_3 \rho_0 + 3 \lambda_2\right)^2}{\left(2 \lambda_2 \rho_0 +1 \right)^3} - \frac{2\lambda_4 \rho_0 + 5 \lambda_3}{\left(2 \lambda_2 \rho_0 +1 \right)^2}\right].
\label{eq_fixed_point_couplings_2}
\end{align}
The equations for the other couplings take the form,
\begin{align}
\lambda_n = C_n(\rho_0,\lambda_2,..,\lambda_n) + A_n(\rho_0,\lambda_2,\lambda_3) \, \lambda_{n+1} + B(\rho_0,\lambda_2) \, \lambda_{n+2}, &&  n \geq 3,
\label{eq:local_pot_recursion}
\end{align}
with $C_n$, $A_n$ and $B_n$ being three functions of the couplings which are determined in Appendix \ref{sec:Local_potential_approximation_fixed_point_coefficients} and given explicitly in \Eqs{eq:A}, \eq{eq:B} and \eq{eq:C} with \eq{eq:X}.

Such a recursive form of the fixed point equation simplifies greatly the search for its parameters. \cred{Indeed}, the coordinates of a fixed point are given by $\rho_0$ and as many $\lambda_n$ as one is able to take into account. This means that if one were to search for solutions of \Eq{eq:effective_pot_fp} in a straightforward way, one would have to search for the correct values of the coupling within a very high dimensional space. This is numerically very demanding and become practically impossible as we take more couplings into account. The recursion relation given in \Eq{eq:local_pot_recursion} enables us to relate $\lambda_{n>2}$ to $\rho_0$ and $\lambda_2$ from the onset. Then there are only two unknown parameters left and it becomes easy to scan all their possible values and find the ones which provide a solution to \Eq{eq:effective_pot_fp}.

$\lambda_{n>2}$ can be related to $\rho_0$ and $\lambda_2$ in a recursive way. First, one can simply isolate $\lambda_3$ from \Eq{eq_fixed_point_couplings_3}. Then, this value of $\lambda_3$ can be inserted in \Eq{eq_fixed_point_couplings_2} and $\lambda_4$ is isolated in the same way. Next the recursion relation \eq{eq:local_pot_recursion} enters. Setting $n = 3$ we get
\begin{align}
\lambda_5 = \frac{\lambda_3 - C_3(\rho_0,\lambda_2,\lambda_3,\lambda_4) - A_3(\rho_0,\lambda_2,\lambda_3) \, \lambda_{4}}{B(\rho_0,\lambda_2)}.
\end{align}
$\lambda_5$ is now a function of $\rho_0$ and $\lambda_2$ only since $\lambda_3$ and $\lambda_4$ are already expressed in terms of these quantities through \Eqs{eq_fixed_point_couplings_3} and \eq{eq_fixed_point_couplings_2}. The next step is to insert $n=4$ in the \Eq{eq:local_pot_recursion} and isolate $\lambda_6$ is the same way as we did with $\lambda_5$. We can keep on going in this way and, in principle express all the $\lambda_n$ in terms of $\rho_0$ and $\lambda_2$. In practice, however we need to stop at some value of $n$. \Ie we need to further approximate $u(\rho)$ by truncating the sum in \Eq{eq:def_potential} at a given power $N$,
\begin{align}
& u(\rho) \cong \lambda_0 + \sum_{n = 2}^{N} \frac{\lambda_n}{n!} (\rho-\rho_0)^n.
\end{align}
Then we can go on applying the recursion relation all the way up to $n = N-2$. Notice that the recursion relation evaluated at $n$ provides and expression for $\lambda_{n+1}$ in terms of $\lambda_2$ and $\rho_0$. The as we increase $n$ we find that $\lambda_{n=3..N}$ are all expressed in terms of $\rho_0$ and $\lambda_2$ when we have reached $n = N-2$. There are still two equations, for $n = N-1$ and $n = N$, which have not been used yet. They fix the last two unknown parameters $\rho_0$ and $\lambda_2$. \cred{Indeed}, with a truncated sum the recursion relation for $n = N-1$ and $n = N$ becomes
\begin{align}
& \lambda_{N-1} = C_{N-1}(\rho_0,\lambda_2,...,\lambda_{N-1}) + A_{N-1}(\rho_0,\lambda_2,\lambda_3) \, \lambda_{N}, \nonumber \\
& \lambda_N = C_N(\rho_0,\lambda_2,...,\lambda_N).
\label{eq:recursion_close}
\end{align}
The additional terms of \Eq{eq:local_pot_recursion} which contain $\lambda_{N+1}$ and $\lambda_{N+2}$ have been set to zero. These two remaining equations are now a system of two equations for the two unknowns $\rho_0$ and $\lambda_2$ only.

\begin{table}[t]
\begin{center}
\begin{tabular}{l | rrr}
 & \multicolumn{1}{l}{Couplings} & \multicolumn{1}{l}{Re-scaled couplings} & \multicolumn{1}{l}{Couplings of \cite{Bervillier:2007rc}} \\
% \hline \hline
\midrule \midrule
$\lambda_1$ & $3.860821804198 \cdot 10^{-3}$ & $0.228628703223$ & \\
$\rho_0$ & $3.064754051849 \cdot 10^{-2}$ & $1.814874604703$ & $1.814 898 403 687$ \\
$\lambda_2$ &  $7.471529410168\hphantom{...10^{-0}}$ & $0.126170700576$ & $0.126 164 421 218$ \\
$\lambda_3$ & $1.045641876778 \cdot 10^{2\hphantom{-}}$ & $0.029818169310$ & $0.029 814 964 767$ \\
$\lambda_4$ & $1.301298120673 \cdot 10^{3\hphantom{-}}$ & $0.006266482270$ & $0.006 262 816 384$ \\
$\lambda_5$ &  $-3.349979877176 \cdot 10^{3\hphantom{-}}$ & $-0.000272419514$ & $-0.000 275 905 516$ \\
$\lambda_6$ & $-8.124166870475 \cdot 10^{5\hphantom{-}}$ & $-0.001115639433$ & \\
$\lambda_7$ &  $-1.327839825548 \cdot 10^{6\hphantom{-}}$ & $-0.000030792128$ & \\
$\lambda_8$ & $2.514985199681 \cdot 10^{9\hphantom{-}}$ & $0.000984868989$ &  \\
$\lambda_9$ & $8.068138754674 \cdot 10^{10\hphantom{.}}$ & $0.000533538044$ & \\
\end{tabular}
\end{center}
\caption{Coordinates of the fixed point. We compute the coefficients of the Taylor expansion of the fixed point potential within the local potential approximation up to order $O(\rho-\rho_0)^9$ and for $d=3$ (first column). We reproduce the right-hand side of Table 1 of Ref.~\cite{Bervillier:2007rc} where the same quantities were computed (third column) and re-scale our results according to \Eq{eq:rescaled_bervillier} (middle column).}
\label{tab:comp_bervillier_me}
\end{table}

All the steps that we just outlined can be performed numerically for a given set of values of $\rho_0$ and $\lambda_2$. Then it is possible to evaluate \Eqs{eq:recursion_close} and check if they are satisfied. If they are not, another set of values of $\rho_0$ and $\lambda_2$ can be chosen and the whole procedure can be repeated until a solution of the whole set of equations is found.
Such a procedure was performed with MATLAB up to order $N = 9$ and for $d=3$. The values of the obtained couplings are shown in the first column of Table (\ref{tab:comp_bervillier_me}). In order to compare with the results of \cite{Bervillier:2007rc}, which are shown in the third column of Table (\ref{tab:comp_bervillier_me}), the field and potential were re-scaled according to
\begin{align}
\hat{\rho} = \frac{d}{\Omega_d} \rho, && \hat{u}(\hat{\rho}) = \frac{d}{\Omega_d} \, u\left(\frac{\Omega_d}{d}\hat{\rho}\right).
\label{eq:rescaled_bervillier}
\end{align}
The re-scaled couplings are shown in the second column of Table (\ref{tab:comp_bervillier_me}). In the calculation of \cite{Bervillier:2007rc} the value $N=32$ was used. Considering the fact that we stopped our calculation at $N=9$, we find a good agreement in between both calculation.

\subsubsection{Critical exponents}

Here we compute the critical exponents of the system and compare our results to \cite{Bervillier:2007rc} as well.
We compute the \RG flow asymptotically close to the fixed point where the \RG flow equations can be linearised.
Then \RG acts on the different couplings simply by re-scaling them.
The re-scaling that is defined in \Eq{eq:rescaled_variables} is not restricted to \RG fixed points. It is made fully general by including a cut-off dependence in $u(\rho) = u_k(\rho)$.
We define
\begin{align}
\delta \rho_0 = \rho_0-\rho_0^*, && \delta \lambda_n = \lambda_n - \lambda_n^*, && \boldsymbol{\delta \lambda} = \left(\delta \lambda_0,\delta \rho_0,\delta \lambda_1,..,\delta \lambda_N\right),
\end{align}
with the starred quantities $\rho_0^*$ and $\lambda_n^*$ being the values the couplings assume at the fixed point which are given in Table (\ref{tab:comp_bervillier_me}).
The right-hand side of the \RG flow equation \eq{eq:flow_local_pot} is a non-linear function of $\delta \lambda_n$ and $\delta \rho$.
By definition it vanishes when $\delta \lambda_n=\delta \rho=0$.
If $\delta \lambda_n$ and $\delta \rho$ are small enough it can be expanded around $\delta \lambda_n=\delta \rho=0$ and all but its linear part can be neglected. Then the flow equation takes the form
\begin{align}
k\partial_k \boldsymbol{\delta \lambda} = \mathcal{M}^* \boldsymbol{\delta \lambda}.
\label{eq:linear_flow}
\end{align}
$\mathcal{M}^*$ is a matrix that depends on the coordinates of the fixed point, $\rho_0^*$ and $\lambda_n^*$. Then \Eq{eq:linear_flow} can be solved by diagonalising $\mathcal{M}^*$. \cred{Indeed}, each eigenvector of $\mathcal{M}^*$, $\mathbf{\text{v}}_n$, obeys
\begin{align}
k \partial_k \mathbf{\text{v}}_n = \mathcal{M}^* \mathbf{\text{v}}_n = \Lambda_n \mathbf{\text{v}}_n.
\end{align}
$\Lambda_n$ is the corresponding eigenvalue. We get $\mathbf{\text{v}}_n(k) = \mathbf{\text{v}}_n(k_{\Lambda}) (k/k_{\Lambda})^{\Lambda_n}$. Then if $\boldsymbol{\delta \lambda}$ is decomposed into the eigenvectors of $\mathcal{M}^*$, $\boldsymbol{\delta \lambda} = \sum_{n=0}^N \alpha_n \mathbf{\text{v}}_n$, we get
\begin{align}
\boldsymbol{\delta \lambda} = \sum_{n=0}^N \mathbf{\text{v}}_n(k_{\Lambda}) (k/k_{\Lambda})^{\Lambda_n}.
\end{align}
We see that only the eigenvectors with negative eigenvalues, $\Lambda_n < 0$ take the flow away from the fixed point. These are the relevant directions. If the \RG flow is initiated close to the basin of attraction (also called the critical surface for non trivial fixed points) of a fixed point, the couplings will flow towards it until they are pulled away again by their component in the relevant direction.
If the \RG flow is initiated exactly on the critical surface, the couplings flow towards their fixed point values and reach them asymptotically as $k \to 0$.

Since everything is re-scaled by the \RG flow the correlation length of the effective theory $\hat{\xi}$, must decrease when the cut-off scale is decreased, $\hat{\xi} = k\xi$.
Therefore, any theory that lies on the basin of attraction of a fixed point must have either a vanishing or an infinite correlation length. If this is not the case $\hat{\xi}$ will depend on $k$.
At a non-Gaussian fixed point we have $\xi = \infty$ since if this was not the case we would have $\xi = 0$ and the fixed point would be Gaussian.
We conclude that the correlation length diverges as we move close to the critical surface. In the case where there is only one relevant 
direction, the critical exponent $\nu$ is defined as
\begin{align}
\xi \sim \left(\delta \lambda_{\text{rel}}\right)^{-\nu}, && \text{as } \delta \lambda_{\text{rel}} \to 0.
\end{align}
$\delta \lambda_{\text{rel}} \equiv \delta \lambda_{\text{rel}}(k_\Lambda )$ is the projection on the relevant direction of the shortest vector joining the critical surface and the point at which the \RG flow is initiated (for $k = k_\Lambda$). It is the physical parameter which one can change to tune the system to its phase transition. This asymptotic form contains an exponent which can be extracted from the linearised \RG flow since we consider the situation where $\delta \lambda_{\text{rel}}$ is very small compared to all the other couplings.

Note that in this case the \RG fixed point is approximately attractive and the couplings flow towards their fixed point values and only $\delta \lambda_{\text{rel}}$ truly matters. We find that the critical properties of the system only depend on the relevant couplings. Since most couplings are irrelevant many different systems share the same fixed point and therefore have the same critical properties. This is universality.

Let us now see how this works in the context of the local potential approximation. The first step is to linearise the \RG flow equations. The beta functions are computed by taking derivatives of \Eq{eq:flow_local_pot} and evaluating them at $\rho = \rho_0$. They are given by,
\begin{align}
& k\partial_k \lambda_0 \equiv  \beta_{\lambda_0} = -d \lambda_0 + \frac{\Omega_d}{d}\frac{1}{2\lambda_2 \rho_0 +1}, \nonumber \\
& k\partial_k \rho_0 \equiv \beta_{\rho_0} = -(d-2) \rho_0 + \frac{\Omega_d}{d} \frac{\frac{2 \lambda_3 \rho_0}{\lambda_2} + 3}{\left(2\lambda_2 \rho_0 +1\right)^2}, \nonumber \\
& k\partial_k \lambda_n \equiv \beta_{\lambda_n} = \lambda_{n+1} \beta_{\rho_0} - d \lambda_n + (d-2) \left(\lambda_{n+1} \rho_0 + n \lambda_n \right) + \frac{\Omega_d}{d} X(n,1).
\label{eq_beta}
\end{align}
$X(n,1)$ is defined in Appendix \ref{sec:Local_potential_approximation_fixed_point_coefficients}, in \Eqs{eq:X}. Then the fixed point equations \eq{eq_fixed_point_couplings_3}, \eq{eq_fixed_point_couplings_2} and \eq{eq:local_pot_recursion} can be written as\footnote{We define $\boldsymbol{\beta} = \left(\beta_{\lambda_0},\beta_{\rho_0},\lambda_2,..\lambda_N\right)$.} $\boldsymbol{\beta}(\boldsymbol{\lambda}^*) = 0$. The linearised flow equations are obtained by inserting $\boldsymbol{\beta}(\lambda) \cong \boldsymbol{\beta}(\lambda^*) + \mathcal{M}^* \boldsymbol{\delta \lambda} = \mathcal{M}^* \boldsymbol{\delta \lambda}$. The matrix elements of $\mathcal{M}^*$ are given by
\begin{align}
\left(\mathcal{M}^*\right)_{i,j} = \left. \frac{\partial \beta_i}{\partial \lambda_j} \right|_{\boldsymbol{\lambda}^*}.
\label{eq:stab_matrix}
\end{align}
They are computed in a straightforward way\footnote{In principle we may  define $\delta u(\rho)$ through $u(\rho) = u^{*}(\rho) + \delta u(\rho)$ and $u^{*}(\rho) = \lambda_0^* + \sum_{s=2}\lambda_s^*/s! (\rho-\rho_0^*)^s$, insert it in \Eq{eq:flow_local_pot} and only keep the linear part in $\delta u(\rho)$. This yields a linear differential equation for $\delta u(\rho)$ from which a set of equations for $\boldsymbol{\delta \lambda}$ can be extracted by inserting \Eq{eq:def_potential}. Such an approach is not wrong, but is not optimal with respect to the convergence of the results as $N\to\infty$. It is better to expand the effective potential around its flowing minimum, $\rho_0^* + \delta \rho_0$.} from \Eqs{eq_beta}.
Using the values given in Table (\ref{tab:comp_bervillier_me}) for $\boldsymbol{\lambda}^*$, the matrix elements of $\mathcal{M}^*$ can be computed explicitly and $\mathcal{M}^*$ can be diagonalised. The eigenvalues that we find are listed in Table (\ref{tab:eig_bervillier_me}). Note that we find only one relevant direction. There is also one vanishing eigenvalue which originates from the inclusion of $\delta \lambda_0$ into $\boldsymbol{\delta \lambda}$. \cred{Indeed}, since this does not represent anything more than a shift in the total energy of the system, one can move in this direction without changing the physics of the fixed point. Finally, note that we find two sets of degenerate eigenvalues which are not in good agreement with the calculation of \cite{Bervillier:2007rc}. Theses degeneracies are lifted as $N$ is increased.

\begin{table}[t]
\begin{center}
\begin{tabular}{rr}
\multicolumn{1}{l}{Eigenvalues} & \multicolumn{1}{l}{Eigenvalues of \cite{Bervillier:2007rc}} \\
\midrule \midrule
$-1.539601891232131$ &  $-1.539499459806173$ \\
$0$ &  \\
$0.6557966422316$ & $0.6557459391933$ \\
$3.1827323507884$ & $3.1800065120592$ \\
$5.9126210155394$ & $5.9122306127477$ \\
$8.7702584075498$ & $8.796092825414\hphantom{0}$ \\
$13.9842288669691$ & $11.798087658337\hphantom{0}$ \\
$13.9842288669691$ & $14.896053175688\hphantom{0}$\\
$24.6148666337971$ & \\
$24.6148666337971$ & 
\end{tabular}
\end{center}
\caption{The eigenvalues of the stability matrix $\mathcal{M}^*$. We insert the couplings of Table (\ref{tab:comp_bervillier_me}) into the stability matrix defined in \Eq{eq:stab_matrix} and compute its eigenvalues. As before we go up to order $O(\rho-\rho_0)^9$ and use $d=3$ (first column). The second column are the values computed in Ref.~\cite{Bervillier:2007rc}.}
\label{tab:eig_bervillier_me}
\end{table}

The correlation length is related to the negative eigenvalue of Table (\ref{tab:eig_bervillier_me}) by computing the two-point correlation function,
\begin{align}
 \langle \phi(\mathbf{x}+\mathbf{r})\phi(\mathbf{x}) \rangle_{\text{c}} & = \int_{\mathbf{p},\mathbf{q}} \text{e}^{-i \left[\mathbf{p}\cdot\left(\mathbf{x}+\mathbf{r}\right) + \mathbf{q}\cdot\mathbf{x}\right]} \left(\frac{\delta^2 \Gamma_{k\to 0}[\phi = \phi_0]}{\delta \phi (\mathbf{p}) \delta \phi (\mathbf{q})}\right)^{-1} \nonumber \\[0.2cm]
& = \int_{\mathbf{p}} \text{e}^{-i \mathbf{p}\cdot \mathbf{r}} \frac{1}{p^2+\left. \frac{\partial^2 U_{k\to 0}}{\partial^2 \phi}\right|_{\phi = \phi_0}}.
\label{eq:correlation_length}
\end{align}
$\phi_0$ is defined through $\left.\delta \Gamma_{k\to 0}/\delta \phi(\mathbf{x})\right|_{\phi = \phi_0} = 0$. The definition of the couplings given in \Eq{eq:def_potential} can be used away from the fixed point. We write
\begin{align}
\langle \phi(\mathbf{x}+\mathbf{r})\phi(\mathbf{x}) \rangle_{c,k} &= \int_{\mathbf{p}}  \frac{\text{e}^{-i \mathbf{p}\cdot \mathbf{r}}}{p^2+ 2 \rho_0(k) \lambda_2(k) k^{2}} = \pi \frac{\text{e}^{-\sqrt{2 \rho_0(k) \lambda_2(k)} k x}}{\sqrt{2 \rho_0(k) \lambda_2(k)} k}.
\label{eq:flowing_correlation}
\end{align}
and identify the flowing correlation length $\xi_k = 1/(k\sqrt{2 \rho_0(k) \lambda_2(k)})$. If we are on the critical surface we have $\rho_0(k\to 0) = \rho^*$ and $\lambda_2(k\to 0) = \lambda_2^*$. Then the physical correlation length $\xi_{k\to\infty}$ is infinite since the divergence coming from the $k$ factor is not cancelled by the \RG flow of $\rho_0(k)$ and $\lambda_2(k)$. On the other hand, if the \RG flow is not initiated on the critical surface, two possibilities arise. If $\rho_0$ is large enough, it grows to infinity as $k\to 0$ in such a way that $\langle \phi \rangle_k = \sqrt{2 k \rho_0(k)}$ saturates to a constant field expectation value. On the other hand, if $\rho_0$ is smaller than its critical value it flows to a constant while $\lambda_2(k)$ flows to infinity and the asymptotic field expectation value vanishes. In both cases, the product $\sqrt{\lambda_1(k) \rho_0(k)}$ grows to infinity as $k$ goes to zero in such a way that the $k$ factor is cancelled out and we recover a finite 
correlation length. See \cite{Adams:1995cv} for an explicit calculation.

If we start close to the critical surface the \RG flow first approaches the fixed point and the irrelevant couplings assume their fixed point values. On the other hand, the relevant coupling scales according to $\delta \lambda_{\text{rel}} \sim k^{\Lambda_{\text{rel}}}$ and grows as $k$ decreases. The closer to the critical surface the \RG flow is initiated, the "longer" it stays close to the fixed point. $\xi_k$ then grows during all that "time" since its dominant $k$-dependence comes from its $1/k$ factor. In this way it can be made arbitrarily large if the initial distance to the critical surface is arbitrarily small.

\Eq{eq:flowing_correlation} only represents the physical correlation function in the limit $k\to 0$. We can not use it to relate the physical correlation length to the distance of the original parameters $\boldsymbol{\delta \lambda}(k_\Lambda)\equiv \boldsymbol{\delta \lambda}$ from the critical surface. We can however re-scale \Eq{eq:correlation_length} with the cut-off scale if we replace the original parameters by their $k$-dependent ones
\begin{align}
& \langle \phi(\mathbf{x}+\mathbf{r})\phi(\mathbf{x}) \rangle_{\text{c}}(\boldsymbol{\delta \lambda}) = k \langle \hat{\phi}(k\mathbf{x}+k\mathbf{r})\hat{\phi}(k\mathbf{x}) \rangle_{\text{c}}(\boldsymbol{\delta \lambda}(k)).
\end{align}
Such an equation is definitely true for $k = k_\Lambda$. However, the change of $\boldsymbol{\delta \lambda}(k)$ with the cut-off scale is precisely defined in such a way that it stays true for all positives values of $k$. Then if $\boldsymbol{\delta \lambda}$ is close to the critical surface we can decrease $k$ on the right hand side until all but the relevant coupling acquire their fixed point value. We have
\begin{align}
& \langle \phi(\mathbf{x}+\mathbf{r})\phi(\mathbf{x}) \rangle_{\text{c}}(\boldsymbol{\delta \lambda}) = k \langle \hat{\phi}(k\mathbf{x}+k\mathbf{r})\hat{\phi}(k\mathbf{x}) \rangle_{\text{c}}(\boldsymbol{\lambda}^*,\delta \lambda_{\text{rel}} \, (k/k_\Lambda)^{\Lambda_{\text{rel}}}).
\end{align}
In principle, such an equation only holds as long a $k$ is small enough that all but the relevant coupling are equal to their fixed point values but large enough that we are still close to the fixed point. However, the smaller $\delta \lambda_{\text{rel}}$ is, the bigger the range of values of $k$ where this is true. In the asymptotic limit $\delta \lambda_{\text{rel}} \to 0$, $k$ is completely free. We can then choose it such that $\delta \lambda_{rel} (k/k_\Lambda)^{\Lambda_{\text{rel}}} = 1$ and extract
\begin{align}
\xi \sim \delta \lambda_{\text{rel}}^{1/\Lambda_{\text{rel}}}, && \nu = -1/\Lambda_{\text{rel}} \cong 0.649518557813480.
\end{align}
This concludes the discussion of the local potential approximation. We have shown how to compute fixed point properties of the critical scalar theory with the local potential approximation. In the case $d=3$ we find a fixed point and compute its scaling exponents. The scaling exponent $\nu$ that was computed is consistent with the Ising universality class. The results of \cite{Bervillier:2007rc} are recovered.

\FloatBarrier
\subsection{Frequency and momentum dependent inverse propagator}
\label{sec:momentum_dependent_propagator}

We will see in Section (\ref{sec:Burgers_functionals}) that in the case of stochastic hydrodynamics the advective derivative of the velocity field leads to a theory with a non-linearity that contains a spatial derivative. In such a context, we do not expect the derivative expansion to produce good results since the momentum dependence of the action plays an essential role.
The derivative expansion was applied to study stationary scaling solutions of the stochastic \KPZ equation in \cite{Canet2005b}. As for the theory of hydrodynamics \KPZ equation contains a derivative in its vertex.
Although the correct scaling exponents were recovered for $d=1$, unphysical values were obtained for $d\geq 2$.

In this section we generalise a method first used in Ref.~\cite{Pawlowski:2003XX} in the context of thermal equilibrium Yang Mills theory. We write a set of \RG fixed point conditions for the second field derivative of the flowing effective action without making any restriction on its momentum or frequency dependence. We will assume that some truncation has been made on the higher order derivatives of the flowing effective action but not specify it here.

The main idea behind these fixed point equations is relatively straightforward.
We impose that the inverse propagator looses all of it's dependence on $k$ and assumes a scaling form once the cut-off is removed, \Eq{eq_gamma2_kto0}. This makes it possible to define the fixed point scaling exponents $\bar{\eta}_i$, the scaling function $g(a)$ and the rescaling factors $\bar{z}_i$.
Then we parametrise the inverse propagator in terms of re-scaled variables, \Eq{eq_param_1}.
In parallel we require that the re-scaled propagator and \RG flow equations, \Eq{eq_flow_1}, depend on the cut-off scale only through the re-scaled variables. This implies that the re-scaled theory does not depend on $k$. It provides a first constraint on the scaling exponents and makes it possible to define the fixed point coupling constant $h$.
Next we require that there be a qualitative difference in between the form of the propagator when $k\to 0$ and $k\to\infty$. This constrains the scaling function $g(a)$.
Finally the normalisation of $g(a)$ is used to constrain the second scaling exponent and the fixed point coupling $h$.
See Table (\ref{tab:momentum_dependent_fp}) where this is summarised.

We return to the velocity field of classical hydrodynamics $\mathbf{v}(\tx)$ and assume that the \RG flow of our theory has reached a fixed point. We define $\Gamma_{k,ij}^{(2)}(\omega,\mathbf{p})$ (with $i,j = 1,2,...,d$) such that
\begin{align}
 & \frac{\delta^2 \Gamma_k[v = 0]}{\delta v_i(\omega',\mathbf{p}') \delta v_j(\op)} 
 = \frac{\Gamma_{k,ij}^{(2)}(\omega,p)}{(2\pi)^{d+1}} \delta(\omega+\omega') \delta\left(\mathbf{p}+\mathbf{p}'\right).
\label{eq:Gamma2}
\end{align}
Because of Galilei and spatio-temporal translation invariance of the fixed points of the theory this particular form for the inverse propagator does not contain any restriction. We have assumed that the effective action has its extremum for $\mathbf{v}(\tx) = 0$ ($\delta \Gamma_k[\mathbf{v}=0]/\delta v_i(\tx) = 0$) which up to a Galilei boost is always true.

Since we are looking for fixed points of the \RG flow we write $\Gamma_{k,ij}^{(2)}(\omega,p)$ in terms of re-scaled dimensionless variables and extract its non universal parts in such a way that we recover a scaling form when $k \to 0$. In this section we set $d=1$ and drop the indices in order to make the notation simpler. The generalisation to the case $d\neq 1$ is straightforward. Since we consider a system which invariant under spatial rotations the different objects of this section do not depend on the sign of $p$.
In order to make the notation simpler we use the short-hand notation\footnote{Note that this is only a short hand notation for $d=1$. When the momentum is a vector $p$ is defined as its norm.} $p=\left|p\right|$. We write
\begin{align}
\Gamma_{k}^{(2)}(\omega,p) = k \bar{z}_1 \left(\frac{p}{k}\right)^{\bar{\eta}_1} & \left\{  g\left[\left(\frac{p}{k}\right)^{\bar{\eta}_2} \left(\frac{\omega \bar{z}_2}{k^2}\right)\right] + \delta Z\left(\frac{\omega \bar{z}_2}{k^2},\frac{p}{k}\right) \right\}.
\label{eq_param_1}
\end{align}
$\bar{\eta}_i$, $\bar{z}_i$ and the scaling function $g(a)$ are defined through the supplementary conditions
\begin{align}
& \lim_{\hat{p} \to \infty} \delta Z(a \hat{p}^{-\bar{\eta}_2},\hat{p}) = 0, && \text{for } a>0,
\label{eq_param_2}
\end{align}
\begin{align}
g(0)=1, & & \left.\frac{dg}{da}\right|_{a=0}=1,
\label{eq_param_3}
\end{align}
and the re-scaled variables
\begin{align}
\hat{p} = p/k, && \hat{\omega} = \omega \, \bar{z}_2 /k^2.
\end{align}
$\bar{z}_{i}$ are $k$-dependent dimensionless re-scaling factors which will be discussed shortly. \Eq{eq_param_2} must be valid for all dimensionless numbers, $a$ and is used to define $g(a)$, $\bar{\eta}_1$ and $\bar{\eta}_2$. It ensures that we recover a scaling solution in the limit $k\to 0$. \Eqs{eq_param_3} define the normalisation of the scaling function. They are arbitrary and have no physical interpretation but are necessary to properly define $\bar{z}_1$ and $\bar{z}_2$. They are the fixed point equivalent of the \RG conditions that we discussed at the very beginning of this Section.
The $k$ pre-factor is extracted in order to make the rest dimensionless. $\delta Z(\hat{\omega},\hat{p})$, $g(a)$, $\bar{\eta}_1$ and $\bar{\eta}_2$ are unknown physical parameters that can be determined by solving the \RG fixed point equations.

At a fixed point $\bar{z}_i$ take a particularly simple form. \cred{Indeed}, taking the limit $k\to 0$ and inserting \Eq{eq_param_2} we obtain
\begin{align}
\lim_{k \to 0} \Gamma_{k}^{(2)}(\omega,p) = k \bar{z}_1 \hat{p}^{\bar{\eta}_1}  g\left(\hat{p}^{\bar{\eta}_2} \hat{\omega}\right).
\label{eq_gamma2_kto0}
\end{align}
In this limit the dependence on the cut-off scale must vanish, see Section (\ref{sec:defgamma}). This is only possible if $\bar{z}_i$ are power laws of the cut-off scale
\begin{align}
\bar{z}_1 = \bar{z}_{10} \, k^{\bar{\eta}_1-1}, && \bar{z}_2 = \bar{z}_{20} \, k^{\bar{\eta}_2+2}.
\label{eq_scaling_zi}
\end{align}
The only free parameters that they contain are the pre-factors $\bar{z}_{i0}$.
As anticipated, we are left with a scaling form for $\Gamma_{k \to 0}^{(2)}(\omega,p)$. $\bar{\eta}_1$ and $\bar{\eta}_2$ are the scaling exponents of the fixed point and $g(a)$ is the scaling function. In particular we can identify the dynamical scaling exponent $z =-\bar{\eta}_2$.

The parametrisation made in \Eq{eq_param_1} corresponds to rewriting the inverse propagator in terms of dimensionless variables, $\hat{p}$ and $\hat{\omega}$ and asking for the re-scaled inverse propagator $\hat{\Gamma}_{k}^{(2)}(\hat{\omega},\hat{p}) \equiv \Gamma_{k}^{(2)}(\hat{\omega},\hat{p})/(k \bar{z}_1)$, to loose its explicit dependence on the cut-off scale. If we were not at a fixed point, $\bar{z}_i$ would have a non-trivial dependence on $k$ and $\delta Z(\hat{\omega},\hat{p})$ would have an additional explicit $k$-dependence.

We are now ready to insert \Eq{eq_param_1} in the flow equation Eq. (\ref{eq:wetterich2}). Since all the cut-off dependence drops out in the limit $k\to 0$, the pre-factor of \Eq{eq_param_1} as well as the scaling function contain no dependence on $k$ (see \Eq{eq_gamma2_kto0}). Then the scale derivative on the left-hand side of the flow equation only acts on the arguments of $\delta Z(\hat{\omega},\hat{p})$,
\begin{align}
 k\partial_k \Gamma_{k}^{(2)}(\omega,p) = k \bar{z}_1 \hat{p}^{\bar{\eta}_1} \left[ \bar{\eta}_2 \, \hat{\omega} \frac{\partial \delta Z}{\partial \hat{\omega}} - \hat{p} \frac{\partial \delta Z}{\partial \hat{p}} \right].
\label{eq:k_derivative_deltaZ}
\end{align}
In order to obtain closed \RG flow equations one introduces a truncation for the flowing effective action $\Gamma_k\brv$. Then \Eq{eq:wetterich2} can be projected onto the flow of the inverse propagator by inserting the truncated effective action in
\begin{align}
 & \frac{1}{2} \left. \frac{\delta^2}{\delta v(\omega,p) \delta v (\omega',{p}')} \text{Tr}\left[ k\partial_k R_k \left(\Gamma_k^{(2)}+R_k\right)^{-1} \right] \right|_{v = 0} \equiv \delta(p + p') \delta(\omega +\omega') \, I_k(\omega,p),
\label{eq:defI}
\end{align}
and require that $\Gamma_{k}^{(2)}(\omega,p)$ obey,
\begin{align}
 k\partial_k \Gamma_{k}^{(2)}(\omega,p) = I_k(\omega,p).
\end{align}
Since we do not restrict the two-point function of $\Gamma_k\brv$ the truncated effective action must contain approximated higher order vertexes.
Then additional equations must be extracted for each unknown term that is included in the truncation and the vertexes of $\Gamma_k\brv$ must be parametrised in the same way as its inverse propagator. \Ie they must depend on the same re-scaled variables, reduce to a scaling form in the limit $k\to 0$ and loose all of their cut-off dependence when re-scaled with the appropriate power of $k$ and $\bar{z}$ pre-factor. Note that we must introduce a new pre-factor for each vertex. Then when this is inserted into \Eq{eq:defI}, we find that $I_k(\omega,p)$ is automatically parametrised in the same way as its components,
\begin{align}
& I_k(\omega,p) = k \bar{z}_3 \, \hat{I}(\hat{\omega},\hat{p}).
\label{eq:def_z3}
\end{align}
We have introduced a third $k$-dependent re-scaling factor $\bar{z}_3$ which is a combination of the re-scaling pre-factors of the inverse propagator $\bar{z}_i$ and of the vertexes of $\Gamma_k[\mathbf{v}]$, $\bar{z}$. Its form can be inferred from the particular flow equation that we use and depends on the truncation of $\Gamma_k\brv$. See Section (\ref{sec:fixed_point_equations}) and \Eq{eq:flow_integrals} with \Eqs{eq:bar_to_notbar} for an example. Note that with the re-scaled truncation inserted $\hat{I}(\hat{\omega},\hat{p})$ contains no explicit dependence on $k$ and that all the re-scaling prefactors $\bar{z}_i$ as well as $\bar{z}$ have been extracted and put together in $\bar{z}_3$. $\hat{I}(\hat{\omega},\hat{p})$ only contains, $g(a)$, $\bar{\eta}_i$ and $\delta Z(\hat{\omega},\hat{p})$.

Putting everything together we write the flow equation as
\begin{align}
 \frac{\text{d}}{\text{d}\hat{p}} \delta Z(a \hat{p}^{-\bar{\eta}_2},\hat{p}) = - \frac{\bar{z}_3}{\bar{z}_1} \frac{\hat{I}(a \hat{p}^{-\bar{\eta}_2},\hat{p})}{\hat{p}^{\bar{\eta}_1+1}}.
\label{eq_flow_1}
\end{align}
The full frequency dependence of $\delta Z(\hat{\omega},\hat{p})$ is taken into account by the fact that \Eq{eq_flow_1} is valid for all values of the dimensionless number $a$.
Eq. (\ref{eq_param_2}) is used as an initial condition which makes it possible to determine $\delta Z(a \hat{p}^{-\bar{\eta}_2},\hat{p})$ if $\bar{\eta}_i$, $\bar{z}_3/\bar{z}_1$ and $g(a)$ are known. The scaling exponents $\bar{\eta}_i$, the scaling functions $g(a)$ and the pre-factors $z_{i0}$ are still undetermined and must be constrained in some other way.

First, one of $\bar{\eta}_i$ can be related to the other by noting that the only term that explicitly depends on $k$ in \Eq{eq_flow_1} is $\bar{z}_3/\bar{z}_1$. It is then apparent that $\bar{z}_3/\bar{z}_1$ does not depend on $k$.
Given \Eqs{eq_scaling_zi} and the fact that there will be a similar equation for the other components of $\Gamma_k[\mathbf{v}]$, we can write that $\bar{z}_3 = \bar{z}_{30} \, k^{\bar{\eta}_3}$. The fact that $\bar{z}_3$ is a combination of $\bar{z}_i$ and $\bar{z}$ implies that $\bar{\eta}_3$ will be a (typically affine) function of $\bar{\eta}_i$ and of the corresponding exponents of the vertexes.
$\bar{z}_1$ and $\bar{z}_3$ must scale with $k$ in the same way in order for all of the explicit $k$-dependence to drop out. Inserting the first of \Eqs{eq_scaling_zi} we get
\begin{align}
 \bar{\eta}_3 = \bar{\eta}_1 - 1,
\label{eq:consistency_0}
\end{align}
which provides a scaling relation in between the different scaling exponents included in the truncation of $\Gamma_k[\mathbf{v}]$ and reduces the number of free parameters by one. Note that an additional re-scaling factor and scaling exponent will be introduced for each vertex that is taken into account in $\Gamma_k[\mathbf{v}]$. However and additional scaling relation will also be generated for each additional flow equation that is used. Hence whatever the truncation we use for $\Gamma_k[\mathbf{v}]$ there will always be only one independent scaling exponent.

Taking into account \Eq{eq:consistency_0} we find that the right-hand side of \eq{eq_flow_1} has a dimensionless $k$-independent pre-factor,
\begin{align}
 h \equiv \frac{\bar{z}_3}{\bar{z}_1} = \frac{\bar{z}_{30}}{\bar{z}_{10}}.
\label{eq:fp_coupling}
\end{align}
$h$ can be interpreted as a fixed point coupling constant. \cred{Indeed}, if $h=0$ we trivially find that $\delta Z(a \hat{p}^{-\bar{\eta}_2},\hat{p})=0$ and the inverse propagator has a scaling form from the onset without being affected by the non-linearity of the theory. Moreover, note that if we do not sit exactly on the \RG fixed point but flow towards it, we will find that $h$ flows with $k$. The coupling defined in \Eq{eq:fp_coupling} is the asymptotic value that the \RG flow reaches when $k\to 0$.

The scaling function, $g(a)$, is determined by a supplementary condition on the large-$k$ limit of $\delta Z(\hat{\omega},\hat{p})$. In this limit the cut-off function $R_k(p)$ is huge. We have an effective theory with a very large mass term. This is very different from the physical theory (when $k\rightarrow 0$) and we can expect the scaling to be different from that of \Eq{eq_gamma2_kto0}. Then the asymptotic form of $\delta Z(a \hat{p}^{-\bar{\eta}_2},\hat{p})$ must be
\begin{align}
 \delta Z(a \hat{p}^{-\bar{\eta}_2},\hat{p}) = - g\left(a\right) + \frac{f(a \hat{p}^{-\bar{\eta}_2},\hat{p})}{\hat{p}^{\bar{\eta}_1}}, && \hat{p} \ll 1,
\label{eq_deltaZ_0}
\end{align}
in order for the physical (small-$k$) scaling to be undone when $k$ is very large. When this is inserted in \Eq{eq_param_1} we find that $g(a)$ is cancelled out ant that the pre-factor of the second term on the right-hand side of \Eq{eq_deltaZ_0}, $\hat{p}^{-\bar{\eta}_1}$, removes the $\bar{\eta}_1$-scaling and replaces it with something else given by $f(\hat{\omega},\hat{p})$,
\begin{align}
 \lim_{k\to\infty} \Gamma_{k}^{(2)}(\omega,p) = k \bar{z}_1 f(\hat{\omega},\hat{p}).
\end{align}

In \Eq{eq_deltaZ_0} $f(a \hat{p}^{-\bar{\eta}_2},\hat{p})\,\hat{p}^{-\bar{\eta}_1}$ contains all the $\hat{p}$-dependence of the asymptotic form of $\delta Z(a \hat{p}^{-\bar{\eta}_2},\hat{p})$.
Simply defining $f(a \hat{p}^{-\bar{\eta}_2},\hat{p})\,\hat{p}^{-\bar{\eta}_1}$ as the asymptotic form of $\delta Z(a \hat{p}^{-\bar{\eta}_2},\hat{p})$ minus its constant (in $\hat{p}$) part is ambiguous. One can however, see that $f(a \hat{p}^{-\bar{\eta}_2},\hat{p})$ is well defined by looking at \Eq{eq_flow_1}. \cred{Indeed}, using the boundary condition given by \Eq{eq_param_2} we can write \Eq{eq_flow_1} as\footnote{Here we assume that this integral is convergent. See Section (\ref{sec:uv_divergent_fixed_points}) where the general case is discussed.}
\begin{align}
 \delta Z(a \hat{p}^{-\bar{\eta}_2},\hat{p}) = h \int_{\hat{p}}^\infty \text{d}y \, \frac{\hat{I}(a y^{-\bar{\eta}_2},y)}{y^{\bar{\eta}_1+1}}.
\label{eq:deltaZ_integrated_general}
\end{align}
We see that the $\hat{p}^{-\bar{\eta}_1}$ pre-factor is already present in the integrand. In the limit $y \ll 1$, $\hat{I}(a y^{-\bar{\eta}_2},y)$ will be simple because the cut-off function is dominating the flow. Let us assume that we have the form
\begin{align}
 \hat{I}(a y^{-\bar{\eta}_2},y) = y^\alpha \left(a_0 + a_1 y + ... \, \right), && y \ll 1.
\label{eq:asymptotic_flow}
\end{align}
The exponent $\alpha$ and the coefficients $a_i$ depend on $a$, $\eta_i$, $g(a)$ and $h$. Because in this limit everything is dominated by the cut-off we expect this dependence to be relatively simple. In the case of Burgers stochastic hydrodynamics we were able to compute $\alpha$ and its relation to $\bar{\eta}_i$ exactly. See Section (\ref{sec:frg_calculation}) and \Eqs{eq:alphas} and \eq{eq:eta12alpha1}.
Since we are talking about the limit $\hat{p}\to 0$ we can introduce $\hat{p}<\epsilon \ll 1$ and brake the integral of \Eq{eq:deltaZ_integrated_general} in two parts: one integral from $\hat{p}$ to $\epsilon$ and one from $\epsilon$ to $\infty$. The we can insert \Eq{eq:asymptotic_flow} into the first part,
\begin{align}
 \delta Z(a \hat{p}^{-\bar{\eta}_2},\hat{p}) & = h \left[\int_{\hat{p}}^{\epsilon} \text{d}y\, \frac{\hat{I}(a y^{-\bar{\eta}_2},y)}{y^{\bar{\eta}_1+1}}+ \int_\epsilon^\infty\text{d}y\, \frac{\hat{I}(a y^{-\bar{\eta}_2},y)}{y^{\bar{\eta}_1+1}}\right] \nonumber \\
& = h \left[ \int_\epsilon^{\infty} \text{d}y \, \frac{\hat{I}(a y^{-\bar{\eta}_2},y)}{y^{\bar{\eta}_1+1}} + \epsilon^{\alpha-\bar{\eta}_1} \left(\frac{a_0}{\alpha-\bar{\eta}_1} + \epsilon \frac{a_1}{\alpha-\bar{\eta}_1+1} + ... \right) \right. \nonumber \\
& \qquad \left. {}- \hat{p}^{\alpha-\bar{\eta}_1} \left(\frac{a_0}{\alpha-\bar{\eta}_1} + \hat{p} \frac{a_1}{\alpha-\bar{\eta}_1+1} + ...\right) \right],
\end{align}
with $\epsilon \ll 1$ small enough that \Eq{eq:asymptotic_flow} applies for $\hat{p} < \epsilon$. We can now make the identifications
\begin{align}
& g(a) = -h \left[ \int_\epsilon^{\infty} \text{d}y \, \frac{\hat{I}(a y^{-\bar{\eta}_2},y)}{y^{\bar{\eta}_1+1}} + \epsilon^{\alpha-\bar{\eta}_1} \left(\frac{a_0}{\alpha-\bar{\eta}_1} + \epsilon \frac{a_1}{\alpha-\bar{\eta}_1+1} + ... \right) \right], \label{eq:eq_for_g} \\
& f(a \hat{p}^{-\bar{\eta}_2},\hat{p}) = - h \, \hat{p}^{\alpha} \left(\frac{a_0}{\alpha-\bar{\eta}_1} + \hat{p} \frac{a_1}{\alpha-\bar{\eta}_1+1} + ...\right).
\label{eq:g_and_f}
\end{align}
We define $f(a \hat{p}^{-\bar{\eta}_2},\hat{p})\,\hat{p}^{-\bar{\eta}_1}$ as the part of $\delta Z(a \hat{p}^{-\bar{\eta}_2},\hat{p})$ from which a factor of $\hat{p}^{-\bar{\eta}_1}$ can be extracted. What is left will not depend on $\hat{p}$ by construction. The scaling function is then constrained by enforcing \Eq{eq:eq_for_g}. Note that in practice the sum on the right-hand side of \Eq{eq:asymptotic_flow} will be truncated at a finite order in $y$. Then it is only an approximation and $\epsilon$ should be taken as small as possible in order to minimise the error. We should therefore take the limit $\epsilon \to 0$ in \Eq{eq:eq_for_g}. This limit will only be finite if the order at which we truncate \Eq{eq:asymptotic_flow} is large enough for all the terms that grow as $\epsilon$ decreases to be taken into account. We need to consider all terms of order smaller than $y^m$ with $m > \bar{\eta}_1-\alpha$.

With this definition of $f(\hat{\omega},\hat{p})$ we can now use \Eq{eq_deltaZ_0} to constrain $g(a)$. We are then left with one of the exponents $\bar{\eta}_i$, one of the pre-factors $z_{i0}$ and the coupling $h$ to determine. We use the normalisation constraints on $g(a)$ given in \Eq{eq_param_3} to do this. Note that $z_{i0}$ only enter \Eqs{eq_flow_1} and \eq{eq_deltaZ_0} through $h$. \Eqs{eq_param_3} therefore constrain the values of the second exponent and of the coupling. The last pre-factor is left undetermined. This is a desirable property since the value of the pre-factor of $\Gamma_k^{(2)}(\omega,p)$ depends on the particular system of units that we use and is fixed by comparing the results with an experiment. It is not a property of an \RG fixed point.

Let us now summarise the method to extract the \RG fixed point properties. See Table (\ref{tab:momentum_dependent_fp}). First we define the different properties of the fixed point $\bar{\eta}_i$, $z_{i0}$, $g(a)$ and $\delta Z(\hat{\omega},\hat{p})$ in \Eqs{eq_param_1}, \eq{eq_param_2}, \eq{eq_param_3} and \eq{eq_scaling_zi}. Next we use the \RG flow equation \Eq{eq:wetterich2} and the initial condition given by \Eq{eq_param_2} to constrain $\delta Z(\hat{\omega},\hat{p})$. This naturally leads to a constraint on one of the two scaling exponents $\bar{\eta}_i$ and to the definition of the fixed point coupling $h$. This constraint is a necessary condition for the fixed point version of the flow equation \eq{eq_flow_1}, to be truly independent of the cut-off scale. Then we constrain $g(a)$ by requiring that the solution of the \RG flow equations be different in the two limits $\hat{p} \ll 1$ and $\hat{p} \gg 1$. Finally we use the normalisation constraints given by \Eqs{eq_param_3} to constrain the values of the remaining scaling exponent and the coupling.

\begin{table}[t]
\begin{center}
\begin{tabular}{p{3.5cm}p{2.3cm}p{2.7cm}c}
Steps & Equations & Free parameters & Fixed parameters \\
\midrule \midrule
Define the different parameters. & \eq{eq_param_1}, \eq{eq_param_2}, \eq{eq_param_3}, \eq{eq_scaling_zi} & $\bar{\eta}_i$, $z_{i0}$, $g(a)$, $\delta Z(\hat{\omega},\hat{p})$ & \\ \midrule
Use the \RG flow equation and the initial condition. & \eq{eq:wetterich2}, \eq{eq_param_2} & $\bar{\eta}_i$, $z_{i0}$, $g(a)$ & $\delta Z(\hat{\omega},\hat{p})$ \\ \midrule
Remove $k$-dependence from the \RG flow equation. & \eq{eq_flow_1}, \eq{eq:consistency_0} & $\bar{\eta}_1$, $z_{10}$, $h$, $g(a)$ & $\bar{\eta}_2$ \\ \midrule
Enforce that the limits $k\to 0$ and $k\to\infty$ be different & \eq{eq_deltaZ_0}, \eq{eq:eq_for_g} & $\bar{\eta}_1$, $z_{10}$, $h$ & $g(a)$ \\ \midrule
Enforce the normalisation constraints of $g(a)$. & \eq{eq_param_3} & $z_{10}$ & $\bar{\eta}_1$, $h$
\end{tabular}
\end{center}
\caption{The different steps that lead to extracting all of the fixed point properties of $\Gamma_k^{(2)}(\omega,p)$. The steps are to be followed from top to bottom. They are outlined in the first column and the relevant equations are given in the second column. The third column lists the parameters that are undetermined at the corresponding step. The fourth column gives the parameters that are fixed by the corresponding step. They can be completely determined if the parameters of the third column are given. We have made the choice of fixing $\bar{\eta}_2$ and expressing $\bar{z}_{20}$ in terms of $h$ at the third step. This is of course arbitrary either one of the $\bar{\eta}_i$, $\bar{z}_{i0}$ could have been chosen.
Note that $z_{10}$ is never fixed since it depends on the particular system of units that we use.}
\label{tab:momentum_dependent_fp}
\end{table}

It is surprising that the two most interesting properties of the fixed point, namely $\bar{\eta}_i$ and $h$ are fixed by the arbitrary normalisation given in \Eqs{eq_param_3}. \cred{Indeed}, it looks like we could choose a different normalisation and get different values of $\bar{\eta}_i$ and $h$. This is of course not the case. A change of the normalisation (as compared to \Eqs{eq_param_3}) of $g(a)$ can be reabsorbed into a redefinition of $\delta Z(\hat{\omega},\hat{p})$, $\hat{\omega}$ and $\hat{p}$.  Then we recover the same set of equations as before.
For example, if we choose $g(1) = b$, we can simply re-define 
\begin{align}
& \hat{p}' = \hat{p} \, b^{1/\bar{\eta}_1}, && \hat{\omega}' = \hat{\omega} \, b^{\bar{\eta}_2/\bar{\eta}_1}, \nonumber \\
& \delta Z'(\hat{\omega}',\hat{p}') = \delta Z(\hat{\omega},\hat{p}) /b, && g'(a) = g(a)/b.
\end{align}
Then we are back to where we started without having touched $\bar{\eta}_i$ or $h$.

Note that in the case of a Gaussian fixed point the theory is quadratic in $\mathbf{v}(\tx)$. Then we simply have $\delta Z(\hat{\omega},\hat{p}) = 0$ and $h=0$ and there is no need to impose the relation given by \Eq{eq:consistency_0} in between $\bar{\eta}_i$ since \Eq{eq_flow_1} is trivially independent of $k$ in this case. If we have a Gaussian theory the \RG flow of the effective average action is trivial. We simply get $k\partial_k\Gamma_k^{(2)}(\omega,p)=0$ and $\Gamma_k^{(n)} = 0$ for $n>2$. Then all one has to do to find a fixed point is to choose a scaling form for $\Gamma_k^{(2)}(\omega,p)$. The exponents and the scaling function are not constrained by the fixed point equations.

\subsubsection{Ultraviolet divergent fixed points}
\label{sec:uv_divergent_fixed_points}

In the previous section we have assumed that the integral on the right-hand side of \Eq{eq:deltaZ_integrated_general} is convergent. This is a self-consistent assumption since if it is not true we can not have \Eq{eq_param_2},
\begin{align}
& \lim_{\hat{p} \to \infty} \delta Z(a \hat{p}^{-\bar{\eta}_2},\hat{p}) = 0.
\label{eq_param_2_bis}
\end{align}
We discuss here the fact that this is only a formal restriction on the form of $\Gamma_k[\mathbf{v}]$. \cred{Indeed}, it is physically perfectly acceptable that
\begin{align}
& \lim_{\hat{p} \to \infty} \delta Z(a \hat{p}^{-\bar{\eta}_2},\hat{p}) = \infty ,
\end{align}
since no fixed point is perfectly realised in nature. There is always a cut-off to make everything finite.
Such a divergence happens when the exact fixed point theory is \UV divergent and it not defined in the limit $k\to \infty$. See Section (\ref{sec:cascades}) where this is discussed in the context of stochastic hydrodynamics and cascades of energy. The \FRG program (start with $\Gamma_{k\to\infty}[\mathbf{v}] = S[\mathbf{v}]$ and progressively lower the cut-off scale to $\Gamma_{k \to 0}[\mathbf{v}]=\Gamma[\mathbf{v}]$) can still be carried out but the initial condition must be taken at an arbitrary but not infinite cut-off scale $\Gamma_{k=\Lambda}[\mathbf{v}] = S[\mathbf{v}]$. All momenta larger than $\Lambda$ are not physical.
In this case we can still find a theory with correlation functions that are asymptotically scale invariant but there will always be a trace of the initial scale $\Lambda$ which restricts the scaling range. The actual value of $\Lambda$ depends on the particular details of the \UV theory. It is not universal. Since we are investigating the scaling regime we will take $\Lambda$ as large as possible.

We refer to such fixed points as \UV divergent and to fixed points where \Eq{eq_param_2_bis} is realised as \UV convergent.
Note that the case $\lim_{\hat{p} \to \infty} \delta Z(a \hat{p}^{-\bar{\eta}_2},\hat{p}) = {const} \neq 0$ is \UV convergent. \cred{Indeed}, one can simply redefine $\bar{\delta Z}(\hat{\omega},\hat{p}) = \delta Z(\hat{\omega},\hat{p}) - {const}$ and $\bar{g}(a) = g(a) + {const}$ and \Eq{eq_param_2_bis} is true for the new variables.

In the case of \UV divergent fixed points \Eq{eq:deltaZ_integrated_general} has to be supplemented with an \UV cut-off which modifies the fixed point conditions. This can be done in two different ways. One can either modify the flow equation \Eq{eq_flow_1} in such a way that the \RG flow does not build up infinities in the \UV.
Or one can abandon the strict fixed point conditions and only require that we are close to an \RG fixed point. We will see that these two options are actually equivalent. In both cases one must keep a finite (although large) \UV cut-off to compute the universal properties of the system, $\bar{\eta}_i$ and $g(a)$. Then one must check that the latter are independent of the former and increase the value of the \UV cut-off if they are not.

In the first case we replace \Eq{eq_flow_1} by
\begin{align}
 \frac{\text{d}}{\text{d}\hat{p}} \delta Z(a \hat{p}^{-\bar{\eta}_2},\hat{p}) = - h \frac{\hat{I}(a \hat{p}^{-\bar{\eta}_2},\hat{p})}{\hat{p}^{\bar{\eta}_1+1}} \, \theta(B-\hp).
\label{eq:modif_flow}
\end{align}
$\theta(x)$ is the usual step function.
This means that the right-hand side of the flow equation \Eq{eq:wetterich2} is truncated for momenta larger than $\Lambda \equiv kB$. 
Then \Eq{eq_param_2_bis} is equivalent to
\begin{align}
& \lim_{\hat{p} \to B} \delta Z(a \hat{p}^{-\bar{\eta}_2},\hat{p}) = 0,
\label{eq:modif_limit}
\end{align}
since $\delta Z(a \hat{p}^{-\bar{\eta}_2},\hat{p})$ is simply constant for $\hat{p}>B$. For any finite value of $B$ \Eq{eq:modif_limit} can always be satisfied.
The original \RG flow equations are then recovered by taking $B \gg 1$.
The fact that we have an \UV divergent fixed point prevents us from taking the limit $B\to\infty$. We must instead choose it to be large enough for the momentum scale that we probe to be correctly reproduced by the modified \RG flow equation. \cred{Indeed}, properties of $\Gamma_k[\mathbf{v}]$ which depend on momentum $p$ and satisfy $k \ll p \ll k B = \Lambda$ will not be sensitive to what is happening at the scale $\Lambda$. This is apparent from the diagrammatic expression of the flow equation of the inverse propagator,
\begin{align}
\begin{aligned}
\begin{tikzpicture}[scale=1]
\draw[color=black, line width = 1.5pt] (1-0.4, 0) circle (1);
\draw[color=black, line width = 1.5pt] (4.7, 0) circle (1);
\path[draw] (-0.6*0.8/0.6-0.4,0) -- (0-0.4,0);
\path[draw] (2-0.4,0) -- (2+0.6*0.8/0.6-0.4,0);
\path[draw] (4.7-0.4243*0.8/0.6,-1-0.4243*0.8/0.6) -- (4.7,-1) -- (4.7+0.4243*0.8/0.6,-1-0.4243*0.8/0.6);
\node[draw,circle, fill=gray!30] (a) at (0-0.4, 0) {3};
\node[draw,circle, fill=gray!30] (b) at (2-0.4, 0) {3};
\node[draw,circle, fill=black] (c) at (1-0.4, 1) {};
\node[draw,circle, fill=gray!30] (d) at (4.7, -1) {4};
\node[draw,circle, fill=black] (e) at (4.7, 1) {};
\node (f) at (1.8+0.8/0.6, 0) {$ \displaystyle-\frac{1}{2}$};
\node (g) at (-1-1.2/0.6, 0) {$ \displaystyle  k\partial_k \Gamma_k^{(2)}(\omega,p)=$};
\end{tikzpicture} \end{aligned} \enspace .
\label{eq:diagramm1}
\end{align}
The thick lines denote $G_{k}=(\Gamma_k^{(2)}+R_k)^{-1}$, the thin lines external momenta and frequencies, the black dots insertions of the derivative $k\partial_{k}R_{k}$ of the regulator and the circled numbers the corresponding derivatives of $\Gamma_k\brv$. The $k\partial_{k}R_{k}$ insertions limit the loop momentum to be close to the cut-off scale. Then all the momenta that enter \Eq{eq:diagramm1} are not far from $k$, $p$ or $p\pm k$. We see that when $k \ll p \ll \Lambda$, $\Lambda$ does not enter the flow equations.

If instead we do not wish to modify the flow equations but prefer to move slightly away from the \RG fixed point we write
\begin{align}
 \delta Z(a \hat{p}^{-\bar{\eta}_2},\hat{p}) = h \int_{\hat{p}}^{B(a;k)} \text{d}y \, \frac{\hat{I}(a y^{-\bar{\eta}_2},y)}{y^{\bar{\eta}_1+1}},
\label{eq:initial_flow_cut-off}
\end{align}
instead of \Eq{eq:deltaZ_integrated_general}. $B(a;k)$ is a function of the cut-off scale that marks the limit in between close-to and far-from-the fixed point in the $(\hat{\omega},\hat{p},k)$ space. It will be defined more precisely below.
The idea here is to make the fixed point conditions weaker. We only assume that the \RG flow takes us close to the fixed point but does not actually stop on it.
This is a more natural situation than simply truncating the flow equations but leads an explicit $k$ dependence in $\delta Z(\hat{\omega},\hat{p})$. This means that we are not at an \RG fixed point and the fixed point conditions do not strictly apply any more.

Let us assume that the \RG flow passes close to one of its fixed points.
We define $\Lambda_1$ and $\Lambda_2$ as the scales in between which $\Gamma_k[\mathbf{v}]$ approximately behaves as a fixed point theory. For $\Lambda_1 < k < \Lambda_2$ the re-scaled parameters of the flowing effective are almost constant. Then a scaling range emerges. \cred{Indeed}, any correlation function takes the form
\begin{align}
 O(\omega,p) = k^{d_O+\eta_O} O(\hat{\omega},\hat{p};\mathbf{g}(k)), && \text{for all }k.
\label{eq:observable_type}
\end{align}
$\mathbf{g}(k)$ are the parameters that characterise $\hat{\Gamma}_k[\hat{\mathbf{v}}]$ and $d_O$ and $\eta_O$ are the canonical and anomalous dimensions of $O(\omega,p)$ respectively. $\mathbf{g}(k)$ change with the cut-off scale precisely in the right way for \Eq{eq:observable_type} to be true. Close to an \RG fixed point we have
\begin{align}
 \mathbf{g}(k) \cong \mathbf{g}^*, && \text{for }\Lambda_1\ll k \ll \Lambda_2.
\end{align}
Then we are free to choose $\hp = p/k =1$ and extract a scaling form for $O(\omega,p)$ as long as $\Lambda_1\ll p \ll \Lambda_2$.
In our case $\hat{\Gamma}_k[\hat{\mathbf{v}}]$ is characterised by an infinite set of parameters since have flowing functions of momentum. We can write this as $\mathbf{g}(k) = \mathbf{g}(\hat{\omega},\hat{p};k)$. Then the close-to-a-fixed-point condition can be written as
\begin{align}
 k\partial_k\mathbf{g}(\hat{\omega},\hat{p};k) \ll \mathbf{g}(\hat{\omega},\hat{p};k).
\label{eq:fixed_point_regions}
\end{align}
This separates the $3d$ space spanned by $(\hat{\omega},\hat{p},k)$ into regions where \Eq{eq:fixed_point_regions} is satisfied and regions where it's not. $B(a;k)$ is defined as the value of $\hp$ on the upper limit of the region where \Eq{eq:fixed_point_regions} is true,
\begin{align}
 B(a;k) = \text{max}\left\{\hp|\, \text{\Eq{eq:fixed_point_regions} is satisfied}\right\}.
\end{align}

We can use the scaling range $\Lambda_1\ll p \ll \Lambda_2$ to define $\bar{\eta}_i$, $g(a)$, $\bar{z}_i$ as before. Then $\delta Z(\hat{\omega},\hat{p})$ is defined through \Eq{eq_param_1}. Note that since we are not exactly at a fixed point anymore it contains an additional explicit $k$-dependence $\delta Z(\hat{\omega},\hat{p}) \to \delta Z(\hat{\omega},\hat{p};k)$.
Outside of the scaling range $\delta Z(\hat{\omega},\hat{p};k)$ changes with $k$ in such a way that the scaling form breaks down. As long as $k$ is restricted to $\Lambda_1 \ll k \ll \Lambda_2$ \Eq{eq_flow_1} is not modified. Only its initial condition \Eq{eq_param_2_bis}, is changed to
\begin{align}
& \lim_{\hp \to B(a;k)} \delta Z(a \hp^{-\bar{\eta}_2},\hp;k) = 0.
\label{eq:new_init_cond}
\end{align}
The picture that emerges is that as $k$ is lowered from $\Lambda_2$ to $\Lambda_1$ $\hp$ increases. This in turn decreases $\delta Z(a \hat{p}^{-\bar{\eta}_2},\hat{p};k)$ and brings it to zero when the \RG flow leaves the fixed point. \Ie for $k=\Lambda_1$. Note that in this regime the system is close to the fixed point. The explicit dependence of $\delta Z(a \hat{p}^{-\bar{\eta}_2},\hat{p};k)$ on $k$ is weak.
For smaller values of the cut-off scale, $\delta Z(a \hat{p}^{-\bar{\eta}_2},\hat{p};k)$ becomes strongly dependent on $k$ again. If $p$ is out of the scaling range $\delta Z(a \hat{p}^{-\bar{\eta}_2},\hat{p};k)$ increases again and the scaling form is destroyed. On the other hand if $p$ is in the scaling range $\delta Z(a \hat{p}^{-\bar{\eta}_2},\hat{p};k)$ stays roughly constant as $\hp$ is increases while $k$ is decreased.

Note that the existence of the scaling range implies
\begin{align}
& \lim_{k\to 0} \delta Z(\hat{\omega},\hat{p};k) \cong 0, && \text{for }\Lambda_1 \ll p \ll \Lambda_2, \nonumber \\
& k\partial_k \delta Z(\hat{\omega},\hat{p};k) \ll \delta Z(\hat{\omega},\hat{p};k), && \text{for }\Lambda_1 \ll k \ll \Lambda_2.
\end{align}
The first equation ensures that we recover scaling in the physical limit $k\to 0$. The initial condition given by \Eq{eq:new_init_cond} and the fact that $\delta Z(a (p/k)^{-\bar{\eta}_2},p/k;k)$ does not change very much for $k<\Lambda_1$ and $p \gg k$ ensure that it is realised. The second equation is related to the fact that we have an approximate fixed point for $\Lambda_1 \ll k \ll \Lambda_2$. It states that $\delta Z(\hat{\omega},\hat{p};k)$ only weakly depends on $k$ close to the fixed point. This equation relies on the fact that $B(a;k)$ depends weakly on $k$. If this is not true then we need to restrict the $(\hat{\omega},\hat{p},k)$ space a little more by replacing $B(a;k)$ by
\begin{align}
 B(a) = \text{min}_k\left[B(a;k)\right].
\end{align}
Then \Eq{eq:initial_flow_cut-off} becomes
\begin{align}
 \delta Z(a \hat{p}^{-\bar{\eta}_2},\hat{p}) = h \int_{\hat{p}}^{B(a)} \text{d}y \, \frac{\hat{I}(a y^{-\bar{\eta}_2},y)}{y^{\bar{\eta}_1+1}}.
\end{align}
This picture provides approximate fixed point conditions. Using $B(a)$ instead of $B(a;k)$ reduces the quality of the approximation but eliminate all the explicit $k$-dependence from the fixed point equations. This reduction of the quality of the approximation is in principle a practical limitation but does not make any difference in the end. \cred{Indeed}, we can assume that the \RG flow come asymptotically close to the fixed point. Then the closer we come to it the bigger $B(a)$ will be. It then acts as a cut-off exactly as in \Eq{eq:modif_flow}. We can then even use $B = \text{min}_a\left[B(a)\right]$. The only strict requirement is that we choose $B\gg 1$.

We remark that for a fixed momentum $p$ we can not lower $k$ below $p/B$. We can only recover physical results if $B \gg 1$ since in this limit there is a large range of momenta $p\gg k$ which are resolved.
Then we recover a scaling form for $\Gamma_k^{(2)}(\omega,p)$ with a restricted scaling range given by $k \ll p \ll k B \equiv \Lambda$.
We see that the \UV cut-off $\Lambda = k B$ is proportional to $k$. In the limit $k\to 0$ the scaling range shrinks to a point and only the largest scales are resolved. However since $B$ is arbitrary it can always be chosen large enough for the momentum scale that we are interested in to be within the range $k \ll p \ll \Lambda$ and the physical properties are that of an \RG fixed point.

Note that a completely different solution to this problem of finding the properties of \UV divergent fixed points would be to reverse the direction of the \RG flow. If one uses a cut-off function that vanishes for $p\ll k$ and diverges for $p\gg k$ \Eq{eq:wetterich2} still applies but the boundaries of the flow are reversed $\Gamma_{k\to0}\brv = S\brv$ and $\Gamma_{k\to\infty}\brv = \Gamma \brv$. The interpretation is a little strange since instead of coarse-graining fluctuations on scales smaller than $k$ we do the inverse. Large scale fluctuations are integrated out first. This does not matter here since we look for fixed points. \Ie theories where there is no \RG flow anyway. The fixed point equations then stay unchanged except for the \Eq{eq_param_2_bis} that becomes,
\begin{align}
& \lim_{\hat{p} \to 0} \delta Z(a \hat{p}^{-\bar{\eta}_2},\hat{p}) = 0.
\end{align}
We see that \UV divergent fixed points will not bother us any more. We may however encounter \IR divergent fixed points where $\delta Z(a \hat{p}^{-\bar{\eta}_2},\hat{p})$ diverges at $\hp = 0$.

\chapter{Burgers Turbulence}
\label{sec:burgers_turbulence}
\ResetAbbrevs{All}

In this section we discuss the stationary states of the stochastic Burgers equation.
We start with a brief introduction to the physics that we wish to describe and set up its mathematical formulation. In Section (\ref{sec:scaling_and_correlation_functions}) we review the results of the literature with a focus on the scaling properties of correlation functions. Next we establish a path integral formulation of the steady state generating functional in Section (\ref{sec:Burgers_functionals}). In particular we compute its action (see \Eqs{eq:Burgersaction} or \Eq{eq:Burgersaction_fourier}). Section (\ref{sec:frg_calculation}) is the heart of the present work. We apply the \FRG and the fixed point equations established in Section (\ref{sec:momentum_dependent_propagator}) to compute the two-point correlation function of the stochastic Burgers equation. We establish an approximation scheme in Section (\ref{sec:approximation_scheme}) and write \RG flow equations for its parameters in Section (\ref{sec:flow_equations}). Next we adapt the fixed point equations of Section (\ref{sec:momentum_dependent_propagator}) to our truncation in Section (\ref{sec:fixed_point_equations}). The different quantities that we have introduced in Sections (\ref{sec:approximation_scheme}) and (\ref{sec:fixed_point_equations}) are related to physical observables in Section (\ref{sec:computing_observables}). Finally the \RG fixed point equations are analysed in detail in Section (\ref{sec:asymptotic_properties_of_the_flow_integrals}) and fixed points as well as their properties are discussed.

\begin{figure}[t]
 \begin{center}
\begin{tabular}{p{5cm} p{1cm} p{5cm}}
\resizebox{!}{4.8cm}{\includegraphics{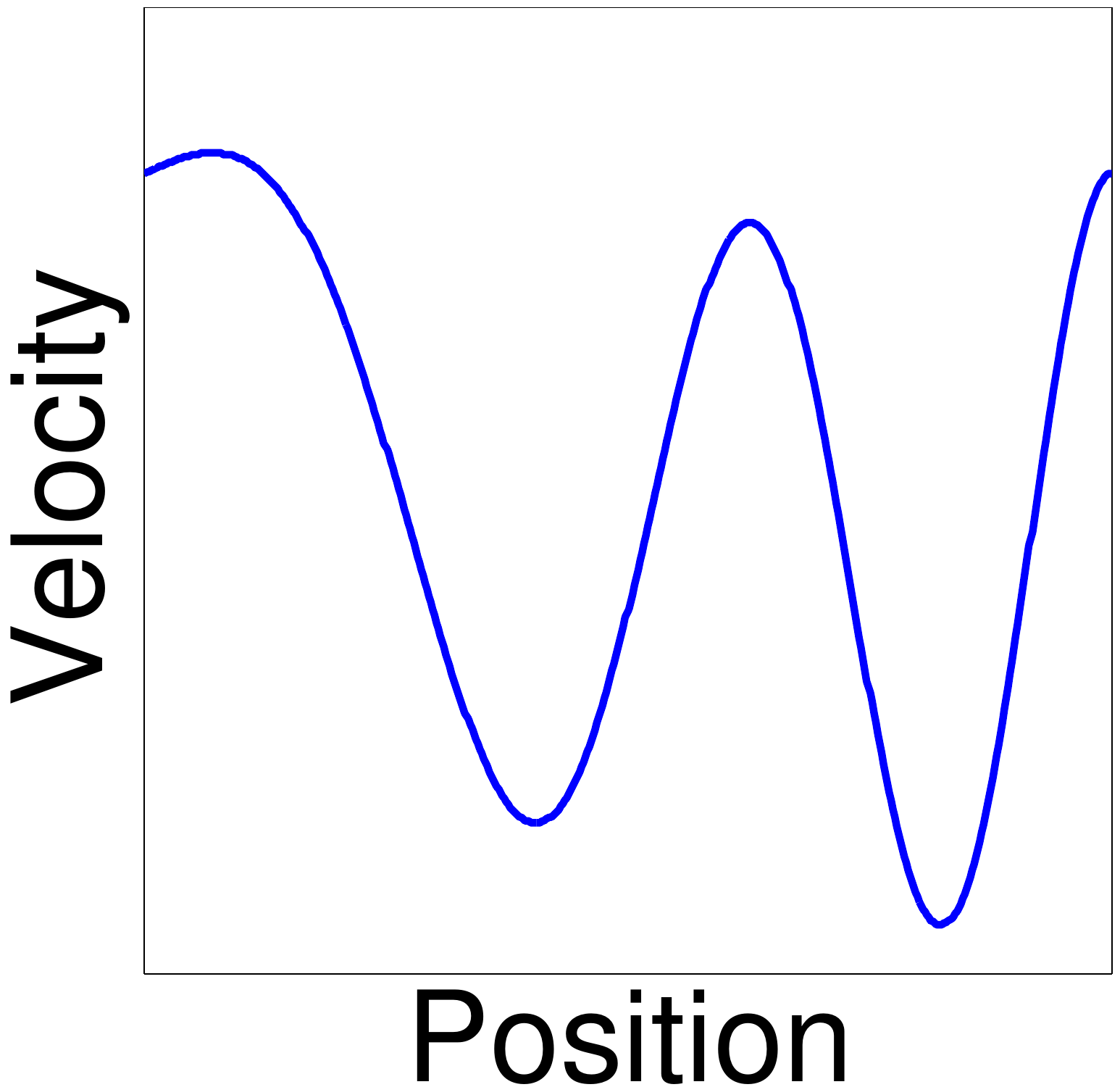}} & & \resizebox{!}{4.8cm}{\includegraphics{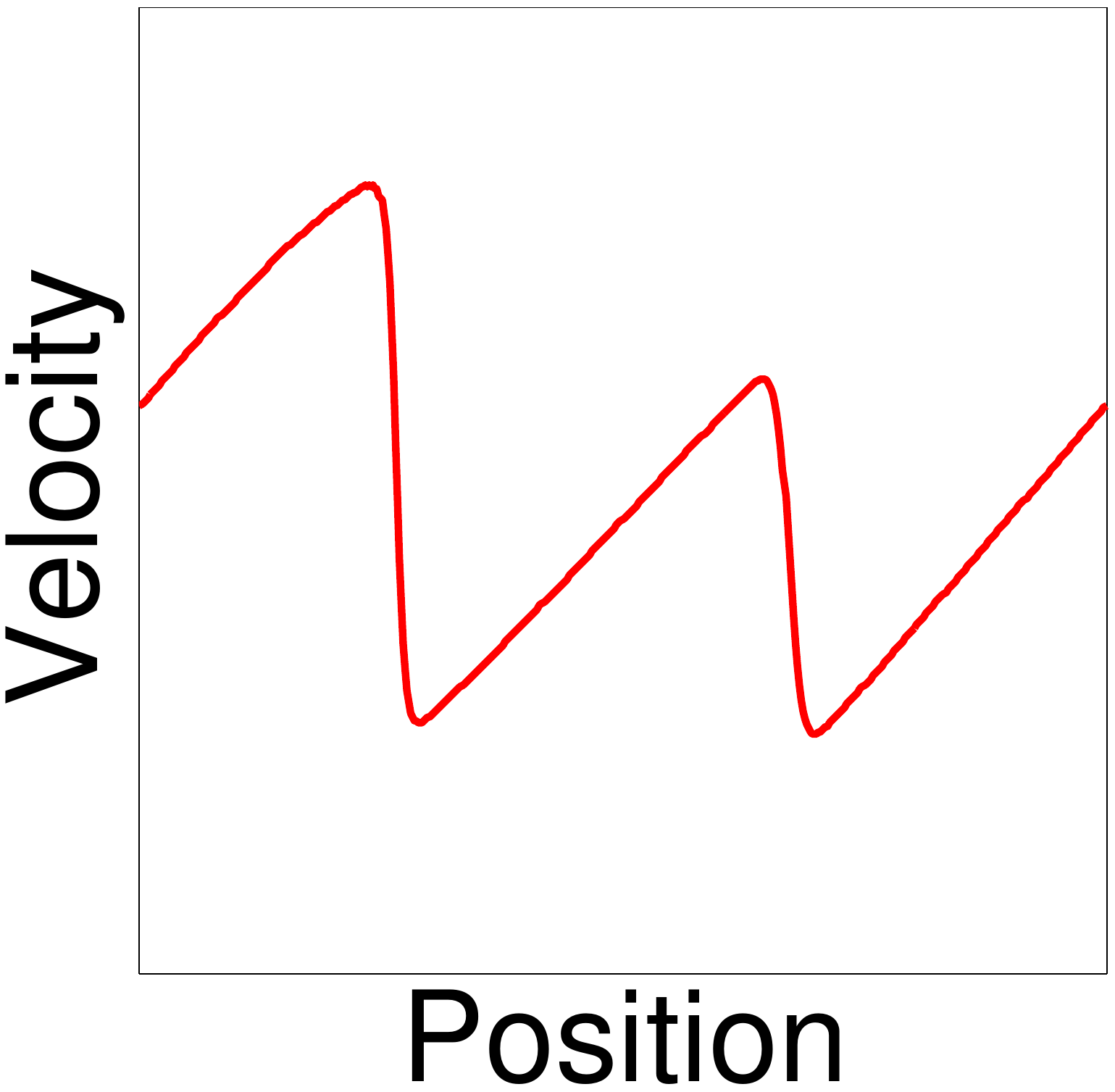}}
\end{tabular}
\end{center}
\caption{A typical velocity profile for Burger's equation in $1d$ at two successive times (left, then right)}
\label{fig:shocks}
\end{figure}

Let us start by discussing the classical turbulence described by Burgers' equation \cite{Burgers1939a},
\begin{align}
% \mathbf{E}(\mathbf{v}(\tx)) \equiv \partial_t \mathbf{v}(\tx) + \left[\mathbf{v}(\tx) \cdot \boldsymbol{\nabla}\right] \mathbf{v}(\tx) - \nu \Delta \mathbf{v}(\tx) = 0.
\mathbf{E}\brv \equiv \partial_t \mathbf{v} + \left[\mathbf{v} \cdot \boldsymbol{\nabla}\right] \mathbf{v} - \nu \Delta \mathbf{v} = 0.
\label{eq:burgers}
\end{align}
$\mathbf{v}(\tx)$ is a space and time dependent velocity field and $\nu$ is the kinematic viscosity (see \cite{Frisch2000a,Bec2007a} for reviews).
Burgers' equation is equivalent to the \NS equation if the equation of state is assumed to impose a constant pressure, $P = const$ \cite{guyon2001a}. It can be interpreted as a model for fully compressible hydrodynamics since there is no pressure to stop two fluid elements from being squeezed together. In the limit $\nu \to 0$, this leads to the appearance of spatial discontinuities in the velocity field after a finite time even when smooth initial conditions are chosen \cite{Grafke2013a,Mesterhazy2013b}.
This is illustrated in Figure (\ref{fig:shocks}) where two shocks are shown to appear from smooth initial conditions for $d=1$.
Such shocks correspond to situations where the velocity field has a strong enough gradient that fluid elements which are moving with a large velocity are advected onto the slow ones too quickly for the latter to get out of the way.
Then the shock propagates through the system and "eats up" the slow particles on the way.
Fluid accumulates at singular points of space.
Note that this picture looks a little different in the frame that it moving with the shock. In this case the whole velocity profile is shifted vertically such that the net velocity is zero. The strong gradient then implies that the velocity changes sign. The picture is then that of two masses of fluid flowing towards each other and colliding at the position of the shock.
For $d>1$ the shocks are not restricted to being points and can have a rich topology \cite{Bec2007a}.
Note that a small but non-vanishing value of $\nu$ leads to shocks which are actually smooth at the dissipation scale, $l \cong \nu/v_{\text{typical}}$. We can see this in Figure (\ref{fig:shocks}). The shocks are actually rounded off because of the finite value of $\nu$.

Burgers' equation is often considered as a toy model for classical hydrodynamics. Given an initial velocity configuration it can be solved analytically in the limit $\nu \to 0$ with the method of characteristics. Moreover, contrarily to the incompressible \NS equation, Burgers' equation can be used to study $1d$ turbulence\footnote{Note that a $1d$ incompressible fluid is trivial.}. This simplifies greatly numerical computations and makes it possible to study stochastic versions of \Eq{eq:burgers}, \cite{Hayot1996a,Bec2007a,Grafke2013a,Mesterhazy2013b}.

Burgers' equation has however many physical applications ranging from the modelling of dust in the early universe to polymers in random media (see Ref.~\cite{Frisch2000a,Bec2007a} and references therein).
The irrotational Burgers equation can be mapped onto the Kardar-Parisi-Zhang (KPZ) equation \cite{Kardar1986a},
\begin{align}
 {\partial_t \theta + \frac{\lambda}{2} \left(\boldsymbol{\nabla} \theta \right)^2 = \nu \nabla^2 \theta,}
\end{align}
with $\mathbf{v} = \boldsymbol{\nabla} \theta$.
The coupling constant $\lambda$ must be set to one if \KPZ equation is to be mapped on Burgers' equation. Indeed, in hydrodynamic theories the non-linearity is part of advective time derivative and can not be re-scaled independently of the partial time derivative.
When \KPZ equation is not used for hydrodynamics, $\lambda$ can be arbitrary and measures the strength of the non-linearity as compared to the dissipative term. In particular when $d\geq2$ one finds a phase transition in the dynamics of the stochastic stationary \KPZ equation. For small values of $\lambda$ the non-linearity is irrelevant and the properties of the steady state are the same as for the linear theory. On the other hand when $\lambda$ is large enough the non linear term becomes relevant and the steady state dynamics is more complicated. See \eg \cite{barabasi1995fractal} for an overview.

The KPZ equation is typically used to describe non-linear interface growth but can also be applied to the dynamics of phase fluctuations in an ultracold Bose gas described by the \SGPE \cite{Carusotto2013a,Altman2013a}, or to directed polymers in random media \cite{Huse1985a,Kardar1987a,Bouchaud1995a}.
Shocks can also appear in the \GPE but, due to the definition of the phase on a compact circle, lead to the creation of (quasi) topological defects, \eg dissolve into soliton trains \cite{Dutton2001a,Kevrekidis2008a}.

In the present work we are mainly interested in out-of-equilibrium stationary states. In order to study such systems we need to include a forcing mechanism. Then the dynamics of the system adapts in order to transfer the energy from the forcing to the dissipation scale and an asymptotic steady state can be reached.
A natural set-up would be to add a deterministic forcing that is periodic in time, for example, to the right-hand side of \Eq{eq:burgers}. We would then need to choose a particular form of such a forcing in space and time before we can extract a particular solution. We are however not really interested in the detailed response of the fluid to a particular forcing because it is expected that averages over space or time will provide smooth and symmetric correlation functions independently of the details of the forcing mechanism. Instead we consider all possible types of forcing and weigh them with a given probability distribution. 
We therefore consider the stochastic Burgers equation,
\begin{align}
\mathbf{E}\brv = \partial_t \mathbf{v} + (\mathbf{v} \cdot \boldsymbol{\nabla}) \mathbf{v} - \nu \Delta \mathbf{v} = \mathbf{f}.
\label{eq:burgers_stochastic}
\end{align}
$\mathbf{f}$ is a force with zero average and Gaussian fluctuations
\begin{align}
\langle \mathbf{f}\rangle = 0, && \langle f_{i}(t,\mathbf{x}) f_{j}(t',\mathbf{x'})\rangle = \delta_{ij}\delta(t-t') F\left(\left|\mathbf{x}-\mathbf{x'}\right|\right).
\label{eq:forcing_corr}
\end{align}
It has the following probability distribution\footnote{We use the definition $\int_{\mathbf{y}} F^{-1}\left(\left|\mathbf{x}-\mathbf{y}\right|\right) \, F\left(\left|\mathbf{y}-\mathbf{x}'\right|\right) = \delta(\mathbf{x}-\mathbf{x}')$ for $F^{-1}\left(\left|\mathbf{x}-\mathbf{y}\right|\right)$.}
\begin{align}
P[\mathbf{f}] = \frac{1}{N} \exp\left[-\frac{1}{2} \int_{\tx,\mathbf{x}'} \mathbf{f}(\tx) \cdot \mathbf{f}(t,\mathbf{x}') \, F^{-1}\left(\left|\mathbf{x}-\mathbf{x'}\right|\right)\right].
\label{eq:forcing_prob}
\end{align}
$N$ is a constant normalisation factor.
We assume that the dynamics of the fluid is ergodic in the sense that we can trade the spatio-temporal mean for the average over the stochastic forcing. Such an averaging procedure makes it possible to define an out-of-equilibrium steady state which is fully invariant under space and time translations. There is no need to take some kind of spatio-temporal averaging to recover universal properties as in the case of a deterministic forcing.
$F\left(\left|\mathbf{x}-\mathbf{x}'\right|\right)$ determines the fluctuations of the forcing and is left undetermined at this point. The integral on the right-hand side of \Eq{eq:forcing_prob} takes a simple form in Fourier space,
\begin{align}
P[\mathbf{f}] = \frac{1}{N} \exp\left[-\frac{1}{2} \int_{\op} \frac{\mathbf{f}(\op) \cdot \mathbf{f}(\mop)}{F\left(p\right)}\right].
\label{eq:forcing_prob_fourier}
\end{align}
We see that the Fourier modes of $\mathbf{f}$ fluctuate independently from each other, follow Gaussian distributions and that their variance is given by
\begin{align}
\sigma^2(\op) = F\left(p\right) \equiv \int_{\mathbf{x}} \text{e}^{i \mathbf{p}\cdot \mathbf{x}} F\left(x\right).
\end{align}
To distinguish different types of forcing we define the exponent $\beta$ as,
\begin{align}
F\left(p\right) \sim p^{\beta}.
\label{eq:forcing_corr_beta}
\end{align}
Hence, $\beta$ determines the degree of non-locality of the forcing. 
For $\beta > 0$ the energy is mainly injected into the \UV modes while for $\beta<0$ the forcing acts on large \IR scales. The case $\beta = 0$ corresponds to a forcing delta correlated in space.

Note that the forcing correlation function that we defined in \Eq{eq:forcing_corr} is not consistent with a potential forcing field. \cred{Indeed}, if there is a function $u(\tx)$ such that $\mathbf{f}(\tx) = \boldsymbol{\nabla}u(\tx)$ the correlation function of the forcing must have the form
\begin{align}
\langle f_{i}(t,\mathbf{x}) f_{j}(t',\mathbf{x'})\rangle = \frac{\partial}{\partial x_i} \frac{\partial}{\partial x_j'} \langle u(\tx) u(t',\mathbf{x}') \rangle.
\end{align}
If we require that the forcing be invariant under space and time translations and rotations we can restrict the correlation function of the potential to satisfy
\begin{align}
\langle u(\op) u(\omega',\mathbf{p}') \rangle = \delta(\omega+\omega') \delta(\mathbf{p}+\mathbf{p}') \, U(p). 
\label{eq:potential_corr}
\end{align}
Then the correlation function of the forcing is
\begin{align}
 \langle f_i(\op) f_j(\omega',\mathbf{p}') \rangle = \delta(\omega+\omega') \delta(\mathbf{p}+\mathbf{p}') \, p_i p_j U(p).
 \label{eq:forcing_corr_potential}
\end{align}
This is consistent with \Eq{eq:forcing_corr} only when $d=1$. Then we can identify $F(p) = p^2 U(p)$. For $d>1$ there is a difference in the tensor structure of the correlation function of \Eq{eq:forcing_corr_potential} as compared to \Eq{eq:forcing_corr}. This originates in the fact that the definition of the forcing mechanism given in \Eqs{eq:forcing_corr} implies that we are forcing vorticity as well as energy into the fluid. The correlation function of the vorticity injection is
\begin{align}
& \langle \left[ \boldsymbol{\nabla} \times \boldsymbol{f}(\tx) \right] \cdot \left[ \boldsymbol{\nabla} \times \boldsymbol{f}(0,\mathbf{0}) \right] \rangle =  \delta(t) \, 2 (1-d) \nabla^2 [F(x)]\neq 0,
\end{align}
where $F'(x) = \left.\text{d}F/\text{d}x\right|_x$.

\section{Scaling and correlation functions}
\label{sec:scaling_and_correlation_functions}

Within such a set-up observables that one can compute are the moments of the velocity field averaged over the different realisations of the forcing. Typically one looks at the velocity increments \cite{Frisch2004a},
\begin{align}
 S_q(\tau,\mathbf{r}) = \left< \left[\left( \mathbf{v}(t+\tau,\mathbf{x}+\mathbf{r}) - \mathbf{v}(t,\mathbf{x}) \right)\cdot\frac{\mathbf{r}}{r}\right]^q \right>,
\end{align}
because of their symmetry properties. \cred{Indeed}, the out-of-equilibrium steady state of the stochastic Burgers equation is expected to be invariant under time and space translations and spatial rotation. Moreover $S_q(\tau,\mathbf{r})$ contain a velocity difference so that they are symmetric under Galilei boosts as well. Additionally, the turbulent state is expected to be invariant under scale transformations for a large class of forcing mechanism. In particular we define the exponents $\zeta_q$ through the equal time structure function,
\begin{align}
 S_q(0,\lambda \mathbf{r}) = \lambda^{\zeta_q} S_q(0,\mathbf{r}), && \text{for } \lambda > 0,
\end{align}
and the exponents $\chi$ and $z$ through the the second order structure function,
\begin{align}
S_2(\tau,\mathbf{r}) = r^{2(\chi-1)} \, g\left(\frac{\tau}{r^z}\right).
\end{align}
Note that we have $\zeta_2 = 2(\chi-1)$. $g(a)$ is the scaling function and is analogous to $g(a)$ defined in Section (\ref{sec:momentum_dependent_propagator}). It is well known from analytical \cite{Forster1977a,Kardar1986a,Medina1989a,Adzhemyan1999a,Berera2001a,Bec2007a,Kloss2012a} as well as numerical \cite{Tang1992a,Ala-Nissila1993a,Hayot1996a,hayot1998dynamical,Castellano1999a,Marinari2000a,Aarao2004a,Ghaisas2006a,Kelling2011a} calculations that $\chi$ and $z$ are related by
\begin{align}
 \chi + z = 2.
\label{eq:chiofeta1andorz}
\end{align}
This can be attributed to Galilean invariance which prohibits an anomalous scaling of the velocity field.
Since the non-linearity of $\mathbf{E}\brv$ is part of the advective derivative of the fluid velocity, it must scale in the same way as the partial time derivative. 
This is only possible if the velocity scales as position divided by time ($\sqrt{S_2(0,\mathbf{r})} \sim r/r^z$), which implies, the relation \eq{eq:chiofeta1andorz} when the exponents are matched.

Note that this relation holds for stationary stochastic \KPZ dynamics as well. In this case the relevant symmetry is not Galilei invariance but infinitesimal tilting of the interface, $\theta(\tx) \to \theta(t,\mathbf{x}+t \lambda \mathbf{v}) + \mathbf{x}\cdot\mathbf{v} - \lambda v^2 /2$.
On the other hand the relation \eq{eq:chiofeta1andorz} does not hold any more if the forcing correlation function is not white in time. \cred{Indeed}, if we make the replacement
\begin{align}
\delta(t-t') \, F(\left|\mathbf{x}-\mathbf{x}'\right|) \to F\left(t-t',\left|\mathbf{x}-\mathbf{x}'\right|\right),
\end{align}
in \Eq{eq:forcing_corr} the forcing correlation function looses its invariance under Galilei transformations\footnote{Under Galilei transformations $\mathbf{x}-\mathbf{x}'$ goes to $\mathbf{x}-\mathbf{x}'-(t-t') \mathbf{v}_0$. This quantity is only invariant if $t=t'$.}. See \cite{Medina1989a} where this case is examined.

The dependence of $\zeta_q$ on the order of the structure function $q$, and more precisely their deviations from a linear behaviour is called intermittency because it has its origin in rare and large events within the hydrodynamic flow \cite{Frisch2004a}. For Gaussian statistics we get $\zeta_q = \zeta_2 \, q/2$. Intermittency is an active topic of research within the incompressible \NS turbulence set-up \cite{ictr2008,Takashi20009a,Chevillard2012,ruelle2012a,song2014,carneiro2014,ruelle2014,Schumacher2014a} and is still not fully understood within Burgers' equation, especially for $d>1$.

In the case $d=1$ Burgers' and \KPZ equations are equivalent and many results are available.
For a forcing correlation function of the type given in \Eq{eq:forcing_corr_beta} and for $\beta > 0$ the forcing fluctuates mainly on small spatial scales and the shocks are overwhelmed. Then there is no intermittency \cite{Hayot1996a,Mesterhazy2013b}. One finds that $\zeta_q = (\chi-1) q$. In this regime perturbative \RG techniques apply \cite{Medina1989a,Frey1994a,Janssen1999a,frey1999scaling} and it was found that
\begin{align}
 \chi = \text{max}\left(\frac{1}{2},-\frac{\beta}{3}+1\right).
 \label{eq:exponent_1d_medina}
\end{align}

When $\beta < 0$ non-linearities become relevant and perturbative \RG techniques fail. The case $\beta < 3$ is however relatively simple. When $\beta$ is small enough the forcing acts on large spatial scales and its realisations are differentiable and smooth enough for shocks to form naturally (see Figure (\ref{fig:shocks})) and propagate through the system. Then the stationary state contains a finite density of shocks and the scaling of the velocity increments in the regime where $\left|r\right|$ is much smaller than the average distance in between shocks but still much larger than the shock size can be estimated by a simple argument \cite{Frisch2000a}. \cred{Indeed}, we can write
\begin{align}
S_q(0,{r}) = P\left(\text{${r}$ passes a shock}\right) \Delta v_{\text{shock}} + P\left(\text{${r}$ does not pass a shock}\right) \left[\left(\partial_x {v}\right) \left|r\right| \right]^q.
\label{eq:non_local_forcing_velocity_increment}
\end{align}
$\Delta v_{\text{shock}}$ is the typical jump in velocity across a shock and $P(\text{event})$ is the probability of 'event'. By '${r}$ passes a shock' we mean that there is a shock in between ${x}$ and ${x}+{r}$\footnote{Remember that we average over many realisations of the forcing such that ${x}$ actually drops out of the final results.}. Since the shocks are discontinuous $\Delta v_{\text{shock}}$ does not depend on $r$. On the other hand if there is no shock the velocity profile is differentiable and the velocity increment is simply proportional to $\left|r\right|^q$.
We compute the average velocity difference by summing the product of the probability that ${r}$ passes over a shock with the corresponding velocity difference and the complementary probability multiplied by the corresponding linearised velocity difference. Note that the velocity difference in the second term of \Eq{eq:non_local_forcing_velocity_increment} is smooth enough to be linear in $\left|r\right|$ by definition. The discontinuous velocity differences are in the first term. In order to compute $\zeta_q$ we estimate the two probabilities as behaving as
\begin{align}
& P\left(\text{${r}$ touches a shock}\right) && \sim \left|r\right|, \nonumber \\
& P\left(\text{${r}$ does not touch a shock}\right) = 1-P\left(\text{${r}$ touches a shock}\right) && \sim 1 - \left|r\right| .
\end{align}
\cred{Indeed}, when $\left|r\right|$ is much bigger than the shock size the latter appear point-like and the probability of finding one with a vector of length $\left|r\right|$ is simply proportional to $\left|r\right|$. We finally keep only the leading (for small $\left|r\right|$) behaviour in the second term and write
\begin{align}
S_q(0,{r}) \sim \Delta v_{\text{shock}} \, \left|r\right| + \left[\left(\partial_x {v}\right)\right]^q \, \left|r\right|^q,
\end{align}
and see that $\zeta_q = \text{min}\left(1,q\right)$. The regime $-3<\beta<0$ is not as well understood. Dimensional analysis and matching with the known scaling for $\beta = 0$ and $\beta = -3$ suggests $\zeta_q = \text{min}\left(-q\beta/3,1\right)$ \cite{Bec2007a}. This assumption seems to be supported by \cite{Hayot1996a}.

Note that for a forcing correlation function of the type given in \Eq{eq:forcing_corr_beta} and for $\beta =2$ an analytic expression for the probability distribution of the single point velocity field was obtained. See \cite{Sasamoto2010a,Corwin2012a} and references therein for detailed discussions.

Much less is known about the cases $d>1$ and apart from \cite{Neate2011a,Choquard2013}, the literature seems to focus exclusively on the case of a potential forcing (as in \Eq{eq:forcing_corr_potential}) and very often directly assumes $\mathbf{v} = \boldsymbol{\nabla}\theta$ in order to simply switch to \KPZ equation. Little is known about $\zeta_q$ and intermittency. The two-point correlation function and its exponents $\chi$ and $z$ are studied in more detail. Much interesting work has been done in this context using a variety of non-perturbative methods.
Most of these concentrate on the case $\beta = 2$, corresponding to white-noise forcing in space.
In this context, predictions for the scaling exponents, scaling functions and upper critical dimension have been made, using, \egc the mode coupling approximation \cite{vanBeijeren1985a,Bouchaud1993a,Frey1996a,Colaiori2001a}, the self-consistent expansion \cite{schwartz1992a,Schwartz2008a}, or the weak-noise scheme \cite{Fogedby2001a,Fogedby2005a,Fogedby2006a}.
The case $\beta < 0$ of a forcing that is concentrated on large scales was tackled in Ref.~\cite{Bouchaud1995a} by means of a replica-trick approach being exact in the limit $d \to \infty$, and bi-fractal scaling of the velocity increments was obtained.
The tails of the probability distribution of velocity differences were addressed in \cite{Polyakov1995a} using an operator product expansion in $d=1$, and in \cite{Gurarie1996a,Grafke2013a} within an instanton approach.
Decaying Burgers turbulence was studied in \cite{Fedorenko2013a}.

We close this section by mentioning the work of \cite{Canet2010a,Canet2011a,Kloss2012a,Kloss2013a} since it is closely related to ours. In these papers the stationary state of the stochastic \KPZ equation is studied for $\beta = 2$ \cite{Canet2010a,Canet2011a,Kloss2012a} with the \FRG. A non-perturbative approximation scheme that respects Galilei invariance is devised and \RG fixed points are found for $d= 1,2,3$. Corresponding scaling exponents as well as scaling functions are estimated. For $d=1$ the exact scaling exponents of \Eq{eq:exponent_1d_medina} are recovered and the obtained scaling function compares very well with the exact result \cite{Sasamoto2010a,Corwin2012a}. For $d\geq2$ the Gaussian fixed point becomes attractive as well and a phase transition occurs.
For general $\beta$ the authors of \cite{Kloss2013a} assume that the noise correlation function takes the form\footnote{Note that this is work about the \KPZ equation. The forcing is a potential field. See \Eq{eq:forcing_corr_potential} where $U(p)$ is defined.},
\begin{align}
 U(p) = D \left(1 + w \, p^{\beta-2}\right),
\end{align}
and look into the cases of $\beta \leq 2$.
It was found  that for $\beta < \beta_\text{c} = 4-d$ there is no stable Gaussian fixed point. When $\beta < \beta_{\text{trans}}(d) \cong 2 -d/2$ the non-local noise is relevant and we get $\chi = (4-d-\beta)/3$. On the other hand for $\beta_\text{c}<\beta<\beta_{\text{trans}}(d)$ the \RG flow is attracted to a fixed point with a local forcing ($w = 0$) and $\chi = (4-d-\beta_{\text{trans}}(d))/3 \cong (4-d)/6$ does not depend on $\beta$ any more. \Eq{eq:exponent_1d_medina} is generalised to
\begin{align}
\chi = \text{max}\left(\frac{4-d-\beta_{\text{trans}}(d)}{3},\frac{4-d-\beta}{3}\right), && \beta_{\text{trans}}(d) \cong \left\{\begin{array}{cc}
 3/2  & \text{for } d = 1\\
 0.863  & \text{for } d = 2\\
 0.100  & \text{for } d = 3 \end{array}\right. .
\label{eq:chi_arbitrary_beta}
\end{align}
See \Eq{eq:KPZ_literature} with $\chi = \eta_1 - d$ for the corresponding values of $\chi$ when $\beta > \beta_{\text{trans}}(d)$.
Note that $\beta_{\text{trans}}(d) \cong 2-d/2$ is a simple estimation based on a linear interpolation in between known results at $d=1$ and $d=4$ \cite{Frey1994a,Janssen1999a}). The values of $\beta_{\text{trans}}(d)$ given in \Eq{eq:chi_arbitrary_beta} are average values from the \KPZ literature (see \cite{Tang1992a,Ala-Nissila1993a,Castellano1999a,Marinari2000a,Aarao2004a,Ghaisas2006a,Kelling2011a} and Table I. of Ref.~\cite{Kloss2012a}.
For $\beta > \beta_\text{c}$ the Gaussian fixed point becomes attractive as well as the local one. We see again a phase transition.
We remark that \cite{Frey1994a,Janssen1999a} tackles this problem perturbatively and postulates most of these results.

\section{Functional treatment}
\label{sec:Burgers_functionals}

In this section we apply the \MSR formalism \cite{Martin1973a,Bausch1976a,Janssen1976a,DeDominicis1978a,Zinnjustin2002a} and express the generating functional of velocity correlation functions as a path integral. We extract the corresponding action functional which is given in \Eqs{eq:Burgersaction} and \eq{eq:Burgersaction_fourier}.
We follow closely the derivation of \cite{Zinnjustin2002a}, chapter 4, where classical Langevin equations are analysed.

We define the generating functional of velocity field correlation functions as
\begin{align}
 Z[\mathbf{J}] = \langle \text{e}^{\int \mathbf{J} \cdot \mathbf{v}} \rangle.
\label{eq:gen_functional_0}
\end{align}
All the velocity field correlation functions can be extracted from $Z[\mathbf{J}]$ by taking derivatives with respect to $\mathbf{J}(\tx)$ before evaluating at $\mathbf{J}(\tx)=0$. \Eq{eq:gen_functional_0} contains an averaging over the stochastic forcing $\mathbf{f}(\tx)$. We write it as
\begin{align}
 Z[\mathbf{J}] = \int \mathcal{D}[\mathbf{f}] \, \text{e}^{\int \mathbf{J} \cdot \mathbf{v}[\mathbf{f}]} P[\mathbf{f}].
 \label{eq:gen_functional_1/2}
\end{align}
$\mathbf{v}[\mathbf{f}]$ is the solution of \Eq{eq:burgers_stochastic} with the realisation $\mathbf{f}$ inserted and $P[\mathbf{f}]$ is the probability density functional of the forcing which is given in \Eqs{eq:forcing_prob} and \eq{eq:forcing_prob_fourier}. Not that we do not worry about initial conditions here because we are interested in the steady state where the initial time is far away in the past, $t_0\to-\infty$.
Next we reformulate \Eq{eq:gen_functional_1/2} by including an integration over $\mathbf{E}(\tx)$ and enforcing the equations of motion with a delta functional,
\begin{align}
 Z[\mathbf{J}] = \int \mathcal{D}[\mathbf{f}] \mathcal{D}[\mathbf{E}] \, \text{e}^{\int \mathbf{J} \cdot \mathbf{v}[\mathbf{E}]} P[\mathbf{f}] \, \delta[\mathbf{E}-\mathbf{f}].
\end{align}
%
%We have assumed that the function $\mathbf{E}[\mathbf{v}]$ that is defined in \Eq{eq:burgers} can be inverted. $\mathbf{v}[\mathbf{E}]$ is this inverse.
We can then change variables from $\mathbf{E}$ to $\mathbf{v}$ under the integral,
\begin{align}
 Z[\mathbf{J}] = \int \mathcal{D}[\mathbf{f}] \mathcal{D}[\mathbf{v}] \, \text{e}^{\int \mathbf{J} \cdot \mathbf{v}} P[\mathbf{f}] \, \delta[\mathbf{E}\brv-\mathbf{f}] \left|\frac{\delta \mathbf{E}\brv}{\delta \mathbf{v}}\right|.
\label{eq_generating_interm}
\end{align}
The Jacobian of the coordinate change is given by
\begin{align}
\left|\frac{\delta \mathbf{E}\brv}{\delta \mathbf{v}}\right| = \text{det}\left[\frac{\partial E_i\brv(\tx)}{\partial v_j(t',\mathbf{x}')}\right].
\label{eq:det}
\end{align}
Note that this is a functional determinant. It is the determinant of an operator which acts on the space of functions of spatio-temporal variables $(\tx)$. It can be included in the weight of the path integral by virtue of the identity $\log\left(\text{det}\left[A\right]\right) = \text{Tr}\left(\log\left[A\right]\right)$. Finally we evaluate the delta function and insert \Eq{eq:forcing_prob} to write
\begin{align}
 Z[\mathbf{J}] = \int \mathcal{D}[\mathbf{v}] \, \text{e}^{- S[\mathbf{v}] + \int \mathbf{J} \cdot \mathbf{v}},
\label{eq:gen_functional_1}
\end{align}
with
\begin{align}
 S[\mathbf{v}] = \frac{1}{2} \int_{\tx,\mathbf{x}'} \mathbf{E}\brv(\tx) \cdot \mathbf{E}\brv(t,\mathbf{x}') \, F^{-1}(\left|\mathbf{x}-\mathbf{x}'\right|) - \int_{\tx} \log\left[\frac{\partial E_i\brv(\tx)}{\partial v_j(t',\mathbf{x}')}\right](\tx,\tx).
\label{eq:action_0}
\end{align}
As for the determinant of \Eq{eq:det} we have here a functional logarithm. It can be interpreted as a power series in its argument. We have now arrived at an action for the steady state of the stochastic Burgers equation. We can however simplify it. \cred{Indeed}, we show in the following that the second term of \Eq{eq:action_0} does not contribute to the steady-state dynamics.

Let us start by giving an expression for the operator that is inside the logarithm in coordinate space,
\begin{align}
 \frac{\partial E_i\brv(\tx)}{\partial v_j(t',\mathbf{x}')} = \delta(t-t') \delta(\mathbf{x}-\mathbf{x}') \left[ \delta_{ij} \left(\partial_{t'} - \nu \nabla^2_{\mathbf{x}'} + v_l(t,\mathbf{x}) \partial_{x'_l} \right) + \frac{\partial v_i(t,\mathbf{x})}{\partial x_j} \right].
\label{eq:jacobian_intermediary}
\end{align}
Defining the time derivative operator and its inverse,
\begin{align}
 D_t(\tx;t',\mathbf{x}') = \delta(t-t') \delta(\mathbf{x}-\mathbf{x}') \, \partial_{t'} \, , && I_t(\tx;t',\mathbf{x}') = \theta(t-t') \, \delta(\mathbf{x}-\mathbf{x}'),
\end{align}
we can extract $D_t$ from \Eq{eq:jacobian_intermediary}.
We write ${\partial E}/{\partial v} = D_t \cdot \left(\mathds{1}+X\right)$ with\footnote{${\partial E}/{\partial v}$ denotes the operator with matrix elements given by \Eq{eq:jacobian_intermediary} in real space. $\mathds{1}$ is the identity operator. $\theta(x)$ is the step function. It is zero if $x<0$ and $1$ if $x>0$.}
\begin{align}
 X(\tx;t',\mathbf{x}') = \theta(t-t') \delta(\mathbf{x}-\mathbf{x}') \left[ \delta_{ij} \left( - \nu \nabla^2_{\mathbf{x}'} + v_l(t,\mathbf{x}) \partial_{x'_l} \right) + \frac{\partial v_l(t,\mathbf{x})}{\partial x_l} \right].
\label{eq:jacobian_intermediary_2}
\end{align}
This form for ${\delta \mathbf{E}\brv}/{\delta \mathbf{v}}$ makes the trace of its logarithm particularly simple. First note that the multiplicative factor that we extracted does not depend on the velocity field and can be absorbed into the normalisation of $Z[\mathbf{J}]$. Furthermore the step function in front of $X$ implies that $X^n$ will be proportional to $\left(t-t'\right)^{n-1}$ in the limit $t \to t'$. Let us illustrate this with a simpler example,
\begin{align}
 A(t,t') = \theta(t-t') \, a(t').
\end{align}
$a(t')$ is a function of $t'$ as in \Eq{eq:jacobian_intermediary_2}. The general case is completely analogous but contains heavy expressions. It is easy to check that
\begin{align}
 A^2(t,t') \equiv \int_\tau \theta(t-\tau) a(\tau) \, \theta(\tau-t') a(t') & = \theta(t-t') \int_0^{t-t'} \text{d}\tau \, a(t-\tau) a(t') \nonumber \\
& \cong \theta(t-t') (t-t') a(t) a(t').
\label{eq:jacobian_intermediary_3}
\end{align}
The last equality was obtained by inserting $a(t-\tau) \cong a(t)$ which is asymptotically exact in the limit $t \to t'$. Using \Eq{eq:jacobian_intermediary_3} it can be shown by recurrence that
\begin{align}
 A^n(t,t') \cong \theta(t-t') \frac{\left(t-t'\right)^{n-1}}{(n-1)!} \, a(t)^n, && \text{as } t \to t'.
\end{align}
We can now insert this into \Eq{eq:action_0} and use the fact that the second term is a trace and is evaluated at $(\tx) = (t',\mathbf{x}')$. Then only the term linear in $X$ contributes. Once the constant contribution arising from $D_t$ is subtracted we get
\begin{align}
 S[\mathbf{v}] = \frac{1}{2} \int_{\tx,\mathbf{x}'} \mathbf{E}\brv(\tx) \cdot \mathbf{E}\brv(t,\mathbf{x}') \, F^{-1}(\left|\mathbf{x}-\mathbf{x}'\right|) - \int_{\tx} X(\tx,\tx).
\label{eq:action_1}
\end{align}
Note that $X(\tx,\tx)$ is not well defined since it contains a delta function which is evaluated at zero and since it is proportional to $\theta(0)$. The value of the step function evaluated at zero is related to the precise definition of the continuum limit of the theory and has to be chosen consistently. See \cite{Lau2007a,canet2011b} or chapter 4 of \cite{Zinnjustin2002a} for a detailed discussion. A natural choice (that we do not make at this point and) which corresponds to a forward discrete time propagation is the It\={o} prescription $\theta(0) = 0$.

The $\delta(\mathbf{0})$ pre-factor looks however more troubling at first but can be shifted into the normalisation of the path integral in the same way as the $\text{Tr}\left[\log\left(D_t\right)\right]$ factor. Performing the trace over the spatial indexes and assuming that the delta distribution has been regularised in some way, we formally write the second term of \Eq{eq:action_1} as
\begin{align}
 \int_{\tx} X(\tx,\tx) = \theta(0) \int_{\tx} \left[ d \left(-\nu \frac{\partial^2 \delta\left(\mathbf{0}\right)}{\partial x_l \partial x_l} + v_l(t,\mathbf{x}) \frac{\partial \delta\left(\mathbf{0}\right)}{\partial x_l}  \right) + \delta\left(\mathbf{0}\right) \frac{\partial v_l(t,\mathbf{x})}{\partial x_l} \right].
\end{align}
We already see that the first term does not depend on $\mathbf{v}$ and only contributes as a shift of $S[\mathbf{v}]$ and that the third term only depends on the velocity field evaluated at the boundary of space $\mathbf{x} \to \infty$. We discard them both. Finally we can see that the middle term actually behaves as the third. \cred{Indeed}, if we write it as
\begin{align}
d \theta(0) \int_{\tx,\mathbf{x}'} \delta(\mathbf{x}-\mathbf{x}') \, v_l(t,\mathbf{x}) \, \frac{\partial \delta\left(\mathbf{x}-\mathbf{x}'\right)}{\partial x_l} = \frac{d\theta(0) }{2} \int_{\tx,\mathbf{x}'} \frac{\partial \left[\delta\left(\mathbf{x}-\mathbf{x}'\right)\right]^2}{\partial x_l} v_l(t,\mathbf{x}),
\end{align}
we can integrate it by parts and perform the $\mathbf{x}'$ integration. It becomes
\begin{align}
d \theta(0) \int_{\tx,\mathbf{x}'} \delta(\mathbf{x}-\mathbf{x}') \, v_l(t,\mathbf{x}) \, \frac{\partial \delta\left(\mathbf{x}-\mathbf{x}'\right)}{\partial x_l} = -\frac{d\theta(0) \delta(\mathbf{0})}{2} \int_{\tx} \frac{\partial v_l(t,\mathbf{x})}{\partial_{x_l}}.
\end{align}
Note that we can avoid the manipulation of the square of a delta function by defining the delta function evaluated at zero as a volume factor, $\delta(\mathbf{0}) = \int_{\mathbf{y}}\, 1/(2\pi)^d$. Then we have
\begin{align}
 \int_{\tx} X(\tx,\tx) = \frac{\theta(0)}{(2\pi)^d} \int_{\tx,\mathbf{y}} \boldsymbol{\nabla}\cdot \mathbf{v}(\tx) + d\left[\nu y^2 + i \, \mathbf{y}\cdot \mathbf{v}(\tx)\right].
\end{align}
The first two terms can be neglected as before. The third term however now vanishes because it contains an integration over all space of an odd function, $\int_{\mathbf{y}} \mathbf{y} = 0$. If one is not comfortable with such manipulations one can always choose to work with It\={o}'s prescription and choose $\theta(0)=0$. It is however interesting to note that the way in which we enforce causality at the microscopic level, \ie choice of the value of $\theta(0)$, does not seem to play a role here. In fact we will never have to choose a value of $\theta(0)$ in this work because we work with a steady state and do not look into response functions.

We finally write the action for the stochastic Burgers equation as
\begin{align}
& S[\mathbf{v}] = \frac{1}{2}\int_{t,\mathbf{x},\mathbf{y}} \mathbf{E}\brv(t,\mathbf{x}) \cdot \mathbf{E}\brv(t,\mathbf{x}') \, F^{-1}(|\mathbf{x}-\mathbf{y}|), \nonumber \\
& \mathbf{E}\brv(\tx) = \partial_t \mathbf{v}(\tx) + \left[\mathbf{v}(\tx) \cdot \boldsymbol{\nabla}\right] \mathbf{v}(\tx) - \nu \Delta \mathbf{v}(\tx),
\label{eq:Burgersaction}
\end{align}
or
\begin{align}
& S[\mathbf{v}] = \frac{1}{2}\int_{\op} \frac{\mathbf{E}\brv(\op) \cdot \mathbf{E}\brv(\mop)}{F(p)}, \nonumber \\
& \mathbf{E}\brv(\op) = \left(i\omega-\nu p^2\right) \mathbf{v}(\op) - i\int_{\omega',\mathbf{p}'} \mathbf{v}(\omega',\mathbf{p}') \left[\mathbf{v}(\omega-\omega',\mathbf{p}-\mathbf{p}') \cdot \mathbf{p}'\right],
\label{eq:Burgersaction_fourier}
\end{align}
in Fourier space.

\Eq{eq:gen_functional_1} together with \Eqs{eq:Burgersaction} or \eq{eq:Burgersaction_fourier} contains the full information on the steady state of Burgers turbulence. It will however be more practical to work with different variables. As in thermal equilibrium one can define the Schwinger functional,
\begin{align}
 W[\mathbf{J}] = \log\left(Z[\mathbf{J}]\right), && \langle v_{i_1}(t_1,\mathbf{x}_1)..v_{i_n}(t_n,\mathbf{x}_n) \rangle_{\text{c}} = \left. \frac{\delta^n W}{\delta J_{i_1}(t_1,\mathbf{x}_1)..\delta J_{i_n}(t_n,\mathbf{x}_n)}\right|_{\mathbf{J}=0},
\end{align}
which generates connected correlation functions of the velocity field. Then we switch from the representation in terms of $\mathbf{J}(\tx)$ to the one in terms of $\langle \mathbf {v}(\tx) \rangle$ by taking the Legendre transform of $W[\mathbf{J}]$,
\begin{align}
 \Gamma[\mathbf{v}] = \sup_{\mathbf{J}}\left[- W[\mathbf{J}] + \int \mathbf{J} \cdot \mathbf{v}\right].
% && \frac{\delta \Gamma[\mathbf{v}]}{\delta v_i(\tx)} = J_i(\tx).
\end{align}
$\Gamma[\mathbf{v}]$ is the \OnePI effective action. It takes into account the fluctuations of the velocity field through an infinite series of \OnePI diagrams. The physical velocity field expectation value is the extremum of $\Gamma[\mathbf{v}]$
\begin{align}
 \left. \frac{\delta \Gamma}{\delta v_i(\tx)}\right|_{\mathbf{v} = \langle \mathbf{v}\rangle} = 0,
\end{align}
and the higher order correlation functions are computed through the higher order derivatives of $\Gamma[\mathbf{v}]$ evaluated at $\mathbf{v} = \langle \mathbf{v}\rangle$. See Section (\ref{sec:1pi_effective_action}) where this is explained in detail.

Note that the action \eq{eq:Burgersaction_fourier} is rather complicated. Its inverse propagator,
\begin{align}
S^{(2)}_{ij} [\mathbf{0}](\op) = \delta_{ij} \, F^{-1}(p) \left(\omega^2 + \nu^2 p^4\right),
\end{align}
contains a non-trivial dependence on momentum through $F^{-1}(p)$ and $S[\mathbf{v}]$ contains two different vertexes which depend on momentum, frequency and spatial indexes in a complicated way. We do not give explicit expressions for $S^{(3)}_{ijl}[\mathbf{0}](\op;\omega',\mathbf{p}')$ and $S^{(4)}_{ijlm}[\mathbf{0}](\op;\omega',\mathbf{p}';\omega'',\mathbf{p}'')$ because \Eq{eq:Burgersaction_fourier} is formally very similar to the ansatz that we will choose for the flowing effective action $\Gamma_k[\mathbf{v}]$ in Section (\ref{sec:approximation_scheme}). The vertexes of $S\brv$ can be obtained by making the replacements
\begin{align}
 &\Gamma_k[\mathbf{v}] \to S[\mathbf{v}], && F_k^{-1}(p) \to F^{-1}(p), && \nu_k(p) \to \nu,
\end{align}
in \Eqs{eq:gamma3} and \eq{eq:gamma4} of Appendix \ref{sec:explicit_expressions_for_the_flow_integrals}. 

Let us close this section by noting that the \MSR formalism is usually used with an additional response field. See \eg \cite{Canet:2006xu,Mejiamonasterio2012a,Tauber2013a} where this is done. Instead of directly performing the integration over the forcing in \Eq{eq_generating_interm} we could have taken the Fourier transform of the delta function,
\begin{align}
 \delta \left[\mathbf{E}\brv - \mathbf{f}\right] = \int \mathcal{D}[\tilde{\mathbf{v}}] \, \text{e}^{i \int \tilde{\mathbf{v}} \cdot \left[ \mathbf{E}\brv-\mathbf{f}\right]},
\end{align}
and performed the $\mathbf{f}$-integration with $\tilde{\mathbf{v}}(\tx)$ left free since the argument of the exponential stays quadratic in the forcing. This has three advantages. First the resulting action
\begin{align}
 S\left[\mathbf{v},\tilde{\mathbf{v}}\right] = \frac{1}{2} \int_{\op} F(p) \, \tilde{\mathbf{v}}(\op) \cdot \tilde{\mathbf{v}}(\mop) - 2 i \, \tilde{\mathbf{v}}(\op) \cdot \mathbf{E}\brv(\mop),
\end{align}
does not contain the inverse of the forcing correlation function any more. \cred{Indeed}, when one integrates over the forcing the $F^{-1}(\left|\mathbf{x}-\mathbf{x}'\right|)$ term in the probability distribution (see \Eq{eq:forcing_prob}) is inverted. Then we can choose $F(p)$ to vanish for certain values of $p$ without having to worry about divergences in the action. Secondly the action that one obtains in terms of $\mathbf{v}(\tx)$ and $\tilde{\mathbf{v}}(\tx)$ has a simpler structure than the one of \Eq{eq:Burgersaction} since it only contains one vertex.
%\cred{Indeed} the action of \Eq{eq:Burgersaction} has a quadratic term that contains two time and four space derivatives, a three point vertex and a four point vertex. See Appendix \ref{sec:integrals} where these are computed explicitly and shown in \Eqs{eq:gamma2_delta}, \eq{eq:gamma2}, \eq{eq:gamma3_delta}, \eq{eq:gamma3}, \eq{eq:gamma4_delta} and \eq{eq:gamma4}. We will see in Section (\ref{sec:flow_equations}) that this complicates calculations notably.
Finally even though it was introduced as a ghost field $\tilde{\mathbf{v}}(\tx)$ can be interpreted as a response field. If one adds a deterministic part to the stochastic forcing of \Eq{eq:burgers_stochastic} $\mathbf{f}(\tx) \to \mathbf{f}(\tx) + \mathbf{\Phi}(\tx)$ on can write the response function as
\begin{align}
 \left. \frac{\delta}{\delta \Phi_i(\tx)} \langle v_j(t',\mathbf{x}') \rangle \right|_{\Phi(\tx)=0}= -i\,\langle v_j(t',\mathbf{x}') \tilde{v}_i(\tx) \rangle.
\end{align}
Note however that the in theory of both fields which is described by $S\left[\mathbf{v},\tilde{\mathbf{v}}\right]$, we must deal with twice as much degrees of freedom which makes it necessary to consider all the objects of the theory as tensors in space as well as field indexes indexes.

\section{Functional renormalisation group calculation}
\label{sec:frg_calculation}

We have seen in Section (\ref{sec:scaling_and_correlation_functions}) that scale invariance is an essential property of Burgers turbulence.
We therefore turn to the \RG and look for fixed point of the field theory defined by \Eq{eq:gen_functional_1} together with \Eq{eq:Burgersaction} or \Eq{eq:Burgersaction_fourier}.
Perturbative \RG methods yield a continuum of non-Gaussian \IR attractive fixed points with exponents given by \Eqs{eq:exponent_1d_medina} and \eq{eq:chiofeta1andorz} for $d=1$ and for $\beta > 0$ \cite{Medina1989a,Frey1994a,Janssen1999a,frey1999scaling}.
Outside of this range non-perturbative effects become important and a more sophisticated approach is necessary.
We turn to the non-perturbative \RG which is outlined in Section (\ref{sec:def_flow_equation}). We take as starting point the exact flow equation \Eq{eq:wetterich2},
\begin{align}
k \partial_k \Gamma_k[\mathbf{v}] 
= \frac{1}{2} \text{Tr}\left[ \left(\Gamma_k^{(2)}[\mathbf{v}]+R_k\right)^{-1} k\partial_k R_k \right],
\label{eq:wetterich_b}
\end{align}
for the effective average action, $\Gamma_k\brv$.
%Since we are interested in scaling phenomena we look for fixed points within the \FRG framework.
In order to find self-similar turbulent configurations we look for \IR fixed points of the flow, \ie for solutions of \eq{eq:wetterich_b} which are scaling in $\omega$ and $p$ in the limit $k\to0$.

Our main goal is to get a handle on the second order velocity increment $S_2(\tau,\mathbf{r})$ of the stochastic Burgers equation and on the exponents $\chi$ and $z$. For this we start by making a non-perturbative approximation for the flowing effective action $\Gamma_k[\mathbf{v}]$.
See Section (\ref{sec:approximation_scheme}).
This approximation contains two arbitrary functions of momentum and is formally similar to $S\brv$.
Next we write \RG flow equations for these two functions in Section (\ref{sec:flow_equations}).
The \RG fixed point equations of Section (\ref{sec:momentum_dependent_propagator}) are adapted to our approximation in Section (\ref{sec:fixed_point_equations}).
The two unknown functions of momentum are related to the physically observable quantities $S_2(\tau,\mathbf{r})$, $\chi$ and $z$ in Section (\ref{sec:computing_observables}). Finally we solve the \RG fixed point equations in both asymptotic limits of momentum much larger and much smaller than the cut-off scale and discuss the properties of the fixed points that we find in Section (\ref{sec:asymptotic_properties_of_the_flow_integrals}).

\subsection{Approximation scheme}
\label{sec:approximation_scheme}

\Eq{eq:wetterich_b} relates the flowing effective action with its second moment and therefore creates an infinite hierarchy of integro-differential equations for all the field derivatives of $\Gamma_k[\mathbf{v}]$.
To allow their solution in practice, truncations are in order.
We have discussed different approach to such truncations in Section (\ref{sec:FRG_fixed_point_equations}) and we choose to truncate the high order correlation functions of the theory in favour of a more precise treatment of the frequency and momentum dependence of the two-point correlation function. \cred{Indeed}, both vertexes of $S[\mathbf{v}]$ contain a non-trivial momentum dependence because of the advective non-linearity of fluid dynamics and through the forcing correlation function $F(p)$. Moreover, the three-point vertex depends on frequency as well as momentum. Momentum dependence is already an essential property of $S\brv$. To not take this into account in the flowing effective action will not give good results. See \cite{Canet2005b} where this was tried with the stochastic \KPZ equation.

Note that the \BMW approach \cite{benitez2012} can not be used here because it relies on evaluating \Eq{eq:wetterich_b} at a constant field while including the momentum (and frequency) dependence of low order correlation functions only.
The strength of the \BMW approximation is that it takes into account the full dependence of the derivatives of $\Gamma_k\brv$ on this constant field as in the case of the local potential approximation.
Here Galilei invariance implies that the derivatives of the effective action behaves as\footnote{This can be seen by applying a Galilei boost, $\mathbf{v} \to \mathbf{v} - \mathbf{u}$.}
\begin{align}
\Gamma_{i_1i_2..i_n}^{(n)}[\mathbf{u}](\omega_1,\mathbf{p}_1;..;\omega_{n-1},\mathbf{p}_{n-1}) = \Gamma_{i_1i_2..i_n}^{(n)}[\mathbf{0}](\omega_1+\mathbf{p}_1\cdot \mathbf{u},\mathbf{p}_1;..;\omega_{n-1}+\mathbf{p}_{n-1}\cdot \mathbf{u},\mathbf{p}_{n-1}),
\end{align}
when $\mathbf{u}(\tx)=\mathbf{u}$ is a constant velocity field. We see that the frequency, momentum and field dependence are all related by Galilei invariance in such a way the \BMW approximation can not be applied. Indeed, Galilei invariance implies that truncating the frequency and momentum dependence of $\Gamma_k\brv$ is equivalent to truncating its dependence on the constant field $\mathbf{u}$.
 
The solutions resulting from our truncation should preserve the symmetries of the underlying theory. For this we make the ansatz 
\begin{align}
  \Gamma_k[\mathbf{v}] & = \frac{1}{2} \int_{\omega,\mathbf{p}} \mathbf{E}_k\brv(\op) \cdot \mathbf{E}_k\brv(\mop) \, F^{-1}_k(p), \nonumber \\
 \mathbf{E}_k\brv(\op) & = \left[ i\omega + \nu_k(p) p^2\right] \mathbf{v}(\omega,\mathbf{p})  - i \int_{\omega',\mathbf{q}} 
 \left[\mathbf{v}(\omega-\omega',\mathbf{p}-\mathbf{q})\cdot \mathbf{q} \right] \, \mathbf{v}(\omega',\mathbf{q}) ,
\label{eq:trunc}
\end{align}
for the effective average action, in terms of the $k$-dependent inverse force correlator $F_{k}^{-1}(p)$ and kinematic viscosity $\nu_{k}(p)$.
$\Gamma_{k}[\mathbf{v}]$ has the same form as the action $S[\mathbf{v}]$ of the underlying Burgers equation \eq{eq:Burgersaction_fourier}, but with the inverse force correlator and the kinematic viscosity allowed to be $k$-dependent. This ensures that no symmetry is broken by the ansatz.
In particular this truncated action functional is manifestly Galilei invariant.
We anticipate that $\nu_{k}(p)$ will become $p$-dependent as a result of the \RG flow because the advective derivative renders the cubic and quartic couplings of the velocity field momentum dependent.

With the above ansatz, we keep a general dependence of the inverse propagator on $p$ while taking into account the $\omega$-dependence in an expansion to first order in the frequency squared,
%\footnote{Note that here and in the following we use $\Gamma^{(2)}_{k}(\omega,\mathbf{p})$ to denote the diagonal elements of $\Gamma^{(2)}_{k,ij}(\omega,\mathbf{p})$ in the spatial indexes. This should not be confused with the $\Gamma^{(2)}_{k}(\omega,\mathbf{p})$ of Section (\ref{sec:momentum_dependent_propagator}) which is equal to $\Gamma^{(2)}_{k,ij}(\omega,\mathbf{p})$ in the special case $d=1$.}
%
\begin{align}
\Gamma^{(2)}_{k,ij}(\omega,\mathbf{p}) & \equiv \delta_{ij} \, \Gamma^{(2)}_{k}(\omega,\mathbf{p}) =  \delta_{ij} \left[ \nu_k(p)^2 p^4 +\omega^2\right]F^{-1}_k(p).
\label{eq:Gprop}
\end{align}
The sole dependence on the norm $p$ reflects the assumed rotational invariance.
Note that isotropy of the problem implies as well that $\delta \Gamma_k[\mathbf{v}=0]/\delta \mathbf{v}=0$. \Ie $\langle \mathbf{v}(\tx)\rangle = 0$.
The truncation of the frequency dependence ensures that the integrand on the right hand side of the flow equation \eq{eq:wetterich_b} is a rational function of $\omega$. We will see that this enables the analytic integration over the frequency variable in the flow equation so that there is no need to explicitly cut off the frequency degrees of freedom. $R_k(\op) = R_k(p)$ is sufficient (see Appendix \ref{sec:explicit_expressions_for_the_flow_integrals} for details).
Finally, note that $\Gamma_k\brv$ is chosen such that the inverse force correlator $F_k^{-1}(p)$ and thus the inverse propagator are diagonal in momentum $p$ and in the field indices $i,j$. 
As a consequence, this truncation is only able to capture solutions of the \RG flow equations where vorticity is injected in the steady state. A correlation function of the form \eq{eq:forcing_corr_potential} can not be accounted for in the classical limit $k\to \infty$ and is not allowed to develop as the cut-off scale is lowered.

$\nu_k(p)$ and $F_k^{-1}(p)$ can be interpreted as an effective viscosity and forcing correlation function that emerges on the large scales when the small scales are integrated out. \cred{Indeed}, one can choose $\nu_{k\to\infty}(p) = \nu$ and $F_{k\to\infty}^{-1}(p) = 1/F(p)$ as initial conditions and compute the whole $k$-dependence of both quantities from the flow equation. For each value of $k$ along the way to $k = 0$, $\Gamma_k[\mathbf{v}]$ can be used as an effective bare action as long as the modes with momentum larger than $k$ are cut off. Such a procedure is not guaranteed to lead to a scale invariant solution when $k \to 0$ since any value can be chosen for $\nu$ and the function $F(p)$ must be specified. Each different choice can lead to a different solution which will most likely still depend on the scales that were chosen at $k\to\infty$. In contrast we will write in Section (\ref{sec:fixed_point_equations}), directly a set of \RG fixed point equations without ever having to specify $\nu_k(p)$ 
and 
$F_k^{-1}(p)$. The equations provide solutions which are scale invariant by construction.

\subsection{Flow equations}
\label{sec:flow_equations}

Before we go on to the \RG fixed point equations let us write flow equations for $\nu_k(p)$ and $F_k^{-1}(p)$. These are the main ingredient of the fixed point equations but can be used as well to compute the full \RG flow. The main difference in between these and the fixed point equations resides in the initial conditions.
We insert the truncated effective action \eq{eq:trunc} into the flow equation \eq{eq:wetterich_b} and define
\begin{align}
I_{k}[\mathbf{v}] \equiv \frac{1}{2} \text{Tr}\left[ \left(\Gamma_k^{(2)}[\mathbf{v}]+R_k\right)^{-1}k\partial_k R_k \right],
\end{align}
as the term on the right-hand side of \Eq{eq:wetterich_b} with the truncation \eq{eq:trunc} inserted. See Section (\ref{sec:def_flow_equation}) and \Eqs{eq:w_times_G} to \eq{eq:proper_definitions_2} for precise definitions. The $k$-dependence of $\nu_k(p)$ and $F_k^{-1}(p)$ can be extracted from the flow equation of $\Gamma^{(2)}_{k}(\omega,\mathbf{p})$ by expanding everything in powers of $\omega^2$. From \Eq{eq:wetterich_b} we get\footnote{We use a definition analogous to the definition of $\Gamma_{k,ij}^{(2)}(\omega,p)$ in \Eq{eq:Gamma2} for $I_{k,ij}^{(2)}(\omega,\mathbf{p})$. See \Eqs{eq:derivatives_1} and \eq{eq:inv_cumulants}.}
\begin{align}
 k \partial_k \Gamma^{(2)}_{k,ij}(\omega,\mathbf{p}) = I_{k,ij}^{(2)}(\omega,\mathbf{p}).
\label{eq:flow_intermediary}
\end{align}
The truncation \eq{eq:trunc} leads to an inverse propagator \eq{eq:Gprop}, that is diagonal in the spatial indexes $ij$. We therefore further project \Eq{eq:flow_intermediary} onto its diagonal part by taking its trace,
\begin{align}
 d \, k \partial_k \Gamma^{(2)}_{k}(\omega,\mathbf{p}) = I_{k,ii}^{(2)}(\omega,\mathbf{p}) \equiv d\, I_{k}^{(2)}(\omega,\mathbf{p}).
\label{eq:flow_integrals_def_with_d}
\end{align}
$I_{k}^{(2)}(\omega,\mathbf{p})$ is computed by taking two field derivatives of $I_{k}[\mathbf{v}]$, evaluating at $\mathbf{v}=0$, taking the trace over the field indexes and dividing by $d$. It has the following diagrammatic representation
\begin{equation}
\begin{aligned}
 \begin{tikzpicture}[scale=1]
\draw[color=black, line width = 1.5pt] (1-0.4, 0) circle (1);
\draw[color=black, line width = 1.5pt] (4.7, 0) circle (1);
\path[draw] (-0.6*0.8/0.6-0.4,0) -- (0-0.4,0);
\path[draw] (2-0.4,0) -- (2+0.6*0.8/0.6-0.4,0);
\path[draw] (4.7-0.4243*0.8/0.6,-1-0.4243*0.8/0.6) -- (4.7,-1) -- (4.7+0.4243*0.8/0.6,-1-0.4243*0.8/0.6);
\node[draw,circle, fill=gray!30] (a) at (0-0.4, 0) {3};
\node[draw,circle, fill=gray!30] (b) at (2-0.4, 0) {3};
\node[draw,circle, fill=black] (c) at (1-0.4, 1) {};
\node[draw,circle, fill=gray!30] (d) at (4.7, -1) {4};
\node[draw,circle, fill=black] (e) at (4.7, 1) {};
\node (f) at (1.8+0.8/0.6, 0) {$ \displaystyle-\frac{1}{2}$};
\node (g) at (-0.7-1.2/0.6, 0) {$ \displaystyle  I_{k}^{(2)}(\omega,\mathbf{p})=$};
\end{tikzpicture} \end{aligned} \enspace .
\label{eq:flow_3}
\end{equation}
The thick lines denote $G_{k}=(\Gamma_k^{(2)}+R_k)^{-1}$, the thin lines external momenta and frequencies, and the black dots insertions of the derivative $k\partial_{k}R_{k}$ of the regulator. The 3- and 4-vertices are given in Appendix \ref{sec:explicit_expressions_for_the_flow_integrals}, in \Eqs{eq:gamma3} and \eq{eq:gamma4}, respectively. The equations for $\nu_k(p)$ and $F_k^{-1}(p)$ are then
\begin{align}
  k\partial_k \left[F^{-1}_k(p) \nu_k(p)^2 p^4 \right]  = I_{k}^{(2)}(0,\mathbf{p}), && 
  k\partial_k F^{-1}_k(p)\,  = \frac{\partial I_{k}^{(2)}(0,\mathbf{p})}{\partial \omega^2}.
\label{eq:flow1}
\end{align}
$I_{k}^{(2)}(0,p)$ and ${\partial I_{k}^{(2)}}/{\partial \omega^2}{(0,p)}$ are computed in a straightforward although lengthy way. We refer to Appendix \ref{sec:explicit_expressions_for_the_flow_integrals} for details on their calculation and simply state the result here
\begin{align}
I_{k}^{(2)}(0,\mathbf{p}) = & -\frac{k^d}{2d} \int_{\Omega} \left[F^{-1}_k(k) F^{-1}_k(q) \nukk \nukq (\nukk+\nukq)\right]^{-1} 
\nonumber \\
& \times \Big[ F^{-1}_k(q)^2 \nukq (\nukq+\nukk) 
   \left[ d(p^2+k^2)-2 k \mathbf{e}_{\mathbf{r}} \cdot \mathbf{p}\right]  
   \nonumber \\
& \quad -\ \theta \left(q^2-k^2 \right) \Big( F^{-1}_k(q)^{2} \nukq (\nukk+\nukq) \left[d(p^2+k^2)-2k\, \mathbf{p}\cdot\mathbf{e}_{\mathbf{r}}\right]
\nonumber\\
&  \quad\ \ + F^{-1}_k(k)^{2} \nukk(\nukk+\nukq) \left[d(p^2+q^2)-2\mathbf{p}\cdot \mathbf{q}\right]
\nonumber\\
&  \quad\ \ + F^{-1}_k(p)^2  \left.\nukp\right.^2 \left[d(k^2+q^2)+2k\, \mathbf{q}\cdot\mathbf{e}_{\mathbf{r}}\right]
\nonumber\\
&  \quad\ \ - F^{-1}_k(p)F^{-1}_k(q) \nukp\nukq  2d \, \mathbf{p} \cdot \mathbf{q} 
\nonumber \\
&  \quad\ \ - F^{-1}_k(p)F^{-1}_k(k) \nukp\nukk 2d \, k\,\mathbf{p}\cdot\mathbf{e}_{\mathbf{r}} \Big) \Big],
\label{eq:explicit_flow_1} \\[1cm]
%\end{align}
%
%\begin{align}
\frac{\partial I_{k}^{(2)}}{\partial {\omega^2}}(0,\mathbf{p}) = & \, \frac{k^d}{2d} \int_{\Omega} \theta\left(q^2-k^2\right)F^{-1}_k(p) \left[F^{-1}_k(k) F^{-1}_k(q) \nukk \nukq (\nukk + \nukq)^{3} \right]^{-1}
\nonumber\\
& \times \Big( 
F^{-1}_k(p) \, \left[(\nukk+\nukq)^{2} - \left. \nukp \right.^2  \right] \left[d(q^2+k^2)+2 k\,\mathbf{q}\cdot\mathbf{e}_{\mathbf{r}}  \right]  
\nonumber \\
& \quad + 
F^{-1}_k(q) \, \nukq \left[\nukk+\nukq + \nukp  \right]  2d \, \mathbf{p} \cdot \mathbf{q} 
\nonumber \\
& \quad + 
F^{-1}_k(k) \, \nukk \left[\nukk+\nukq + \nukp \right]  2d \, k\,\mathbf{p}\cdot \mathbf{e}_{\mathbf{r}}
\Big).
\label{eq:explicit_flow_2}
\end{align}
We have used the sharp cut-off function $R_{k}(p) =  k^d z_1\tilde R_{k}(p)$, with
\begin{align}
\tilde R_{k}(p) \left\{ \begin{array}{ll}
=0 & \text{if } p \ge k \\
\to\infty & \text{if } p < k
\end{array}\right. ,
\end{align}
which makes it possible to analytically perform the radial part of the momentum integration in $I_{k}^{(2)}(\omega,\mathbf{p})$. We are left with its angular part, $\int_{\Omega}$ which is defined through\footnote{For $d=1$, this reduces to $\int_{\Omega} f(p) = [f(p)+f(-p)]/(2\pi)$.} $\int_{\vec{p}} = \int_0^\infty p^{d-1} dp \int_{\Omega}$. The vector $\mathbf{e}_{\mathbf{r}}$ which appears in \Eqs{eq:explicit_flow_1} and \eq{eq:explicit_flow_2} is of unit length and contains the remaining angular integration. It points in the direction defined by the angle $\Omega$.
We define as well $\mathbf{q} = \mathbf{p} - k \mathbf{e}_{\mathbf{r}}$ and the short-hand notation $\nuk{p}{}=\nu_{k}(p)p^{2}$. Finally note that the dependence on the full vector $\mathbf{p}$ is only apparent. \cred{Indeed}, the trace over the spatial indexes makes everything isotropic. It can be check in \Eq{eq:explicit_flow_appendix_0} that $I_{k}^{(2)}(\omega,\mathbf{p})$ only depends on $p$ and the square of the frequency $\omega^2$.

\subsection{Fixed point equations}
\label{sec:fixed_point_equations}

We now apply the fixed point conditions of Section (\ref{sec:momentum_dependent_propagator}) to the truncation \eq{eq:trunc}.
We follow precisely the reasoning of Section (\ref{sec:momentum_dependent_propagator}) and relate the equations that we write here to the corresponding equations of Section (\ref{sec:momentum_dependent_propagator}). We give short explanations of the most important features the fixed point equations here and reffer to Section (\ref{sec:momentum_dependent_propagator}) where everything is explained in great detail.

Since we have truncated the frequency dependence of $\Gamma^{(2)}_{k,ij}(\omega,\mathbf{p})$ the equations of Section (\ref{sec:momentum_dependent_propagator}) can be simplified as well. The form \eq{eq:Gprop} for the inverse propagator implies that the scaling function is
\begin{align}
 g(a) = 1 + a^2.
\end{align}
Note that it is no longer possible to impose \Eqs{eq_param_3} on the derivative of $g(a)$. The normalisation of the scaling function is however a matter of convention. We choose instead
\begin{align}
g(0)=1, & & \left.\frac{dg}{d(a^2)}\right|_{a=0}=1.
\end{align}
The parametrisation made in \Eq{eq_param_1} applied to \Eq{eq:Gprop} can be written as
\begin{alignat}{3}
 &\Gamma^{(2)}_{k}(0,p) && =  k^{d} z_1 \, \hat{p}^{\eta_1} \left[1 + \delta Z_1\left(\hat{p}\right) \right], 
\nonumber \\
 &\left.\frac{\partial \Gamma^{(2)}_{k}}{\partial {\omega^{2}}}\right|_{(\omega^2=0,p)} && =  k^{d-4} z_2 \, \hat{p}^{\eta_2} \left[1 + \delta Z_2\left(\hat{p}\right) \right].
\label{eq:parametrization}
\end{alignat}
and
\begin{align}
& \hat{p} = \frac{p}{k}, & 
& \hat{\omega} = \frac{1}{k^2} \sqrt{\frac{z_2}{z_1}} \omega, & 
& \hat{\mathbf{v}}(\hat{\omega},\hat{\mathbf{p}}) = k^{d+1} {\mathbf{v}}(\op).
\label{eq:dim_variables}
\end{align}
The parameters $\nu_k(p)$ and $F_k^{-1}(p)$ are easily related to the ones we just introduced. Comparing \Eqs{eq:Gprop} and \eq{eq:parametrization} we can directly write
\begin{alignat}{3}
& \nu_k(p) & & = \sqrt{z_1/z_2}\, \hat{p}^{(\eta_1-\eta_2-4)/2}  \sqrt{\frac{1+\delta Z_1(\hat{p})}{1+\delta Z_2(\hat{p})}}, 
\nonumber \\
& F^{-1}_k(p) & & = k^{d-4} z_2\, \hat{p}^{\eta_2} \left[1+\delta Z_2(\hat{p})\right].
\label{eq:nuOm_2_dZ}
\end{alignat}

As for the normalisation of the scaling function we have chosen slightly different definitions for the scaling pre-factors $z_i$ and exponents $\eta_i$ as compared to Section (\ref{sec:momentum_dependent_propagator}). We have the following correspondences
\begin{align}
& \eta_1 = \bar{\eta}_1, && z_1 = \bar{z}_1, \nonumber \\
& \eta_2 = \bar{\eta}_1 +\bar{\eta}_2, && z_2 = \bar{z}_1 \bar{z}_2.
\label{eq:bar_to_notbar}
\end{align}
Finally we get an additional truncation on the $\hat{\omega}$ behaviour of $\delta Z(\hat{\omega},\hat{p})$ (which is defined in \Eq{eq_param_1}) because of the frequency truncation. It is constrained to the form
\begin{align}
 \delta Z(\hat{\omega},\hat{p}) = \delta Z_1(\hat{p}) + \hat{\omega}^2 \, \hat{p}^{\eta_1-\eta_1} \, \delta Z_2(\hat{p}),
\end{align}
with \Eq{eq_param_2} becoming $\delta Z_i(\hat{p}\to \infty) = 0$.

The parametrisation \eq{eq:parametrization} of the inverse propagator is made in such a way that the latter has the two following fixed point properties. First when it is rescales with $z_1 k^d$ and expressed in terms of the re-scaled variables of \Eqs{eq:dim_variables} is loses completely its explicit dependence on the cut-off scale $k$. Secondly the property $\delta Z_i(\hp\to\infty) = 0$ ensures that in the physical limit $k\to 0$ the inverse propagator assumes a scaling form,
\begin{align}
& \Gamma^{(2)}_{k\to0}(\omega,\mathbf{p}) = k^d z_1 \, \hat{p}^\eta_1 \left(1+\hat{p}^{\eta_2-\eta_1} \, \hat{\omega}\right) = k^{d-\eta_1} z_1 \, p^{\eta_1} + k^{d-\eta_2-4} z_2 \, p^{\eta_2} \, \omega^2.
\label{eq:scaling_form_fourier}
\end{align}
As in \Eqs{eq_gamma2_kto0} and \eq{eq_scaling_zi} this expression is independent of $k$ only if
\begin{align}
&z_1 = z_{10} \, k^{\eta_1-d}, & 
& z_2 = z_{20} \, k^{\eta_2+4-d}.
\label{eq:zi}
\end{align}

We now express $I_k^{(2)}(\op)$ in terms of the re-scaled variables \eq{eq:dim_variables} and the non-scaling part of the inverse propagator $\delta Z_i(\hat{p})$,
\begin{align}
& I_{k}^{(2)}[0]\left(\omega,p\right) = {k^d} \sqrt{\frac{z_2}{z_1}} \left[\hat{I}^{(2)}_1(\hat{p}) + \hat{\omega}^2 \hat{I}^{(2)}_2(\hat{p}) + \mathcal{O}\left(\hat{\omega}^4\right)\right].
\label{eq:flow_integrals}
\end{align}
The flow integrals $\hat{I}^{(2)}_i(\hat{p})$, are obtained by inserting \Eqs{eq:nuOm_2_dZ} in \Eqs{eq:explicit_flow_1} and \eq{eq:explicit_flow_2}, dividing out all the dimensionless pre-factors $z_i$ and the dimensionfull powers of $k$ and expanding in powers of $\hat{\omega}^2$.

Note that $\hat{I}^{(2)}_i(\hat{p})$ contain the scaling exponents $\eta_i$ and the functions $\delta Z_i(\hat{p})$ but the cut-off scale as well as the re-scalling pre-factors $z_i$ have been completely extracted and are part of the pre-factors of \Eq{eq:flow_integrals}. Explicit expressions are given in Appendix \ref{sec:rescaled_flow_integrals} in \Eqs{eq:rescaled_flow_appendix_1} and \eq{eq:rescaled_flow_appendix_2} with \eq{eq:F11} to \eq{eq:F24} inserted.
We can now compare this with \Eq{eq:def_z3} and identify $\bar{z}_3 = \sqrt{z_2/z_1}$. Since there are only two free parameters in the truncation \eq{eq:trunc}, no additional parameter enters the flow equations and $\bar{z}_3$ only depends on $z_i$. \Eqs{eq:consistency_0} and \eq{eq:fp_coupling} become
\begin{align}
 3\eta_1 & = \eta_2+4+2d,
\label{eq:consistency}
\end{align}
and
\begin{align}
h & \equiv \sqrt{\frac{z_2}{z_1}} \frac{1}{z_1} .
\label{eq:defh}
\end{align}
These equations arrise form the requirement that there be no explicit cut-off dependence at the fixed point. Indeed, taking the cut-off derivative of \Eqs{eq:parametrization} and taking into acount \Eqs{eq:zi} we find that the flow equation of the inverse propagator \eq{eq:flow_intermediary} can be written as
\begin{align}
\frac{\text{d}\delta Z_i\left(\hat{p}\right)}{\text{d}\hat{p}} = -h \frac{\hat{I}^{(2)}_i(\hat{p})}{\hat{p}^{\eta_i+1}},
% && \text{or} && \delta Z_i\left(\hat{p}\right) = h \int_{\hat{p}}^\infty \text{d}y \frac{\hat{I}^{(2)}_i(y)}{y^{\eta_i+1}}.
\label{eq:flow2}
\end{align}
for $i = 1,2$. The $h$ factor that arrises is defined in terms of $z_i$ in \Eq{eq:defh}. \Eq{eq:consistency} has to be satisfied in order for $h$ not to have any explicit cut-off dependence. We see that we have only one free scaling exponent.

$h$ is the fixed point coupling of the theory.
We can see this by inserting the dimensionless re-scaled variables into the truncated effective action we find $\Gamma_k[\mathbf{v}] = \hat{\Gamma}_k[\hat{\mathbf{v}}]/h$. This shows that when $h\to 0$ all the fluctuations are damped and only the extremum of $\hat{\Gamma}_k[\hat{\mathbf{v}}]$ contributes to the path integals. The theory is classical in the sense that its fluctuations are Gaussian. On the other hand when $h$ is large all the fluctuations are important and we have a strongly interacting theory. Note that in the case of a Gaussian fixed point $h=0$ and there is no need to impose the relation \eq{eq:consistency} in between $\eta_i$. Instead we find $\delta Z_i(\hp)=0$ and the inverse propagator assumes a scaling form from the onset. The scaling exponents are then completely free.

Finally the constraint on the $k\to\infty$ limit of the fixed point equations \eq{eq_deltaZ_0} are now
\begin{align}
\delta Z_i(\hat{p}\rightarrow 0) = -1 + \frac{f_i(\hat{p})}{\hat{p}^{\eta_i}}.
\label{eq:small_p}
\end{align}
\Eqs{eq:small_p} are the consequence of the requirement that the fixed point effective action be qualititatively different in the two limits $k\to 0$ and $k\to\infty$. Indeed, we can see by inserteing them in \Eqs{eq:parametrization} that the form \eq{eq:small_p} for $\delta Z_i(\hp)$ is such that the scaling in terms of $\eta_i$ is removed and replaced by the functions $f_i(\hp)$.

\subsection{Computing observable quantities}
\label{sec:computing_observables}

Before we move on to the solutions of the \RG flow equations we relate $\eta_i$, $h$ $\nu_k(p)$ and $F_k^{-1}(p)$ to the physically observable quantities $S_2(\tau,\mathbf{r})$, $\chi$ and $z$.
The physical quantities are only related to the flowing quantities in the limit $k\to 0$. We can however compute a scale dependent velocity increment from the truncated effective action \eq{eq:trunc}.
This one reduces to the physical quantity when the cut-off is removed $k\to 0$.

First we express the second order velocity increment in terms of $\nu_k(p)$ and $F_k^{-1}(p)$. For this we write it in terms of the Fourier modes of the velocity field. Using \Eqs{eq:inverse_gamma} and \eq{eq:Gamma2_intro_inverse} we can write
\begin{align}
& \langle v_i(\omega,\mathbf{p}) v_j(\omega',\mathbf{p}') \rangle = \delta(\omega+\omega') \delta(\mathbf{p}+\mathbf{p}') \delta_{ij} \, \frac{(2\pi)^{d+1}}{F_k^{-1}(p)\left[\omega^2+\nu_k(p)^2p^4\right]}.
\label{eq:fourier_correlation_function}
\end{align}
Then we take the double Fourier transform of \Eq{eq:fourier_correlation_function} and write
\begin{align}
 \langle v_i(t+\tau,\mathbf{x}+\mathbf{r}) v_j(\tx) \rangle_k & = \, \delta_{ij} \int_{\op} \text{e}^{i\left(\omega \tau -\mathbf{p}\cdot \mathbf{r} \right)} \frac{1}{F_k^{-1}\left(\omega^2 + \nu_k(p)^2 p^4\right)} \nonumber \\[0.3cm]
& = \frac{\delta_{ij}}{2} \int_{\mathbf{p}} \text{e}^{-i\mathbf{p}\cdot\mathbf{r}} \frac{\text{e}^{-\nu_k(p)p^2 \left|\tau\right|}}{F_k^{-1}(p)\nu_k(p)p^2}.
\label{eq:1sts_residue}
\end{align}
We have used the residue theorem in order to perform the frequency integration and obtain the second equality.
We finally get
\begin{align}
 S_{2,k}(\tau,\mathbf{r}) = \bigintsss_{\mathbf{p}} \frac{1-\exp\left[-i\mathbf{p}\cdot\mathbf{r}-\nu_k(p)p^2 \left|\tau\right|\right]}{F_k^{-1}(p)\nu_k(p)p^2}.
\end{align}
Inserting \Eqs{eq:nuOm_2_dZ} and taking the limit $k\to0$ we can write
\begin{align}
 S_{2}(\tau,\mathbf{r}) = \sqrt{\frac{z_2}{z_1}} \frac{k^2}{z_2} \bigintss_{\hat{\mathbf{p}}} \frac{1-\exp\left[{-i\hat{\mathbf{p}}\cdot k\mathbf{r}}{-\sqrt{\frac{z_1}{z_2}} \, \hat{p}^{(\eta_1-\eta_2)/2} \, k^2\, \left|\tau\right|}\right]}{\hat{p}^{(\eta_1+\eta_2)/2}}.
\end{align}
The apparent $k$-dependence can be removed by changing the variable integration to $\mathbf{p}$ and inserting \Eqs{eq:zi}. Since $k$ is a free parameter we can chose it to be $k = 1/r$ and get
\begin{align}
 S_{2}(\tau,\mathbf{r}) = r^{(\eta_1+\eta_2-2d)/2} \, \sqrt{\frac{z_{20}}{z_{10}}} \frac{1}{z_{20}} \bigintss_{\hat{\mathbf{p}}}{ \frac{1-\exp\left[{-i\hat{{p}}_z}{-\sqrt{\frac{z_{10}}{z_{20}}}\,(r\hat{p})^{(\eta_1-\eta_2)/2} \, \left|\tau\right|}\right]}{\hat{p}^{(\eta_1+\eta_2)/2}}}.
\label{eq:scaling_0}
\end{align}
$\hat{p}_z$ is the $z$-component of $\hat{\mathbf{p}}$. From this we can identify
\begin{align}
& \chi = \frac{\eta_1+\eta_2-2d}{4}+1, & z = \frac{\eta_1-\eta_2}{2}.
\label{eq:eta_to_chi}
\end{align}
Inserting \Eq{eq:consistency} this can be written as
\begin{align}
& \chi = \eta_1-d, & z = -\eta_1+2+d.
\label{eq:eta_to_chi_new}
\end{align}
Note that we recover \Eq{eq:chiofeta1andorz}, $\chi + z = 2$, from the fixed point equations. This is the first non-trivial result that emerges form our approach.

Let us finally note that \Eq{eq:scaling_0} can be slightly simplified by using \Eqs{eq:defh} and \eq{eq:consistency} to express $z_2$ and $\eta_2$ in terms of $z_1$, $h$ and $\eta_1$,
\begin{align}
S_{2}(\tau,r) = \frac{r^{2(\eta_1-d-1)}}{z_{10}^2} \, \hat g\left(\frac{\tau}{r^{d+2-\eta_1}}\frac{1}{z_{10}}\right),
&& 
\hat g(x) = \frac{d}{h} \bigintsss_{\hat{\mathbf{p}}} \frac{1-\exp\left[{-{\hat{p}^{d+2-\eta_1}x}/{h }}{-i \hat{p}_z}\right] }{\hat{p}^{2\eta_1-2-d}}.
\end{align}
We see that even if all the fixed point parameters are known $z_{10}$ stays undetermined and can be fixed by comparing $S_{2}(\tau,r)$ to an experiment.

\Eq{eq:fourier_correlation_function} can be used to compute the kinetic energy of the system
\begin{align}
E_{k,\text{kin}} \equiv \frac{1}{2} \int_{\mathbf{x}} \langle v^2(\tx) \rangle_k = \frac{\mathcal{V}}{2} \int_{\op} \frac{1}{F_k^{-1}(p)\left[\omega^2+\nu_k(p)^2 p^4\right]} = \frac{\mathcal{V}}{4} \int_{\mathbf{p}} \frac{1}{F_k^{-1}(p)\nu_k(p)p^2}.
\label{eq:ekin_burgers}
\end{align}
$\mathcal{V} = \int_{\mathbf{x}}$ is the volume of the system. As in \Eq{eq:1sts_residue} we used the residue theorem to perform the frequency integration. We can as well identify the kinetic energy spectrum from \Eq{eq:ekin_burgers},
\begin{align}
 \epsilon_{k,\text{kin}}(\mathbf{p}) \equiv \frac{1}{2} \int_\omega \langle \mathbf{v}(\op)\cdot\mathbf{v}(\mop)\rangle_{k} = \frac{1}{2} \frac{1}{F_k^{-1}(p)\nu_k(p)p^2}.
\end{align}
which in the limit $k\to0$ scales as
\begin{align}
 \epsilon_{0,\text{kin}}(\mathbf{p}) = \epsilon_{\text{kin}}(\mathbf{p}) \sim p^{-(\eta_1+\eta_2)/2} = p^{d+2-2\eta_1},
\label{eq:kin_spectrum_exponent_burgers}
\end{align}
at the \RG fixed point. See Sections (\ref{sec:cascades}) or (\ref{sec:super_fluid_turbulence_and_non_thermal_fixed_points}) for a discussion of its physical interpretation.

\subsection{Asymptotic properties of the flow integrals}
\label{sec:asymptotic_properties_of_the_flow_integrals}

In this section we investigate the solutions of the fixed point equations.
In particular we look into the properties of $\hat{I}_{i}^{(2)}(\hat{p})$ for very large and very small arguments and solve the fixed point euqations in the assymptotic regimes. We consider the consequences of \Eqs{eq:flow2} with $\eta_{i}$ related through \Eq{eq:consistency} without imposing the conditions given by \Eqs{eq:small_p} yet. $\eta_{1}$ and $h$ are to be considered as free parameters first. The idea is to solve the differential equations for arbitrary values of $\eta_{i}$ and $h$ and to choose the values for which Eqs.\eq{eq:small_p} are true later. Note that the following discussion only applies to non Gaussian fixed points where $h\neq 0$ and \Eq{eq:consistency} is mandatory. We start by writing our flow integrals as\footnote{Note that the exponents of the following expression contain the Kronecker delta $\delta_{d1}$. \cred{Indeed}, cancellations in the leading pre-factors of $F_{1,3}(\hat{p})$, $F_{2,3}(\hat{p})$ and $F_{2,4}(\hat{p})$ occur for $d=1$ and the sub-leading terms must be taken into account.}
\begin{align}
\hat{I}^{(2)}_1(\hat{p}) 
&= \hat{p} \, F_{1,1}(\hat{p}) 
\nonumber\\
& \qquad + \hat{p}^{(\eta_1+\eta_2)/2+2} \sqrt{1+\delta Z_1(\hat{p})} \, \sqrt{1+\delta Z_2(\hat{p})} \, F_{1,2}(\hat{p})
\nonumber\\
& \qquad + \hat{p}^{\eta_1+\eta_2+2\,\delta_{d1}} \left[1+\delta Z_1(\hat{p})\right] \left[1+\delta Z_2(\hat{p})\right] F_{1,3}(\hat{p}), 
\label{eq:rescaled_flow_1}
\end{align}
\begin{align}
 \hat{I}^{(2)}_2(\hat{p}) 
 &= \hat{p}^{(\eta_1+\eta_2)/2+2} \sqrt{1+\delta Z_1(\hat{p})} \, \sqrt{1+\delta Z_2(\hat{p})} \, F_{2,1}(\hat{p}) 
 \nonumber \\
 &\qquad + \hat{p}^{\eta_2+2} \left[1+\delta Z_2(\hat{p})\right] F_{2,2}(\hat{p}) 
 \nonumber \\
 &\qquad + \hat{p}^{\eta_1+\eta_2 + 2\,\delta_{d1}} \left[1+\delta Z_1(\hat{p})\right] \left[1+\delta Z_2(\hat{p})\right] F_{2,3}(\hat{p})
 \nonumber \\
 &\qquad + \hat{p}^{2\eta_2+2 \, \delta_{d1}} \left[1+\delta Z_2(\hat{p})\right]^2 F_{2,4}(\hat{p}).
\label{eq:rescaled_flow_2}
\end{align}
Explicit expressions for the functions $F_{i,j}(\hat{p})$ are given in \Eqs{eq:F11} to \eq{eq:F24} of Appendix \ref{sec:rescaled_flow_integrals}. These functions were defined in such a way that they are analytic and non vanishing at $\hat{p}=0$. They can be Taylor expanded $F_{i,j}(\hat{p}\to 0) = F_{i,j}(0) + F_{i,j}'(0) \, \hat{p} + \text{O}(\hat{p}^2)$. Expressions (in terms of $\eta_1$ and $\delta Z_i(1)$) for the first term of their Taylor expansions are given in \Eqs{eq:Fij_0_1d} and \eq{eq:Fij_0_Dd} of Appendix \ref{app:small_p}. Similarly, their large $\hp$ asymptotic behaviour can be computed and turn out to be power laws. Details on how to compute their exponents and pre-factors are given Appendix (\ref{app:big_p}) and explicit expressions are given in \Eqs{eq:F11_big_p} to \eq{eq:F23_big_p} and \eq{eq:F12_big_p} to \eq{eq:F22_big_p}.

\subsubsection{Scaling limit (\texorpdfstring{$p\ll k$}{p<<k})}
\label{sec:scaling_limit_p_ll_k}

We start by looking at the limit $\hp \to 0$. Since all but the the constant terms of $F_{i,j}(\hp)$ are sub-dominant in the limit $\hat{p}\to0$ the asymptotic form of $\hat{I}^{(2)}_{i}(\hat{p}\ll1)$ only depends on $F_{i,j}(0)$. This makes apparent the asymptotic behaviour of our flow integrals as a sum of power laws multiplied by combinations of $\left[1+\delta Z_{i}(\hat{p}\ll 1)\right]$ and $F_{i,j}(0)$. We see that different asymptotic behaviours of $\delta Z_{i}(\hat{p}\ll1)$ and values of $\eta_{i}$ lead to different terms dominating. The true dominating terms must be determined self-consistently. The asymptotic form of \Eqs{eq:flow2} is
\begin{align}
\frac{\delta Z_1}{\text{d}\hat{p}}(\hp) \cong & -\frac{h}{\hp^{\eta_1}} \left\{ F_{1,1} + F_{1,2} \, \hp^{(\eta_1+\eta_2)/2+1} \sqrt{\left[1+\delta Z_1(\hp)\right] \left[1+ \delta Z_2(\hp)\right]} \right. \nonumber \\
& \qquad \left. + F_{1,3} \, \hp^{\eta_1+\eta_2+2\delta_{d1}-1} \left[1+ \delta Z_1(\hp)\right] \left[1+ \delta Z_2(\hp) \right] \vphantom{\sqrt{\left[1+\delta Z_1(\hp)\right] \left[1+ \delta Z_2(\hp)\right]}}\right\}, \nonumber \\
\frac{\delta Z_2}{\text{d}\hat{p}}(\hp) \cong & -\frac{h}{\hp^{\eta_2}} \left\{ F_{2,1} \, \hp^{(\eta_1+\eta_2)/2+1} \, \sqrt{\left[1+ \delta Z_1(\hp) \right]\left[1+\delta  Z_2(\hp)\right] }  \right. \nonumber \\
& \qquad  \left. + F_{2,2} \, \hp^{\eta_2+1} \left[1+\delta Z_2(\hp) \right] + F_{2,3} \, \hp^{\eta_1+\eta_2 +2\delta_{d1}-1}\, \left[1+\delta Z_1(\hp)\right] \left[1+\delta Z_2(\hp)\right] \right. \nonumber \\
& \qquad  \left. + F_{2,4}\, \hp^{2\eta_2+2\delta_{d1}-1} \left[1+\delta Z_2(\hp)\right]^2 \right\},
\label{eq:asymptotic_eq}
\end{align}
with the short-hand notation $F_{i,j} = F_{i,j}(0)$. The asymptotic behaviour of $\delta Z_{i}(\hp)$ can be extracted from these equations with the method of dominant balance \cite{Bender1999}. This is however a lengthy process which requires the solution of 12 different sets of coupled differential equations since each combinations of terms must be considered separately. We instead make the further assumption that both functions behave as power laws at small momenta. More precisely, we will use the form
\begin{align}
 \delta Z_{i}(\hp \to 0) \cong c_i + \frac{A_i}{\alpha_i-\eta_i} \hp^{\alpha_i-\eta_i},
 \label{eq:def_alphas}
\end{align}
which is consistent with \Eq{eq:small_p} only when $c_i = -1$ and $f_i(\hat{p}) \cong A_i \hat{p}^{\alpha_i}/(\alpha_i-\eta_i)$. Note that this is a self consistent assumption. \cred{Indeed}, it gives $\delta Z_i(\hp\to0) \sim \hp^{\text{min}(0,\alpha_i-\eta_i)}$ which implies that we have power laws on both sides of \Eq{eq:asymptotic_eq}. Moreover since this expression is motivated by the differential equation (\ref{eq:flow2}), $\alpha_i$ is really defined through\footnote{We use the notation $\delta Z_i'(\hat{p}) = \text{d} \delta Z_i(\hat{p})/\text{d}\hat{p}$.} $\delta Z_i'(\hp\to0) \sim \hp^{\alpha_i-\eta_i-1}$ and $c_i$ is an integration constant. Then the case $\alpha_i -\eta_i = 0$ is to be understood as $\delta Z_{i}(\hp \to 0) \sim c_i + A_i \log(\hp)$.

We can now insert this ansatz into \Eq{eq:asymptotic_eq} and match the exponents on both sides. The asymptotic form $\delta Z_i(\hp\to0) \sim \hp^{\text{min}(0,\alpha_i-\eta_i)}$ implies that we need to consider the four cases given by $\alpha_i-\eta_i < 0$ and $\alpha_i-\eta_i > 0$ separately if we want to extract values of $\alpha_i$ that provide solutions to \Eqs{eq:flow2}. Note that the cases $\alpha_i-\eta_i = 0$ correspond to logarithmic divergences of $\delta Z_i(\hat{p})$ and should be treated separately as well.

\begin{figure}[t]
\begin{center}
\begin{tikzpicture}[scale=5]
\path[draw,dashed,line width=20/9pt] (-2/9,2/9) -- (0,4/9);
\path[draw,line width=20/9pt] (-31/90,1/10) -- (-2/9,2/9);
\path[draw,line width=20/9pt] (0,4/9) -- (1/15,23/45);
\draw[fill,white] (-3/9,4.5/9) rectangle (6/9,4.5/9+1/60);
\draw[fill,white] (-3/9-1/60,-2.5/9) rectangle (-3/9,4.5/9);
\draw (-3/9,-2.5/9) rectangle (6/9,4.5/9);
\path[draw,line width=10/9pt,color=green] (-2/9,1/9) -- (5/9,1/9);
\draw[fill,color=blue] (-2/9,1/9) circle (0.15/9);
\path[draw,color=blue, line width=20/9pt] (-2/9,1/9) -- (-2/9,-2/9);
\draw[fill,color=red] (5/9,1/9) circle (0.15/9);
\draw[fill,color=blue] (-2/9,-2/9) circle (0.15/9);
\draw[fill,color=black] (0,0) circle (0.15/9);
\draw[fill,color=black] (0,1/9) circle (0.15/9);
\path[draw,color=black, line width=20/9pt] (0,0) -- (0,1/9);
\path[] (-2/9,-2.5/9) node [below] {$-2$} -- (1.5/9,-2.5/9) node [below] {\begin{tabular}{c}  \\ $\alpha_2$ \end{tabular}} -- (0,-2.5/9) node [below] {$0$} -- (2/9,-2.5/9) node [below] {$2$} -- (4/9,-2.5/9) node [below] {$4$} -- (6/9,-2.5/9) node [below] {$6$};
\path[] (-1/3,-2/9) node [left] {$-2$} -- (-1/3,0) node [left] {$0$} -- (-1/3,1/9) node [left] {\begin{tabular}{cc} $\alpha_1$ & \end{tabular}} -- (-1/3,2/9) node [left] {$2$} -- (-1/3,4/9) node [left] {$4$};
\end{tikzpicture}
\end{center}
\caption{A graphical representation of the values of the bare exponents $\alpha_1$ and $\alpha_2$, defined in \Eqs{eq:def_alphas} in blue and red for $d=1$ and black and red for $d \neq 1$, which characterise the non-Gaussian fixed points.
These different solutions are connected by a thin horizontal green line which marks the solutions of \Eqs{eq:alphas_2nd_case_equation} where \Eqs{eq:small_p} can not be satisfied. Note that it stops at the upper black dot when $d\neq 1$ and at the upper blue for $d=1$. It connects the solutions of \Eq{eq:eq_alphas} in all dimensions.
The scaling exponents of the original stochastic Burgers equation, as described by $S\brv$, \Eqs{eq:Burgersaction} or \eq{eq:Burgersaction_fourier}, are related by $\alpha_1 = \alpha_2 +4$ and are shown as a black line. 
The blue dot at $(\alpha_1,\alpha_2) = (1,-2)$ corresponds to the fixed point investigated in \cite{Kloss2012a,Medina1989a}, while the top half of the blue line (for $\alpha_1 >0$) represents the set of points found in \cite{Medina1989a} for different types of forcing, corresponding to a bare action with exponents given by the dotted black line.
In all dimensions we find a continuum of fixed points, and a further, new fixed point that may arise if the fluid is forced on large scales (red dot). Along the blue line we find that the scaling exponent satisfies $\eta_1 = 2-\alpha_1/2$ while it is $\eta_1 = 2-\alpha_1/2+d$ along the black line. $\eta_2$ is related to $\eta_1$ through \Eq{eq:consistency}.}
\label{fig:alphas}
\end{figure}
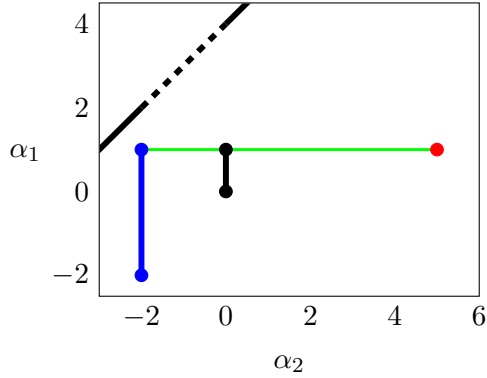

\begin{itemize}
\item We first consider the case of both $\alpha_i-\eta_i < 0$. Then the constants $c_i$ are sub-dominant and do not need to be taken into account on the right-hand-side of \Eqs{eq:asymptotic_eq}. Note that this case also arises when both $c_i = -1$, \ie at solutions of the full \RG fixed-point equations. We see that, $\eta_{i}$ do not appear any more in the exponents. $\alpha_{i}$ must satisfy
\begin{align}
\alpha_1
&  =  \text{min}\left(1,\frac{\alpha_1}{2}+\frac{\alpha_2}{2}+2,\alpha_1+\alpha_2+2\delta_{d1}\right),
 \nonumber \\
\alpha_2
& = \text{min}\left(\frac{\alpha_1}{2}+\frac{\alpha_2}{2}+2,\alpha_2+2,\alpha_1+\alpha_2+2\delta_{d1},2\alpha_2+2\delta_{d1}\right).
\label{eq:eq_alphas}
\end{align}
This implies that $\alpha_{i}$ are restricted to
\begin{align}
&\left(\alpha_1,\alpha_2\right) \in \left(\left[-2\delta_{d1},1\right],-2\delta_{d1}\right), \nonumber \\
&\left(\alpha_1,\alpha_2\right) = \left(1,5\right).
\label{eq:alphas}
\end{align}
See Figure (\ref{fig:alphas}) where these solution are plotted in red and black (red and blue for $d=1$) vertical lines and big dots.
This asymptotic solutions can only arise if $\alpha_i-\eta_i < 0$ or $c_{i}  = -1$.

\item Next we consider the case $\alpha_1-\eta_1 < 0$ and $\alpha_2-\eta_2 > 0$. In this case, matching the exponents on both sides of \Eqs{eq:asymptotic_eq} leads to
\begin{align}
 \alpha_1
&  =  \text{min}\left(1,\frac{\alpha_1}{2}+\frac{\eta_2}{2}+2,\alpha_1+\eta_2+2\delta_{d1}\right),
 \nonumber \\
\alpha_2
& = \text{min}\left(\frac{\alpha_1}{2}+\frac{\eta_2}{2}+2,\eta_2+2,\alpha_1+\eta_2+2\delta_{d1},2\eta_2+2\delta_{d1}\right).
\label{eq:alphas_2nd_case_equation}
\end{align}
Then we get
\begin{align}
& \alpha_1 \in \left\{ \begin{array}{cc} \left\{1\right\} & \text{if } \eta_2 > -2\delta_{d1} \\
\left]-\infty,1\right] & \text{if } \eta_2 = -2\delta_{d1} \\
\varnothing & \text{if } \eta_2 < -2\delta_{d1} \end{array} \right. ,
\label{eq:alphas_2nd_case}
\end{align}
and $\alpha_2$ is a function of $\alpha_1$ and $\eta_2$ given by the second line of \Eqs{eq:alphas_2nd_case_equation}. Inserting \Eq{eq:consistency} we can express $\eta_2$ in terms of $\eta_1$ in \Eq{eq:alphas_2nd_case} and check that we always get solutions that satisfy $\alpha_1-\eta_1 < 0$. However second condition for this case to be realised $\alpha_2-\eta_2 > 0$ restricst the solution further. \cred{Indeed}, we have $\alpha_2-\eta_2 > 0$ only for $\eta_2 \in \left]-2\delta_{d1},5\right[$. This in turn implies that the only consistent solutions are
\begin{align}
 & \eta_2 \in \left]-2\delta_{d1},5\right[, \nonumber \\
& \alpha_1 = 1 , \nonumber \\
& \alpha_2 = \text{min}\left(\frac{5}{2}+\frac{\eta_2}{2},\eta_2+2,\eta_2+1+2\delta_{d1},2\eta_2+2\delta_{d1}\right).
\end{align}
These solutions are plotted on Figure (\ref{fig:alphas}) as a horizontal thin green line. Note that they seem to connect the different solutions of \Eqs{eq:eq_alphas}.
\item The same analysis can be done for all the other cases: $\alpha_1-\eta_1 > 0$, $\alpha_2-\eta_2 < 0$, both $\alpha_i-\eta_i > 0$ and logarithmic divergences. None of these cases has a consistent solution.
\end{itemize}

The second case $\alpha_1-\eta_1 < 0$ and $\alpha_2-\eta_2 > 0$ provides an asymptotic solutions to the differential equation (\ref{eq:flow2}). However, it requires that
\begin{align}
-1+\delta Z_2(\hp \to 0) \sim {const} \neq 0.
\end{align}
If ${const} = 0$ then the power law, $-1+\delta Z_2(\hp \to 0) \sim \hp^{\alpha_2-\eta_2}$ becomes dominant again and we are back to the first case. In other words we have found two sets of solutions of the differential equations (\ref{eq:flow2}) but only one of them can contain solutions of \Eqs{eq:small_p} as well. We focus on these solutions in the remainder of this Section.

The above results suggest that the allowed combinations ($\alpha_1,\alpha_{2})$ correspond to the different possible non-Gaussian fixed points of the theory.
For the different dimensions, we find, for the scaling exponents relevant in the limit $p\ll k$, a connected interval for $\alpha_{1}$ and an additional point at $\left(\alpha_1,\alpha_2\right) = \left(1,5\right)$.

Having identified the different fixed points of the \RG flow we can now study their properties. In principle this can be done by numerically solving \Eqs{eq:flow2} for different values of $\eta_1$ and $h$ and tuning both parameters to satisfy \Eqs{eq:small_p}. Note that \Eq{eq:consistency} makes $\eta_2$ a simple function of $\eta_1$. We instead look a little deeper into the analytic properties of the fixed point equations. \cred{Indeed}, we show now that the value of the scaling exponents $\eta_i$ at the different fixed points can be inferred by looking at the $\hp \ll 1$ limit of \Eqs{eq:flow2}.

Along the lines of fixed point given by $(\alpha_1,\alpha_2) = (]-2\delta_{d1},1[,-2\delta_{d1})$ and shown in red and black in Figure (\ref{fig:alphas}) we can relate the physical scaling exponents $\eta_i$ to the \UV exponent $\alpha_1$ in a simple way.
Inserting \Eqs{eq:def_alphas} and keeping only the dominating terms we have seen that the right hand side of \Eqs{eq:asymptotic_eq} is a linear combination of monomials. Since we assume that $\alpha_i-\eta_i < 0$ the exponents of these monomials only depend on $\alpha_i$.
In fact the different exponents are given by the different arguments of the minimum functions on the right-hand side of \Eqs{eq:eq_alphas} minus $(1+\eta_i)$ (see \Eqs{eq:monomials} in Appendix \ref{sec:equations_for_ai}). We see that once we have chosen values of $\alpha_i$ from the solutions of \Eqs{eq:eq_alphas} we can compare the different exponents and identify the leading monomial in the limit $\hp \to 0$.
It then becomes possible to match the pre-factors on both sides of \Eqs{eq:asymptotic_eq} and extract equations containing both $A_i$. These are given in Appendix \ref{sec:equations_for_ai} in \Eqs{eq:eq_ai_first_dD} to \Eqs{eq:eq_ai_last_dD} and \Eqs{eq:eq_ai_first_1d} to \Eqs{eq:eq_ai_last_1d} for $d=1$ with the short-hand notation, $a_i = A_i (\alpha_i-\eta_i)$.
Contrarily to \Eqs{eq:eq_alphas} these are not closed and can not be solved independently of \Eqs{eq:flow2}.
However, for the interval sets of $\alpha_{i}$ marked by the black and blue lines in Figure (\ref{fig:alphas}), the ratio of the equation for $A_i$, \Eqs{eq:aclosed_1d} and \eq{eq:aclosed_Dd}, only contains $\alpha_1$ and $\eta_i$ and makes it possible to relate both quantities.
Taking the ratio of \Eqs{eq:aclosed_1d} and using the fact that $F_{1,3}=F_{2,4}$ gives
\begin{align}
\frac{a_1}{a_2} \frac{\eta_1-\alpha_1}{\eta_2 +2} = \frac{a_1}{a_2},
\label{eq:ratioaiEqs1d}
\end{align}
for $d=1$ and the ratio of \Eqs{eq:aclosed_Dd} gives
\begin{align}
\frac{a_1}{a_2} \frac{\eta_1-\alpha_1}{\eta_2} = \frac{a_1}{a_2},
\label{eq:ratioaiEqsDd}
\end{align}
for $d\neq1$.
We see that $a_i$ drop out of these equations and that, inserting \Eq{eq:consistency}, $\eta_1$ can be related to $\alpha_1$,
\begin{align}
\eta_1 = \left\{\begin{array}{ll}
2-\alpha_1/2, & d=1\\
2-\alpha_1/2+d, & d\neq 1
\end{array}\right. .
\label{eq:eta12alpha1}
\end{align}
Specifically, for $d = 1$, we get $3/2<\eta_1<3$ and, for $d \neq 1$, $3/2+d<\eta_1<2+d$. These intervals are marked by dark grey shading in Figure (\ref{fig:eta_range}). 
%This regime, for $d\not=1$, is disjunct with the regime where a direct energy cascade may occur.
%Only for $d=1$, it allows for such a cascade, but does not require it.
In view of Refs.~\cite{Janssen1999a,Kloss2013a}, we note that we find an upper bound to the regime of allowed $\eta_{1}$.
Specifically, \Eq{eq:eta_to_chi_new} implies that above this bound, the dynamical critical exponent would become negative.
It is not possible to relate $\eta_1$ to $\alpha_i$ at the other solutions of \Eq{eq:eq_alphas} in a similar way. At these points, the full solution of \Eq{eq:flow2} is necessary to verify the existence of the RG fixed points and extract the values of the $\eta_{i}$.

Note that there is a qualitative difference in between the intervals of fixed points, $(\alpha_1,\alpha_2) = (]-2\delta_{d1},1[,-2\delta_{d1})$ and their end points $(\alpha_1,\alpha_2) = (\{-2\delta_{d1},1\},-2\delta_{d1})$. Along the blue and black lines of Figure (\ref{fig:alphas}) only one of the monomials of the right-hand side of \Eq{eq:asymptotic_eq} dominates while at the end points two terms contribute to the asymptotic behaviour (see Appendix \ref{sec:equations_for_ai} and \Eqs{eq:monomials}). As a result \Eq{eq:eta12alpha1} does not apply at the end points since the corresponding equations do not simplify. The values of $\eta_i$ is still undetermined.
However, under the assumption that $\eta_1$ is a continuous function of $\alpha_1$, one obtains $\eta_1 = 2-\alpha_1/2 + d - \delta_{d1}$ also at the end points. 
We remark that the calculation reported in Ref.~\cite{Medina1989a} provides a continuum of fixed points, with  $\eta_1 = 2 - \alpha_1/2$ and $0<\alpha_1 <1$, and an additional fixed point with $\eta_1 = 3/2$, the end-point value.

Finally let us remark that none of these resulting small-$\hat{p}$ scalings corresponds to that of the action $S\brv$ for Burgers' equation.
In $S\brv$, $\nu$ is $p$-independent, which implies that the ratio of {$\Gamma_{k \to \infty}^{(2)}(0,p)$ and $\partial \Gamma_{k \to \infty}^{(2)}/\partial \omega^2 (0,p)$} scales as $p^4$, see \Eq{eq:Gprop}. In the bare limit $k\to\infty$ we have
\begin{align}
\Gamma_{k \to \infty}^{(2)}(0,\mathbf{p}) \cong \frac{A_1}{\alpha_1-\eta_1} \, p^{\alpha_1},
& & \left. \frac{\partial \Gamma_{k \to \infty}^{(2)}}{\partial \omega^{2}}\right|_{(0,\mathbf{p})} \cong \frac{A_2}{\alpha_2-\eta_2}  \, p^{\alpha_2}.
\label{eq:G_ktoinf}
\end{align}
Hence, taking the ratio of \Eqs{eq:G_ktoinf} and requiring it to scale as in the bare case gives $\alpha_1 = \alpha_2 +4$.
Analogously, \Eq{eq:forcing_corr_beta} implies that {$\partial \Gamma_{k \to \infty}^{(2)}/\partial \omega^2 (0,p)$} scales as $p^{-\beta}$ and thus that $\alpha_2 = - \beta$.
The resulting possible combinations $(\alpha_{1},\alpha_{2})=(4-\beta,-\beta)$ are marked by the black (dashed/solid) line in Figure (\ref{fig:alphas}).
As expected, there is no choice of the forcing exponent $\beta$ that makes the stochastic Burgers equation sit at a non-Gaussian RG fixed point for all values of $k$.

Finally note that the relatively large value of $\alpha_2=5$ at the isolated fixed point suggests that it is realised for a highly non local forcing, $\beta = -5$.

\subsubsection{Scaling limit (\texorpdfstring{$p\gg k$}{p>>k})}
\label{sec:scaling_limit_p_gg_k}

In the opposite limit of vanishing cut-off all fluctuations are integrated out and the full effective theory emerges. 
We have seen in Section (\ref{sec:uv_divergent_fixed_points}) that two possibilities arise. Either $\delta Z_i(\hat{p} \to \infty) = 0$ or $\delta Z_i(\hat{p} \to \infty) = \infty$.
In the first case, once the cut-off scale is sent to zero, see \Eq{eq:parametrization}, one finds fixed points with correlations given by scaling functions across all momentum scales.
If, on the other hand, $\lim_{\hat{p}\to\infty} \delta Z_i(\hat{p}) = \infty$, an \RG fixed point can only exist if the scaling range is restricted to momenta smaller than some upper cut-off $\Lambda$. 
Then scaling only arises within the range of physical momenta and $\delta Z_i(1\ll \hat{p}< \Lambda/k) \cong 0$. 
In this situation of an \UV divergent fixed point, the theory is not well defined exactly at the fixed point but the latter can be approached arbitrarily by choosing $\Lambda$ accordingly large. See Section (\ref{sec:uv_divergent_fixed_points}) for a detailed discussion.

\begin{figure}[t]
\begin{subfigure}[b]{0.5\textwidth}
\centering
\begin{tikzpicture}[scale=0.7]
\draw[line width=2pt,color = blue] (-0.05,3.1) -- (7/5,1/5) -- (3,5) -- (5.05,9.1);
\draw[line width=2pt,color = black,dashed] (-0.050,7.1) -- (3,1) -- (5,7) -- (6.05,9.1);
\draw[line width=0.5pt,color = red] (0,0) -- (6,6);
\draw[fill,color = white] (-0.1,0) rectangle (0,9);
\draw[fill,color = white] (6,0) rectangle (6.15,9.15);
\draw[fill,color = white] (0,-0.05) rectangle (6,0);
\draw[fill,color = white] (0,9) rectangle (6,9.15);
\draw (0,0) rectangle (6,9);
\path[] (0,0)  node [below] {$0$} -- (1,0) node [below] {$1$} -- (2,0) node [below] {$2$} -- (3,0) node [below] {$3$} -- (3.5,0) node [below,align=center] {\vphantom{$3$}\\$\eta_1$} -- (4,0) node [below] {$4$} -- (5,0) node [below] {$5$} -- (6,0) node [below] {$6$};
\path[] (0,1) node [left] {$1$} -- (0,2) node [left] {$2$} -- (0,3) node [left] {$3$} -- (0,4) node [left] {$4$} -- (0,4.5) node [left] {$\beta_1$ \hphantom{$4$}} -- (0,5) node [left] {$5$} -- (0,6) node [left] {$6$} -- (0,7) node [left] {$7$} -- (0,8) node [left] {$8$} -- (0,9) node [left] {$9$};
\end{tikzpicture}
\end{subfigure}
\begin{subfigure}[b]{0.4491\textwidth}
\centering
\begin{tikzpicture}[scale=0.2589]
\draw[line width=2pt,color = blue] (-7.05,-10.05) -- (-9/2,-15/2) -- (3,5) -- (9,11) -- (10.05,37/3+1/15);
\draw[line width=2pt,color = black,dashed] (-7,-12) -- (-11/2,-21/2) -- (5,7) -- (9+3/80,37/3+0.05);
\draw[line width=0.5pt,color = red] (-7,-7) -- (10,10);
\draw[fill,color = white] (-7.15,-12) rectangle (-7,37/3);
\draw[fill,color = white] (-7,-12.1) rectangle (10,-12);
\draw[fill,color = white] (-7,37/3) rectangle (10,37/3+0.1);
\draw[fill,color = white] (10,-12) rectangle (10.15,37/3+0.15);
\draw (-7,-12) rectangle (10,37/3);
\path[draw] (-7,-12) -- (-6,-12) node [below] {$-6$} -- (-3,-12) node [below] {$-3$} -- (0,-12) node [below] {$0$} -- (3,-12) node [below] {$3$} -- (6,-12) node [below] {$6$} -- (9,-12) node [below] {$9$} -- (10,-12);
\path[draw] (-7,-12) node [left] {$-12$} -- (-7,-9) node [left] {$-9$} -- (-7,-6) node [left] {$-6$} -- (-7,-3) node [left] {$-3$} -- (-7,3) node [left] {$3$} -- (-7,6) node [left] {$6$} -- (-7,9) node [left] {$9$} -- (-7,12) node [left] {$12$} -- (-7,37/3);
\path[] (1.5,-12) node [below,align=center] {\vphantom{$0$}\\$\eta_2$} -- (-7,0) node [left] {$\beta_2$ {$0$}};
\end{tikzpicture}
\end{subfigure}
\caption{Plots of $\beta_1$ (left) and $\beta_2$ (right) with respect to $\eta_1$, $\eta_2$ respectively for $d=1$ (blue solid lines) and $d=3$ (black dotted lines). $\eta_i$ is plotted on both plots as a thin red line. We see that we can only have $\beta_i-\eta_i<0$ for a finite range of $\eta_1$. As always $\eta_2$ is related to $\eta_1$ through \Eq{eq:consistency}.}
\label{fig:beta}
\end{figure}
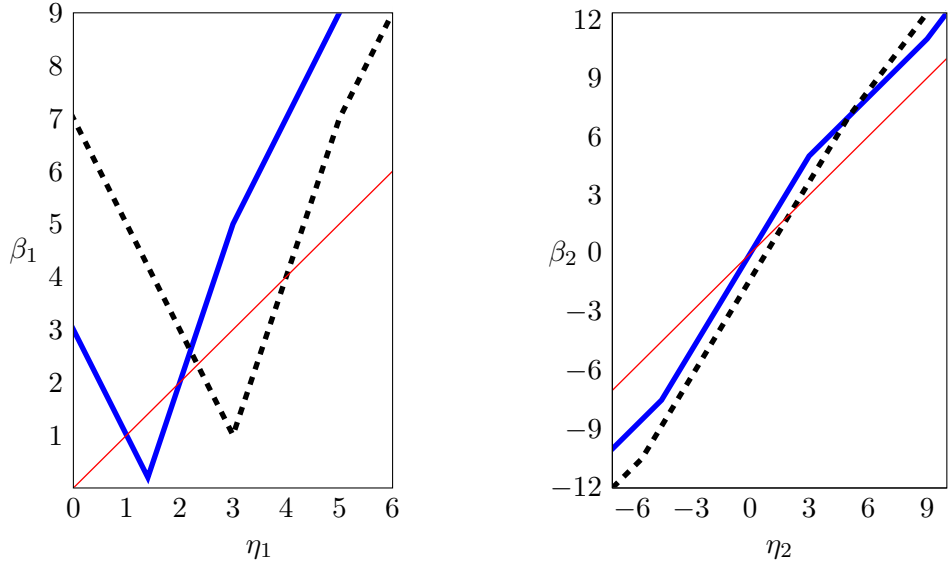

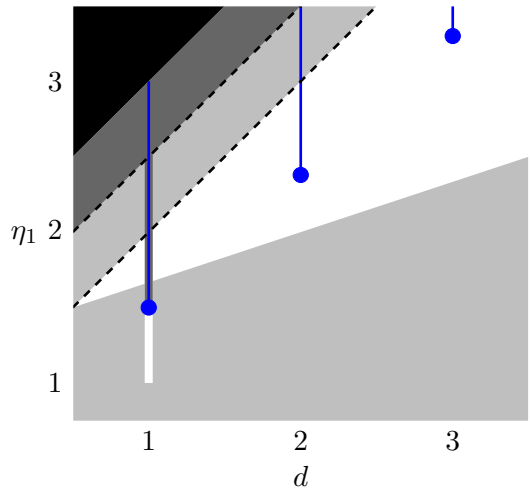
\begin{figure}[h!t]
\begin{center}
\begin{tikzpicture}[scale=2]
\path[fill,color=gray!50] (1/2,3/2) -- (5/2,7/2) -- (1/2,7/2) -- cycle;
\path[fill,color=gray!50] (1/2,3/2) -- (7/2,5/2) -- (7/2,3/4) -- (1/2,3/4) -- cycle;
\path[fill,color=black] (1/2,5/2) -- (3/2,7/2) -- (1/2,7/2) -- cycle;
%\shade[shading=axis,bottom color=blue!50,top color=transparent!0,shading angle=45]
%  (1,3/2) -- (2,2.379) -- (3,3.3) -- (3.2172,7/2) -- (11/4,7/2) -- (1,7/4) -- cycle;  
%\path[draw,color=blue, line width=3pt] (1,3/2) -- (2,2.379) -- (3,3.3) -- (3.2172,7/2);
\path[draw,color=white, line width=3pt] (1,2) -- (1,1);
\path[draw,color=gray!120, line width=3pt] (1,3/2) -- (1,2.8);
\path[fill,color=gray!120, line width=2pt] (1/2,2) -- (2,7/2) -- (3/2,7/2) -- (1/2,5/2) -- cycle;
\path[draw,color=blue, line width=1pt] (1,3/2) -- (1,3);
%\path[draw,line width=1pt,color=blue,decoration={shade change, start shade = 100, end shade = 0},decorate] (1,3/2) -- (1,3.5);
\path[draw,color=blue, line width=1pt] (2,2.379) -- (2,3.5);
\path[draw,color=blue, line width=1pt] (3,3.3) -- (3,3.5);
\draw[fill,color=blue] (1,3/2) circle (0.05);
\draw[fill,color=blue] (2,2.379) circle (0.05);
\draw[fill,color=blue] (3,3.300) circle (0.05);
\draw[dashed,line width=1pt] (1/2,3/2) -- (5/2,7/2);
\draw[dashed,line width=1pt] (1/2,2) -- (2,7/2);
\draw[fill,color=white] (1/2,1/2) rectangle (1/2-1/40,7/2);
\draw[fill,color=white] (1/2,7/2) rectangle (7/2+2/40,7/2+2/40);
\draw[fill,color=white] (7/2,3/4) rectangle (7/2+2/40,7/2+2/120);
\path[] (1,3/4) node [below] {$1$} -- (2,3/4) node [below,text width=1cm,align=center] {$2$ \\ $d$} -- (3,3/4) node [below] {$3$};
\path[] (1/2,1) node [left] {$1$} -- (1/2,2) [left] node {$\eta_1$ $2$} -- (1/2,3) [left] node {$3$};
\end{tikzpicture}
\end{center}
\caption{The range of values of $\eta_{1}$ (defined in \Eqs{eq:parametrization} and related to $\chi$ and $z$ in \Eqs{eq:eta_to_chi_new}), which correspond to UV convergent non-Gaussian fixed-points for different spatial dimensions $d$ (white area).
The white vertical stripe at $d=1$ only applies to this single dimension.
In the regions shaded in light and dark grey any potential fixed point is UV divergent such that the RG flow integrals must be regularised in the UV.
The blue dots correspond to the average literature values for the exponent $\eta_{1}$, \cite{Tang1992a,Ala-Nissila1993a,Castellano1999a,Marinari2000a,Aarao2004a,Ghaisas2006a,Kelling2011a}.
Their values, as given in Table I. of Ref.~\cite{Kloss2012a}, are given in \Eq{eq:KPZ_literature}. They correspond to a delta correlated forcing in the KPZ framework, \ie the exponent $\beta$, defined in \Eq{eq:forcing_corr_beta}, is equal to two.
The vertical blue lines marks the range of possible exponents found in \cite{Kloss2013a} for different forcings, with the exponent $\beta<2$.
It is not known where this set of fixed points ceases to exist.
The dark-grey shaded area marks the range of values of $\eta_1$ that we find at the sets $(\alpha_{1},\alpha_{2})$ shown as the blue and black solid lines at $\alpha_{2}=-2$ and $0$ in Figure (\ref{fig:alphas}), respectively.
It continues accordingly also for $\eta_{1}>7/2$.
In between the two dotted lines there is a direct cascade of kinetic energy. 
In the black area at the left top, the dynamical critical exponent $z$ is negative.}
\label{fig:eta_range}
\end{figure}

Here, we consider \UV finite fixed points and take the limit $\hat{p}\to \infty$ in the flow integrals. 
The boundary condition $\delta Z_i(\hat{p}\rightarrow \infty)=0$ then allows us to write \Eqs{eq:flow2} in the integral form 
\begin{align}
& \delta Z_i\left(\hat{p}\right)
= h \int_{\hat{p}}^\infty \text{d}y \, \frac{\hat{I}^{(2)}_i(y)}{y^{\eta_i+1}}.
\label{eq:flow3}
\end{align}
For $\hat{p}\gg 1$, also the integration variable $y$ exceeds $1$ by far such that we can approximate $\delta Z_i(y)=0$ in the integrals $\hat{I}^{(2)}_i(y)$. 
The flow integrals can be further approximated by keeping only their leading term as $\hat{p} \to \infty$, 
\begin{align}
\hat{I}^{(2)}_i(\hat{p}\to\infty) \sim \hat{p}^{\beta_{i}}.
\label{eq:large_p}
\end{align}
The exponents $\beta_{i}$ are determined in Appendix \ref{app:big_p}. We give the final result here,
\begin{align}
& \beta_1 = \left\{ 
 \begin{array}{ll} 
 4-2\delta_{d1}-2\eta_1+d  &\mbox{if}\ (6+3d-2\delta_{d1})/5  \geq \eta_1  \\[0.1cm]
 3\eta_1-2d-2  & \mbox{if}\ (6+3d-2\delta_{d1})/5 < \eta_1 \leq d+2 \\[0.1cm]
 2\eta_1-d  &\mbox{if}\ \eta_1 > d+2 
\end{array} \right. ,
\label{eq:beta1_text}
\\[0.5cm]
& \beta_2  = \left\{ \begin{array}{ll} 
 3(\eta_1 -d-2)  &\mbox{if}\  \eta_1 \leq d/2  \\[0.1cm]
5\eta_1-4d-6 &\mbox{if}\  d/2 < \eta_1 \leq d+2 \\[0.1cm]
3\eta_1-2d -2 &\mbox{if}\  d+2 < \eta_1 \leq d+2 + 2 \delta_{d1} \\[0.1cm]
4\eta_1-3d-4 -2\delta_{d1} &\mbox{if}\  d+2 + 2 \delta_{d1} < \eta_1 
\end{array} \right. .
\label{eq:beta2_text}
\end{align}
These are piece-wise affine functions of $\eta_1$. They are plotted for $d=1$ (blue solid lines) and $d=3$ (black dashed lines) in Figure (\ref{fig:beta}) as an example.
Note that these equations are strictly valid only when $\beta_i-\eta_i<0$ since they were computed with the assumption $\delta Z_i(\hp \to \infty) =0$.
In order to obtain finite integrals on the right hand side of \Eqs{eq:flow3}, it is necessary that $\beta_i < \eta_i$. We see from Figure (\ref{fig:beta}) that this requirement restricts $\eta_1$ (and $\eta_2$) to a finite range. A careful analysis of \Eqs{eq:beta1_text} and \eq{eq:beta2_text} shows that this range is given by
\begin{align}
\left. \begin{array}{cc} 
d=1: & d \\ 
d\neq 1: & ({d+4})/{3} \end{array} \right\} < \eta_1 < d+1 ,
\label{eq:range}
\end{align}
which is shown as the white area in Figure (\ref{fig:eta_range}).

As we discussed in the beginning of this section \Eq{eq:range} limits the range of \UV convergent fixed points. Outside of the white area of Figure (\ref{fig:eta_range}) the flow integrals need to be regularised in the \UV before they can be used to extract fixed point properties. The introduction of such an \UV cut-off introduces a scaling range where the fixed point properties are realised. In such cases the fixed point theory is not well defined since it contains an infinity in the \UV . It can however be approached arbitrarily close by taking a large enough value of the \UV cut-off.

Figure (\ref{fig:eta_range}) contains the range of \UV convergent fixed points as well as the values of $\eta_1$ that can be computed form the literature on \KPZ equation.
The blue dots correspond to a potential\footnote{Note that this is work about the \KPZ equation. The forcing is a potential field. See \Eq{eq:forcing_corr_potential} where $U(p)$ is defined.} delta correlated in space, \ie $\beta = 2$. These values of $\eta_1$ are average values from the \KPZ literature (see \cite{Tang1992a,Ala-Nissila1993a,Castellano1999a,Marinari2000a,Aarao2004a,Ghaisas2006a,Kelling2011a} and Table I. of Ref.~\cite{Kloss2012a}). They are given by,
\begin{align}
&\eta_{1}=3/2 && \text{for }d=1,\nonumber \\
&\eta_{1}=2.379(15) && \text{for }d=2,\nonumber \\
&\eta_{1}=3.300(12)&& \text{for }d=3.
\label{eq:KPZ_literature}
\end{align}
The vertical blue lines are the fixed points found in \cite{Kloss2013a} for values of $\beta > \beta_{\text{trans}}(d)$. See the discussion at the end of Section (\ref{sec:scaling_and_correlation_functions}) and \Eq{eq:chi_arbitrary_beta} for more details.
%The vertical blue line at $d=1$ are the fixed points found in \cite{Medina1989a} which correspond to $0\leq\beta\leq 2$. The authors of \cite{Janssen1999a,Kloss2013a} found that as long as $\beta$ is large enough the \RG flow is attracted to the "local" fixed point with $\beta=2$. Then the average literature values apply. However when beta is lowered below a given $\beta_{\text{trans}}(d)$ the non-local fixed point is realised and we have $\eta_1 = (4+2d-\beta)/3$. The value of $\beta_{\text{trans}}(d)$ is determined by the fact that $\eta_1$ is a continuous function of $\beta$. For very small values of $\beta$, we have $\eta_1 = (4+2d-\beta)/3$ and $\eta_1$ is very large. As $\beta$ is increased $\eta_1$ decreases until it becomes equal to the corresponding average literature value which are given for $d=1,2$ and $3$ in Figure (\ref{fig:eta_range}). Then $\eta_1$ stays constant as $\beta$ is increased further. The situation is the same as for $d=1$ (See \Eq{eq:exponent_1d_medina}) apart from the fact $\beta_{\text{trans}}(d\neq1)$ is not known exactly.
The question of weather these lines continue for arbitrarily small (and negative) values of $\beta$ (\ie arbitrarily large values of $\eta_1$) is not addressed in \cite{Kloss2013a} and only mentioned briefly in \cite{Janssen1999a}. It is however clear that something must happen because as $\beta$ is decreased the relation\footnote{Again, see the discussion at the end of Section (\ref{sec:scaling_and_correlation_functions}) and \Eq{eq:chi_arbitrary_beta} where the results of \cite{Kloss2013a} are summarised.} $\eta_1 = (4+2d-\beta)/3$ implies that the dynamical critical exponent $z$ decreases and even reaches zero for $\beta = -d-2$. The black area of Figure (\ref{fig:eta_range}) corresponds to negative values of $z$.

We see that the local \KPZ fixed points (blue dots) are \UV convergent.
For $d=1$ the analysis of Section (\ref{sec:scaling_limit_p_ll_k}) is in good agreement with the results of \cite{Kloss2013a}.
We find a continuum of fixed point with $\eta_1 > 3/2$ and an additional fixed point at the edge of this continuum. Strictly speaking we can not infer the value of $\eta_1$ at the blue dots, but it is reasonable to assume that it is a continuous function of $\alpha_1$. Then we obtain $\eta_1 = 3/2$ on the blue dot at $d=1$. The same structure was found in \cite{Kloss2013a}. There the value $\eta_1 = 3/2$ corresponds to the singular short-range fixed point which is realised for $\beta \geq 3/2$. And the values $\eta_1 > 3/2$ to the set of non-local fixed points with $\beta < 3/2$.

For $d\neq1$ however \Eq{eq:eta12alpha1} provides fixed points which are \UV divergent. Moreover, based to the different values of $\eta_1$ that they find, the authors of \cite{Kloss2013a} seem to find more fixed points than we do. Our calculation has however the advantage of explicitly excluding any fixed point with $z<0$.
We explain these discrepancies by the fact that our truncation is not consistent with \KPZ equation for $d\neq 1$ since it can not describe an irrotational fluid. Even though they have overlapping values of $\eta_1$, the fixed points that we find are most likely different from the ones of \cite{Kloss2013a} which are purely irrotational.

\subsubsection{Implications for driving and turbulent cascades}
\label{sec:cascades}

The bounds \eq{eq:range} on $\eta_1$ have distinct physical interpretations: 
While the lower bound can be expressed as a regularity condition on the type of forcing that is sampled by the stochastic process, the upper bound marks the onset of a direct energy cascade.

\paragraph{Locality of the forcing}\hspace{0pt}\\

Let us discuss first the lower bound at $\eta_1 = (d+4)/3$ and $\eta_1 = 1$ for $d=1$. As the forcing is a Gaussian random variable, it follows the probability distribution
\begin{align}
 P_{k}[\mathbf{f}] = \frac{1}{N} \exp\left[-\frac{1}{2}\int_{\omega,\mathbf{p}} \left|\mathbf{f}(\omega,\mathbf{p})\right|^2 F_k^{-1}(p)\right].
\end{align}
$N$ is a normalisation factor.
This distribution implies that the probability of a spatially local force field $\mathbf{f}(t,\mathbf{x}) \sim \delta(\mathbf{x}-\mathbf{x}_{f})$ can be finite if $\int_{\mathbf{p}} F_k^{-1}(p)$ is finite and is necessarily zero otherwise.
Furthermore, for the latter integral to be finite, the critical exponent at an \UV finite fixed point needs to fulfil $\eta_1<(d+4)/3$, as one finds by inserting the parametrisation \Eq{eq:nuOm_2_dZ} for $F_{k}^{-1}$. 
We conclude that in the lower grey shaded area of Figure (\ref{fig:eta_range}),  $\eta_1 < (d+4)/3$, local Gaussian forcing of the type $\mathbf{f}(t,\mathbf{x})\sim \delta(\mathbf{x}-\mathbf{x}_{f})$ is included while it is suppressed for $\eta_1 > (d+4)/3$.
We have shown above that \UV finite non-Gaussian fixed points require $\eta_1 > (d+4)/3$.
Hence, for an RG fixed point to be \UV finite, the forcing needs to be sufficiently regular in space-time, specifically $\lim_{p\to\infty}\left|f(\omega,\mathbf{p})\right| = \lim_{\omega\to\infty}\partial_\omega\left|f(\omega,\mathbf{p})\right|=0$.

Note that, in the case $d=1$, the regularity condition is modified. 
The fixed points are \UV finite above the relatively lower limit $\eta_1 > 1$.
This is a consequence of the fact that in $d=1$ dimension, point-like shocks are stable solutions. 
Applying a force such as $\mathbf{f}(t,\mathbf{x}) \sim \delta(\mathbf{x}-\mathbf{x}_{s})$ is comparable to inserting a shock at the position $\mathbf{x}_{s}$.

%\hspace{0pt}\newline
\paragraph{Energy cascades}\hspace{0pt}\\

The upper bound, $\eta_1 = d+1$, can be related to the appearance of a direct cascade of energy.
We briefly sketch the argument leading to this result in the following.
We look at the kinetic energy spectrum, $\epsilon_{\text{kin}}(q)$, which is defined through the kinetic energy density of the system
\begin{align}
E_{\text{kin}} \equiv \frac{1}{2} \int_{\mathbf{x}} \langle v^2(\tx) \rangle = \mathcal{V} \int_{\mathbf{q}} \epsilon_{\text{kin}}(q),
&& \epsilon_{\text{kin}}(q) = \frac{1}{2} \langle \mathbf{v}(t,\mathbf{q}) \cdot \mathbf{v}(t,-\mathbf{q})\rangle.
\label{eq:kin_spectrum}
\end{align}
$\mathcal{V} = \int_{\mathbf{x}}$ is the volume of the system.
$\epsilon_{\text{kin}}(q)$ characterises the distribution of kinetic energy over the different Fourier modes of the stationary state.
Note that we consider a uniform system and abuse the notation by defining,
\begin{align}
\langle \mathbf{v}(t,\mathbf{q}) \cdot \mathbf{v}(t,\mathbf{q}')\rangle \equiv \delta(\mathbf{q}+\mathbf{q}') (2\pi)^d \, \langle \mathbf{v}(t,\mathbf{q}) \cdot \mathbf{v}(t,-\mathbf{q})\rangle .
\label{eq:abuse_notation}
\end{align}
By definition the kinetic energy density as well as the kinetic energy spectrum are time-independent in a stationary state.
Then one can compute the time derivative of $\epsilon_{\text{kin}}(p)$, insert the equation of motion \eq{eq:burgers} and equate this to zero. We find
\begin{align}
\partial_t \epsilon_{\text{kin}}(q) = - \epsilon_{\nu}(\mathbf{q}) + \epsilon_{f}(\mathbf{q}) + \epsilon_{\text{adv}}(\mathbf{q}) = 0,
\label{eq:steady_state}
\end{align}
with\footnote{The notation of \Eq{eq:abuse_notation} is used here as well for $\langle \mathbf{f}(t,\mathbf{q}) \cdot \mathbf{v}(t,-\mathbf{q}) \rangle$.}
\begin{align}
& \epsilon_{\nu}(\mathbf{q}) =  \nu q^2 \, \langle \mathbf{v}(t,\mathbf{q}) \cdot \mathbf{v}(t,-\mathbf{q})\rangle, \nonumber \\
&\epsilon_{f}(\mathbf{q}) = \frac{1}{2} \left[ \langle \mathbf{f}(t,\mathbf{q}) \cdot \mathbf{v}(t,-\mathbf{q}) \rangle + \langle \mathbf{v}(t,\mathbf{q}) \cdot \mathbf{f}(t,-\mathbf{q}) \rangle \right],
\end{align}
and
\begin{align}
\epsilon_{\text{adv}}(\mathbf{q}) =  \frac{i}{2} \int_{\mathbf{p}} \mathbf{p} \cdot \left\{ \langle \mathbf{v}(t,\mathbf{q}-\mathbf{p}) \left[\mathbf{v}(t,\mathbf{p}) \cdot \mathbf{v}(t,-\mathbf{q}) \right]\rangle + \langle \mathbf{v}(t,-\mathbf{q}-\mathbf{p}) \left[\mathbf{v}(t,\mathbf{p}) \cdot \mathbf{v}(t,\mathbf{q})\right]\rangle\right\}.
\end{align}
Note that we use the equivalent of the notation of \Eq{eq:abuse_notation} for the three-point function here (see \Eq{eq:def_av_v^2}). This makes possible the study of fluxes of energy at the steady state. $\epsilon_{\nu}(\mathbf{q})$ and $\epsilon_{f}(\mathbf{q})$ represent the energy dissipation and injection rates respectively while $\epsilon_{\text{adv}}(\mathbf{q})$ is the amount of kinetic energy that arrives at the Fourier mode $\mathbf{q}$ from the other modes. $\epsilon_{f}(\mathbf{q})$ is hard to estimate but $\epsilon_{\nu}(\mathbf{q})$ and $\epsilon_{\text{adv}}(\mathbf{q})$ are both expressed in terms of velocity field correlation functions and can be computed from the effective average action. See Section (\ref{sec:1pi_effective_action}) and \Eqs{eq:inverse_gamma} and \eq{eq:three_point_function}. $\epsilon_{f}(\mathbf{q})$ is then determined by \Eq{eq:steady_state}. In terms of $F_k^{-1}(p)$ and $\nu_k(p)$ we have,
\begin{align}
\epsilon_{k,\nu}(\mathbf{q}) = \frac{d\nu}{2} \frac{1}{F_{k}^{-1}(q)\nu_{k}(q)},
\end{align}
and
\begin{align}
 \epsilon_{k,\text{adv}}(\mathbf{q})
 & = \frac{1}{8}\int_{\mathbf{p}} \frac{1}{\nuk{p}{}+\nuk{q}{}+\nuk{\left|\mathbf{p}-\mathbf{q}\right|}{}} \left[ \frac{dp^2 + \mathbf{p} \cdot \left(\mathbf{q}-\mathbf{p}\right)}{F_k^{-1}(\left|\mathbf{p}-\mathbf{q}\right|)F_k^{-1}(p) \, \nuk{\left|\mathbf{p}-\mathbf{q}\right|}{}\nuk{p}{}} \right. \nonumber \\
& \qquad \left. + \frac{p^2-\mathbf{p} \cdot \mathbf{q}}{F_k^{-1}(q)F_k^{-1}(p) \, \nuk{q}{} \nuk{p}{}} + \frac{\mathbf{p} \cdot (\mathbf{q}-\mathbf{p}) - d \mathbf{p} \cdot \mathbf{q}}{F_k^{-1}(\left|\mathbf{p}-\mathbf{q}\right|)F_k^{-1}(q) \, \nuk{\left|\mathbf{p}-\mathbf{q}\right|}{} \nuk{q}{}} \right] \nonumber \\
& \qquad + \frac{1}{8}\int_{\mathbf{p}} \frac{1}{\nuk{p}{}+\nuk{q}{}+\nuk{\left|\mathbf{p}+\mathbf{q}\right|}{}} \left[ \frac{dp^2-\mathbf{p} \cdot \left(\mathbf{p}+\mathbf{q}\right)}{F_k^{-1}(\left|\mathbf{p}+\mathbf{q}\right|)F_k^{-1}(p) \, \nuk{\left|\mathbf{p}+\mathbf{q}\right|}{}\nuk{p}{}} \right. \nonumber \\
& \qquad \left. + \frac{p^2+\mathbf{p} \cdot \mathbf{q}}{F_k^{-1}(q)F_k^{-1}(p) \, \nuk{q}{} \nuk{p}{}} + \frac{d \mathbf{p} \cdot \mathbf{q}-\mathbf{p} \cdot (\mathbf{p}+\mathbf{q})}{F_k^{-1}(\left|\mathbf{p}+\mathbf{q}\right|)F_k^{-1}(q) \, \nuk{\left|\mathbf{p}+\mathbf{q}\right|}{} \nuk{q}{}} \right].
\label{eq:energy_transfer}
\end{align}
The physical correlation functions are recovered in the limit $k\to0$. As in the case of \Eqs{eq:explicit_flow_1} and \eq{eq:explicit_flow_2} we use the short-hand notation $\nuk{p}{}=\nu_{k}(p)p^{2}$. We see that we are able to identify the kinetic energy transport kernel. Since our system is isotropic all the quantities of this section only depend on the norm of momenta. We define the energy transport kernel $E_{k,\text{trans}}(p,q)$ through,
\begin{align}
E_{k,\text{adv}}(q) \equiv q^{d-1} \int_{\Omega} \epsilon_{k,\text{adv}}(\mathbf{q}) = \int_{0}^{\infty} \text{d}p \, E_{k,\text{trans}}(p,q),
\end{align}
which is invariant under rotations by construction.
We have simply averaged $\epsilon_{k,\text{adv}}(\mathbf{q})$ over all the directions in which $\mathbf{q}$ may point and multiplied by the surface factor $q^{d-1}$. $E_{k,\text{adv}}(q)$ represents the total kinetic energy transfer to the momentum shell of radius $q$.
$E_{k,\text{trans}}(p,q)$ is then the double angular average of the integrand of \Eq{eq:energy_transfer} multiplied by two surface factors $(qp)^{d-1}$.
It measures the transfer of kinetic energy from the momentum shell $p$ to $q$. It characterises the flux of energy across the different scales of the system.

In the limit $k\to0$, $E_{0,\text{trans}}(p,q)\equiv E_{\text{trans}}(p,q)$ is a function of $h$ and $\eta_1$ only.
Evaluating $E_{\text{trans}}(p,q)$ in this way, a direct energy cascade, \iec transport which is local in momentum space on a logarithmic scale, can be identified numerically for $d+1<\eta_1<d+3/2$.
In this regime, $E_{\text{trans}}(p,q)$ is non-vanishing only for $p\cong q$ (locality), positive for $p<q$ and negative for $p>q$ (positive directionality), and $E_{\text{trans}}(p,q)\simeq-E_{\text{trans}}(p,-q)$ (balance of driving and dissipation, \iec inertial turbulent transport).

Note that it is natural to have a direct cascade requiring an \UV regulator:
Physically, a cascade is realised only in a given inertial range. 
For example, in $3d$ Kolmogorov turbulence, energy is injected on the largest scales and transported to smaller scales by the non-linear dynamics which leads to larger eddies feeding into smaller ones. 
The kinetic energy is dissipated into heat once it reaches the end of the inertial range set by the viscosity. 
At an RG fixed point, the inertial range by definition extends over all momenta.
Hence, the \UV cut-off of the dissipation scale is absent.
As a result, energy in a direct cascade is transported to infinitely large momenta, leading to a \UV divergence of the fixed-point theory.

\chapter{Ultra-cold Bose Gases}
\label{sec:ultracold_bose_gases}
\ResetAbbrevs{All}

In this section we consider the dynamics of dilute gases of bosons with a contact interaction.
We start by summarising relevant results concerning out-of-equilibrium steady states in closed systems. We introduce the concept of \NTFPs and their scaling properties in Section (\ref{sec:super_fluid_turbulence_and_non_thermal_fixed_points}). In Section (\ref{sec:driven_dissipative_gross_pitaevskii_equation}) we introduce the \SGPE as a model for a dilute gas of Bosons in contact with external reservoirs of particles and energy. The inclusion of driving and dissipation into the closed system enables the mapping of such a system onto the stochastic Burgers equation which was introduced in Part \ref{sec:burgers_turbulence}. Finally we assume that both approaches, the closed \NTFP and the open \SGPE, describe the same out-of-equilibrium steady state in Section (\ref{sec:resultsSWT}) and apply the results of Section (\ref{sec:frg_calculation}) in order to extract non-trivial scaling relations in the context of out-of-equilibrium Bose gases at a \NTFP. In particular we show that \Eq{eq:chiofeta1andorz} which 
originates in the Galilei invariance of the hydrodynamic theory has a dual expression in terms of the scaling exponents of the Bose gas in Section (\ref{sec:gallilee_invariance_and_kolmogorov_scaling}) and \Eq{eq:kappa_and_z}, and we use the values of $\chi$ and $z$ known form the \KPZ literature for $d=1$, $2$ and $3$ to compute anomalous correlation to the scaling of the compressible kinetic energy spectrum of the Bose gas, see Section (\ref{sec:resultsAcoustic}), \Eqs{eq:anomalousLit} and Figure (\ref{fig:anomalous_simulations}).

In order to describe the ultra-cold Bose gas we consider a field theory defined by the action
\begin{align}
S = \int_{\tx} \, \left[\frac{i}{2} \left( \psi^* \partial_t \psi - \psi \partial_t \psi^* \right) - \frac{1}{2m} \vec{\nabla}\psi \cdot \vec{\nabla}\psi^* + \mu \, \psi \psi^* - \frac{g}{2} \, \left(\psi \psi^* \right)^2\right],
\label{eq:action_gpe}
\end{align}
and bosonic commutation relations. Here and in the following we set $\hbar = 1$.
$g$ is the interaction constant which is proportional to the boson-boson s-wave scattering length, $m$ is the mass of the particles and $\mu$ is the chemical potential.
Outside of thermal equilibrium there is no thermodynamic reservoir and $\mu$ is not precisely defined. There is however a fixed average density of particles which is fixed by a the Lagrange multiplier $\mu$, $\langle n \rangle = f(\mu)$. Then calling $\mu$ "chemical potential" is an abuse of language. What we actually mean is the inverse of $f$, $\mu \equiv f^{-1}(\langle n \rangle)$.
The particular form of the interaction potential that we consider $g \left(\psi \psi^* \right)^2$ applies in the dilute limit where the inter-particle distance is much larger than their scattering length. Only the contact interaction remains.

The full quantum dynamics of the theory can be expressed as a path integral as in Section (\ref{sec:Burgers_functionals}) by means of the Schwinger-Keldysh formalism \cite{Schwinger1961a,KadanoffBaym1962a,Keldysh1964a,Rammer2007a,kita2010}.
At the mean field level the dynamics follow directly from the Euler-Lagrange equations and are described by the \GPE \cite{Gross1961a,Pitaevskii1961a} (see \cite{Gasenzer2009a} for an overview),
\begin{align}
 i \partial_t \psi = \left[-\frac{1}{2m} \nabla^2 - \mu + g \left|\psi \right|^2 \right] \psi.
 \label{eq:gpe}
\end{align}
Because of the Bose condensation the low energy modes are strongly occupied and classical fluctuations are much stronger than quantum fluctuations. Then it becomes meaningful to make a semi-classical approximation where the quantum fluctuations are handled approximately (see \eg \cite{Berges:2007ym,Brewczyk2007a,Blakie2008a,Polkovnikov2010a,davis2013a}). In particular the c-field methods achieve this by introducing an \UV cut-off which separates the phase space into a classical (low energy, high occupation number) and quantum (high energy, small occupation number) regions.

Within this set-up one can write a Fokker--Planck equation for the time evolution of the Wigner quasi-probability distribution (see \eg \cite{Walls1994a,Gardiner2004a}) and project it on the classical region. When the terms containing three field derivatives of the Wigner distribution are neglected, the time evolution of the different correlation functions can be expressed in terms of stochastic differential equations.
Then the field expectation value in the classical region is described in terms of the mean field equation (\GPE) with additional collision integrals to model interactions with the quantum region which is assumed to be at thermal equilibrium. See \cite{Blakie2008a,davis2013a} and references therein for an overview.
When the classical region is highly occupied one can neglect its coupling with the quantum region. Quantum fluctuations then are included in the mean field description by sampling initial conditions from the initial Wigner distribution and integrating the otherwise deterministic \GPE. Averaging over the initial conditions enables the calculation of fluctuating observables. This is the truncated-Wigner approximation \cite{Steel1998b,Sinatra:GPE}. It gives good results as long as the low energy modes of the theory are strongly occupied while the high energy are not. See \eg \cite{Sinatra2002,Polkovnikov2003a} where the domain of validity of this approximation is discussed.

Note that requiring strong occupation numbers for the low energy modes is somewhat incompatible with the diluteness assumption made on the interaction of the bosons. The physics that we describe here is in the middle region where both approximations are valid. We have enough particles for the semi-classical approximation to hold but not to much for the contact interaction to loose its applicability.

\section{Super-fluid turbulence and non-thermal fixed points}
\label{sec:super_fluid_turbulence_and_non_thermal_fixed_points}

As in Section (\ref{sec:burgers_turbulence}) we are mainly interested here in non-equilibrium steady states.
We focus, in particular, on \NTFPs. See \cite{Berges:2008wm,Berges:2008sr,Scheppach:2009wu} or \cite{Vinen2006a,Nowak:2013juc} and references therein for overviews about quantum turbulence and \NTFPs.
%Such fixed points arise when conservation laws prevent the rapid thermalisation of a system.
Such fixed points arise when an initial inhomogeneous distribution of conserved charges prevents the rapid thermalisation of the system. \cred{Indeed}, if the initial state of the system contains spatially separated regions of positive and negative charge both parts need to mix before the uniform thermal state can be reached.
The differently charged particles need to cross the whole volume of the system before they can meet and annihilate.
We will see later in this Section that the conserved quantity that prevents the thermalisation in the case of the ultra-cold Bose gas is the angular momentum and that the entities that need to meet in order to annihilate are opposite sign vortices.

\begin{figure}[t]
\begin{center}
\includegraphics[width=0.8\textwidth]{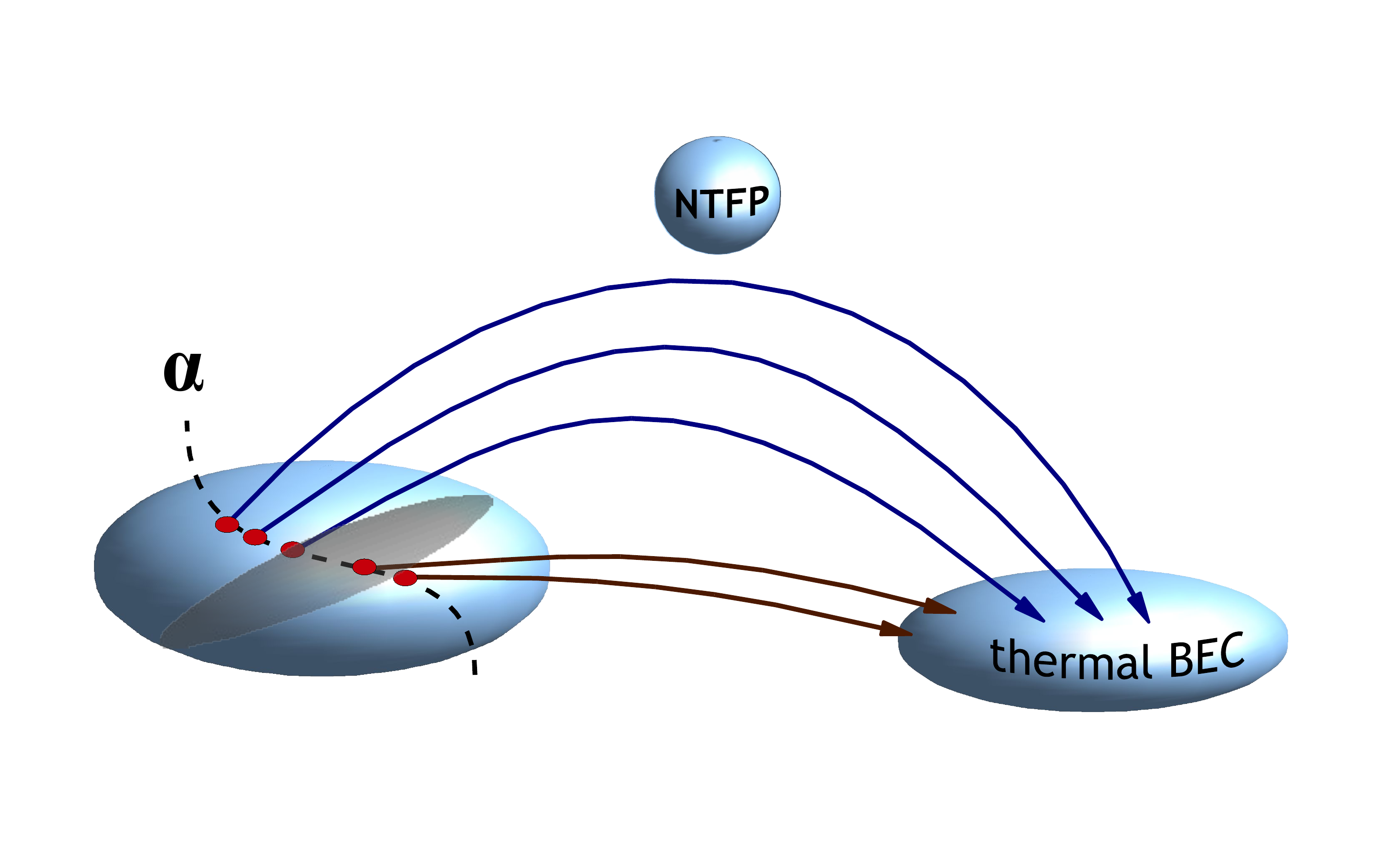}
\end{center}
\caption{Two very different paths to thermal equilibrium \cite{Nowak:2012gd}. The initial states are taken from the region on the left-hand side. The latter is seperated in two regions. If the system is initiated on the right of the grey separation it relaxes to thermal equilibrium. On the other hand if we give it initial conditions from the left it is attracted to the NTFP.}
\label{fig:NTFP}
\end{figure}

Another way of thinking of \NTFPs is in terms of dynamical systems. Let us assume that the state of the system during the out-of-equilibrium dynamics can be characterised by a vector in $\mathds{R}^n$. Then the only stable and attractive fixed point of the time evolution map corresponds to thermal equilibrium.
However, if there is another fixed point with only a few repulsive directions and if the initial state of the system lies very close to the basin of attraction of such a fixed point then the time evolution will lead towards it and stay there for as long as it takes for the relevant directions to take over and bring the system to thermal equilibrium.
We see that two scenarios emerge. Either the initial state of the system is far enough from the basin of attraction of the \NTFP and it relaxes directly to its thermal state or it is attracted to the \NTFP in the initial stage of its time evolution and stays there for a very long time before its thermalisation (see Figure (\ref{fig:NTFP})). A striking property that arises in both cases is that the state of the system during the (quasi-)stationary state is not sensitive to the particular choice of the initial conditions. We expect the properties of the system to be universal. In particular we will see that physical quantities exhibit scale invariance at \NTFPs as in the case of driven-dissipative classical turbulence (see Section (\ref{sec:scaling_and_correlation_functions})).

Such \NTFPs were identified by means of a strong-wave-turbulence analysis on the basis of \TwoPI dynamic field equations \cite{Berges:2008wm,Berges:2008sr,Scheppach:2009wu} as well as semi-classical field simulations \cite{Berges:2008mr,Nowak:2010tm,Gasenzer:2011by,Nowak:2011sk,Schmidt:2012kw,Schole:2012kt,Nowak:2012gd,Karl:2013mn,Nowak:2013juc,Berges:2013eia,Gasenzer:2013era,Karl:2013kua,Berges:2013lsa} within the truncated-Wigner approximation. Before we discuss the outcome of these calculations let us introduce the observables that we use to probe the system.

As in the case of Burgers turbulence we focus on two-point correlation functions. The time ordered Greens function\footnote{$\mathcal{T}$ is the time ordering operator. It moves the operator with the largest time argument to the left.},
\begin{align}
 G(\tau,\mathbf{r}) = \mathcal{T}\left(\begin{array}{cc} \langle \psi(t+\tau,\mathbf{x}+\mathbf{r}) \psi(t,\mathbf{x}) \rangle_{\text{c}} & \langle \psi(t+\tau,\mathbf{x}+\mathbf{r}) \psi^\dagger(t,\mathbf{x}) \rangle_{\text{c}} \\
\langle \psi^\dagger(t+\tau,\mathbf{x}+\mathbf{r}) \psi(t,\mathbf{x}) \rangle_{\text{c}} & \langle \psi^\dagger(t+\tau,\mathbf{x}+\mathbf{r}) \psi^\dagger(t,\mathbf{x}) \rangle_{\text{c}} \end{array} \right),
\end{align}
contain all the two body information. It is usually decomposed,
\begin{align*}
& G(\tau,\mathbf{r}) = F(\tau,\mathbf{r}) - \frac{i}{2} \, \text{sign}(\tau) \rho(\tau,\mathbf{r}),
\end{align*}
in term of the statistical and spectral functions $F(\tau,\mathbf{r})$ and $\rho(\tau,\mathbf{r})$ which have direct physical interpretations (see \eg \cite{Berges:2004yj} for more details). Here and in the following we describe the steady-state properties of the system. The correlation functions therefore only depend on the relative coordinates $(\tau,\mathbf{r})$. In terms of field expectation values $F(\tau,\mathbf{r})$ and $\rho(\tau,\mathbf{r})$ can be written as,
\begin{align}
F_{ab}(\tau,\mathbf{r}) = \frac{1}{2}\langle \left\{\psi_a(t+\tau,\mathbf{x}+\mathbf{r}), \psi_b(t,\mathbf{x})\right\} \rangle_{\text{c}}, &&
\rho_{ab}(\tau,\mathbf{r}) = i\langle \left[\psi_a(t+\tau,\mathbf{x}+\mathbf{r}), \psi_b(t,\mathbf{x})\right]\rangle,
\end{align}
with $\psi_1 = \psi$, $\psi_2 = \psi^\dagger$ and $[\cdot,\cdot]$ and $\{\cdot,\cdot\}$ the commutator and anti-commutator of two fields respectively. Note that the statistical function $F(\tau,\mathbf{r})$ which we just introduced has nothing to do with the forcing correlation function of \Eq{eq:forcing_corr}. We will not talk about the forcing correlation function in this section. $F(\tau,\mathbf{r})$ is always the statistical function.

The kinetic energy of the system can be computed from $F(\tau,\mathbf{r})$,
\begin{align}
E_{\text{kin}} \equiv -\frac{1}{2m} \int_\mathbf{x} \langle \psi^{\dagger}(\tx) \nabla^2\psi(\tx) \rangle = -\frac{\mathcal{V}}{2m}\operatorname{Tr}\left(\nabla^2 F(0,\mathbf{0})\right).
\end{align}
$\mathcal{V} = \int_{\mathbf{x}}$ is the volume of the system. $E_{\text{kin}}$ can be decomposed into its Fourier modes in order to define the kinetic energy spectrum. Taking the Fourier transform of the complex field $\psi$ we can write\footnote{We have inserted $\langle \psi^\dagger(t,\mathbf{p}) \psi(t,\mathbf{p}') \rangle \sim \delta(\mathbf{p}+\mathbf{p}')$ which is a consequence of the invariance of the system under space translations. See Section (\ref{sec:notations_and_conventions}), \Eq{eq:def_av_v^2} for a precise definition.}
\begin{align}
{E_{\text{kin}} \equiv \frac{\mathcal{V}}{2m} \int_{\mathbf{p}} p^2 \, \langle \psi^\dagger(t,\mathbf{p}) \psi(t,-\mathbf{p}) \rangle \equiv \mathcal{V} \int_{\mathbf{p}} \epsilon_{\text{kin}}(\mathbf{p})}.
\end{align}
We have defined
\begin{align}
\epsilon_{\text{kin}}(\mathbf{p}) & = \frac{p^2}{2m} \langle \psi^\dagger (t,\mathbf{p}) \psi (t,-\mathbf{p}) \rangle = \frac{1}{2m} \int_{\mathbf{x}} \text{e}^{i\mathbf{p}\cdot\mathbf{x}} \, \langle \boldsymbol{\nabla}\psi(t,\mathbf{x}) \cdot \boldsymbol{\nabla}\psi^\dagger(t,\mathbf{0})\rangle \nonumber \\
& = -\frac{1}{2m} \int_{\mathbf{x}} \text{e}^{i\mathbf{p}\cdot\mathbf{x}} \, \nabla^2F(0,\mathbf{x}).
\label{eq:epsilon_kin_bose}
\end{align}
Physically $\epsilon_{\text{kin}}(\mathbf{p})$ is the amount of kinetic energy stored in the Fourier mode $\mathbf{p}$. It can be used as a measure of the energy content of the different scales of the system.
We define the particle density spectrum in a completely analogous way. The number of particles is
\begin{align}
{N = \int_\mathbf{x} \langle \psi^{\dagger}(\tx) \psi(\tx) \rangle \equiv \mathcal{V} \int_{\mathbf{p}} \langle \psi^\dagger(t,\mathbf{p}) \psi(t,-\mathbf{p}) \rangle \equiv \mathcal{V} \int_{\mathbf{p}} n(\mathbf{p})}.
\end{align}
We see that both spectra are related by $\epsilon_{\text{kin}}(\mathbf{p}) = p^2 n(\mathbf{p})/(2m)$.

We now briefly summarise the results of Ref.~\cite{Scheppach:2009wu}. We will build upon these in Section (\ref{sec:resultsSWT}).
Stationary scaling solutions for the statistical and spectral two-point correlators,
\begin{align}
F\left(s^z \omega,s\mathbf{p}\right) = s^{-2-\kappa} F\left(\omega,\mathbf{p}\right),
&& \rho\left(s^z \omega,s\mathbf{p}\right) = s^{-2+\eta} \rho\left(\omega,\mathbf{p}\right),
\label{eq:scaling_bose}
\end{align}
respectively, were predicted by means of a non-perturbative wave-turbulence analysis of the \TwoPI dynamic equations for these correlation functions.

The critical behaviour is characterised by the exponents $\kappa$ and $\eta$, as well as the dynamical exponent $z$.
$\eta$ is an anomalous critical exponent which determines the deviation of the spectral scaling from the free behaviour.
In Ref.~\cite{Scheppach:2009wu} two possible solutions were found, corresponding to different strong-wave-turbulence cascades, with scaling exponents
\begin{align}
 \kappa_{\mathrm{P}} = d+2z-\eta_{\mathrm{P}}, 
&& \kappa_{\mathrm{Q}} = d+z-\eta_{\mathrm{Q}},
\label{eq:NTFPkappa}
\end{align}
between $\kappa$, $\eta$, $z$, and $d$.
$(\kappa_{\mathrm{P}},\eta_{\mathrm{P}})$ correspond to an energy cascade while $(\kappa_{\mathrm{Q}},\eta_{\mathrm{Q}})$ reflect a quasi-particle cascade in the wave turbulent system.
Both represent \NTFPs of the non-equilibrium Bose gas.
The scaling of the statistical correlation function $F$ implies scaling of particle density and kinetic energy spectra $n(\mathbf{p})=\int d\omega F(\omega,\mathbf{p})$ and $\epsilon_{\text{kin}}(\mathbf{p}) = p^2 n(\mathbf{p})$:
\begin{align}
\epsilon_{\text{kin}}(\mathbf{p}) \sim p^{-\xi}, && n(\mathbf{p})\sim p^{-\xi-2}, && \text{with}\quad \xi=\kappa-z.
\label{eq:nscaling2PI}
\end{align}
Assuming $\eta_{\mathrm{Q}}=0$, the distribution $n_{\mathrm{Q}}(\mathbf{p}) \sim p^{-d-2}$ corresponds, for $d=2,\,3$ to the scaling of the flow field $v\sim r^{-1}$ with the distance $r$ from a vortex core \cite{Nore1997a,Nore1997b} and, equivalently, of a random distribution of vortices \cite{Nowak:2010tm,Nowak:2011sk}, as we will discuss in more detail in Section (\ref{sec:resultsAcoustic}).
Such an \IR divergence can be interpreted as an inverse cascade of particles.
This plays an important role in the equilibration and condensation process \cite{Svistunov1991a,Semikoz1997a,Micha2003a} after a strong cooling quench in a Bose gas \cite{Berges:2012us,Nowak:2012gd} since the cascade builds up the Bose condensate by accumulating particles on the largest scales.

In Refs.~\cite{Nowak:2010tm,Nowak:2011sk,Schmidt:2012kw,Schole:2012kt,Nowak:2013juc}, the above non-thermal fixed points of the dilute super-fluid gas were discussed in the context of topological defect formation and super-fluid turbulence.
A key result is that nearly degenerate Bose gases in $d=2,3$ dimensions, quenched parametrically close to the Bose-Einstein condensation (in $d=2$ Berezinsky--Kosterlitz--Thouless) transition, can evolve quickly to a quasi-stationary state exhibiting critical scaling \cite{Nowak:2011sk} and slowing-down behaviour \cite{Schole:2012kt}.
The critical scaling exponents $\xi$ of the kinetic energy spectra $\epsilon_{\text{kin}}(\mathbf{p})\sim p^{-\xi}$ corroborated the predictions $\xi_\mathrm{Q} = d-\eta_\mathrm{Q}$ of the strong-wave-turbulence analysis of Ref.~\cite{Scheppach:2009wu} for a quasi-particle cascade, with a very small value of $\eta_\mathrm{Q}$. 
Within the numerical precision it was found that $\xi_\mathrm{Q}=2$ in $d=2$ and $\xi_\mathrm{Q}=3$ in $d=3$ \cite{Nowak:2011sk}.
These exponents turned out to be related to randomly distributed vortices and (large) vortex rings occurring during the approach of the critical state \cite{Nowak:2010tm,Nowak:2011sk}.
%The critical scaling exponents $\zeta$ of the single-particle momentum spectra $n(p)\sim p^{-2}\epsilon_{\text{kin}}(p)\sim p^{-\zeta}$ corroborated the predictions $\zeta_\mathrm{Q} = d+2-\eta_\mathrm{Q}$ of the strong-wave-turbulence analysis of Ref.~\cite{Scheppach:2009wu} for a quasi-particle cascade, with a very small value of $\eta_\mathrm{Q}$. Within the numerical precision it was found that $\zeta_\mathrm{Q}=4$ in $d=2$ and $\zeta_\mathrm{Q}=5$ in $d=3$ \cite{Nowak:2011sk}. These exponents turned out to be related to randomly distributed vortices and (large) vortex rings occurring during the approach of the critical state \cite{Nowak:2010tm,Nowak:2011sk}.

\section{Driven-dissipative Gross--Pitaevskii equation}
\label{sec:driven_dissipative_gross_pitaevskii_equation}

In order to make contact with classical turbulence and the results of Section (\ref{sec:burgers_turbulence}) we include driving and dissipation into the description of the Bose gas given by \Eqs{eq:action_gpe} and \eq{eq:gpe}. Physically this amounts to coupling the Bose gas to an external reservoir which can absorb particles and energy.
A well known example of such system are \EPCs in solid-state systems as well as in ultra cold atomic gases (see \eg \cite{Carusotto2013a} for a review). It was found experimentally \cite{weisbuch1992a,Kasprzak2006a,lai2007a,deng2007a,amo2009a,lagoudakis2008a,Lagoudakis2009a,Amo2011a,hivet2012a} as well as theoretically \cite{Bolda2001,Carusotto2004,ciuti2005,Szymanska2006,Keeling2008,Pigeon2011} that such systems undergoes Bose condensation and exhibit vortices and super-fluid turbulent dynamics similar to the isolated case even though the steady state is maintained by the competition of driving and dissipation.

The driven-dissipative super-fluid dilute Bose gas can be described in an a similar way as the isolated gas. See \eg \cite{wouters2009a,wouters2010c} for applications of the c-field methods to \EPCs. In the semi-classical approximation we can simply replace the \GPE by the \SGPE,
\begin{align}
 i \partial_t \psi = \left[-\left(\frac{1}{2m}-i\nu\right) \nabla^2 - \mu + g \left|\psi \right|^2 \right] \psi+\zeta.
\label{eq:GPE}
\end{align}
We allow for the necessary dissipation, loss, and gain of energy and particles by allowing $\mu=\mu_{1}+i\mu_{2}$ and $g=g_{1}-ig_{2}$ to become complex, including an effective particle gain or loss $\mu_{2}$, as well as two-body interaction and loss parameters $g_{1,2}$. 
The diffusion term $\propto \nu$ is generated through the coarse graining of high-frequency modes. It was first proposed empirically and shown to provide a good model of the \EPC dynamics \cite{Wouters2010a,Wouters2010b}. It was derived from first principles in \cite{Sieberer2013a,Sieberer2013b,Tauber2013a} using \RG techniques.
$\zeta$ is a Gaussian, delta-correlated white noise that satisfies
\begin{align}
\langle \zeta(\tx) \rangle = 0, && \langle\zeta^{*}(t,\mathbf{x})\zeta(t',\mathbf{x}')\rangle=\gamma \, \delta(t-t')\delta(\mathbf{x}-\mathbf{x}').
\end{align}
It accounts for the noise induced by the driving mechanism. The overall driving intensity in turn is given by $\mu_2$. When $\mu_2>0$ the steady state is maintained by the competing of the pumping of individual particles into the system and the two body losses that are taken into account by $g_2$.

We have seen in the previous section that scaling at \NTFPs is the outcome of a cascade of particles or energy. In a closed system this can not go on forever. The system eventually thermalises. During the cascade process the system is however quasi-stationary. The cascade is maintained because the non-thermal distribution of particles and energy allows for depletions and over-occupations of the different Fourier modes of the system which act as sinks and sources. During the particles cascade the condensate is not fully formed yet and there is still a lot of room in the zero mode. We have a sink in the \IR. In order to understand the source in the \UV we must look a little more into the early the time evolution of the system, \ie before it reaches the \NTFP. \cred{Indeed}, when the far-from-equilibrium initial conditions contain a few very highly occupied \IR modes, the field expectation value initially oscillates in time and space. This oscillating background then drives the occupation of the higher 
energy modes just like parametric resonances in classical physics \cite{case1996a,fossen2011parametric}. See \cite{Berges:2004yj} for a detailed discussion of the parametric resonance mechanism and \cite{Schole:2012kt,Nowak:2013juc} for details on early time dynamics of the Bose gas. Then when the system reaches the \NTFP there are a lot of particles in the \UV which act as a source for the cascade towards the \IR. Note that this source of particles is in the \UV as compared to the zero mode. Actually it is at intermediate scales, at momentum of about one half of the inverse healing length\footnote{The healing length is the scale which one can use to make \Eq{eq:gpe} dimensionless. If the distance is measured in units of $\xi = (2m \mu)^{-1/2}$ then the pre-factor of the kinetic term is equal to the chemical potential and can be scaled into a dimensionless interaction parameter $\hat{g} = g m \xi^{2-d}$.} \cite{Nowak:2011sk,Nowak:2012gd}. There is still a lot of depleted modes in the (far) \UV for energy to 
accumulate there. What is observed in numerical simulations (\cite{Nowak:2013juc} and references therein) is that the large number of particles that cascade to the \IR looses most of its energy to a small number of particles which constitute the \UV part of the spectrum. In this way while particles are cascading to the \IR, energy is simultaneously cascading to the \UV.

On the other hand, an \EPC contains built-in driving and dissipation such that cascades can be sustained forever. The driving mechanism is incoherent pumping of particles. It is simply proportional to the local (in momentum space) particle spectrum. The dissipation happens through two-body losses which is non-linear. It is proportional to the product of two spectra at different momenta. Then if there is a Bose condensate most of the dissipated particles will be lost to the condensate since it is macroscopically occupied as compared to the rest of the spectrum. We see that once more we find a sink in the \IR. It then becomes reasonable to assume that the cascades in both open an closed systems are a properties of the underlying dynamics instead of the detailed driving mechanism. In the rest of this section we assume that this is the case. We will study the scaling properties of the non-equilibrium steady state of the \EPC and compare them to results known in the context of \NTFPs in the 
closed system.

Super-fluid turbulence \cite{Vinen2006a,Maurer1998,Walmsley2007a,Walmsley2008a,Nazarenko2001a,Araki2002a,Kobayashi2005a,Volovik2004a,Kozik2009a,Tsubota2008a,Tsubota2010a,Numasato2010a} manifests itself in self-similar field configurations in the domain of long-wavelength hydrodynamic excitations.
The hydrodynamic formulation of the \SGPE results by introducing the parametrisation $\psi=\sqrt{n}\exp[i\theta]$ in terms of the fluid density $n$ and velocity fields {$\mathbf{v}=m^{-1}\boldsymbol{\nabla}\theta$}.
The phase angle $\theta$ then obeys a Langevin equation of the \KPZ type which is equivalent to Burger's equation \eq{eq:burgers_stochastic} for the curl-free velocity field $\mathbf{v}$, under the  condition that $\mathbf{f}=m^{-1}\boldsymbol{\nabla} U$, with a random potential field $U$.
We therefore relate the \SGPE to the stochastic Burgers equation \eq{eq:burgers_stochastic}.
\cred{Indeed}, the hydrodynamic decomposition of the complex field, {$\psi = \sqrt{n}\exp\{i \theta\}$}, makes it possible to write \Eq{eq:GPE} as
\begin{alignat}{3}
& \partial_t \theta + \frac{1}{2m}\left( \boldsymbol{\nabla}\theta \right)^2 - \nu \nabla^2 \theta &&= U, \nonumber \\
& \partial_t n + \frac{1}{m} \boldsymbol{\nabla} \cdot \left( n \boldsymbol{\nabla}\theta \right) &&= S.
\end{alignat}
This is formally similar to the equations that arise from the conservative GPE, with the addition that the continuity equation is in-homogeneous and that the \KPZ equation has a non-zero dissipative term,
\begin{alignat}{3}
& U &&= \frac{1}{4 m \sqrt{n}} \boldsymbol{\nabla} \cdot \left(\frac{\boldsymbol{\nabla}n}{\sqrt{n}}\right) + \frac{\nu}{n}\boldsymbol{\nabla}n \cdot \boldsymbol{\nabla}\theta  +\mu_1 - g_1 n - \frac{\operatorname{Re}(\zeta \text{e}^{-i\theta})}{\sqrt{n}}, \nonumber \\
& S && = \nu \sqrt{n} \boldsymbol{\nabla} \cdot \left(\frac{\boldsymbol{\nabla}n}{\sqrt{n}}\right) - 2 \nu n \left(\boldsymbol{\nabla}\theta\right)^2 - 2\mu_2 n - 2 g_2 n^2 + 2 \sqrt{n} \operatorname{Im}(\zeta \text{e}^{-i\theta}).
\end{alignat}
These equations are coupled non-linear Langevin equations.
If the fluctuations of the field amplitude are sub-dominant the former can be decoupled by assuming that $U$ plays the role of the potential of the stochastic forcing {$\mathbf{f} = m^{-1}\boldsymbol{\nabla}U$}, with noise correlator
\begin{align}
\langle U(\omega,\mathbf{p}) U(\omega',\mathbf{p}') \rangle 
= \delta(\omega+\omega')\, \delta\left(\mathbf{p}+\mathbf{p'}\right) \, u(\omega,\mathbf{p}).
\label{eq:forcingKPZ}
\end{align}
This describes particles being injected and removed as amplitude fluctuations, such that the system reaches a state where they can be described by a (not necessarily thermal) distribution and feed energy to the phase fluctuations.
%Note that, contrarily to the correlations of $\zeta$, we do not require $U$ to be delta correlated in space.
Burgers' equation is obtained by setting {$\mathbf{v} = m^{-1}\boldsymbol{\nabla}\theta$}. Note that a very similar mapping was first introduced in \cite{Altman2013a}.
% and then in \cite{mathey2014}.

The kinetic energy spectrum can be written in therms of $\theta$ and $n$ by means of the density and phase representation.
It is decomposed into three parts,
\begin{alignat}{4}
&E_{\text{kin}} && = \frac{\rho}{2m} \int_\mathbf{x} \langle \left(\boldsymbol{\nabla}\theta\right)^2 \rangle 
&& + \frac{1}{2m} \int_\mathbf{x} \langle \frac{\left(\boldsymbol{\nabla}n\right)^2}{4 n} \rangle && + \frac{1}{2m}\int_\mathbf{x} \, \langle \delta n \left(\boldsymbol{\nabla}\theta\right)^2\rangle \nonumber \\
& && = E_{\text{hydro}} &&+ E_{\text{quantum}} &&+ E_{\text{exchange}}.
\label{eq:decomposed_energy}
\end{alignat}
The amplitude of $\psi$ is separated into $n = \langle n \rangle + \delta n \equiv \rho  + \delta n$. 
At sufficiently low energies, the average value of the amplitude is much larger than its fluctuations and the major contribution to the kinetic energy is $E_{\text{hydro}}$. 
Then,
\begin{align}
 E_{\text{kin}} \cong \frac{\rho}{2m} \int_{\mathbf{x}} \, \langle \left(\boldsymbol{\nabla}\theta\right)^2\rangle \equiv \mathcal{V} \int_\mathbf{p} \epsilon_{\text{kin}}(\mathbf{p}),
\end{align}
where $\mathcal{V}$ is the volume of the system. 
Hence $\epsilon_{\text{kin}}(\mathbf{p})$ can be related to the two-point correlation function of the Burgers fluid,
\begin{align}
 \epsilon_{\text{kin}}(\mathbf{p}) = \frac{m \rho}{2} \int_{\omega} \, \langle \mathbf{v}(\omega,\mathbf{p}) \cdot \mathbf{v}(-\omega,-\mathbf{p})\rangle .
 \label{eq:kin_spectrum_burgers}
\end{align}

\section{Strong wave and quantum turbulence}
\label{sec:resultsSWT}

In the following we discuss results of the present work.
We exploit the mapping of the \SGPE onto the stochastic Burgers equation which was introduced in Section (\ref{sec:driven_dissipative_gross_pitaevskii_equation}) to extract information on quantum turbulence in dilute Bose gases as described by the \GPE.

\subsection{Galilei invariance and Kolmogorov scaling}
\label{sec:gallilee_invariance_and_kolmogorov_scaling}

The kinetic energy spectrum of the ultra-cold Bose gas is expressed in terms of the velocity field of Burgers' equation in \Eq{eq:kin_spectrum_burgers}.
%The latter is shown in \Eq{eq:ekin_burgers}. As always the physical observables are obtained in the limit $k\to0$.
One can relate the scaling exponents of the ultra-cold Bose gas $\kappa$, $\eta$ and $z$ to the exponents of Burgers turbulence $\chi$ and $z$ by computing the scaling exponent of the kinetic energy spectra in both set-ups.
In the case of the Bose gas we insert \Eqs{eq:scaling_bose} into \Eq{eq:epsilon_kin_bose} and get,
\begin{align}
 \epsilon_{\text{kin}}(\mathbf{p}) \sim p^{z-\kappa}.
\label{eq:kin_spectrum_exponent_bose}
\end{align}
For the Burgers fluid we insert \Eqs{eq:parametrization} into \Eq{eq:kin_spectrum_burgers}, take the limit $k\to0$ and get \Eq{eq:kin_spectrum_exponent_burgers}, $\epsilon_{\text{kin}}(\mathbf{p}) \sim p^{2+d-2\eta_1}$.
By matching this scaling with \Eq{eq:kin_spectrum_exponent_bose} we conclude that,
\begin{align}
z-\kappa = d+2- 2\eta_1.
\label{eq:kappa_to_eta}
\end{align}

In the case of the Burgers fluid, inserting \Eq{eq:consistency} into \Eqs{eq:eta_to_chi} gives \Eqs{eq:eta_to_chi_new}. In particular $z = 2+d-\eta_1$.
If we assume that the dynamical critical exponent $z$ is the same for the Bose gas and classical turbulence \Eq{eq:kappa_to_eta} turns into $\kappa = \eta_1$.
Equivalently, \Eqs{eq:eta_to_chi_new} provide,
\begin{align}
\kappa + z = d+2.
\label{eq:kappa_and_z}
\end{align}
This is a non trivial relation in between the scaling exponents of the ultra-cold Bose gas. It has its origin in Galilei invariance through \Eq{eq:chiofeta1andorz} which reduces by one the number of free scaling exponents.

This can thus be used to eliminate $z$ from \Eqs{eq:NTFPkappa} and write
\begin{align}
\kappa_{\mathrm{P}} = d+4/3-\eta_{\mathrm{P}}/3,
&& \kappa_{\mathrm{Q}} = d+1-\eta_{\mathrm{Q}}/2.
\label{eq:NTFPkappawithz}
\end{align}
In turbulence theory, one considers the scaling of the radial kinetic energy distribution $E(p) = p^{d-1} \int_\Omega \epsilon_{\text{kin}}(\mathbf{p}) \sim p^{d-1}\epsilon_{\text{kin}}(\mathbf{p})$. \cred{Indeed}, this observable is isotropic by construction. It measures the energy content of the momentum shell of radius $p$.
Combining the above results, one finds that the direct energy and inverse particle cascades have radial single-particle kinetic energy distributions 
\begin{align}
 E_{\mathrm{P}}(p) \sim p^{-5/3+2\eta_{\mathrm{P}}/3},
 && E_{\mathrm{Q}}(p) \sim p^{-1+\eta_{\mathrm{Q}}},
\label{eq:radialEPQ}
\end{align}
respectively.
We find that, for the direct energy cascade, the strong-wave-turbulence scaling \cite{Scheppach:2009wu} of $E_{\mathrm{P}}(p)$ is equivalent to the classical Kolmogorov law \cite{Kolmogorov1941a,*Kolmogorov1941b,*Kolmogorov1941c,Frisch2004a}, with an intermittency correction $2\eta_{\mathrm{P}}/3$.
Kolmogorov-$5/3$ scaling has been reported to be possible in a super-fluid both experimentally \cite{Maurer1998,Walmsley2007a,Walmsley2008a}, and in simulations \cite{Araki2002a,Kobayashi2005a} of the \GPE.

Given the relation \eq{eq:radialEPQ} of the scaling laws with hydrodynamics and topological and geometric properties of the superfluid gas, we call the exponents $-5/3$ and $-1$ canonical while the effects of fluctuations are captured by the anomalous corrections $2\eta_{\mathrm{P}}/3$ and $\eta_{\mathrm{Q}}$, respectively.  

%=======================================================================
\subsection{Acoustic turbulence in a super-fluid}
\label{sec:resultsAcoustic}

Let us return to the \KPZ dynamics.
In order to make contact with scaling in acoustic turbulence in a super-fluid, we insert the average literature values for $\eta_{1}$ computed within the \KPZ framework into \Eq{eq:kin_spectrum_exponent_burgers} or equivalently into \Eq{eq:kin_spectrum_exponent_bose} with \Eq{eq:kappa_to_eta} inserted. These are given in Table I. of Ref.~\cite{Kloss2012a} and Figure (\ref{fig:eta_range}). They correspond to a forcing potential field delta-correlated in space. Cf.~\Eq{eq:potential_corr} with $U(p)=1$. Note that modulo the difference in the tensor structure of $\langle f_i(\tx) f_j(t',\mathbf{x}')\rangle$ this corresponds to $\beta = 2$ in \Eq{eq:forcing_corr_beta}.
We obtain $\epsilon_{\text{kin}}(\mathbf{p}) \sim p^{-\xi}$, with
\begin{align}
&\xi=0, & & \mbox{for}\ d = 1, \nonumber \\
&\xi=0.758(30), & & \mbox{for}\ d = 2, \nonumber \\
&\xi=1.600(24), & & \mbox{for}\ d = 3.
\label{eq:epsilon_scaling}
\end{align}
These results can be compared with scaling behaviour observed in acoustic turbulence in ultra-cold Bose gases, as summarised in the following.

Results related to the quantum turbulence discussed in the previous section were obtained for a $1$-dimensional Bose gas in Ref.~\cite{Schmidt:2012kw}. 
There, the relation between critical scaling of the single-particle momentum spectrum and the appearance of solitary wave excitations was pointed out.
It was found that this spectrum, as for a thermal quasi-condensate, has a Lorentzian shape if the solitons are distributed randomly in the system, with the width of the Lorentzian being related to the mean density of solitons.
The latter is in general different from and independent of the thermal coherence length of a gas with the same density and energy.
The kinetic energy spectrum, in the regime of momenta larger than the Lorentzian width, correspondingly shows a momentum scaling $\epsilon_{kin}(\mathbf{p}) \sim p^{2} n(\mathbf{p}) \sim p^0$.
This, in turn, is in full agreement with the above result quoted in \Eq{eq:epsilon_scaling}, corresponding to a white-noise forcing, \ie $\beta=2$.
The power law is consistent with that occurring in the single-particle spectrum of a random distribution of grey and black solitons in a one-dimensional Bose gas.

The fixed points found in Ref.~\cite{Kloss2012a}, at which the exponents \eq{eq:epsilon_scaling} apply, describe critical dynamics according to the \KPZ equation describing, \eg the unbounded propagation of an interface moving with coordinates $(\theta,\mathbf{x})$ in a two-component statistical system. 
On the contrary, the \KPZ equation derived for the phase angle $\theta$ of the complex field $\psi$ evolving according to the GPE, see Section (\ref{sec:driven_dissipative_gross_pitaevskii_equation}), is subject to the additional constraint that the range of angles $0 < \theta \leq 2\pi$ is compact.
This constraint plays an important role if the phase excitations are large enough to allow for (quasi) topological defects.
Hence, one can not expect the predictions \eq{eq:epsilon_scaling} to necessarily match the scalings occurring when defects such as vortices are present.  

We now show that the scalings \eq{eq:epsilon_scaling} are present at the \NTFP.
Note that, while the strong-wave-turbulence prediction \Eqs{eq:nscaling2PI} and \eq{eq:NTFPkappa}, $\xi_\mathrm{Q}=d-\eta_\mathrm{Q}\cong d$, is consistent with vortices dominating the infrared behaviour of the single-particle spectrum \cite{Nowak:2011sk}, it does not apply to the $1d$ case where there are no vortex defects. \cred{Indeed}, $\xi_\mathrm{Q}=d=1$ is by $1$ larger than the exponent $\xi=0$ appearing in the Lorentzian distribution at large momenta.
However, also in $d=2$ and $d=3$, a scaling $\xi_{c}\simeq d-1$ appears as a result of kink-like structures and longitudinal, compressible sound excitations. See Section (\ref{sec:kinetic_energy_spectrum_decomposition}) for definitions of the different components of the kinetic energy spectrum.
In Ref.~\cite{Nowak:2011sk}, it was demonstrated that the single-particle spectrum of the compressible component of the super-fluid turbulence can show power-law behaviour, with an  exponent  $\epsilon_{\text{comp}}(\mathbf{p})\sim p^{-d+1}$. The exponent $\xi_{c}\simeq d-1$ is present at the \NTFP but it appears in the compressible part of the kinetic energy spectrum which is sub-dominant as compared to its incompressible component\footnote{Remember that we are dealing with a particle cascade towards the \IR. We consider the case of $p$ being very small.}, $\epsilon_{\text{inc}}(\mathbf{p})\sim p^{-d}$.
This power-law with exponent $\xi_c = d-1$, was ascribed to sound wave turbulence on the background of the vortex gas, in particular to the density depressions remaining for some time in the gas after a vortex and an anti-vortex have mutually annihilated \cite{Berloff2004a}, cf.~Fig.~15 of Ref.~\cite{Nowak:2011sk}.
%To make contact with the scalings \eq{eq:epsilon_scaling}, we note that, while the strong-wave-turbulence prediction $\xi_\mathrm{Q}=d$ is consistent with vortices dominating the infrared behaviour of the single-particle spectrum \cite{Nowak:2011sk}, it does not apply to the $1d$ case where there are no vortex defects, since $\zeta_\mathrm{Q}=d+2=3$ is by $1$ larger than the exponent $\zeta=2$ appearing in the Lorentzian distribution at large momenta. However, also in $d=2$ and $d=3$, a scaling $\zeta_{c}\simeq d+1$ appears as a result of kink-like structures and longitudinal, compressible sound excitations. In Ref.~\cite{Nowak:2011sk}, it was demonstrated that the single-particle spectrum of the compressible component can show power-law behaviour, with an  exponent  $n_{c}(\mathbf{p})\sim p^{-\zeta_{c}}$. This power-law was ascribed to sound wave turbulence on the background of the vortex gas, in particular to the density depressions remaining for some time in the gas after a vortex and an anti-vortex have 
mutually annihilated \cite{Berloff2004a}, cf.~Fig.~15 of Ref.~\cite{Nowak:2011sk}.

%We compare the predictions \eq{eq:epsilon_scaling} with the scalings found in \cite{Nowak:2011sk}.
%Using $\zeta=2\eta_{1}-d=\xi+2$ one finds
%
%\begin{align}
%&\zeta=2, & & \mbox{for}\ d = 1, \nonumber \\
%&\zeta=2.758(30), & & \mbox{for}\ d = 2, \nonumber \\
%&\zeta=3.600(24), & & \mbox{for}\ d = 3.
%\label{eq:n_scaling}
%\end{align}
%

\begin{figure}
%\begin{FPfigure}

\centering
\begin{subfigure}{0.9\textwidth}
\includegraphics[width=0.9\textwidth]{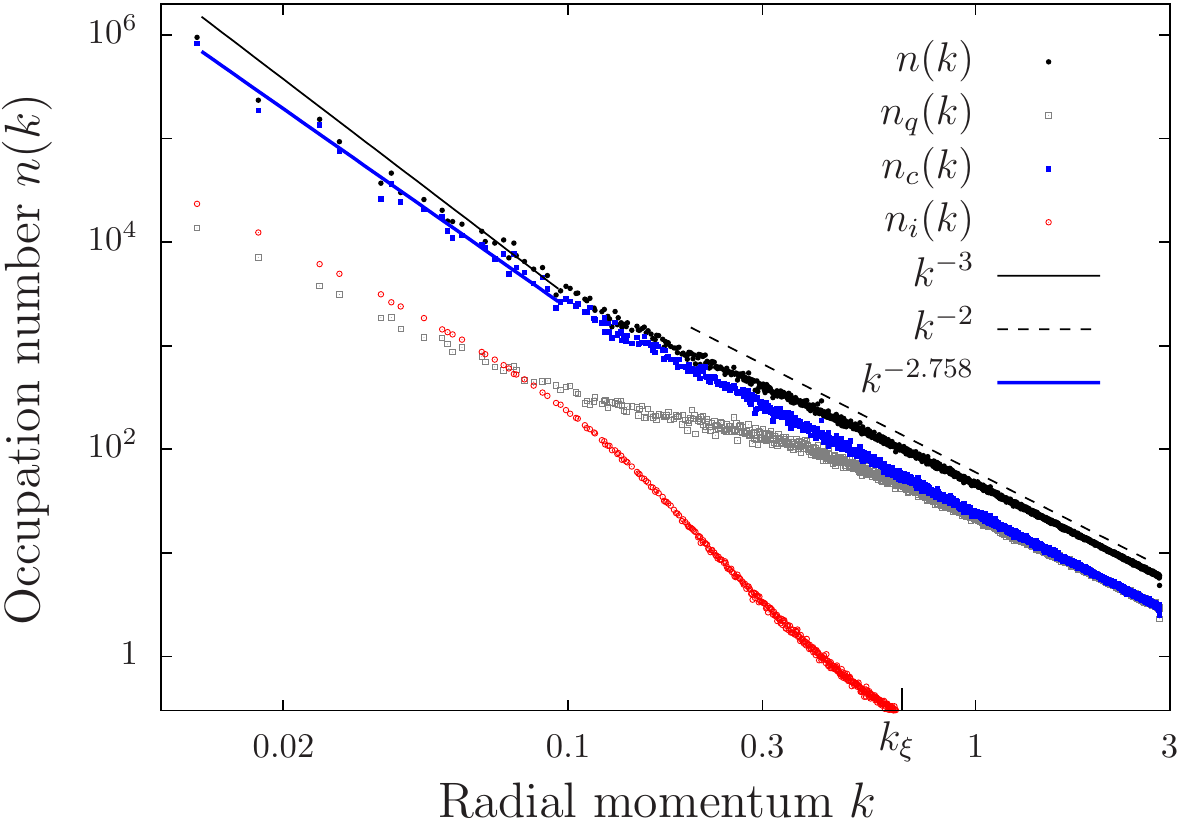}
\end{subfigure}

\begin{subfigure}{0.9\textwidth}
\includegraphics[width=0.9\textwidth]{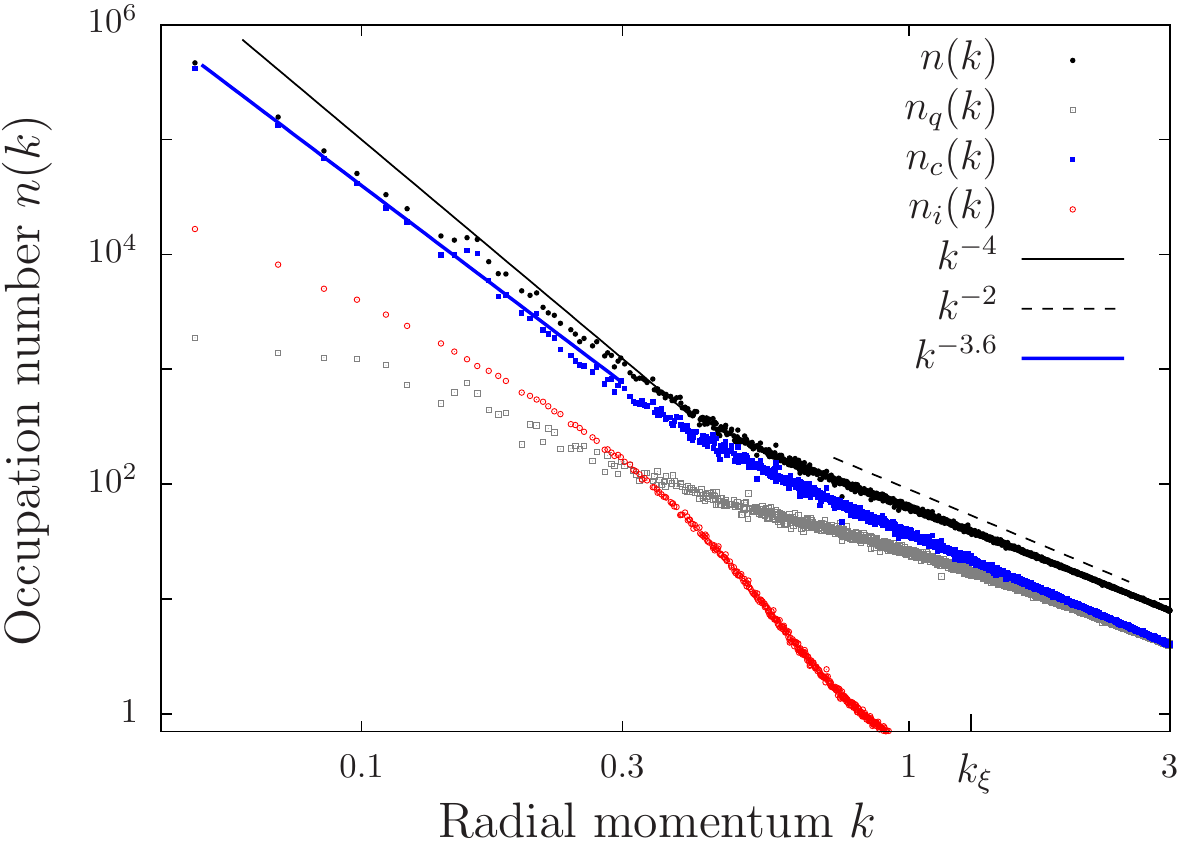}
\end{subfigure}

\caption{Acoustic turbulence in a $d=2$ (top) and $d=3$ (bottom panel) dilute super-fluid Bose gas. Shown are occupation number spectra of the late-stage evolution of a closed system after an initial quench, briefly after the last vortex-antivortex pair ($d=2$) or vortex ring ($d=3$) has disappeared, cf.~Fig.~15 of Ref.~\cite{Nowak:2011sk}.
The blue solid lines correspond to the predictions obtained in the present work, given in \Eq{eq:epsilon_scaling}. Note that $n(\mathbf{p}) \sim p^{-\xi-2}$. The black solid  lines mark the canonical scaling, $n(\mathbf{p}) \sim p^{-d-1}$. 
The latter scaling is expected on geometrical grounds, \eg results from randomly distributed plane-wave density depression waves or solitons \cite{Nowak:2011sk}.
It is emphasised that the comparatively large deviations ($\sim p^{-2.758}$ for $d=2$ and $\sim p^{-3.600}$ for $d=3$) are found independently and contrast the small deviation from the canonical scalings \eq{eq:radialEPQ} of quantum turbulent spectra in the presence of vortices.
The radial momentum is given in lattice units, $k = [2\sum_{i=1}^d \sin^2(2\pi n_i a/L)]^{1/2}$, with $n_i \in \mathbb{Z}$, $-L/(2a) \leq n_i \leq L/(2a)$,  $a$ being the grid constant and $L$ its side length. 
$k_{\xi} = 2\sin(\pi/(2\xi))$ is the momentum corresponding to the inverse healing length.}
\label{fig:anomalous_simulations}
%\end{FPfigure}
\end{figure}

If we assume that the scaling of \Eq{eq:epsilon_scaling} only applies to the incompressible spectra instead of the full $\epsilon_{\text{kin}}(\mathbf{p})$ we can modify \Eq{eq:nscaling2PI} to
%defining an anomalous exponent $\eta$ by requiring that the kinetic energy spectra scale as
%
\begin{align}
\epsilon_{\text{comp}}(\mathbf{p}) \sim p^{-\xi_c} \sim p^{z-\kappa + 1}.
\end{align}
The exponent is greater by one than in \Eq{eq:nscaling2PI} because these are sub-dominant excitations. Then comparing this with \Eqs{eq:epsilon_scaling} and assuming that the particle cascade is realised (\ie inserting the second of \Eqs{eq:NTFPkappa}) we get
\begin{align}
&\eta = 0, & & \text{for $d=1$}, \nonumber \\
&\eta = 0.242(30), & & \text{for $d=2$}, \nonumber \\
&\eta = 0.400(14), & & \text{for $d=3$}.
\label{eq:anomalousLit}
\end{align}
For $d=1$, the  scaling \eq{eq:epsilon_scaling} corresponds to that of the Lorentzian of a random soliton gas as discussed above.
Furthermore, within the numerical precision, the power laws seen in Fig.~15 of Ref.~\cite{Nowak:2011sk} are found to be consistent with the values \eq{eq:epsilon_scaling}. 
We reproduce the data in Figure (\ref{fig:anomalous_simulations}), comparing it with the \IR scaling exponents \eq{eq:epsilon_scaling} for $d=2, 3$ (blue solid lines). 
The figures show the occupation number spectrum $n(\mathbf{p}) \sim p^{-2}\epsilon_{\text{kin}}(\mathbf{p})$ obtained from a numerical simulation of the GPE.
The particular scalings occur shortly after the decay of the last topological excitations, \ie the last vortex-antivortex pair for $d=2$ or vortex ring for $d=3$. 
At the time the picture is taken, the compressible excitations dominate and their scaling exponent can be measured. 
The relatively large anomalous predictions of \Eq{eq:anomalousLit}  fit the data very well.
It was found numerically in \cite{Nowak:2011sk} that the particle cascade is \cred{indeed}, realised in the \IR.

We remark that the grey solitary-wave excitations as well as the density depressions remaining after vortex-anti-vortex annihilation are consistent with the absence of the compactness constraint on $\theta$ in the \KPZ equation.
The soliton gas can be dominated by grey solitons \cite{Schmidt:2012kw} which imply only weak density depressions at the position of the phase jump.
The weaker the depression, the smaller the phase kink and the less relevant the compactness of the range of possible $\theta$.
Similarly the density depressions leading to $\xi_{c}\simeq d-1$ do not require the phase to vary over the full circle.
Hence, we expect in these cases that \KPZ predictions for critical exponents apply also to the \GPE, as defects do not play a role.

\newpage
\thispagestyle{empty}

\chapter{Conclusions and Outlook}
\ResetAbbrevs{All}

In this thesis we have studied steady states of driven-dissipative dynamics. We investigated the classical turbulence described by the stochastic Burgers equation and the stationary quantum turbulence described by the \GPE.

The major focus of this work is on the stochastic Burgers equation. There we have applied the \FRG in order to look for non-perturbative scaling solutions. We established an approximation scheme which takes into account the necessary momentum dependence and is expressed in terms of effective parameters.
We then proceeded to write a set of \RG fixed point equations in order to look directly for scaling solutions. The fixed point equations were thoroughly studied analytically and solved in both asymptotic regimes of infinite and vanishing re-scaled momentum.

In all spatial dimension a continuum of fixed points was found and the corresponding scaling exponents were extracted. These are constrained within a given range that depends on the number of spatial dimensions. Moreover, we find an additional isolated fixed point as well.

For $d=1$ our results compare well with the literature and even suggest the existence of another yet unknown fixed point.
For $d\neq1$ however, the range of vales that we find for the scaling exponents is smaller than the one that appears in related work about \KPZ equation \cite{Kloss2013a}.
This case is however only equivalent to the dynamics of Burgers' equation when the velocity field is constrained to be the gradient of a potential.
There is not much literature about Burgers' equation with vorticity. Authors almost exclusively assume that the velocity field is the gradient of a potential and switch to the related \KPZ problem (see \eg \cite{Frisch2000a,Bec2007a}).
%For $d\neq1$ however there is not much literature to compare to because authors almost exclusively assume that the velocity field is the gradient of a potential and switch to the related \KPZ problem (see \eg \cite{Frisch2000a,Bec2007a}).
On the other hand, our approximation is devised to mimic $3d$ incompressible \NS turbulence where no such potential is introduced.
It contains a forcing which is not the gradient of a potential. \Ie vorticity is being pumped into the system.

Our results indicate that there is a qualitative difference in between the potential and rotational systems. This can be further investigated by using a more sophisticated approximation scheme which is able to capture both situations. However it might be simpler and more straightforward to simply simulate the stochastic Burgers equation and numerically measure the scaling exponents.
For a forcing correlation function with exponent $\beta = 2$ the two-point correlation function of the velocity field scales with and exponent $\eta_1 \cong 2.379$ ($d=2$) and $\eta_1\cong 3.300$ ($d=3$) when the flow is constrained to be potential. On the other hand if the forcing and therefore the velocity field contains a vortical component our study predicts $3.5 \leq \eta_1 \leq 4$ ($d=2$) and $4.5 \leq \eta_1 \leq 5$ ($d=3$) (see Table (\ref{tab:numerical_prediction})).

\begin{table}[ht]
\begin{center}
\begin{tabular}{ccc}
$d$ & Potential forcing & Vortical forcing \\
\midrule \midrule
$2$ & $\eta_1 \cong 2.379$ & $3.5 \leq \eta_1 \leq 4$ \\
$3$ & $\eta_1 \cong 3.300$ & $4.5 \leq \eta_1 \leq 5$
\end{tabular}
\end{center}
\caption{Predictions to be checked numerically for $\beta = 2$. The first column is the dimension of space, the second shows the approximate value that $\eta_1$ assumes when the forcing as well as the velocity field are constrained to be potential and the third gives the predictions of this work which apply when the forcing is pumping vorticity into the system.}
\label{tab:numerical_prediction}
\end{table}

The full numerical solution of the fixed point equations remains as a goal beyond the scope of this thesis.
It is emphasised that the \RG fixed point equations introduced here go far beyond previous approaches.
% and are not a standard approach to finding \FRG fixed points with an included momentum dependence.
Experience based on \cite{Pawlowski:2003XX} suggested that the fixed point equations could be solved iteratively. First $\delta Z_i(\hp)=0$ and arbitrary values of $h$ and $\eta_1$ would be inserted in the right-hand side of the fixed point equations \Eq{eq:flow2}. This would then provide a first estimation for its left-hand side as well as values for $\eta_1$ and $h$. This would then have been inserted back into the right-hand side of \Eq{eq:flow2}. Such a procedure was successfully implemented in the case of thermal equilibrium Yang-Mills theory \cite{Pawlowski:2003XX} but converged immediately to the Gaussian fixed point in our case. Indeed, inserting $\delta Z_i(\hp)=0$ in \Eq{eq:flow2} returned $h=0$ and $\delta Z_i(\hp)=0$.
In the case of stochastic hydrodynamics the vertexes of the action are locked to its quadratic part because of Galilei invariance. This makes it impossible to truncate the propagator (assume $\delta Z_i(\hp)=0$) at the intermediate steps of the calculation.
Since it is analytic our analysis of the asymptotic fixed point equations provide insight into their solutions without having to approach them numerically.
In particular we have learned that they have multiple solutions from which the correct ones need to be selected.

Our \RG study of the Burgers and \KPZ problems opened a new interesting view on aspects of quantum turbulence.
Out-of-equilibrium steady states of super-fluid Bose gases were related to the classical problem of the stochastic Burgers equation by considering an ultra-cold Bose gas coupled to external reservoirs of particles and energy.
Viscosity is an essential property of stochastic hydrodynamics even in the limit $\nu \to 0$. In order to include this in the description of the Bose gas the \GPE was upgraded to the \SGPE which contains driving and dissipation encoded in its complex parameters. Then making the density and phase decomposition of the average wave function of the Bose gas $\psi = \sqrt{n}\,\text{e}^{i\theta}$ one can write a KPZ-like equation for the phase variable. Once more we could not apply our results on Burgers' equation because the dynamics of the phase do not contain vorticity. We could however use results from the \KPZ literature to make predictions about the scaling properties of \NTFPs in ultra-cold Bose gases.

The first step towards these predictions was to assume that the cascade dynamics and its scaling exponents are a property of the underlying \GPE dynamics and do not result from the particular driving and dissipation mechanism. Then the predictions that we make apply to quasi stationary non-equilibrium states of closed systems (\NTFP) while our set-up is truly stationary by virtue of the balance of driving and dissipation.

From there we were able to make two predictions. First we related a well known relation in between the scaling exponents of the stochastic \KPZ equation to its dual in terms of the exponents of the Bose gas.
This provides an additional constraint in between the different scaling exponents of the Bose gas. In the case of a direct cascade of energy this can be used to identify a Kolmogorov $-5/3$ scaling of the kinetic energy spectrum and its anomalous correction at the \NTFP.
Note that a similar scaling form was pointed out in \cite{Scheppach:2009wu} in the case of an inverse particle cascade where the dynamical critical exponent $z$ does not play a role. Both cascades are expected to differ qualitatively through their vortex distribution.

Secondly we could use precise estimations of the scaling exponents of the stochastic \KPZ equation found in the literature to compute the anomalous corrections to the values of the scaling exponents of the compressible part of the kinetic energy spectrum of the Bose gas. On the way this provided an insight to the nature of the \KPZ dynamics. When we use the density and phase decomposition to translate the \GPE to the \KPZ equation we get an additional constraint on $\theta$ because we are dealing with an angle. \Ie $\theta$ is a compact field. This seemingly innocent constraint is actually crucial since quantum turbulence is basically made out of vortices which are the loci of phase jumps.
In such a situation there is no chance for the predictions of the \KPZ literature to be any good. However the decomposition of the kinetic energy spectrum of the Bose gas makes it possible to separate the different contributions to the kinetic energy spectrum.
Then the dominant contribution of the topological excitations can be discarded and we can look at the sub-dominant ones. For this reason we have compared the predictions of the \KPZ literature to the scaling of the compressible kinetic energy spectrum computed from the far-from-equilibrium closed \GPE. We find an excellent agreement (see Figure (\ref{fig:anomalous_simulations})). This seems to indicate that we were right to assume that both steady states (in the quasi-stationary closed system or in the stationary open system) can be described in terms of the same \NTFP.

As a last remark we note that this suggests that there is a qualitative difference in the scaling exponents of the traditional \KPZ equation and of its compactified version. \cred{Indeed}, if $\theta$ is interpreted as an angle instead of a height then vortices become available.
Our experience with \GPE hints towards the fact that if the traditional \KPZ two-point correlation function scales as\footnote{We introduce the \KPZ increment $\Sigma_2(\tau,\mathbf{r}) = \langle \left[\theta(t+\tau,\mathbf{x}+\mathbf{r})-\theta(t,\mathbf{x})\right]^2\rangle$.},
\begin{align}
\Sigma_2(\lambda^z \,\tau,\lambda \, \mathbf{r}) = \lambda^{2\chi} \, \Sigma_2(\tau,\mathbf{r}),
\end{align}
then its compact version scales as
\begin{align}
\Sigma_2(\lambda^z \,\tau,\lambda \, \mathbf{r}) = \lambda^{2\chi+1} \, \Sigma_2(\tau,\mathbf{r}).
\end{align}
In the \GPE however the density vanishes at vortex cores. Hence even if the phase is non-analytic at a vortex the wave function stays smooth because of the vanishing density.
It is not clear if vortex solutions can exist without such a regularisation.

\renewcommand\appendixpagename{\usekomafont{disposition}Appendices}

\begin{appendix}
\addappheadtotoc
\appendixpage

\let\cleardoublepage\clearpage
\chapter{Local Potential Approximation, Fixed Point Coefficients}
\label{sec:Local_potential_approximation_fixed_point_coefficients}
\ResetAbbrevs{All}

In this section we give explicit expressions for the fixed point coefficients $C_n$, $A_n$ and $B_n$ defined in Section (\ref{sec:local_pot_fixed_point}), \Eq{eq:local_pot_recursion},
\begin{align}
\lambda_n = C_n(\rho_0,\lambda_2,..,\lambda_n) + A_n(\rho_0,\lambda_2,\lambda_3) \, \lambda_{n+1} + B(\rho_0,\lambda_2)\, \lambda_{n+2}, &&  n \geq 3.
\label{eq:local_pot_recursion_app}
\end{align}
The coupling $\lambda_n$, defined in \Eq{eq:def_potential} is the $n$-th derivative of $u(\rho)$ evaluated at $\rho=\rho_0$. In order to establish \Eq{eq:local_pot_recursion_app} we start with the fixed point equation \Eq{eq:effective_pot_fp}, differentiate it $n$ times and evaluate it at $\rho = \rho_0$. We get
\begin{align}
\lambda_n = \frac{d-2}{d} \left( \lambda_{n+1} \rho_0 + n \lambda_n \right) - \frac{\Omega_d}{d^2} \frac{\text{d}^{n-1}}{\text{d}^{n-1}\rho}\left( \frac{2 \frac{\text{d}^3 u}{\text{d}^3\rho}\rho + 3 \frac{\text{d}^2u}{\text{d}^2\rho}}{\left[2 \frac{\text{d}^2 u}{\text{d}^2\rho}\rho + \frac{\text{d}u}{\text{d}\rho}+1\right]^2} \right|_{\rho = \rho_0}.
\label{eq:fp_corff_interm0}
\end{align}
In the following we will manipulate \Eq{eq:fp_corff_interm0} in order to extract all the terms that contain $\lambda_{n+1}$ and $\lambda_{n+2}$. Then $C_n$, $A_n$ and $B_n$ can be extracted in a straightforward way. We start by using the fact that\footnote{This identity is true for any couple of differentiable functions and can be proven by recurrence on $n$.}
\begin{align}
\frac{\text{d}^n (f \, g)}{\text{d}^n x} = \sum_{s = 0}^{n} \frac{n!}{s! (n-s)!} \, \frac{\text{d}^s f}{\text{d}^s x} \, \frac{\text{d}^{n-s} g}{\text{d}^{n-s} x},
\label{eq_derivative_of_a_product}
\end{align}
in order to expand the derivative on the right-hand side of \Eq{eq:fp_corff_interm0}. We get
\begin{align}
\lambda_n &= \frac{d-2}{d} \left( \lambda_{n+1} \rho_0 + n \lambda_n \right) \nonumber \\
& \quad {}- \frac{\Omega_d}{d^2} \sum_{s = 0}^{n-1} \frac{(n-1)!}{s! (n-1-s)!} \left( 2 \lambda_{n+2-s} \rho_0 + \left[3+2(n-1-s) \right] \lambda_{n+1-s}\right) \, X(s,2).
\label{eq_fixed_point_couplings_1}
\end{align}
We have defined
\begin{align}
X(n,m) & \equiv \frac{\text{d}^n}{\text{d}^n\rho}\left(\frac{1}{\left[2 \frac{\text{d}^2 u}{\text{d}^2\rho}\rho + \frac{\text{d}u}{\text{d}\rho}+1\right]^m}\right|_{\rho = \rho_0} \nonumber \\
& = -m \frac{\text{d}^{n-1}}{\text{d}^{n-1}\rho}\left(\frac{2 \frac{\text{d}^3 u}{\text{d}^3\rho}\rho + 3\frac{\text{d}^2u}{\text{d}^2\rho}}{\left[2 \frac{\text{d}^2 u}{\text{d}^2\rho}\rho + \frac{\text{d}u}{\text{d}\rho}+1\right]^{m+1}}\right|_{\rho = \rho_0}.
\label{def_X}
\end{align}
We see straight away that the first two terms of the sum on the right-hand side of (\ref{eq_fixed_point_couplings_1}), for $s=1$ and $2$, contain $\lambda_{n+2}$ and $\lambda_{n+1}$. Looking at Eq. (\ref{def_X}), we see that the coupling with the largest $n$ that appears in $X(n,m)$ is $\lambda_{n+2}$. Therefore, neither $\lambda_{n+2}$ nor $\lambda_{n+1}$ appear in the summands of the right-hand side of \eq{eq_fixed_point_couplings_1} if $2 \leq s \leq n-2$. We extract the terms that contain $\lambda_n$ or $\lambda_{n+1}$ and write
\begin{align}
& \lambda_n = \frac{d-2}{d} \left( \lambda_{n+1} \rho_0 + n \lambda_n \right) \nonumber \\
& \qquad {}- \frac{\Omega_d}{d^2} \left\{ 2 \lambda_{n+2} \rho_0 + \left[3+2(n-1) \right] \lambda_{n+1}\right\} \, X(0,2) \nonumber \\
& \qquad {}- \frac{\Omega_d}{d^2} (n-1)\left\{ 2 \lambda_{n+1} \rho_0 + \left[3+2(n-2) \right] \lambda_{n}\right\} \, X(1,2) \nonumber \\
& \qquad {}- \frac{\Omega_d}{d^2} \sum_{s = 2}^{n-2} \frac{(n-1)!}{s! (n-1-s)!} \left\{ 2 \lambda_{n+2-s} \rho_0 + \left[3+2(n-1-s) \right] \lambda_{n+1-s}\right\} \, X(s,2) \nonumber \\
& \qquad {}- \frac{\Omega_d}{d^2} \left( 2 \lambda_{3} \rho_0 + 3 \lambda_{2}\right) X(n-1,2).
\label{eq:fp_corff_interm1}
\end{align}
We see that a structure similar to \Eq{eq:local_pot_recursion_app} is starting to emerge. We still have to extract the factors of $\lambda_{n+1}$ and $\lambda_{n+2}$ from the last term of \Eq{eq:fp_corff_interm1} before the expression can be rearranged in to the desired form. We apply once more \Eq{eq_derivative_of_a_product} on $X(n-1,2)$ and write
\begin{align}
X(n-1,2) = -2\sum_{s = 0}^{n-2} \frac{(n-2)! \, X(s,3)}{s! (n-s-2)!} \left(2 \lambda_{n+1-s} \rho_0 + \left[2(n-s) -1\right]\lambda_{n-s} \right).
\label{eq:fp_corff_interm2}
\end{align}
It is now clear that $\lambda_{n+2}$ does not appear in $X(n-1,2)$ while $\lambda_{n+1}$ only appear in the first term ($s=0$) of the sum of \Eq{eq:fp_corff_interm2}. We can now put everything together and write
\begin{align}
C_n &=  \frac{d-2}{d} \, n \, \lambda_n  + 2 \, \frac{\Omega_d}{d^2} \frac{ 2 \lambda_{3} \rho_0 + 3 \lambda_{2}}{\left( 2 \lambda_2 \rho_0 +1 \right)^3} \left\{n\left[3+2(n-2) \right] \lambda_{n}\right\} \nonumber \\[0.2cm]
& \qquad {}- 2 \, \frac{\Omega_d}{d^2} \frac{\left( 2 \lambda_{3} \rho_0 + 3 \lambda_{2}\right)^2}{\left( 2 \lambda_2 \rho_0 +1 \right)^4} \left[3 (n-2) \left\{2 \lambda_n \rho_0 + \left[3+2(n-3) \right]\lambda_{n-1} \right\} \right] \nonumber \\[0.2cm]
& \qquad {}+ \frac{\Omega_d}{d^2} \sum_{s = 2}^{n-2} \frac{(n-2)! \left( 4 \lambda_{3} \rho_0 + 6 \lambda_{2}\right)}{s! (n-s-2)!} \left\{2 \lambda_{n+1-s} \rho_0 + \left[3+2(n-2-s) \right]\lambda_{n-s} \right\} \, X(s,3) \nonumber \\[0.2cm]
 & \qquad {}- \frac{\Omega_d}{d^2} \sum_{s = 2}^{n-2} \frac{(n-1)!}{s! (n-1-s)!} \left\{ 2 \lambda_{n+2-s} \rho_0 + \left[3+2(n-1-s) \right] \lambda_{n+1-s}\right\} \, X(s,2),
\label{eq:C}
\end{align}
\begin{align}
A_n = \frac{d-2}{d} \rho_0 + \frac{\Omega_d}{d^2} \left[4 n \, \rho_0 \, \frac{ 2 \lambda_{3} \rho_0 + 3 \lambda_{2}}{\left(2 \lambda_2 \rho_0 +1 \right)^3} - \frac{3+2(n-1)}{\left(2 \lambda_2 \rho_0 +1 \right)^2} \right],
\label{eq:A}
\end{align}
and
\begin{align}
B = - 2 \, \frac{\Omega_d}{d^2} \rho_0 \frac{1}{\left(2\lambda_2 \rho_0 +1\right)^2}.
\label{eq:B}
\end{align}
Finally, $X(n,m)$ can be computed recursively. We have:
\begin{align}
& X(0,m) = \frac{1}{\left(2 \lambda_2 \rho_0 +1 \right)^m}, \nonumber \\
& X(n,m) = -m \sum_{s=0}^{n-1} \frac{(n-1)!}{s! (n-1-s)!} \left[ 2 \lambda_{3+s} + \left(2 +3s \right) \lambda_{2+s} \right] X(n-1-s,m+1).
\label{eq:X}
\end{align}

\chapter{Flow Integrals}
\label{sec:explicit_expressions_for_the_flow_integrals}
\ResetAbbrevs{All}

In this appendix we give details on the computation of the flow integrals of \Eqs{eq:explicit_flow_1} and \eq{eq:explicit_flow_2}.
We use the sharp cut-off $R_{k}(p) = k^d z_1(k)\tilde R_{k}(p)$, with
\begin{align}
\tilde R_{k}(p)  
\left\{ \begin{array}{ll}
=0 & \text{if } p \ge k \\
\to\infty & \text{if } p < k
\end{array}\right. .
\end{align}
See Section (\ref{sec:def_flow_equation}). As a result we can use the identity
\begin{align}
& \frac{1}{\Gamma_k^{(2)}+R_k} k\frac{\partial R_k}{\partial k} \frac{1}{\Gamma_k^{(2)}+R_k} = 2k^2 \delta(p^2-k^2) \frac{1}{\Gamma_k^{(2)}},
\label{eq:identity}
\end{align}
to evaluate the flow integrals.
To proceed, we read off, from the ansatz \eq{eq:trunc}, the two-point function
\begin{align}
 & \frac{\delta^2 \Gamma_k[v = 0]}{\delta v_i(\omega',\vec{p}') \delta v_j(\omega,\vec{p})} 
 = \frac{\delta_{ij}}{(2\pi)^{d+1}} \delta(\omega+\omega') \delta\left(\mathbf{p}+\mathbf{p}'\right) \, \Gamma_{k}^{(2)}(\omega,\mathbf{p}),
\label{eq:gamma2_delta}
\end{align}
with $\Gamma_{k}^{(2)}(\omega,\mathbf{p})$ defined in \eq{eq:Gprop} and given by
\begin{align}
\Gamma^{(2)}_{k}(\omega,\mathbf{p}) & =  \left[ \nu_k(p)^2 p^4 +\omega^2\right]F^{-1}_k(p),
\label{eq:gamma2}
\end{align}
the 3-vertex
\begin{align}
&  \frac{\delta^3 \Gamma_k[v=0]}
{\delta v_i(\omega',\mathbf{p}') \delta v_j(\omega,\mathbf{p}) \delta v_l(\omega'',\mathbf{q})} 
= \frac{\delta(\omega+\omega'+\omega'')\, \delta\left(\mathbf{p}+\mathbf{p}'+\mathbf{q}\right)}{(2\pi)^{2(d+1)}} \,\Gamma_{k;ijl}^{(3)}(\omega',\mathbf{p}';\omega,\mathbf{p}),
\label{eq:gamma3_delta}
\end{align}
with
\begin{align}
\Gamma_{k;ijl}^{(3)}(\omega',\mathbf{p}';\omega,\mathbf{p})
  = & -F^{-1}_k(\left|\mathbf{p}+\mathbf{p}'\right|) \left(\omega +\omega' +i\nu_k(\left|\mathbf{p}+\mathbf{p}'\right|) \left|\mathbf{p}+\mathbf{p}'\right|^2\right) 
\left[\delta_{il}p'_j + \delta_{jl}p_i \right] 
\nonumber\\[0.1cm]
& -\ F^{-1}_k(p') \left(-\omega' +i\nu_k(p') \, p'^2 \vphantom{\left|\mathbf{p}+\mathbf{p}'\right|^2}\right) 
\left[\delta_{ij}p_l-\delta_{il}(p'_j+p_j)\right] 
\nonumber \\[0.1cm]
& -\ F^{-1}_k(p) \left( -\omega +i\nu_k(p) \, p^2 \vphantom{\left|\mathbf{p}+\mathbf{p}'\right|^2}\right) 
\left[\delta_{ij}p'_l -\delta_{jl}(p'_i+p_i)\right],
\label{eq:gamma3}
\end{align}
and the 4-vertex
\begin{align}
 \frac{\delta^4 \Gamma_k[v=0]}{\delta v_i(\omega,\mathbf{p}) \delta v_j(\omega',\mathbf{p}')
  \delta v_l(\omega'',\mathbf{q}) \delta v_m(\omega''',\mathbf{q'})} &= \nonumber \\[0.1cm]
& \hspace{-2cm} \frac{\delta(\omega+\omega'+\omega''+\omega''') \delta\left(\mathbf{p}+\mathbf{p}'+\mathbf{q}+\mathbf{q}'\right)}{(2\pi)^{3(d+1)}} \, \Gamma_{k,ijlm}^{(4)}(\mathbf{p},\mathbf{p}',\mathbf{q}),
\label{eq:gamma4_delta}
\end{align}
with
\begin{align}
 \Gamma_{k,ijlm}^{(4)}(\mathbf{p},\mathbf{p}',\mathbf{q}) = & \, F_k^{-1}(\left|\mathbf{p}+\mathbf{p}'\right|) 
\big[\left(\delta_{mi} p_j + \delta_{mj} p'_i \right) \left(p_l + p'_l + q_l\right) -\left(\delta_{li} p_j + \delta_{lj} p'_i \right) q_m\big] 
\nonumber \\[0.1cm]
&  {}+ F_k^{-1}(\left|\mathbf{p}+\mathbf{q}\right|)
\big[\left(\delta_{im} p_l+\delta_{lm}q_i\right) \left(p_j + p'_j + q_j\right) - \left(\delta_{ij}p_l  + \delta_{jl} q_i\right)p'_m \big] 
\nonumber \\[0.1cm]
&  {}+ F_k^{-1}(\left|\mathbf{p}'+\mathbf{q}\right|)
\big[\left(\delta_{jm}p'_l +\delta_{lm}q_j\right)\left( p_i + p'_i + q_i \right) - \left(\delta_{ij} p'_l  +\delta_{il} q_j\right)p_m \big].
\label{eq:gamma4}
\end{align}
Using these, the flow integral \eq{eq:flow_3} can be written as
\begin{align}
& I_{k}^{(2)}[0](\omega,\mathbf{p}) 
 = - \frac{k^{2}}{d} \bigintsss_{\omega',\mathbf{p}'} \frac{\delta(p'^2-k^2)}{\Gamma_{k}^{(2)}(\omega',\vec{p}')}
 \Big[ {\Gamma_{k;iijj}^{(4)}(\mathbf{p},-\mathbf{p},\mathbf{p}')}   \nonumber \\[0.1cm]
& \quad {}-2 \, \frac{\theta \left[ (\vec{p}'-\vec{p})^2 -k^2\right]}{\Gamma_{k}^{(2)}(\omega'-\omega,\vec{p}'-\vec{p})}
{\Gamma_{k;ijl}^{(3)}(-\omega',-\vec{p}';\omega'-\omega,\mathbf{p}'-\vec{p})} \, \Gamma_{k;jil}^{(3)}(\omega-\omega',\vec{p}-\vec{p}';\omega',\vec{p}') 
\Big].&
\label{eq:I}
\end{align}
The theta function arises because of the sharp cut-off.
Since the cut-off diverges for $p<k$, the propagator $(\Gamma_k^{(2)}+R_k)^{-1}$ vanishes in this regime.
This is irrelevant in the diagram depending on the 4-point vertex, cf.~\eq{eq:flow_3},  where the single appearing propagator carries a $\partial_{k}R_{k}$ insertion and is thus evaluated at $p^{2}=k^{2}$. \Eq{eq:I} is divided by $d$ in order to be consistent with \Eq{eq:flow_integrals_def_with_d}.
%Because of rotational symmetry, $\delta^2 I_{k}/\delta v_{i}^2[0](\omega,\mathbf{p})$ does not depend on $i$.
%Hence, taking the trace of $\delta^2 I_{k}/\delta v_{i} \delta v_{j}[0](\omega,\mathbf{p})$ and dividing it by $d$ makes the  integrals rotationally invariant.

The index contractions in \Eq{eq:I} are lengthy but straightforward. We find,
\begin{align}
& \Gamma_{k;ijl}^{(3)}(-\omega',-\mathbf{p}';\omega'-\omega,\mathbf{p}'-\mathbf{p}) \, \Gamma_{k;jil}^{(3)}(\omega-\omega',\mathbf{p}-\mathbf{p}';\omega',\mathbf{p}') = \nonumber \\[0.1cm]
& \qquad \hphantom{+ }\Gamma_k^{(2)}(\mathbf{p},\omega) F^{-1}_k(p) \left[ 2(d-1) p'^2 + d p^2 - 2(d-1)\mathbf{p}' \cdot \mathbf{p} \right] \nonumber \\[0.1cm]
& \qquad {}+ \Gamma_k^{(2)}(\mathbf{p}',\omega') F^{-1}_k(p') \left[ 2(d-1) p^2 + d p'^2 - 2(d-1)\mathbf{p}' \cdot \mathbf{p} \right] \nonumber \\[0.1cm]
& \qquad {}+ \Gamma_k^{(2)}(\mathbf{p}'-\mathbf{p},\omega' -\omega) F^{-1}_k(\left|\mathbf{p}'-\mathbf{p}\right|) \left[ d p'^2 + d p^2 - 2\mathbf{p}' \cdot \mathbf{p} \right] \nonumber \\[0.1cm]
& \qquad {}+ 2dF^{-1}_k(p') F^{-1}_k(p) \left[ \omega' \omega - \nu_k(p')p'^2 \nu_k(p)p^2\right] \mathbf{p}'\cdot \mathbf{p} \nonumber \\[0.1cm]
& \qquad {}+ 2d F^{-1}_k(p) F^{-1}_k(\left|\mathbf{p}-\mathbf{p}'\right|) \nonumber \\
& \qquad \qquad \times \left[\omega (\omega-\omega') - \nu_k(p)p^2 \nu_k(\left|\mathbf{p}-\mathbf{p}'\right|)\left|\mathbf{p}-\mathbf{p}'\right|^2\right] \mathbf{p}\cdot\left(\mathbf{p}-\mathbf{p}'\right) \nonumber \\[0.1cm]
& \qquad {}+ 2dF^{-1}_k(p') F^{-1}_k(\left|\mathbf{p}-\mathbf{p}'\right|) \nonumber \\
& \qquad \qquad \times \left[\omega' (\omega'-\omega) - \nu_k(p')p'^2 \nu_k(\left|\mathbf{p}-\mathbf{p}'\right|)\left|\mathbf{p}-\mathbf{p}'\right|^2\right] \mathbf{p}'\cdot\left(\mathbf{p}'-\mathbf{p}\right),
\label{eq:gamma3_contraction}
\end{align}
and
\begin{align}
\Gamma_{k;iijj}^{(4)}(\mathbf{p},-\mathbf{p},\mathbf{p}') = \, & F^{-1}_k\left( \left| \mathbf{p}-\mathbf{p}' \right| \right) \left[ -2 \, \mathbf{p} \cdot \mathbf{p}' + d(p^2+p'^2) \right] 
\nonumber\\[0.1cm]
&  +\ F^{-1}_k\left( \left| \mathbf{p}+\mathbf{p}' \right| \right) \left[ 2 \, \mathbf{p} \cdot \mathbf{p}' + d(p^2+p'^2) \right].
\label{eq:gamma4_contraction}
\end{align}

Once \Eqs{eq:gamma3_contraction} and \eq{eq:gamma4_contraction} are inserted into \Eq{eq:I}, it becomes apparent that as a result of the truncation of the frequency dependence of the inverse propagator \eq{eq:Gprop}, the integrand is a rational function of $\omega'$. The $\omega$ integration can be done analytically.
Introducing the short-hand notation $\nuk{p}{}=\nu_{k}(p)p^{2}$, etc., a straightforward application of the residue theorem gives
\begin{align}
 \int_{\omega} \frac{1}{\omega^2+\nu_k(p)^2 p^4}  & = \frac{1}{2 \nu_k(p) p^2}, \nonumber \\[0.2cm]
\int_{\omega} \frac{A+B\omega+C\omega^2}{\left[\omega^2+(\nuk{p}{})^2\right]\left[(\omega - f)^2+(\nuk{q}{})^2\right]} & = \frac{ A \left(\frac{\nuk{q}{}}{\nuk{p}{}} +1 \right) + B \, f + C \left[\nuk{q}{}\left( \nuk{q}{} + \nuk{p}{}\right) + f^2 \right] }{2\nuk{q}{}\left[f^2+\left(\nuk{q}{} + \nuk{p}{}\right)^2\right]},
\end{align}
for $A$, $B$ and $C\in \mathds{R}$ and $\nuk{q}{}$, $\nuk{p}{}>0$. The right-hand side of \Eq{eq:I} can be cast into a linear combination of such integrals for appropriate values of $A$, $B$ and $C$.
We finally obtain
\begin{align}
& I_{k}^{(2)}(\omega,\mathbf{p}) 
= -\frac{k^2}{d} \int_{\mathbf{q},\mathbf{r}} 
\frac{{(2\pi)^{d}\delta(r^2-k^2) \,  \delta(\mathbf{p}-\mathbf{q}-\mathbf{r})}}{F^{-1}_k(q) F^{-1}_k(r) \nukq \nukr [\omega^2 + (\nukq + \nukr)^2]}
\nonumber \\
& \times \Big[ F^{-1}_k(q)^2 \nukq \left[\omega^2 + (\nukq + \nukr)^2\right]
\left[d(p^2+r^2) - 2 \mathbf{p} \cdot \mathbf{r}\right]
\nonumber \\
& \quad {}- \theta \left( q^2-k^2\right) \Big( 
F^{-1}_k(q)^2 \nukq \left[\omega^2 + (\nukq + \nukr)^2\right]
\left[d(p^2+r^2)-2\mathbf{p}\cdot\mathbf{r}\right] 
\nonumber \\
& \quad\ {}+ F^{-1}_k(r)^2 \nukr \left[\omega^2 + (\nukq + \nukr)^2\right] 
\left[d(p^2+q^2)-2\mathbf{p} \cdot \mathbf{q}\right] 
\nonumber \\
& \quad\ {}+ F^{-1}_k(p)^{2}  \left(\nukq + \nukr\right) \left[\omega^2 + \nukps\right]
\left[d (q^2 + r^2) + 2\mathbf{q} \cdot \mathbf{r} \right]
\nonumber \\
& \quad\ {}+ F^{-1}_k(p)F^{-1}_k(q) \nukq \left[ \omega^2 -  \nukp\left(\nukq + \nukr\right) \right] 
2d\,\mathbf{p}\cdot \mathbf{q}
\nonumber \\
& \quad\ {}+ F^{-1}_k(p)F^{-1}_k(r) \nukr \left[ \omega^2 -  \nukp\left(\nukq + \nukr\right) \right] 
2d\,\mathbf{p}\cdot \mathbf{r}
\Big)\Big].
\label{eq:explicit_flow_appendix_0}
\end{align}
Note that the terms $\propto d$ and independent of $d$ originate from contractions of the types $\delta^{ij}\delta^{ji}p^k  q^k$ and $\delta^{ij}  \delta^{jl}  p^i  q^l$, respectively.
We can integrate radially, $\int_{\vec{r}} = \int_0^\infty r^{d-1} dr \int_{\Omega}$, which, for $d=1$, reduces to $\int_{\Omega} f(r) = [f(r)+f(-r)]/(2\pi)$.
{The delta distributions allow to set $\mathbf{p}' = k \mathbf{e}_{\mathbf{r}}$ and $\mathbf{q} = \mathbf{p}-k\mathbf{e}_{\mathbf{r}}$, with $|\hat{\mathbf{r}}|=|\mathbf{e}_{\mathbf{r}}|=1$.} 
Expansion in powers of $\omega^2$ gives \Eqs{eq:explicit_flow_1} and \eq{eq:explicit_flow_2}.
\begin{align}
& I_{k}^{(2)}(0,\mathbf{p}) 
= -\frac{k^d}{2d} \int_{\Omega} \frac{1}{F^{-1}_k(k) F^{-1}_k(q) \nukk \nukq (\nukk+\nukq)} 
\nonumber \\
& \times \Big[ F^{-1}_k(q)^2 \nukq (\nukq+\nukk) 
   \left[ d(p^2+k^2)-2 k \mathbf{e}_{\mathbf{r}} \cdot \mathbf{p}\right]  
   \nonumber \\
& \quad -\ \theta \left(q^2-k^2 \right) \Big( F^{-1}_k(q)^{2} \nukq (\nukk+\nukq) \left[d(p^2+k^2)-2k\, \mathbf{p}\cdot\mathbf{e}_{\mathbf{r}}\right]
\nonumber\\
&  \quad\ \ + F^{-1}_k(k)^{2} \nukk(\nukk+\nukq) \left[d(p^2+q^2)-2\mathbf{p}\cdot \mathbf{q}\right]
\nonumber\\
&  \quad\ \ + F^{-1}_k(p)^2  \left.\nukp\right.^2 \left[d(k^2+q^2)+2k\, \mathbf{q}\cdot\mathbf{e}_{\mathbf{r}}\right]
\nonumber\\
&  \quad\ \ - F^{-1}_k(p)F^{-1}_k(q) \nukp\nukq  2d \, \mathbf{p} \cdot \mathbf{q} 
\nonumber \\
&  \quad\ \ - F^{-1}_k(p)F^{-1}_k(k) \nukp\nukk 2d \, k\,\mathbf{p}\cdot\mathbf{e}_{\mathbf{r}} \Big) \Big].
\end{align}
and
\begin{align}
& \left. \frac{\partial I_{k}^{(2)}}{\partial {\omega^2}}\right|_{(0,\mathbf{p})}
= \frac{k^d}{2d} \int_{\Omega} \frac{\theta\left(q^2-k^2\right)F^{-1}_k(p)}{F^{-1}_k(k) F^{-1}_k(q) \nukk \nukq (\nukk + \nukq)^{3}}
\nonumber\\
& \times \Big( F^{-1}_k(p) \, \left[(\nukk+\nukq)^{2} - \left. \nukp \right.^2  \right] \left[d(q^2+k^2)+2 k\,\mathbf{q}\cdot\mathbf{e}_{\mathbf{r}}  \right]  
\nonumber \\
& \quad {}+ F^{-1}_k(q) \, \nukq \left[\nukk+\nukq + \nukp  \right]  2d \, \mathbf{p} \cdot \mathbf{q} 
\nonumber \\
& \quad {}+ F^{-1}_k(k) \, \nukk \left[\nukk+\nukq + \nukp \right]  2d \, k\,\mathbf{p}\cdot \mathbf{e}_{\mathbf{r}}   
\Big).
\end{align}

\chapter{Re-scaled Flow Integrals}
\label{sec:rescaled_flow_integrals}
\ResetAbbrevs{All}

In this section we write explicit expression for the re-scaled flow integrals which are defined in \Eq{eq:flow_integrals} and \Eqs{eq:rescaled_flow_1} and \eq{eq:rescaled_flow_2}. In Sections (\ref{app:small_p}) and (\ref{app:big_p}) we study their asymptotic behaviour for very small and very large arguments respectively.

The re-scaled flow integrals are obtained from \Eqs{eq:explicit_flow_1} and \eq{eq:explicit_flow_2} by inserting the fixed point parametrisation \eq{eq:parametrization} and the re-scaled variables \eq{eq:dim_variables}. We make the replacements given by
\begin{align}
& p = k \, \hp, & & \omega = k^2 \sqrt{\frac{z_1}{z_2}} \hat{\omega}, \nonumber \\
 & \nu_k(p) = \sqrt{z_1/z_2}\, \hat{p}^{(\eta_1-\eta_2-4)/2}  \sqrt{\frac{1+\delta Z_1(\hat{p})}{1+\delta Z_2(\hat{p})}}, & & F^{-1}_k(p) = k^{d-4} z_2\, \hat{p}^{\eta_2} \left[1+\delta Z_2(\hat{p})\right].
\end{align}
The re-scaled flow integrals take the form
\begin{align}
\hat{I}^{(2)}_1(\hat{p}) &= \hat{p} \, F_{1,1}(\hat{p}) + \hat{p}^{(\eta_1+\eta_2)/2+2} S_1(\hat{p}) \, S_2(\hat{p}) \, F_{1,2}(\hat{p}) \nonumber \\[0.1cm]
 & \qquad + \hat{p}^{\eta_1+\eta_2+2\,\delta_{d1}} [S_1(\hat{p}) \, S_2(\hat{p})]^2 F_{1,3}(\hat{p}), 
\label{eq:rescaled_flow_appendix_1}
\end{align}
and
\begin{align}
 \hat{I}^{(2)}_2(\hat{p}) 
 &= \hat{p}^{(\eta_1+\eta_2)/2+2} S_1(\hat{p}) \, S_2(\hat{p}) \, F_{2,1}(\hat{p}) 
 + \hat{p}^{\eta_2+2} S_2(\hat{p})^2 F_{2,2}(\hat{p}) 
 \nonumber \\[0.1cm]
 & \qquad + \hat{p}^{\eta_1+\eta_2 + 2\,\delta_{d1}} [S_1(\hat{p}) \, S_2(\hat{p})]^2 F_{2,3}(\hat{p})
 + \hat{p}^{2\eta_2+2 \, \delta_{d1}} S_2(\hat{p})^4 F_{2,4}(\hat{p}).
\label{eq:rescaled_flow_appendix_2}
\end{align}
We have introduced the short-hand notation $S_i(\hp) = \sqrt{1+ \delta Z_i(\hp)}$.
The functions $F_{i,j}(\hat{p})$ are given by\footnote{Note that the exponents of the following expression contain the Kronecker delta $\delta_{d1}$. \cred{Indeed}, cancellations in the leading pre-factors of $F_{1,3}(\hat{p})$, $F_{2,3}(\hat{p})$ and $F_{2,4}(\hat{p})$ occur for $d=1$ and the sub-leading terms must be taken into account.},
\begin{align}
 F_{1,1}(\hat{p}) =& \frac{1}{2d\,\hat{p}} \int_{\Omega} \Big\{\theta \left(\hat{q}^2-1\right) 
\hat{q}^{-(\eta_1+\eta_2)/2} 
\frac{S_2(1)^2}{S_1(\hat{q})S_2(\hat{q})} \left[d (\hat{p}^2 + \hat{q}^2)-2\hat{\mathbf{p}}\cdot\hat{\mathbf{q}}\right] \nonumber \\
& \quad -\theta \left(1-\hat{q}^2\right) \hat{q}^{\eta_2} \frac{S_2(\hat{q})^2}{S_1(1)S_2(1)}  \left[d(\hat{p}^2+1)-2 \, \hat{\mathbf{p}} \cdot\mathbf{e}_{\mathbf{r}} \right]  \Big\},
\label{eq:F11}
\end{align}
\begin{align}
%\\[0.5cm]
 F_{1,2}(\hat{p}) = & -\hat{p}^{-2} \int_{\Omega} \theta \left(\hat{q}^2-1\right) \T \left\{\frac{\hat{\mathbf{p}} \cdot \hat{\mathbf{q}}}{S_1(1)S_2(1)} + \frac{\hat{\mathbf{p}} \cdot\mathbf{e}_{\mathbf{r}}}{\hat{q}^{(\eta_1+\eta_2)/2} S_1(\hat{q})S_2(\hat{q})} \right\},
\label{eq:F12}
\end{align}
\begin{align}
%\\[0.5cm]
F_{1,3}(\hat{p}) =&\ \frac{\hat{p}^{-2\delta_{d1}}}{2d} \int_{\Omega} \theta \left(\hat{q}^2-1 \right) \T
\frac{d(\hat{q}^2+1) + 2 \hat{\mathbf{q}} \cdot\mathbf{e}_{\mathbf{r}}}{\hat{q}^{(\eta_1+\eta_2)/2} S_1(\hat{q})S_2(\hat{q})S_1(1)S_2(1)},
\label{eq:F13}
\end{align}
\begin{align}
%\\[0.5cm]
F_{2,1}(\hat{p}) =&\  \hat{p}^{-2} \int_{\Omega} \theta \left(\hat{q}^2-1\right)\left.\T \right.^3 \left\{\frac{\hat{\mathbf{p}} \cdot \hat{\mathbf{q}}}{S_1(1)S_2(1)} +\ \frac{\hat{\mathbf{p}} \cdot\mathbf{e}_{\mathbf{r}}}{\hat{q}^{(\eta_1+\eta_2)/2} S_1(\hat{q})S_2(\hat{q})}\right\},
\label{eq:F21}
\end{align}
\begin{align}
%\\[0.5cm]
 F_{2,2}(\hat{p}) =&\ \hat{p}^{-2} \int_{\Omega} \theta \left( \hat{q}^2-1\right) \left.\T \right.^2 
 \left\{ \frac{\hat{\mathbf{p}} \cdot \hat{\mathbf{q}}}{S_1(1)S_2(1)} +\ \frac{\hat{\mathbf{p}} \cdot\mathbf{e}_{\mathbf{r}}}{\hat{q}^{(\eta_1+\eta_2)/2} S_1(\hat{q})S_2(\hat{q}) } \right\},
\label{eq:F22}
\end{align}
\begin{align}
%\\[0.5cm]
F_{2,3}(\hat{p}) = &- \frac{\hat{p}^{-2\delta_{d1}}}{2d} \int_{\Omega} \theta \left( \hat{q}^2-1 \right) \left.\T \right.^3 \frac{d(\hat{q}^2+1) + 2 \hat{\mathbf{q}} \cdot\mathbf{e}_{\mathbf{r}}}{\hat{q}^{(\eta_1+\eta_2)/2} S_1(\hat{q})S_2(\hat{q}) S_1(1)S_2(1)},
\label{eq:F23}
\end{align}
\begin{align}
%\\[0.5cm]
F_{2,4}(\hat{p}) =& F_{1,3}(\hat{p}),
\label{eq:F24}
\end{align}
With $\mathbf{e}_{\mathbf{r}}$ the unit vector pointing in the direction of $\Omega$, $\hat{\mathbf{q}} = \hat{\mathbf{p}} - \mathbf{e}_{\mathbf{r}}$ and the short-hand notation $(\nukk+\nukq)^{-1} = k^{-2} (z_2/z_1)^{1/2} \T$.
These functions were defined in such a way that they are analytic and non vanishing at $\hat{p}=0$. They can be Taylor expanded $F_{i,j}(\hat{p}\to 0) = F_{i,j}(0) + F_{i,j}'(0) \, \hat{p} + \text{O}(\hat{p}^2)$. Their asymptotic behaviour determines the asymptotic form of the flow integrals and is studied in Sections (\ref{app:small_p}) and (\ref{app:big_p}).

\section{Scaling limit (\texorpdfstring{$p\ll k$}{p<<k})}
\label{app:small_p}

Here we give expressions for $F_{i,j}(0)$. They are computed in a straightforward way by Taylor expanding the integrands of \Eqs{eq:F11} to (\ref{eq:F24}) up to leading order and performing the angular integration.
In the following, we use the notation $\delta Z_i'(1) = \left. \text{d} \delta Z_i/\text{d}\hat{p} \right|_{\hat{p}=1}$. 
Such terms arise in $F_{1,1}(0)$ and $F_{2,1}(0)$ because the respective integrands vanishes at $\hat{p}=0$ and the Taylor expansions of $\delta Z_i(\hat{\mathbf{q}})$ around $\hat{q}=1$ enter the leading term.
For $d=1$ we get
\begin{align}
 & F_{1,1}(0) = \frac{1}{{8}\pi S_1^3 S_2} \left[2(\eta_1-1)S_1^2 S_2^2 - S_2^2 \delta Z_1'(1) + S_1^2\delta Z_2'(1)\right],\nonumber  \\ % rechecked 
 & F_{1,2}(0) = -\frac{1}{8 \pi S_1^4 S_2^2} \left[4(\eta_1-1) S_1^2 S_2^2 + S_2^2 \delta Z_1'(1) + S_1^2 \delta Z_2'(1) \right], \nonumber  \\
 & F_{1,3}(0) = F_{2,4}(0) = \frac{1}{{8}\pi S_1^3 S_2}, \nonumber  \\ %rechecked
 & F_{2,1}(0) = \frac{1}{{32}\pi S_1^6} \left[4(\eta_1-1)S_1^2 S_2^2 +S_2^2\delta Z_1'(1)+S_1^2\delta Z_2'(1)\right], \nonumber  \\ %rechecked
 & F_{2,2}(0) = \frac{1}{16\pi S_1^5 S_2} \left[4(\eta_1-1)S_1^2 S_2^2 + S_2^2 \delta Z_1'(1) + S_1^2 \delta Z_2'(1) \right], \nonumber  \\
 & F_{2,3}(0) = -\frac{S_2}{{32} \pi S_1^5}. \nonumber  \\ %rechecked
% & F_{2,4}(0) = \frac{1}{{8}\pi S_1^3 S_2}. %rechecked
\label{eq:Fij_0_1d}
\end{align}
And for $d\neq1$ we have
\begin{align}
 & F_{1,1}(0) = \frac{2^d\Omega_d \Gamma(d/2)^2}{{16}\pi (d-1)! \, S_1^3 S_2} \left[2(\eta_1-1) S_1^2 S_2^2 - S_2^2 \delta Z_1'(1){+}S_1^2\delta Z_2'(1)\right], \nonumber \\% This was rechecked and I find +1 instead of -3.
 & F_{1,2}(0) = -\frac{\Omega_d}{8 d S_1^4 S_2^2} \left[4(\eta_1-1)S_1^2 S_2^2 + S_2^2 \delta Z_1'(1)+ S_1^2 \delta Z_2'(1)\right], \nonumber\\
 & F_{1,3}(0) = F_{2,4}(0) = \frac{\Omega_d (d-1)}{{4}dS_1^3 S_2}, \nonumber \\ %rechecked
 & F_{2,1}(0) = \frac{\Omega_d}{{32}d S_1^6} \left[4(\eta_1-1)S_1^2 S_2^2+S_2^2\delta Z_1'(1)+S_1^2\delta Z_2'(1)\right], \nonumber \\ %rechecked
 & F_{2,2}(0) = \frac{\Omega_d}{16 d S_1^5 S_2} \left[4(\eta_1-1)S_1^2 S_2^2+S_2^2\delta Z_1'(1)+S_1^2\delta Z_2'(1)\right], \nonumber \\
 & F_{2,3}(0) = -\frac{\Omega_d(d-1)S_2}{{16}dS_1^5}. \nonumber \\ %rechecked
% & F_{2,4}(0) = \frac{\Omega_d (d-1)}{{4}dS_1^3 S_2}. %rechecked
\label{eq:Fij_0_Dd}
\end{align}
We have introduced the short-hand notation $S_i = S_i(1)$. This leads to the following asymptotic form for the re-scaled flow integrals,
\begin{align}
\hat{I}^{(2)}_1(\hp) & \cong   F_{1,1} + F_{1,2} \, \hp^{(\eta_1+\eta_2)/2+1} \sqrt{\left[1+\delta Z_1(\hp)\right] \left[1+ \delta Z_2(\hp)\right]} \nonumber \\[0.1cm]
& \qquad + F_{1,3} \, \hp^{\eta_1+\eta_2+2\delta_{d1}-1} \left[1+ \delta Z_1(\hp)\right] \left[1+ \delta Z_2(\hp) \right] \vphantom{\sqrt{\left[1+\delta Z_1(\hp)\right] \left[1+ \delta Z_2(\hp)\right]}}, \nonumber \\[0.2cm]
\hat{I}^{(2)}_2(\hp) & \cong  F_{2,1} \, \hp^{(\eta_1+\eta_2)/2+1} \, \sqrt{\left[1+ \delta Z_1(\hp) \right]\left[1+\delta  Z_2(\hp)\right] }  \nonumber \\
& \qquad  + F_{2,2} \, \hp^{\eta_2+1} \left[1+\delta Z_2(\hp) \right] + F_{2,3} \, \hp^{\eta_1+\eta_2 +2\delta_{d1}-1}\, \left[1+\delta Z_1(\hp)\right] \left[1+\delta Z_2(\hp)\right] \nonumber \\[0.1cm]
& \qquad  + F_{2,4}\, \hp^{2\eta_2+2\delta_{d1}-1} \left[1+\delta Z_2(\hp)\right]^2,
\label{eq:asymptotic_rescaled_flow_integrals}
\end{align}
with the short-hand notation $F_{i,j} = F_{i,j}(0)$. \Eqs{eq:asymptotic_eq} follow by inserting this in \Eqs{eq:flow2}.

\section{Scaling limit (\texorpdfstring{$p\gg k$}{p>>k})}
\label{app:big_p}

In this subsection, the asymptotic behaviour of the integrals $\hat{I}^{(2)}_{1,2}(\hat{p})$ for $\hat{p}\gg1$ is derived from the respective dependence of the integrals $F_{i,j}(\hat{p})$ given in \Eqs{eq:F11} to \eq{eq:F24}. We discuss this for each $F_{i,j}(\hat{p})$ separately, taking into account spherical symmetry.

We start by making a simplification which is valid only in the asymptotic limit $\hp \gg 1$ and for \UV convergent fixed points $\delta Z_i(\hp \to \infty )=0$. For $\hat{p}^2 \gg 1$, then also $\hat{q} \sim \hat{p}$, \iec $\hat{q}^2 - 1 > 0$, and we can approximately set the theta functions and, since $\delta Z_{i}(\hat{q}\to \infty) = 0$, also the $S_{i}(\hat q)$ to one. Separating out the leading UV scaling, $F_{i,j}(\hat p\to\infty) \sim \hat p^{\gamma_{i,j}}$, we write the $F_{i,j}(\hat{p})$ in the form
\begin{align}
 F_{i,j}(\hat{p}) = \hat{p}^{\gamma_{i,j}} \int_{\Omega} f_{i,j}(1/\hat{p},\hat{\mathbf{p}} \cdot \mathbf{e}_{\mathbf{r}}/\hat{p}).
\label{eq:deff}
\end{align}
The $f_{i,j}$ are finite and non-vanishing at $1/\hat{p} = 0$. 

Note that, in \Eqs{eq:F12}, \eq{eq:F21} and \eq{eq:F22}, different terms can be leading in the UV such that the above definition of the $f_{i,j}$ and $\gamma_{i,j}$ depends on the values of the $\eta_{1,2}$. Moreover, the denominator of $\T$ in \Eqs{eq:F12}--\eq{eq:F24} contains a divergence if $\eta_1-\eta_2 > 0$ in which case an additional factor $\hat{p}^{(\eta_2-\eta_1)/2}$ appears. This can be seen by recalling the definition $T(\hat q) = (z_1/z_2)^{1/2} k^{2}(\nukk+\nukq)^{-1} $, which gives (recall $\hat q=|\hat{\mathbf{p}}-\mathbf{e}_\mathbf{r}|$) the large-$\hat{p}$ asymptotic behaviour
\begin{align}
T(|\hat{\mathbf{p}}-\mathbf{e}_\mathbf{r}|) 
& \cong \left[\left(\hat{p}^{2} + 1 
-{2}\hat{p} \frac{\mathbf{p}}{p}\cdot\mathbf{e}_{\mathbf{r}}\right)^{(\eta_1-\eta_2)/4}
+\frac{S_1(1)}{S_2(1)}\right]^{-1} 
\nonumber \\
& \cong \left\{ \begin{array}{ll} 
\hat{p}^{-(\eta_1-\eta_2)/2} & \text{if } \eta_1-\eta_2 > 0 \\ S_2(1)/S_1(1) & \text{if } \eta_1-\eta_2 < 0 \end{array} \right. .
\label{eq:Tq}
\end{align}
Having identified the leading scaling behaviour, the integrals can be computed in the limit $\hat p\to \infty$ by neglecting sub-leading contributions to the integrands. 
We can approximate $f_{i,j}(1/\hat{p},\hat{\mathbf{p}} \cdot \mathbf{e}_{\mathbf{r}}/\hat{p}) \cong f_{i,j}(0,\hat{\mathbf{p}} \cdot \mathbf{e}_{\mathbf{r}}/\hat{p})$ in the integrands and perform the angular integration which gives, for those integrals where $f_{i,j}(0,y)$ does not depend on $y=\hat{\mathbf{p}} \cdot \mathbf{e}_{\mathbf{r}}/\hat{p}$, a surface factor $\Omega_d = \int_{\Omega} = d \pi^{d/2}[(2\pi)^d \Gamma(d/2+1)]^{-1}$. The asymptotic behaviour of the integrals $F_{1,1}(\hat{p})$, $F_{1,3}(\hat{p})$, and $F_{2,3}(\hat{p})$ can be derived in this way. The result is ($S_i \equiv S_i(1)$)
\begin{align}
F_{1,1}(\hat{p}\to&\,\infty) 
\simeq \Omega_d S_2^2 \left[\delta_{d1}/2+(d-1)/d\right] \hat{p}^{1-2 \delta_{d1}} \hat p^{-(\eta_1+\eta_2)/2},
\label{eq:F11_big_p}
\\[0.5cm]
%\end{align}
%
%\begin{align}
F_{1,3}(\hat{p}\to&\,\infty) 
 \simeq  \frac{\Omega_d}{2 S_1 S_2} \hat{p}^{2-2\delta_{d1}} \left\{ \begin{array}{ll} \hat{p}^{-\eta_1} & \text{if } \eta_1>\eta_2 \\[0.1cm] 
\hat{p}^{-\eta_1}(1+S_1/S_2)^{-1} & \text{if } \eta_1 = \eta_2 \\[0.1cm]
 \hat{p}^{-(\eta_1+\eta_2)/2} (S_2/S_1) & \text{if } \eta_1<\eta_2 \end{array} \right.,
\label{eq:F13_big_p}
\\[0.5cm]
%\end{align}
%
%and
%
%\begin{align}
F_{2,3}(\hat{p}\to&\,\infty) \simeq  -\frac{\Omega_d}{2S_1S_2} \hat{p}^{2-2\delta_{d1}} \left\{ \begin{array}{ll} \hat{p}^{-2\eta_1+\eta_2} & \text{if } \eta_1>\eta_2 \\[0.1cm]
 \hat{p}^{-\eta_1} \left(1+S_1/S_2\right)^{-3} & \text{if } \eta_1 = \eta_2 \\[0.1cm] 
\hat{p}^{-(\eta_1+\eta_2)/2} \left({S_2/S_1}\right)^{3} & \text{if } \eta_1<\eta_2 \end{array} \right. .
\label{eq:F23_big_p}
\end{align}

The calculation of the asymptotic behaviour of the integrals \eq{eq:F12}, \eq{eq:F21} and \eq{eq:F22} can become more involved. Two possibilities arise. If $\eta_1+\eta_2 \geq -2$, the asymptotic behaviour is determined in the same way as for $F_{1,1}$, $F_{1,3}$ and $F_{2,3}$. However, for $\eta_1+\eta_2 < -2$ the leading term of $f_{i,j}(\epsilon\to0,\hat{\mathbf{p}} \cdot \mathbf{e}_{\mathbf{r}}/\hat{p})$ is proportional to $\hat{\mathbf{p}} \cdot \mathbf{e}_{\mathbf{r}}/\hat{p}$, and thus vanishes under the angular integral. In this case, the asymptotically leading term is obtained by expanding  $yT(\hat q)\equiv yT(\hat p,y)$ to order $y^{2}$ before the limit $\hat{p}\to \infty$ is taken and the term that is linear in $y$ is neglected. This ensures that we only consider terms that contribute to the angular integration. One can check that truncating at order $y^2$ does not affect the asymptotic behaviour. \cred{Indeed} $y$ enters through the combination $\hat{p}^2-2\hat{\mathbf{p}} \cdot \mathbf{e}_{\mathbf{r}} 
= (1-2y/\hat{p})\hat p^2$. We see that the term of order $y^n$ is multiplied by $1/\hat p^n$ and can only dominate in the asymptotic regime if all the lower order terms are irrelevant.

We discuss the procedure for $F_{1,2}(\hat{p})$ and state the results for the two remaining integrals $F_{2,1}(\hat{p})$ and $F_{2,2}(\hat{p})$. To simplify the derivation we use that $(\eta_{1}+\eta_{2})/2=2\eta_{1}-2-d$ from \Eq{eq:consistency}. We start by approximating $\delta Z_{i}(\hat{q})\simeq0$, $\theta \left( \hat{q}^2-1\right)=1$ in \Eq{eq:F12}, which gives, defining $\epsilon=1/\hat{p}$ such that $\hat{\mathbf{p}} \cdot \mathbf{e}_{\mathbf{r}} = y/\epsilon$,
\begin{align}
& F_{1,2}(\hat{p}) \cong \bigintsss_{\Omega} \left[\frac{\epsilon y-1}{S_1 S_2} - \hat{q}^{-2\eta_1+2+d} \epsilon  y\right] \T ,
\end{align}
with $\hat{q} = \sqrt{1+\epsilon^2-2\epsilon y}/\epsilon$. We factor out $\epsilon^{-2\eta_1+2+d}$ from $\hat{q}^{2\eta_1+2+d}$ in the second term:
\begin{align}
& F_{1,2}(\hat{p}) \cong \bigintsss_{\Omega} \, \T \left[ \frac{\epsilon y -1}{S_1 S_2}
\vphantom{y(1+\epsilon^2-2\epsilon y)^{(-2\eta_1+2+d)/2}} - \epsilon^{-d-1+2\eta_1}  y(1+\epsilon^2-2\epsilon y)^{(-2\eta_1+2+d)/2} \right] .
\label{eq:befored+2}
\end{align}
The asymptotic behaviour of $T(\hat q)$ is determined by the sign of $-(\eta_{1}-\eta_{2})/2=\eta_{1}-2-d$, see \Eq{eq:Tq}. For both signs, different $\eta_{1}$ will render either of the terms in \Eq{eq:befored+2} dominating for large $\hat p$ ($\epsilon\to0$).

1. $\eta_1<d+2$, $T(\hat{q}\to\infty) \sim \hat{p}^{\eta_1-d-2}\,$: We write $\T = \epsilon^{-\eta_1+2+d} \, \twidlT$ such that $\tilde{T}(\epsilon\to 0) = 1$ and
\begin{align}
& F_{1,2}(\hat{p}) \cong \bigintsss_{\Omega} \left[\epsilon^{-\eta_1+d+2} \frac{\epsilon y-1}{S_1 S_2} - \epsilon^{\eta_1+1} (1+\epsilon^2-2\epsilon y)^{(-2\eta_1+2+d)/2} y\right] \twidlT .
\label{eq:before(d+1)_2}
\end{align}
There are three sub-cases to be distinguished: (a) For $2\eta_1<d+1$, the second term, which provides an extra scaling factor $\epsilon^{\eta_1+1}$, is dominant. Then the leading-power exponent defined in \Eq{eq:deff} reads $\gamma_{1,2}=-\eta_1-1$, and the integrand is
\begin{align}
& f_{1,2}(\epsilon,y) = \left[\epsilon^{-2\eta_1+d+1}\frac{\epsilon y-1}{S_1 S_2} -(1+\epsilon^2-2\epsilon y)^{(-2\eta_1+2+d)/2} y \right] \twidlT .
\end{align}
The leading term $f_{1,2}(0,y)= -y$ does not contribute to the angular integral. Taking the sub-leading factors into account by expanding  to second order in $y$,
\begin{align}
 f_{1,2}(\epsilon,y) & \cong  -\twidlT \Big( y [1+\epsilon^2]^{(-2\eta_1+2+d)/2}  \nonumber\\[0.1cm]
&\qquad {}+ y^2 \epsilon [1+\epsilon^2]^{(-2\eta_1+d)/2} \left[ 2\eta_1 -2-d - (\eta_1-2-d)(1+\epsilon^2)^{(-\eta_1+2+d)/2} \, \twidlT \right] \nonumber \\[0.1cm]
&\qquad {}+ \epsilon^{-2\eta_1+d+1}/(S_1S_2) \Big\{ 1-\epsilon y \left[1 + (\eta_1-2-d) (1+\epsilon^2)^{(-\eta_1+d)/2} \, \twidlT \right] \nonumber \\[0.1cm]
&\qquad \quad {}+ \epsilon^2 y^2 (\eta_1-2-d) (1+\epsilon^2)^{(-\eta_1-2+d)/2} \, \twidlT \nonumber \\[0.1cm]
&\qquad \qquad \times \left[ (\eta_1-2-d)(1+\epsilon)^{(-\eta_1+2+d)/2} \, \twidlT  +(d-\eta_1)/2 + 1+ \epsilon^2 \vphantom{(-\eta_1+2+d)(1+\epsilon)^{(-\eta_1+2+d)/2} \, \twidlT}\right] \Big\} \Big),
\end{align}
we find that two terms are competing, giving rise to a further case distinction: If $\eta_1 < d/2$, the contributions proportional to $\epsilon^{-2\eta_1+d+1}$ are sub-leading and the quadratic term in $y$ dominates. In turn, if $\eta_1 > d/2$, the term that does not depend on $y$ dominates. Both must be taken into account if $\eta_1 = d/2$. As a result,
\begin{align}
& {f}_{1,2}(\epsilon\to0,y) 
 \simeq -\ y -\frac{\epsilon}{S_1 S_2} \left\{ \begin{array}{ll} y^2  \, \eta_1 S_1 S_2 &  \eta_1 < d/2 \\[0.1cm]
  \left(1+ y^2 d S_1 S_2/2 \right) &  \eta_1 = d/2 \\[0.1cm]
 \epsilon^{-2\eta_1+d} &  d/2 < \eta_1
< (d+1)/2 \end{array} \right.,
\end{align}
and, after angular integration, $\int_{\Omega} y^{2} = \Omega_d/d$,
\begin{align}
& F_{1,2}(\hat{p}\to\infty) \simeq  -\frac{\Omega_d}{S_1 S_2} \hat{p}^{-2} \left\{ \begin{array}{ll} \hat{p}^{-\eta_1} \, \eta_1 S_1 S_2/d & \eta_1 < d/2 \\[0.1cm] 
 \hat{p}^{-d/2} \left(1+S_1 S_2 /2 \right) & \eta_1 = d/2 \\[0.1cm] 
\hat{p}^{\eta_1-d} & d/2 < \eta_1
< (d+1)/2 \end{array} \right. .
\label{eq:asymptotic1}
\end{align}
(b) For $2\eta_1 = d+1$, both terms under the integral \eq{eq:before(d+1)_2} are equally important. We obtain $\gamma_{1,2} = -(d+3)/2$ and
\begin{align}
& f_{1,2}(\epsilon,y) = \left[\frac{\epsilon y-1}{S_1 \, S_2} - (1+\epsilon^2-2\epsilon y)^{1/2} y \right]\twidlT.
\end{align}
The relevant contribution is $f_{1,2}(0,y) = - y - (S_1S_2)^{-1}$ while  the terms of order $y^2$ are sub-dominant. 
As a result, the asymptotics \eq{eq:asymptotic1} is supplemented with
\begin{align}
& F_{1,2}(\hat{p}\to\infty) \simeq  - \frac{\Omega_d}{S_1S_2} \hat{p}^{-(d+3)/2} &  \mbox{if}\ \eta_1 = (d+1)/2 .
\label{eq:F12caseb}
\end{align}
(c) For $(d+1)/2<\eta_1<d+2$, the leading terms are  interchanged. From \Eq{eq:before(d+1)_2}, one finds $\gamma_{1,2} = \eta_1-d-2$ and
\begin{align}
& f_{1,2}(\hat{p}) = \left[\frac{\epsilon y-1}{S_1 S_2}  - \epsilon^{2\eta_1-d-1}(\epsilon^2+1-2\epsilon y)^{(d+2-2\eta_1)/2} y\right] \twidlT.
\end{align}
We find $f_{1,2}(0,y) = -(S_1S_2)^{-1}$, and, together with relation \eq{eq:F12caseb}, the last case of the asymptotics \eq{eq:asymptotic1} reads
\begin{align}
& F_{1,2}(\hat{p}\to\infty) \simeq  -\frac{\Omega_d}{S_1 S_2} \hat{p}^{-2} \, \hat{p}^{\eta_1-d}, &&  \mbox{for}\ d/2 < \eta_1 < d+2 .
\label{eq:asymptotic3}
\end{align}

2. $\eta_1 \geq d+2$,  $T(\hat{q}\to \infty) \sim const\,$: In this case, $\hat{q}^{-\eta_1+2+d}$ does not diverge for $\epsilon \to 0$, such that no powers of $1/\hat p$ arise from $\T$. Again, two competing terms in  \Eq{eq:befored+2} require the distinction of three sub-cases. However, for $\eta_1 \geq d+2$, the term proportional to $\epsilon^{-d-1+2\eta_1}$ is always sub-dominant and can be neglected. The term proportional to $(S_{1}S_{2})^{-1}$ in \Eq{eq:befored+2} is dominant, such that $\gamma_{1,2} = 0$ and
\begin{align}
& f_{1,2}(\epsilon,y) = \left[\frac{\epsilon y-1}{S_1 S_2} -\epsilon^{2\eta_1-d-1}(1+\epsilon^2-2\epsilon y)^{(d+2-2\eta_1)/2} y\right] \T.
\end{align}
Taking the limit $f_{1,2}(\epsilon\to0,y)$ and performing the angular integral one obtains the final asymptotics
\begin{align}
& F_{1,2}(\hat{p}\to\infty) \simeq  -\frac{\Omega_d}{S_1 S_2} \hat{p}^{-2} \left\{ \begin{array}{ll} \hat{p}^{-\eta_1} \, \eta_1 S_1 S_2/d & \eta_1 < d/2 \\[0.1cm] 
 \hat{p}^{-d/2} \left(1+S_1 S_2 /2 \right) & \eta_1 = d/2 \\[0.1cm] 
\hat{p}^{\eta_1-d} &  d/2 < \eta_1 < d+2 \\[0.1cm] 
 \hat{p}^{2} \, (1+S_1/S_2)^{-1} &  \eta_1 = d+2 \\[0.1cm] 
 \hat{p}^{2} S_2/S_1 &  d+2 < \eta_1 \end{array} \right. .
\label{eq:F12_big_p}
\end{align}
Using analogous arguments we find
\begin{align}
& F_{2,1}(\hat{p}) \simeq \frac{\Omega_d}{S_1 S_2} \hat{p}^{-6} \left\{ \begin{array}{ll} \hat{p}^{\eta_1-2d} S_1 S_2 (\eta_1+2+d)/(2d) & \eta_1 < d/2 \\[0.1cm] 
 \hat{p}^{-3d/2} \left[1+ S_1 S_2 (3d+4)/(4d) \right] & \eta_1 = d/2 \\[0.1cm] 
\hat{p}^{3\eta_1-3d} &  d/2 < \eta_1 < d+2 \\[0.1cm] 
 \hat{p}^{6} \left(1+S_1/S_2\right)^{-3} &  \eta_1 = d+2 \\[0.1cm] 
 \hat{p}^{6} (S_2/S_1)^3 &  d+2 < \eta_1 \end{array} \right.\!\!,
\label{eq:F21_big_p}
\\[0.5cm]
& F_{2,2}(\hat{p}) \simeq \frac{\Omega_d}{S_1 S_2} \hat{p}^{-6} \left\{ \begin{array}{ll} \hat{p}^{\eta_1-2d} \, S_1 S_2 (5\eta_1-8-4d)/d & \eta_1 < d/2 \\[0.1cm] 
 \hat{p}^{-3d/2} \left[1 - S_1 S_2 (16+3d)/(2d) \right] & \eta_1 = d/2 \\[0.1cm] 
 \hat{p}^{3\eta_1-3d} &  d/2 < \eta_1 < d+2 \\[0.1cm] 
 \hat{p}^{6} \left(1+S_1/S_2\right)^{-3} &  \eta_1 = d+2 \\[0.1cm] 
 \hat{p}^{6} (S_2/S_1)^3 &  d+2 < \eta_1 
 \end{array} \right. \!\!.
\label{eq:F22_big_p}
\end{align}
The resulting expressions for the UV leading behaviour of the integrals $F_{i,j}(\hat{p})$ can be inserted back into \Eqs{eq:rescaled_flow_1} and \eq{eq:rescaled_flow_2} in order to compute the asymptotic behaviour of $\hat{I}_{1,2}^{(2)}(\hat{p})$. Each case needs to be considered separately. With \Eq{eq:consistency}, we find that
\begin{align}
  \hat{I}_{i}^{(2)}(\hat{p}\gg 1) \sim \hat{p}^{\beta_i},
\end{align}
with
\begin{align}
& \beta_1 = \left\{ 
 \begin{array}{ll} 
 4-2\delta_{d1}-2\eta_1+d  &\mbox{if}\ (6+3d-2\delta_{d1})/5  \geq \eta_1  \\[0.1cm]
 3\eta_1-2d-2  & \mbox{if}\ (6+3d-2\delta_{d1})/5 < \eta_1 \leq d+2 \\[0.1cm]
 2\eta_1-d  &\mbox{if}\ \eta_1 > d+2 
\end{array} \right. ,
\label{eq:beta1}
\\[0.5cm]
& \beta_2  = \left\{ \begin{array}{ll} 
 3(\eta_1 -d-2)  &\mbox{if}\  \eta_1 \leq d/2  \\[0.1cm]
5\eta_1-4d-6 &\mbox{if}\  d/2 < \eta_1 \leq d+2 \\[0.1cm]
3\eta_1-2d -2 &\mbox{if}\  d+2 < \eta_1 \leq d+2 + 2 \delta_{d1} \\[0.1cm]
4\eta_1-3d-4 -2\delta_{d1} &\mbox{if}\  d+2 + 2 \delta_{d1} < \eta_1 
\end{array} \right. .
\label{eq:beta2}
\end{align}
See Figure (\ref{fig:beta}) where $\beta_1$ and $\beta_2$ are plotted with respect to $\eta_1$ and $\eta_2$ respectively for $d=1$ and $d=3$.
The integrals on the right-hand side of \Eq{eq:flow3} converge if $\beta_i < \eta_i$, corresponding to the allowed range \eq{eq:range} for $\eta_{1}$.

\chapter{Equations for \texorpdfstring{$A_i$}{Ai}}
\label{sec:equations_for_ai}
\ResetAbbrevs{All}

In this section we write down explicit equations containing the pre-factors of \Eqs{eq:def_alphas}, $A_i$. We use a slightly different notation and replace \Eqs{eq:def_alphas} by
\begin{align}
 \delta Z_{i}(\hp \to 0) \cong c_i + a_i \, \hp^{\alpha_i-\eta_i}, & & \text{as } \hp \to 0.
 \label{eq:small_p_appendix}
\end{align}
This corresponds to the short-hand notation $a_i = A_i (\alpha_i-\eta_i)$.
We consider here cases where either $c_i = -1$ or $\alpha_i-\eta_i < 0$.
Then when we insert \Eqs{eq:small_p_appendix} into the re-scaled flow integrals \Eqs{eq:asymptotic_rescaled_flow_integrals} an take the limit $\hp \to 0$ we can neglect all the $1+c_i$ factors since they are sub-dominant as compared to the $a_i\hp^{\alpha_i-\eta_i}$. We get,
\begin{align}
\hat{I}^{(2)}_1(\hp) & \cong  F_{1,1} + F_{1,2} \, \sqrt{a_1 a_2} \, \hp^{(\alpha_1+\alpha_2)/2+1} + F_{1,3} \, a_1 a_2 \, \hp^{\alpha_1+\alpha_2+2\delta_{d1}-1}, \nonumber \\[0.2cm]
\hat{I}^{(2)}_2(\hp) & \cong  F_{2,1} \, \sqrt{a_1 a_2} \, \hp^{(\alpha_1+\alpha_2)/2+1} + F_{2,2} \, a_2 \, \hp^{\alpha_2+1} \nonumber \\[0.1cm]
& \qquad  + F_{2,3} \, a_1 a_2 \, \hp^{\alpha_1+\alpha_2 +2\delta_{d1}-1} + F_{2,4}\, a_2^2 \, \hp^{2\alpha_2+2\delta_{d1}-1}.
\label{eq:monomials}
\end{align}
We find linear combinations of monomials.
When $c_i = -1$ or $\alpha_i-\eta_i < 0$ the exponents of these monomials only depend on $\alpha_i$. We see that once we have chosen values of $\alpha_i$ from the solutions of \Eqs{eq:eq_alphas} we can compare the different exponents and identify the leading monomial in the limit $\hp \to 0$.
It then becomes possible to match the pre-factors on both sides of \Eqs{eq:asymptotic_eq} and extract equations containing both $a_i$.

Each case of \Eqs{eq:alphas} must be considered separately. Moreover the case $d=1$ must also be considered separately since the exponents are different than for $d\neq 1$.
We start with $d=1$.

\begin{itemize}
\item If $(\alpha_1,\alpha_2) = (1,5)$ \Eqs{eq:monomials} can be simplified to
\begin{align}
\hat{I}^{(2)}_1(\hp) \cong \, F_{1,1}, && 
\hat{I}^{(2)}_2(\hp) \cong \, F_{2,1} \, \sqrt{a_1 \, a_2} \, \hp^{4}. \nonumber
\end{align}
The terms proportional to $F_{1,1}$ and $F_{2,1}$ are dominating.
The corresponding equations are
\begin{align}
 & a_1(\eta_1-1) = h \, F_{1,1}, \nonumber \\
 & a_2(\eta_2-5) = h \sqrt{a_1 a_2} \, F_{2,1}.
 \label{eq:eq_ai_first_1d}
\end{align}

\item If $-2<\alpha_1 < 1$ and $\alpha_2 = -2$, the dominating terms are the ones that are proportional to $F_{1,3}$ and $F_{2,4}$,
\begin{align}
\hat{I}^{(2)}_1(\hp) \cong \, F_{1,3} \, a_1 \, a_2 \, \hp^{\alpha_1-1}, &&
\hat{I}^{(2)}_2(\hp) \cong \, F_{2,4} \, a_2^2 \, \hp^{-3}, \nonumber
\end{align}
and thus
\begin{align}
 a_1 (\eta_1 - \alpha_1) &= h a_1 a_2 \, F_{1,3}, \nonumber \\
 a_2 (\eta_2+2) &= h a_2^2 \, F_{2,4}.
\label{eq:aclosed_1d}
\end{align}

\item If $\alpha_1 =-2$ and $\alpha_2 = -2$, the term proportional to $F_{1,3}$ still dominates $\hat{I}_1^{(2)}(\hat{p})$, but the term proportional to $F_{2,3}$ is of the same order as the one proportional to $F_{2,4}$.
Taking both into account gives
\begin{align}
\hat{I}^{(2)}_1(\hp) \cong \, F_{1,3} \, a_1 \, a_2 \, \hp^{-3}, &&
\hat{I}^{(2)}_2(\hp) \cong \, \left[F_{2,3} \, a_1 \, a_2 + F_{2,4}\, a_2^2 \right] \hp^{-3}, \nonumber
\end{align}
and
\begin{align}
& a_1 (\eta_1+2) = h a_1 a_2 \, F_{1,3}, \nonumber \\
& a_2 (\eta_2+2) = h \left[ a_1 a_2 \, F_{2,3}+ a_2^2 \, F_{2,4}\right].
\end{align}

\item Finally, for $\alpha_1 =1$ and $\alpha_2 = -2$ the dominating terms are proportional to $F_{1,1}+ a_1 a_2 F_{1,3}$ and $F_{2,4}$, \ie
\begin{align}
\hat{I}^{(2)}_1(\hp) \cong \, F_{1,1} + F_{1,3} \, a_1 \, a_2, &&
\hat{I}^{(2)}_2(\hp) \cong \, F_{2,4}\, a_2^2 \, \hp^{-3}, \nonumber
\end{align}
and
\begin{align}
& a_1 (\eta_1-1)=  h \left[ F_{1,1} + a_1 a_2 \, F_{1,3}\right], \nonumber \\
& a_2 (\eta_2+2)= h a_2^2 \, F_{2,4}.
\label{eq:eq_ai_last_1d}
\end{align}
\end{itemize}

We now discuss the cases $d\neq 1$.

\begin{itemize}
 \item If $(\alpha_1,\alpha_2) = (1,5)$, we obtain
\begin{align}
\hat{I}^{(2)}_1(\hp) \cong \, F_{1,1} &&
\hat{I}^{(2)}_2(\hp) \cong \, F_{2,1} \, \sqrt{a_1 a_2} \, \hp^{4}, \nonumber
\end{align}
and
\begin{align}
 a_1 (\eta_1-1) &= h \, F_{1,1},\nonumber \\
 a_2 (\eta_2-5) &= h \sqrt{a_1 a_2} \, F_{2,1}.
 \label{eq:eq_ai_first_dD}
\end{align}
\item If $0<\alpha_1<1$ and $\alpha_2 = 0$, we get
\begin{align}
\hat{I}^{(2)}_1(\hp) \cong \, F_{1,3} \, a_1 a_2 \, \hp^{\alpha_1-1}, &&
\hat{I}^{(2)}_2(\hp) \cong \, F_{2,4} \, a_2^2 \, \hp^{-1}, \nonumber
\end{align}
and
\begin{align}
 a_1 (\eta_1-\alpha_1) &= h a_1 a_2 \, F_{1,3}, \nonumber \\
 a_2 \, \eta_2 &= h a_2^2 \, F_{2,4}.
\label{eq:aclosed_Dd}
\end{align}
\item If $\alpha_1 = \alpha_2 = 0$, we find
\begin{align}
\hat{I}^{(2)}_1(\hp) \cong \, F_{1,3} \, a_1 a_2 \, \hp^{-1}, &&
\hat{I}^{(2)}_2(\hp) \cong \, \left[ F_{2,3} \, a_1 a_2  + F_{2,4}\, a_2^2 \right] \hp^{-1}. \nonumber
\end{align}
and the equations are
\begin{align}
& a_1 \, \eta_1 = h a_1 a_2 \, F_{1,3}, \nonumber \\
& a_2 \, \eta_2 = h \left[ a_1 a_2 \, F_{2,3}+a_2^2 \, F_{2,4}\right].
\end{align}
\item Finally when $\alpha_1 = 1$ and $\alpha_2 = 0$, one obtains
\begin{align}
\hat{I}^{(2)}_1(\hp) \cong \, F_{1,1} + F_{1,3} \, a_1 a_2 , &&
\hat{I}^{(2)}_2(\hp) \cong \, F_{2,4}\, a_2^2 \, \hp^{-1}. \nonumber
\end{align}
with
\begin{align}
 a_1 (\eta_1-1) &= h \left[ F_{1,1} + a_1 a_2 \, F_{1,3}\right], 
 \nonumber \\
 a_2 \, \eta_2 &= h a_2^2 \, F_{2,4}.
 \label{eq:eq_ai_last_dD}
\end{align}
\end{itemize}

Note that all of these equations depend explicitly on $F_{i,j}$ and $h$. $F_{i,j}$ are shown in \Eqs{eq:Fij_0_1d} and \eq{eq:Fij_0_Dd} and are non-linear functions of $\eta_i$, $\delta Z_i(1)$ and $\delta Z_i'(1)$ and $h$ is related to the full functions $\delta Z_i(\hp)$ through \Eqs{eq:flow2}. This means that the asymptotic behaviour of $\delta Z_i(\hp)$ which is described by $a_i$ is coupled to the rest of the functions. $h$ can be eliminated from these equations by taking their ratio but $a_i$ stay coupled to $\delta Z_i(1)$ and $\delta Z_i'(1)$ through $F_{i,j}$.

For the cases of \Eqs{eq:aclosed_1d} and \eq{eq:aclosed_Dd} the ratio of the two equations simplifies greatly because $F_{1,3}=F_{2,4}$ cancels out. This leads to a relation in between $\eta_i$ and $\alpha_1$ which is discussed in the main text, at the end of Section (\ref{sec:scaling_limit_p_ll_k}). $a_i$ however drop out as well so that we can not say that the asymptotic behaviour of $\delta Z_i(\hp)$ is decoupled form the rest of the functions.

\chapter{Energy spectrum decomposition}
\label{sec:kinetic_energy_spectrum_decomposition}
\ResetAbbrevs{All}

In this section we decompose the kinetic energy spectrum of the ultra-cold Bose gas and give definitions for its different components,
\begin{align}
\epsilon_{\text{kin}}(\mathbf{p}) = \epsilon_{\text{inc}}(\mathbf{p}) + \epsilon_{\text{comp}}(\mathbf{p}) + \epsilon_{\text{quant}}(\mathbf{p}).
\end{align}
See \cite{Nowak:2011sk} (which we follow closely here) and reference therein.
Our starting point is the kinetic energy density of the Bose gas,
\begin{align}
%\int_\mathbf{p} \epsilon_{\text{kin}}(p)
\frac{E_{\text{kin}}}{\mathcal{V}} = \frac{1}{2m} \langle \boldsymbol{\nabla}\psi \cdot \boldsymbol{\nabla}\psi^\dagger\rangle =   \frac{1}{2m} \left[\langle \left(\sqrt{n} \, \boldsymbol{\nabla}\theta \right)^2\rangle + \langle \left(\boldsymbol{\nabla}\sqrt{n}\right)^2 \rangle\right],
\end{align}
expressed in terms of the density and the phase of the complex field $\psi = \sqrt{n}\, \text{e}^{i\theta}$. We see already that there are two parts to the kinetic energy spectrum. The first contains the fluctuations of $\mathbf{w} = \sqrt{n}\boldsymbol{\nabla}\theta$ and the second of $\boldsymbol{\nabla}\sqrt{n}$. The first can be associated with classical hydrodynamics because it contains the fluctuations of the phase while the second is a result of the quantum nature of the system. We define,
\begin{align}
\epsilon_{\text{quant}}(\mathbf{p}) = \frac{1}{2m}\int_{\mathbf{x}} \text{e}^{i\mathbf{p}\cdot\mathbf{r}} \, \langle \boldsymbol{\nabla}\sqrt{n}(t,\mathbf{x}+\mathbf{r}) \cdot \sqrt{n}(t,\mathbf{x}) \rangle.
\end{align}
In order to further decompose the kinetic energy spectrum the vector field $\mathbf{w}$ is decomposed into
\begin{align}
\mathbf{w} = \mathbf{w}_{\text{inc}} + \mathbf{w}_{\text{comp}}, && \text{with } \boldsymbol{\nabla}\cdot\mathbf{w}_{\text{inc}} = 0.
\end{align}
$\mathbf{w}_{\text{inc}}$ is a divergence-less field analogous to the velocity field of incompressible hydrodynamics. $\mathbf{w}_{\text{comp}}$ is defined as the difference of $\mathbf{w}$ and $\mathbf{w}_{\text{inc}}$.
We define
\begin{align}
& \epsilon_{\text{inc}}(\mathbf{p}) = \frac{1}{2m}\int_{\mathbf{x}} \text{e}^{i\mathbf{p}\cdot\mathbf{r}} \, \langle \mathbf{w}_{\text{inc}}(t,\mathbf{x}+\mathbf{r}) \cdot \mathbf{w}_{\text{inc}}(t,\mathbf{x}) \rangle, \nonumber \\
& \epsilon_{\text{comp}}(\mathbf{p}) = \frac{1}{2m}\int_{\mathbf{x}} \text{e}^{i\mathbf{p}\cdot\mathbf{r}} \, \langle \mathbf{w}_{\text{comp}}(t,\mathbf{x}+\mathbf{r}) \cdot \mathbf{w}_{\text{comp}}(t,\mathbf{x}) \rangle.
\end{align}
Note that there is in principle an additional term arising from the product of $\mathbf{w}_{\text{inc}}$ and $\mathbf{w}_{\text{comp}}$.
This term is expected to be small because it accounts for cross-correlations in between very different physical processes. See \cite{Nowak:2011sk} where this was found numerically.
Incompressible excitations contain conservative hydrodynamic flows. They do not remove matter from the system.
For example the phase of a pure vortex solutions is a linear function of the angle around the position of the vortex\footnote{$m$ is the circulation of the vortex and $\boldsymbol{e}_\phi$ is the unit vector along the azimuthal direction.}, $\boldsymbol{w} = m\sqrt{n}/r \, \boldsymbol{e}_\phi$. Then we have $\boldsymbol{\nabla}\cdot \boldsymbol{w} = \sqrt{n} \nabla^2 \theta = 0$ which is incompressible.
On the other hand $\mathbf{w}_{\text{comp}}$ contains the rest of the hydrodynamics: pressure dynamics, sound waves, dissipative processes, etc.

\chapter{Notation and conventions}
\label{sec:notations_and_conventions}
\ResetAbbrevs{All}

In this section we introduce some notation and conventions that will be useful later on. This section is meant to give a precise meaning to definitions that may be a little fuzzy in the main text and to serve as a reminder for the reader. It is not meant to be read all at once, but rather to be consulted when in need.

\begin{itemize}
 \item Vectors are noted in boldface $\mathbf{v} = \left(v_1,v_2,..,v_d\right)$ and their components are in normal font $v_i$. Lower case indices ($i,j,k,...$) represent spatial indices and run from $1$ to $d$. Repeated indices are to be summed over and the norm of a vector is in normal font without index,
\begin{align}
 v = \sqrt{v_i \, v_i} = \sqrt{\sum_{i=1}^d v_i^2} \, .
\end{align}

\item Real space differential operators are defined as follows
\begin{align}
& \boldsymbol{\nabla} = \left(\frac{\partial}{\partial x_1},\frac{\partial}{\partial x_2},..,\frac{\partial}{\partial x_d}\right),
& & \nabla^2 = \frac{\partial^2}{\partial x_i \partial x_i},
& & \Delta = \left( \nabla^2,\nabla^2,..,\nabla^2\right).
\end{align}
Note that we distinguish the Laplacian $\nabla^2$, from the vector Laplacian $\Delta$, by the object on which they act. The former acts on scalars while the latter on vectors.

\item We use the following normalisation for the Fourier transformations
\begin{align}
&\mathbf{v}(t,\mathbf{x}) = \int \frac{\mathrm{d}\omega\,\mathrm{d}^{d}p}{(2\pi )^{d+1}} \, \text{e}^{i\left(\omega t- \mathbf{p} \cdot \mathbf{x}\right)} \, \mathbf{v}(\op), & & \mathbf{v}(\op) = \int\mathrm{d}t\,\mathrm{d}^{d}x \, \text{e}^{-i\left(\omega t- \mathbf{p} \cdot \mathbf{x}\right)} \, \mathbf{v}(t,\mathbf{x}).
\label{eq:def_fourier}
\end{align}

\item We use the short-hand notation
\begin{align}
&\int_{t,\mathbf{x}}=\int\mathrm{d}t\,\mathrm{d}^{d}x , && \int_{\omega,\mathbf{p}}=\frac{1}{(2\pi )^{d+1}}\int\mathrm{d}\omega\,\mathrm{d}^{d}p.
\end{align}
Moreover when the integral is the same whether its integrand is expressed in real or Fourier space we write, \eg
\begin{align}
 \int_{\tx} \mathbf{v}(\tx) \cdot \mathbf{J}(\tx) = \int_{\op} \mathbf{v}(\op) \cdot \mathbf{J}(\mop) = \int \mathbf{v} \cdot \mathbf{J}.
\end{align}
Finally the symbol $\Omega$ is used for angular integrations in Fourier space,
\begin{align}
\int_{\mathbf{p}} = \int_0^{\infty} \, \text{dp} \, p^{d-1} \, \int_{\Omega}.
\end{align}
Note that $\int_{\Omega}$ contains a $1/(2\pi)^d$ factor.

\item Angular brackets are used to denote averages with respect either to a stochastic forcing, to quantum fluctuations or both depending on the context. The $c$ index denotes connected correlation functions. If the moments of the field $\mathbf{v}(\tx)$ are generated by
\begin{align}
 & Z[\mathbf{J}] \equiv \langle \text{e}^{\int \mathbf{J} \cdot \mathbf{v}} \rangle ,\nonumber \\
& \langle v_{i_1}(t_1,\mathbf{x}_1) \cdot .. \cdot v_{i_n}(t_n,\mathbf{x}_n) \rangle = \frac{\delta^n Z[\mathbf{J}=0]}{\delta J_{i_1}(t_1,\mathbf{x}_1) .. \delta J_{i_n}(t_n,\mathbf{x}_n)},
\end{align}
then the connected correlation functions are generated by its logarithm,
\begin{align}
 & W[\mathbf{J}] \equiv \log\left(Z[\mathbf{J}]\right) ,\nonumber \\
& \langle v_{i_1}(t_1,\mathbf{x}_1) \cdot .. \cdot v_{i_n}(t_n,\mathbf{x}_n) \rangle_{\text{c}} = \frac{\delta^n W[\mathbf{J}=0]}{\delta J_{i_1}(t_1,\mathbf{x}_1) .. \delta J_{i_n}(t_n,\mathbf{x}_n)}.
\end{align}
In particular we have
\begin{align}
\langle v_i(t,\mathbf{x}) v_j(t',\mathbf{x}') \rangle_{\text{c}} = \langle v_i(t,\mathbf{x}) v_j(t',\mathbf{x}') \rangle - \langle v_i(t,\mathbf{x}) \rangle \langle v_j(t',\mathbf{x}') \rangle .
\end{align}

\item We will encounter angular brackets with a $k$ index $\langle X \rangle_k$. This stands for a correlation function that is computed from the flowing effective action $\Gamma_k[\mathbf{v}]$ (see Section (\ref{sec:defgamma})). It only represents a physical correlation function in the limit where the cut-off is removed,
\begin{align}
\langle X \rangle_{k\to 0} = \langle X \rangle.
\end{align}

\item We will always consider systems that are invariant under space and time translations.
The correlation functions will therefore only depend on relative spatio-temporal arguments.
In Fourier space this translates to
\begin{align}
& \langle v_{i_1}(\omega_1,\mathbf{p}_1) .. v_{i_n}(\omega_n,\mathbf{p}_n) \rangle = \delta(\omega_1+..+\omega_n) \delta(\mathbf{p}_1+\mathbf{p}_n) (2\pi)^{d+1}  \nonumber \\[0.1cm]
& \qquad \times \langle v_{i_1}(\omega_1,\mathbf{p}_1) .. v_{i_n}(-(\omega_1 + .. + \omega_n),-(\mathbf{p}_1 + .. + \mathbf{p}_n)) \rangle .
\label{eq:def_av_v^2}
\end{align}
We have slightly abused the notation here. Strictly speaking
\begin{align}
\langle v_{i_1}(\omega_1,\mathbf{p}_1) .. v_{i_n}(-(\omega_1 + .. + \omega_n),-(\mathbf{p}_1 + .. + \mathbf{p}_n)) \rangle
\end{align}
is not well defined since it is proportional to $\delta(\mathbf{0})$.
We use instead \Eq{eq:def_av_v^2} as a definition of
\begin{align}
\langle v_{i_1}(\omega_1,\mathbf{p}_1) .. v_{i_n}(-(\omega_1 + .. + \omega_n),-(\mathbf{p}_1 + .. + \mathbf{p}_n)) \rangle.
\end{align}
Note that we have extracted a factor of $(2\pi)^{d+1}$.

\item We deal with a lot of functional differentiation. We write the Taylor expansion of a functional $\Gamma[\mathbf{v}]$ as
\begin{align}
 \Gamma[\mathbf{v}] = \sum_{n = 1}^\infty \int_{\omega_1,\mathbf{p}_1;..;\omega_n,\mathbf{p}_n} \frac{v_{i_1}(\omega_1,\mathbf{p}_1) \cdot .. \cdot v_{i_n}(\omega_n,\mathbf{p}_n)}{n!} \, \Gamma_{i_1i_2..i_n}^{(n)}(\omega_1,\mathbf{p}_1;..;\omega_n,\mathbf{p}_n),
\label{eq:cumulants}
\end{align}
which is consistent with
\begin{align}
 \Gamma_{i_1i_2..i_n}^{(n)}(\omega_1,\mathbf{p}_1;..;\omega_n,\mathbf{p}_n) = (2\pi)^{n(d+1)} \, \left.\frac{\delta^n \Gamma}{\delta v_{i_1}(\omega_{1},\mathbf{p}_{1}) .. \delta v_{i_n}(\omega_{n},\mathbf{p}_{n})} \right|_{\mathbf{v}=0}.
 \label{eq:derivatives_1}
\end{align}
%
%$\Gamma_k[\mathbf{v}]$ is the flowing effective action which depends on the velocity field $\mathbf{v}(\tx)$. It will be defined in Section (\ref{sec:defgamma}). All the relations in this section apply of course to all functionals.

\item We use the short-hand notation $\delta(\mathbf{x}) = \Pi_{i=1}^d \delta(x_i)$ for delta distributions with vectors as arguments.

\item As for correlation functions most of the functional derivatives that we deal with are invariant under space and time translations. This implies the following property for the derivatives of physical quantities
\begin{align}
\Gamma_{i_1i_2..i_n}^{(n)}(\omega_1,\mathbf{p}_1;..;\omega_n,\mathbf{p}_n)  & = (2\pi)^{d+1} \, \delta\left(\mathbf{p}_1+ .. + \mathbf{p}_n\right) \delta\left(\omega_1+ .. + \omega_n\right) \nonumber \\
& \qquad \times \Gamma_{i_1i_2..i_n}^{(n)}(\omega_1,\mathbf{p}_1;..;\omega_{n-1},\mathbf{p}_{n-1}).
\label{eq:inv_cumulants}
\end{align}
Note that we used the same notation $\Gamma^{(n);i_1,..,i_n}$, in \Eqs{eq:cumulants} and \eq{eq:inv_cumulants}. There is no risk of confusion since the number of spatial variable is $n$ and $n-1$ respectively.

\item Straightforward definitions for functional operators and operations in between them are taken in real space,
For operators $A$ and $B$ with matrix elements that depend on space and time we have,
\begin{align}
& {AB}(\tx;t',\mathbf{x}') = \int_{t'',\mathbf{x}''} {A}(\tx;t'',\mathbf{x}'') {B}(t'',\mathbf{x}'';t',\mathbf{x}'), \nonumber \\
& {AA^{-1}}(\tx;t',\mathbf{x}') = \int_{t'',\mathbf{x}''} {A}(\tx;t'',\mathbf{x}'') {A}^{-1}(t'',\mathbf{x}'';t',\mathbf{x}') \equiv \delta(t-t') \delta(\mathbf{x}-\mathbf{x}'),\nonumber \\
& \text{Tr}\left({A}\right) =  \int_{\tx} A(\tx;\tx).
\end{align}
The corresponding operations in Fourier space are defined by requiring them to be consistent with \Eqs{eq:def_fourier},
\begin{align}
 A(\op,\omega',\mathbf{p}') \equiv \int_{\tx;t',\mathbf{x}'} \text{e}^{-i\left(\omega t + \omega' t' - \mathbf{p} \cdot \mathbf{x}- \mathbf{p}' \cdot \mathbf{x}'\right)} A(\tx;t',\mathbf{x}').
\end{align}
We have,
\begin{align}
& {AB}(\op;\omega',\mathbf{p}') = \int_{\omega'',\mathbf{p}''} {A}(\op;-\omega'',-\mathbf{p}'') {B}(\omega'',\mathbf{p}'';\omega',\mathbf{p}'), \nonumber \\
& {AA^{-1}}(\op;\omega',\mathbf{p}') = (2\pi)^{d+1} \delta(\omega+\omega') \delta(\mathbf{p}+\mathbf{p}'), \nonumber \\
& \text{Tr}\left({A}\right) =  \int_{\op} {A}(\op;-\omega-\mathbf{p}).
\end{align}
Note that ${A}^{-1}(\omega,\mathbf{p};\omega',\mathbf{p}')$ is defined as
\begin{align}
 A^{-1}(\op,\omega',\mathbf{p}') \equiv \int_{\tx;t',\mathbf{x}'} \text{e}^{-i\left(\omega t + \omega' t' - \mathbf{p} \cdot \mathbf{x}- \mathbf{p}' \cdot \mathbf{x}'\right)} \, A^{-1}(\tx;t',\mathbf{x}'),
\label{eq:def_a-1_fourier}
\end{align}
while ${A}^{-1}(\tx;t',\mathbf{x}')$ is defined through ${AA^{-1}}(\tx;t',\mathbf{x}') = \delta(t-t') \delta(\mathbf{x}-\mathbf{x}')$. The asymmetry is only apparent since \Eq{eq:def_a-1_fourier} is equivalent to ${AA^{-1}}(\op;\omega',\mathbf{p}') = (2\pi)^{d+1} \delta(\omega+\omega') \delta(\mathbf{p}+\mathbf{p}')$.

In particular
\begin{align}
 & \frac{\delta^2 \Gamma_k[v = 0]}{\delta v_i(\omega',\mathbf{p}') \delta v_j(\omega,\mathbf{p})} 
 \equiv \delta_{ij} \Gamma_{k}^{(2)}(\omega,\mathbf{p}) \frac{\delta(\omega+\omega') \delta\left(\mathbf{p}+\mathbf{p}'\right)}{(2\pi)^{d+1}},
\label{eq:Gamma2_intro}
\end{align}
implies
\begin{align}
 & \left(\frac{\delta^2 \Gamma_k[v = 0]}{\delta v_i(\omega',\mathbf{p}') \delta v_j(\omega,\mathbf{p})} \right)^{-1}
 = \frac{\delta_{ij}}{\Gamma_{k}^{(2)}(\omega,\mathbf{p})} \frac{\delta(\omega+\omega') \delta\left(\mathbf{p}+\mathbf{p}'\right)}{(2\pi)^{d+1}}.
\label{eq:Gamma2_intro_inverse}
\end{align}
Note that we use two different notations for the inverse of an operator. Compare \Eqs{eq:def_a-1_fourier} and \eq{eq:Gamma2_intro_inverse}. They are completely equivalent.

\item The identity operator is given by
\begin{align}
& \mathds{1}(\tx;t',\mathbf{x}') = \delta(t-t') \delta(\mathbf{x}-\mathbf{x}') \, \delta_{ij} ,\nonumber \\
& \mathds{1}(\op;\omega',\mathbf{p}') = (2\pi)^{d+1} \delta(\omega+\omega') \delta(\mathbf{p}+\mathbf{p}') \, \delta_{ij}.
\end{align}

\item The two-point correlation function of the stochastic forcing $\mathbf{f}(\tx)$ is defined as
\begin{align}
\langle f_{i}(t,\mathbf{x}) f_{j}(t',\mathbf{x'})\rangle = \delta_{ij} \, \delta(t-t') F\left(\left|\mathbf{x}-\mathbf{x'}\right|\right).
\label{eq:force_corr_intro}
\end{align}
We define its inverse $F^{-1}_{lj}(t,\mathbf{x};t',\mathbf{x}')$ such that
\begin{align}
\int_{t'',\mathbf{x}''} \langle f_{i}(t,\mathbf{x}) f_{l}(t'',\mathbf{x}'')\rangle \, F^{-1}_{lj}(t'',\mathbf{x}'';t',\mathbf{x'}) = \delta_{ij} \delta(t-t') \delta(\mathbf{x}-\mathbf{x}'),
\label{eq:inverse_forcing_intro}
\end{align}
and get
\begin{align}
 F^{-1}_{ij}(t',\mathbf{x};t',\mathbf{x'}) = \delta_{ij} \, \delta(t-t') F^{-1}\left(\left|\mathbf{x}-\mathbf{x'}\right|\right).
\end{align}
$F^{-1}(x)$ is then defined through
\begin{align}
 \int_{\mathbf{x}''} F\left(\left|\mathbf{x}-\mathbf{x}''\right|\right) F^{-1}\left(\left|\mathbf{x}''-\mathbf{x'}\right|\right) = \delta(\mathbf{x}-\mathbf{x}').
\end{align}
Note that this definition does not imply $F^{-1}(x) = 1/F(x)$ in real space. Invariance under spatial translation makes this however true in Fourier space,
\begin{align}
& F(p) \equiv \int_{\mathbf{x}} \text{e}^{i\mathbf{p}\cdot \mathbf{x}} F(x) = \frac{1}{F^{-1}(p)}.
\end{align}

\item At an \RG fixed point observables assume a scaling form,
\begin{align}
 O(\lambda^z t;\lambda x) = \lambda^{\eta}  O(t,x),
\end{align}
for a generic space-time dependent observable $O(t,x)$ and $\lambda > 0$. We are then free to choose $\lambda x = 1$ and write
\begin{align}
 O(t,x)  = x^{\eta} \, O\left(\frac{t}{x^z},1\right) \equiv x^\eta \, g\left(\frac{t}{x^z},1\right).
\end{align}
We use $g(a)$ as a generic scaling function. It may have a different physical meaning depending on the context but is always defined as $g(a) = O\left(a,1\right)$ for an appropriate observable, $O(t,x)$.

\item The time ordering operator is defined in a standard way. Applied to a product of time dependent operators it orders them from left to right with decreasing times. In particular we have
\begin{align}
 \mathcal{T}\left[\psi_{i}(t_1) \psi_{j}(t_2)\right] = \psi_{i}(t_1) \psi_{j}(t_2) \, \theta(t_1-t_2) + \psi_{j}(t_2) \psi_{i}(t_1) \, \theta(t_2-t_1).
\end{align}

\item A primed function of a single argument is its derivative $F'(x) = \text{d}F(x)/\text{d}x$, $\delta Z_i'(1) = \left. \text{d} \delta Z_i/\text{d}\hat{p} \right|_{\hat{p}=1}$.

\item The vorticity of a $3d$ vector field $\mathbf{v}(\tx)$ is given by the pseudo-vector
\begin{align}
 \mathbf{w}(\tx) = \boldsymbol{\nabla} \times \mathbf{v}(\tx).
\end{align}
It can be physically interpreted as the local rotation of the vector field. If $w(\tx)\neq 0$, a fluid element that is following $\mathbf{v}(\tx)$ and located at $\mathbf{x}$ rotates around an axis which passes through $\mathbf{x}$ and is parallel to $\mathbf{w}(\tx)$. Its angular velocity is $w(\tx)/2$.

\item The symbol $\cong$ means "approximately equal" or "asymptotically equal",
\begin{align}
f(x) \cong g(x) \, \text{as } x\to a \, \Leftrightarrow \, \lim_{x\to a}f(x)/g(x) = 1,
\end{align}
depending on the context.

\item The symbol $\sim$ means "scales as",
\begin{align}
 f(x) \sim x^\alpha \, \text{as } x\to a \, \Leftrightarrow \, \lim_{x\to a}f(x) \, x^{-\alpha} = const \neq 0 .
\end{align}
$f(x) \sim x^\alpha$ differs from $f(x) \cong x^\alpha$ by the fact that with $\sim$ we do not say anything about the pre-factor of the power law.

\end{itemize}

\chapter{List of Abbreviations}
\label{sec:list_of_abbreviations}
\ResetAbbrevs{All}

\section{Acronyms}

\begin{itemize*}
\item \OnePI
\item \TwoPI
\item \BMW
\item \SGPE
\item \EPC
\item \FRG
\item \GPE
\item \IR
\item \KPZ
\item \MSR
\item \NS
\item \NTFP
\item \RG
\item \UV
\end{itemize*}

\section{Short-hand notations}

\begin{multicols}{2}
\begin{itemize}
\item $\int_{tx} = \int \text{d}t \text{d}^dx$
\item $\int_{\op} = {(2\pi)^{-d-1}} \int \text{d}\omega \text{d}^dp$
\item $\mathcal{D}\left[\mathbf{v}\right] = \Pi_{\op} \text{d}^dv(\op)$
\item $\delta(\mathbf{x}) = \Pi_{i=1}^d \delta(x_i)$
\item $\Omega_d = \int_{\Omega} = d \pi^{d/2}[(2\pi)^d \Gamma(d/2+1)]^{-1}$
\item $\mathbf{q} = \mathbf{p} - k e_{\mathbf{r}}$
\item $\nuk{p}{}=\nu_{k}(p)p^{2}$
\item $F_{ij} = F_{ij}(0)$
\item $S_i(\hp) = \sqrt{1+ \delta Z_i(\hp)}$
\item $S_i = S_i(1)$
\item $\T = k^2 \sqrt{z_1/z_2}/{(\nukk+\nukq)}$
\item $a_i = A_i (\alpha_i-\eta_i)$
\item $y=\hat{\mathbf{p}} \cdot \mathbf{e}_{\mathbf{r}}/\hat{p}$
\end{itemize}
\end{multicols}
\end{appendix}
\backmatter %use with scrbook
\begin{footnotesize}
\bibliography{Mybib}
\bibliographystyle{utcaps_modif}
\end{footnotesize}

\end{document}